\pdfoutput=1
\documentclass[11pt, oneside]{mnthesis}

\usepackage{vcfp}

\begin{document}
\bibliographystyle{hunsrt} % style of bibliography
%%%%%%%%%%%%%%%%%%%%%%%%%%%%%%%%%%%%%%%%%%%%%%%%%%%%%%%%%%%%%%%%%%%%%%%%%%%%%%%%

%%%%%%%%%%%%%%%%%%%%%%%%%%%%%%%%%%%%%%%%%%%%%%%%%%%%%%%%%%%%%%%%%%%%%%%%%%%%%%%%
% Title and other sections that come before the body of the document
%%%%%%%%%%%%%%%%%%%%%%%%%%%%%%%%%%%%%%%%%%%%%%%%%%%%%%%%%%%%%%%%%%%%%%%%%%%%%%%%
%%%%%%%%%%%%%%%%%%%%%%%%%%%%%%%%%%%%%%%%%%%%%%%%%%%%%%%%%%%%%%%%%%%%%%%%%%%%%%%%
% title.tex - Set up the beginning of thesis.
%%%%%%%%%%%%%%%%%%%%%%%%%%%%%%%%%%%%%%%%%%%%%%%%%%%%%%%%%%%%%%%%%%%%%%%%%%%%%%%%
% For a  PhD give the command \phd. Default is masters
%\degree (normally Doctor of Philosophy or Master of Science)
%\initials (normally Ph.D. or M.S.)
\ms % use if for a Master of Science thesis
%\phd % use if for a Ph.D. dissertation
%\draft

\title{\bf A Higher-Order Abstract Syntax Approach to the\\
           Verified Compilation of Functional Programs }
\author{Yuting Wang}
\campus{University of Minnesota} 
\program{Computer Science and Engineering} 
\degree{DOCTOR OF PHILOSOPHY}
\director{Gopalan Nadathur} 

% Optionally specify the month and year.
\submissionmonth{December} % defaults to current month.
\submissionyear{2016} % defaults to current year.

%Comment out below on final copy
\abstract{%%%%%%%%%%%%%%%%%%%%%%%%%%%%%%%%%%%%%%%%%%%%%%%%%%%%%%%%%%%%%%%%%%%%%%%%%%%%%%%%
% abstract.tex: Abstract
%%%%%%%%%%%%%%%%%%%%%%%%%%%%%%%%%%%%%%%%%%%%%%%%%%%%%%%%%%%%%%%%%%%%%%%%%%%%%%%%

\noindent This thesis concerns the verified compilation of \emph{functional
programming languages}. Functional programming languages,
or \emph{functional languages} for short, provide a high degree of
abstraction in programming and their mathematical foundation makes
programs written in them easy to analyze and to be proved
correct. Because of these features, functional languages are playing
an increasingly important role in modern software
development. However, there is a gap that must be closed before we can
derive the full benefits of verifying programs written in functional
languages. Programs are usually verified with regard to the
computational models underlying the functional languages, while the
execution of programs proceeds only after they are transformed by
compilers into a form that is executable on real hardware.
To get programs verified end-to-end, the compilers must
also be proved correct.
%% Compilers are complex pieces of software and
%% their verification is therefore too large and error-prone a task to
%% carry out effectively by hand. Fortunately, significant strides have
%% been taken in recent years towards providing computer-based support
%% for verifying compilers.

Significant strides have been taken in recent years towards
the goal of verifying compilers. However, much of the attention in
this context has been on imperative programming languages. The
verification of compilers for functional languages poses some
additional challenges.
A defining characteristic of such languages is that they treat
functions as first-class objects. In describing the compilation of
programs written in functional languages and in reasoning about this
compilation, it is therefore necessary to treat functions as data
objects. In particular, we need to provide a logically correct
treatment of the relationship between the arguments of a function and
their use within its body, i.e., the \textit{binding structure} of the
function. Most existing proof systems for formal verification provide
only very primitive support for reasoning about binding structure. As
a result, significant effort needs to be expended to reason about
substitutions and other aspects of binding structure and this
complicates and sometimes overwhelms the task of verifying compilers
for functional languages.

We argue that the implementation and verification of
compilers for functional languages are greatly simplified by
employing a higher-order representation of syntax known as
\emph{Higher-Order Abstract Syntax} or \emph{HOAS}. The underlying
idea of \HOAS is to use a meta-language that provides a built-in and
logical treatment of binding related notions. By embedding the
meta-language within a larger programming or reasoning framework, it
is possible to absorb the treatment of binding structure in the object
language into the meta-theory of the system, thereby greatly
simplifying the overall implementation and reasoning processes.

We develop the above argument in this thesis. In particular, we
present and demonstrate the effectiveness of an approach to the
verified implementation of compiler transformations for functional
programs that exploits HOAS.
In this approach, transformations on functional programs are
first articulated in the form of rule-based relational specifications.
These specifications are rendered into programs in \LProlog, a
language that is well-suited to encoding rule-based relational
specifications and that supports an \HOAS-style treatment of formal
objects such as programs.
Programs in \LProlog serve both as specifications and as executable
code.
One consequence of this is that the encodings of compiler
transformations serve directly as their implementations.
Another consequence is that they can be input to the theorem
proving system \Abella that provides rich capabilities for reasoning
about such specifications and thereby for proving their correctness as
implementations.
The \Abella system also supports the use of the HOAS approach.
Thus, the \LProlog language and the \Abella system together constitute
a framework that can be used to test out the benefits of the HOAS
approach in verified compilation.
We use them to implement and verify a compiler for a representative
functional programming language that embodies the transformations that
form the core of many compilers for such languages.
In both the programming and the reasoning phases, we
show how the use of the HOAS approach significantly simplifies the
representation, manipulation, analysis and reasoning of binding
structure.

Carrying out the above exercise revealed some missing capabilities in
the \Abella system.
At the outset, it was possible to reason about only a subset
of \LProlog programs using the system. Some compiler transformations
required the use of features not available in this subset.
Another limitation was that it did not support the ability to reason
about polymorphic specifications, thereby leading to a loss of
modularity in programs and in reasoning.
We have addressed these issues as well in this thesis.
In particular, we have developed the theoretical underpinnings for
introducing polymorphism into \Abella and for treating the full range
of \LProlog specifications.
These ideas have also been implemented to yield a new version
of \Abella with the additional capabilities.

%% It is a non-trivial task to carry out the above exercise, especially
%% the part about verifying compiler. We have therefore made some
%% enhancements to the \Abella system to meet our need. The version of
%% \Abella we started with was only able to reason about a limited class of \LProlog
%% specifications. We have removed this limitation so that the full
%% logical structure of
%% \LProlog specifications can now be used to encode compiler
%% transformations. We have also incorporated a form of polymorphism
%% into \Abella which enables the sharing of code and proofs, thereby
%% enhancing the modularity of our implementations and proofs of compiler
%% transformations.

%%%%%%%%%%%%%%%%%%%%%%%%%%%%%%%%%%%%%%%%%%%%%%%%%%%%%%%%%%%%%%%%%%%%%%%%%%%%%%%%
}
\words{331}    % number of words in the abstract
\copyrightpage % Do you want copyright protection?
\acknowledgements{%%%%%%%%%%%%%%%%%%%%%%%%%%%%%%%%%%%%%%%%%%%%%%%%%%%%%%%%%%%%%%%%%%%%%%%%%%%%%%%%
% acknowledge.tex: Acknowledgements
%%%%%%%%%%%%%%%%%%%%%%%%%%%%%%%%%%%%%%%%%%%%%%%%%%%%%%%%%%%%%%%%%%%%%%%%%%%%%%%%

Many people have provided me help and support during my
doctoral studies. I would like to take this opportunity to express my 
gratitude to them.

First and foremost, I would like to thank my advisor Gopalan Nadathur
for his mentorship throughout my stay at the University of Minnesota,
which helped me better understand what research in computer science is
all about and how to conduct such research. The dissertation would
not have been in its current state without his guidance during the
thesis research,
his patient examination and revision of the thesis drafts, and his
constant encouragement throughout the project. I look forward to
continuing to collaborate with him in the future.

I would like to thank Kaustuv Chaudhuri for his guidance and
collaboration during my years as a doctoral student. I did two summer
internships in the Parsifal team in Inria Saclay under his
supervision.
% GN You should usually leave the success for others to gauge.
% These internships were very fruitful:
The extensions to the Abella system described in the
thesis were either direct results of the internships or partially
inspired by the work done during that time. To the extent that this
work was successful, I owe a great deal to the inspiring
discussions I had with Kaustuv and to his supportive attitude
towards the research I was interested in doing.

I am grateful to Dale Miller for hosting me in the Parsifal team during
my internships and to the other (former or present) members and visitors of
the Parsifal team with whom I interacted and learned a great deal
from, including but not limited to Anupam Das, Taus Brock-Nannestad,
Danko Ilik, Quentin Heath, Tomer Libal, Siddhartha Prasad, Zakaria
Chihani, Beniamino Accattoli and Nicolas Guenot.

My interest in exploring the use of the higher-order abstract syntax
approach in verified compilation originated from discussions with
Michael Whalen. Although the project took quite a different path in
the end, Michael has expressed continuous support and interest in my
work, for which I am thankful.

Special thanks to Andrew Gacek for creating the original version of
the Abella system and for patiently explaining to me the details of the
implementation that I have periodically had questions about. I have also
learned a great deal of the underlying theory of Abella via
interactions with him in the past few years.

I am honored to have had Professors Eric Van Wyk, Stephen McCamant and
Karel Prikry on my thesis committee. Comments especially from Eric and
Stephen have helped me improve the quality of my dissertation. 
Special thanks to Eric for taking care of the programming
languages lab and providing useful suggestions for being an effective
researcher. I am also grateful to Wayne Richter for having served on my
thesis proposal committee and for teaching the courses that provided
me an introduction to mathematical logic.

I have benefited from discussions with Olivier Savary
Belanger, Brigitte Pientka, Derek Dreyer, Chung-Kil Hur and Georg
Neis. I would like to thank them all for the patience with which they
have explained aspects of their work and for the insights into it that
they have thereby provided. Conversations with all of them have helped
me understand the big picture in the various areas this thesis relates to.

Finally, I would like to thank the students and friends I worked
and studied with at the University of Minnesota, especially Dan
DaCosta, Mary Southern, Ted Kaminski, Kevin Williams, Dawn Michaelson,
Dongjiang You, Ming Zhou, Peng Liu, Shuo Chang, Xiang Cao, Ziqi Fan,
Yuan Li and Yin Liu.

Work on this dissertation has been partially supported by the National
Science Foundation through the grants CCF-0917140 and CCF-1617771.
Support has also been received from a Doctoral Dissertation Fellowship
and a Grant in Aid of Research from the University of
Minnesota. Opinions, findings and conclusions or recommendations that
are expressed in this dissertation should be understood as mine. In
particular, the do not necessarily reflect the views of the National
Science Foundation.

%%%%%%%%%%%%%%%%%%%%%%%%%%%%%%%%%%%%%%%%%%%%%%%%%%%%%%%%%%%%%%%%%%%%%%%%%%%%%%%%
}
\dedication{To my parents and grandparents
}

% Use a special preface
%\extra{\input{preface}}

% The \beforepreface command actually causes insertion of the title, 
% abstract, signature, and copyright pages into the new document.
\beforepreface 

% Define the text to go before the table of contents
\figurespage
\tablespage

% The \afterpreface command actually causes insertion of the
% contents, list of figures, etc. into the new document.
\afterpreface            
%%%%%%%%%%%%%%%%%%%%%%%%%%%%%%%%%%%%%%%%%%%%%%%%%%%%%%%%%%%%%%%%%%%%%%%%%%%%%%%%

%%%%%%%%%%%%%%%%%%%%%%%%%%%%%%%%%%%%%%%%%%%%%%%%%%%%%%%%%%%%%%%%%%%%%%%%%%%%%%%%

%%%%%%%%%%%%%%%%%%%%%%%%%%%%%%%%%%%%%%%%%%%%%%%%%%%%%%%%%%%%%%%%%%%%%%%%%%%%%%%%
% Now lets include the body of the document...
%%%%%%%%%%%%%%%%%%%%%%%%%%%%%%%%%%%%%%%%%%%%%%%%%%%%%%%%%%%%%%%%%%%%%%%%%%%%%%%%
%%%%%%%%%%%%%%%%%%%%%%%%%%%%%%%%%%%%%%%%%%%%%%%%%%%%%%%%%%%%%%%%%%%%%%%%%%%%%%%
% intro.tex: Introduction to the thesis
%%%%%%%%%%%%%%%%%%%%%%%%%%%%%%%%%%%%%%%%%%%%%%%%%%%%%%%%%%%%%%%%%%%%%%%%%%%%%%%%
\chapter{Introduction}
\label{ch:intro}

The focus of this thesis is on the verified compilation of functional
programming languages. Functional programming languages, or
\emph{functional languages} for short, form a class of programming
languages that has been the subject of significant research and that
%GN note added later: don't need references here, but you do need them
%in section 1 where indicated.
% include references to OCaml, Haskell, the paper by Minsky, F#,
% Swift, Clojure, Racket in the cite part.
is now seeing increasing use in research and industrial settings.
One reason for the growing popularity of these languages is the high
degree of abstraction they offer both in the representation of data
and in the way computations can be expressed. Their close
correspondence to mathematical formalisms such as the lambda calculus
also makes it easier to analyze programs and even to prove their
correctness. There is, however, a catch to these kinds of behavioral
analyses. On the one hand, to derive benefit from the structure of the
language that is used, the analyses of programs have to be based on
the computational model underlying the language. On the other hand,
the execution of a program proceeds only after it has been translated
into a form that can be run directly on actual hardware. Thus, to
leverage the benefits of using functional languages in obtaining
assurances of the correctness of a program, it is necessary to ensure
that the translation process does not change its meaning, i.e. to
verify the compiler.

The importance of compiler verification to overall program correctness
has been long recognized and interest in the topic dates back
almost to the advent of programming languages (e.g.,
see~\cite{mccarthy67}). However, the actual task of verifying a
compiler is complex, tedious and error-prone and so few such efforts
were undertaken in the early years. The availability of sophisticated
tools for mechanizing reasoning in recent times has changed this
situation dramatically: there have been several well-documented and
successful efforts to verify compilers for programming languages that
are actually used in practice. Much of this work has been aimed at
conventional imperative programming languages such as C or
Java---see \cite{dave03sigsoft} for a survey on this topic. Because
of the high-level of abstraction that functional languages support, the
verification of compilers for these languages must deal with some
problems that do not arise relative to imperative
languages. One particular source of these problems is that functional
languages treat functions as first-class objects, allowing
them to be passed as arguments and returned as values. As a
consequence, compilers for these languages must pay careful attention
to the {\it structure} of functions and, more specifically, to the
connection between the arguments of a function and their use within
the body. In fact, the initial phases of compilation of functional
languages typically involve the explicit manipulation of this kind of
structure that we refer to as the {\it binding structure}
of expressions. To prove the correctness of these phases, it becomes
necessary also to validate such manipulation of binding
structure. Experience has shown this to be a complex task. In fact, it
is now well appreciated that without specialized tools or techniques
the cost of analyzing, manipulating and reasoning about binding
structure can overwhelm the compilation and verification processes.

The need to treat binding structure is not limited to the domain of
compiler verification. This kind of need arises in the context of a
variety of systems that are geared towards the formal treatment of
objects such as programs, formulas, proofs and types. One of the
approaches that has been proposed for simplifying the programming and
reasoning tasks in these contexts is that of \emph{higher-order
  abstract syntax} or
\emph{HOAS}\cspc\cite{miller87slp,pfenning88pldi}. The underlying idea
in this approach is to use a meta-language for representing formal
objects that provides a built-in, logical means for capturing binding
related notions. By embedding this kind of a meta-language within a
larger programming or reasoning framework, it is possible to absorb
the treatment of binding structure in the object language into the
meta-theory of the system, thereby greatly simplifying the overall
implementation and reasoning processes.  Of course, the choice of
meta-language must be properly calibrated so as to ensure that these
benefits actually flow in practice.  There have been two successful
realizations of this approach that possess this characteristic. One of
these approaches, that is embodied in the Twelf
system~\cite{pfenning99cade} and its successor
Beluga~\cite{pientka10ijcar}, is based on the use of a dependently
typed lambda calculus~\cite{harper87lics} to encode and to reason
about formal objects. The second approach, which has resulted in the
specification and
% In addition to the book and the tutorial paper, use also the
% original references here: Andrews system description paper for
% Abella and the NM87 paper for LProlog.
programming language \LProlog~\cite{miller12proghol,nadathur88iclp} and
the reasoning system \Abella~\cite{gacek08ijcar,baelde14jfr}, uses a
predicate logic over simply typed lambda terms to realize similar
capabilities.

Our objective in this thesis is to show that the \HOAS approach can be
used to significantly simplify the implementation and verification of
compilers for functional programming languages.
Towards this end, we show how the second realization of the \HOAS
approach that is described above can be utilized to benefit in this
task.
Specifically, we show how some of the complex transformations involved
in compiling functional programs can be elegantly encoded in
\LProlog.
We then show how these implementations can be effectively reasoned
about using the Abella system.
In carrying out these tasks, we expose a methodology that we believe
to be broadly applicable in this domain.
In the course of this work, we have discovered ways to strengthen
the Abella system so as to make it more suitable to such
applications.
We present these ideas as well in this thesis.

We elaborate on the broad ideas described above in this introductory
chapter, as a prelude to their technical development in the rest of
the thesis. Section\cspc\ref{sec:intro_mtv} motivates the general
compiler verification endeavor. Section\cspc\ref{sec:intro_vcfl_problems}
discusses the specific difficulties related to the treatment of
binding structure that arise in the verified compilation of functional
languages. Section\cspc\ref{sec:intro_binding} exposes some of the
approaches that have been proposed and used for dealing with these
issues; this discussion also introduces the \HOAS approach.
In Section\cspc\ref{sec:intro_vcfl_hoas}, we expand on the idea of
using the \HOAS approach to simplify verified compilation of
functional programs and we present the  specific goals for this thesis
in this setting. Section\cspc\ref{sec:contributions}
summarizes the contributions that we make through this
thesis. Section\cspc\ref{sec:thesis_overview} concludes this chapter
by explaining how each of the following chapters fit into
the overall dissertation.

\section{Program Correctness and Compiler Verification}
\label{sec:intro_mtv}

With the increasing reliance of modern society on software systems,
the correct operation of such systems has become a major concern. A
commonly used method for gaining confidence in software behavior has
been the idea of testing. In this approach, programs are run
repeatedly under systematically varying conditions and their results
are checked against the expected outcomes. While testing has been used
successfully in many situations, it also has a fundamental limitation:
at its very best, testing can only provide \emph{evidence} for the
correctness of a program, never a water-tight guarantee for the
absence of bugs in it.
There are many safety critical software systems for which it
is in fact necessary to have a guarantee of correctness. The only
way in which such an assurance can be provided is by formally
verifying the properties of software using principles of mathematical
reasoning.

The above considerations have led to a large amount of effort being
invested towards developing
approaches for formally verifying the correctness of programs. Over the
years, there has been a convergence on two main ideas in realizing the
overall verification goal. First, there has been an emphasis on
developing programming languages that make it possible to
describe computations at a level at which they are amenable to
mathematical analysis. Second, methods have been developed for
utilizing the structure of these high-level languages to
support the process of reasoning about programs written in them.
There has been considerable success in realizing these two
ideas and they have indeed provided the basis for verifying the
properties of many non-trivial programs.

Functional languages are a particular class of high-level programming
languages that have been investigated in this context. The structure
of these languages is best understood in comparison with the more
commonly used \emph{imperative programming languages} such as C,C++
and Java. The latter class of languages views computation as the result
of the repeated alteration of state, which is given by the values
stored in memory cells and the location of control; the building
blocks for programs within this paradigm are \emph{statements} or
\emph{commands} for modifying program state. In contrast, programs in
functional languages are given almost entirely by expressions,
computation consisting essentially of evaluating these
expressions. This yields a more abstract view of programming, a view
in which the focus is on describing what has to be done rather than
how it should be carried out. The common foundation of functional
programming is the $\lambda$-calculus, a mathematical system that
possesses well-understood logical
properties\cspc\cite{church32am}. Because of the mathematical basis of
functional languages that also implies freedom from lower-level
implementation details, programs written in these languages are more
concise and easier to reason about than those written in imperative
programming languages. Despite the abstract nature of these languages,
research over the last three decades has shown that programs written in
them can be translated into a machine understandable form that can be
executed with considerable efficiency. As a consequence of these
results, functional languages are starting to play an
increasingly important role in modern software practice. Indeed, there
%GN You need references here. For OCaml, use the Minsky reference in
%addition to something about the language.
are a variety of languages within this paradigm such as Common Lisp,
Scheme, Racket, Clojure, Erlang, Haskell, Standard ML, OCaml, Scala,
Swift and F\# that are being used today in a range of academic,
industrial and commercial settings (\eg, see
\cite{minsky11queue,felleisen03theteachscheme,armstrong03phd,fsharp30}).

As noted above, one of the benefits of high-level programming
languages is that they make it easier to reason about program
behavior. However, there is a gap that must be closed before we can
derive the full benefits of verifying programs written in these
languages. On the one hand, to take advantage of the structure of the
language in arguments about program correctness, it is necessary to
reason within the high-level computational model associated with the
language. On the other hand, in order to actually execute programs in
these languages, they must be translated by an intervening
process known as \emph{compilation} into code at a lower level that can be run
directly on a machine. Now, it is possible for this compilation process to
be buggy, leading thereby to executable versions of programs
whose behavior is different from the ones intended of the
original versions. Traditionally, the consensus has been to assume the
correctness of compilers, with confidence in this assessment being
built over time and repeated use. However, in an absolute sense, this
approach is no different from testing. To obtain the full force of
verification, it is necessary also to establish the correctness of the
compilers that are used.

The importance of proving the correctness of compilers to the program
verification endeavor has long been recognized and there has also been
much work aimed at understanding how this can be done; a survey of the
early efforts appears, for example, in
\cite{dave03sigsoft}.  Compilers are complex pieces of software
and their verification is therefore too large, tedious and error-prone
a task to carry out effectively by hand. This has hampered several
past attempts at showing its practicality. However, significant
strides have been taken in recent years towards providing
computer-based support for the theorem proving task and this has
dramatically altered the situation. Indeed, there has been a
mushrooming of efforts related to formal compiler verification using
theorem proving systems such as
Isabelle\cspc\cite{paulson94book}, HOL\cspc\cite{nipkow02book} and
Coq\cspc\cite{coq02manual}. One of the more successful exercises in
this direction has been the CompCert project that has developed and
verified a compiler for a large subset of the C language using the Coq
theorem prover\cspc\cite{leroy09cacm}. Such efforts have also provided
an impetus to more ambitious projects such as the Verified Software
Toolchain project\cspc\cite{appel14book} related to overall program
verification.

A majority part of the existing work on verified compilation has been
devoted to verifying compilers for imperative programming
languages. In this thesis, we focus on verified compilation for
functional programming languages.  With the increasingly important
role functional programming languages play in software development,
their verified compilation has also been a research topic on the rise
in recent years. Notable work in this area includes the Lambda
Tamer project\cspc\cite{chlipala07pldi,chlipala10popl}, a verified
compiler for a subset of Standard ML called
CakeML\cspc\cite{kumar14popl} and the Pilsner
project\cspc\cite{neis15icfp}. There are some new difficulties that
arise in the verified compilation of functional languages in contrast
to that for imperative programming languages. We discuss this task and
some of the issues involved in it in the next section.

\section{Verified Compilation of Functional Languages}
\label{sec:intro_vcfl_problems}

One important source of power in functional languages is that they
allow functions to be treated as data objects. Functions can be
embedded in data structures or functions themselves, passed as
parameters to other functions and returned as results of computations.
These feature are used often in applications, leading to concise
programs that are easy to understand and to reason about.

Although the presence of the above features provides high-level
abstraction capabilities, it also makes compilation of functional
programs a more complex undertaking. The traditional approach to
compiling functional languages is to put the source programs
through a series of transformations that replace program constructs
for supporting the high-level features with simpler devices and
eventually render the original programs into a form to which
compilation techniques that are well-known from the context of
imperative programs can be applied.
In describing these transformations and in reasoning about them, it is
therefore necessary to have a clear and flexible way of treating
functions as objects.
The difficulty with objectifying functions is that they have inputs or
{\it arguments} and it is necessary for compilers to build in an
understanding of the logical relation that these arguments have with
their occurrences within the body of the function; we will refer to
this logical relation as the \textit{binding structure} of
the function.
Given a particular expression, a compiler for a functional language
must be able to differentiate between
variables bound by some arguments and variables not bound by any
argument, \ie \emph{free variables}, in a function valued expression.
The compiler should also not distinguish between expressions that
differ only in the names used for their bound variables and it should
be able to realize substitution over expressions in a way does not
replace bound variables and also avoids the inadvertent capture of free
variables in the expression being substituted.
Manipulating objects with binding structure adds an additional layer
of complexity to compilers for functional languages on top of the
existing complexity of program transformations common to all
compilers.
This complexity percolates also into the process of reasoning about
the correctness of the transformations.
Most programming languages and theorem-proving systems
provide only rudimentary support for manipulating and reasoning about
binding structure.
The person implementing and verifying the compiler must therefore
build an additional infrastructure for treating binding related
issues.
This effort is, in a sense, orthogonal to the main focus in compiler
verification and, without suitable support, it has often been known to
overwhelm the real objective.

To concretely expose the complexity of dealing with binding structure
in implementing and verifying compilers for functional programs, let
us consider a typical transformation in such compilers known as
% GN Use an earlier reference that is more definitive (look at the
% Wand paper for something that has this character). People who work
% with fp compilers will be offended if you use the minamide paper as
% your only reference here.
\emph{closure
  conversion}~\cite{reynolds72acm,appel1992,adams86cc,wand94popl,minamide96popl}.
This transformation replaces each expression that has a functional
structure with a version of that expression that is parameterized by
its free variables paired with an environment that provides binding
for these variables; such a combination is referred to as a
\emph{closure}. Closure conversion is an important step in the
compilation process partly because it enables further simplification
steps; for example, it enables the elimination of nested functions
because a closed expression can be moved out to the top level of a
program without changing its meaning. To realize the closure
conversion transformation, it is necessary to identify all the free
variables of an expression, to transform this collection into an
environment and to add an environment as an extra parameter to the
expression.
For example, when closure conversion is applied to the following
pseudo-code in a functional language
\begin{tabbing}
\qquad\=\quad\=\kill
\>$\kwd{let}\app x = 2\app \kwd{in}\app \kwd{let}\app y = 3\app \kwd{in}$\\
\>\>$\kwd{fun}\app z \rightarrow z + x + y$
\end{tabbing}
it will yield the code fragment
\begin{tabbing}
\qquad\=\quad\=\kill
\>$\kwd{let}\app x = 2\app \kwd{in}\app \kwd{let}\app y = 3\app \kwd{in}$\\
\>\>$\clos{(\kwd{fun}\app z\app e \rightarrow z + e.1 + e.2)}{(x,y)}$
\end{tabbing}
We write $\clos{F}{E}$ here to represent a closure whose function part
is $F$ and environment part is $E$. Further, we represent an
environment as a tuple and we write $e.i$ to indicate the selection of
the $i$-th element of a tuple $e$. Intuitively, a closure stands for
the application of its function part to its environment, but an
application that is ``suspended'' until the closure is provided with
a value for its ``actual'' argument which, in the case of the example
considered, is $z$.

A key part of implementing the closure conversion transformation is,
as we have indicated above, the identification of the free variables of
an expression and the replacement of these variables by indexed
selections from a new environment parameter that is added to the
expression. Carrying out these steps obviously requires us to perform
a non-trivial analysis of the binding structure of expressions.
This can be a complicated task in an environment that does not support an
explicit representation of binding structure. This difficulty gets
further amplified when we have to reason about the correctness of the
transformation: properties about binding structure that are implicit
in the representation and treated through user code must be explicitly
proved in the reasoning process. For example, to make the intuition that
closure conversion preserves the meaning of programs formal, we need
to explicitly prove a theorem that describes this property in a
reasoning system. To formalize the meaning of programs we often need a
formal notion of substitution. As a part of proving the meaning
preservation theorem, we will need to show that the substitution
operations do not change the function part of a closure. This property
is implied by the description of closure conversion because
closures generated by it must be \emph{closed}, \ie they must not contain
free variables, and substitution has no effect on closed terms. To
play out this argument in our proof, we will need to formally
describe the property of being a closed term and then show that
substitutions have no effect on such terms. If the implementation and
theorem-proving framework do not have an effective way to deal with
binding structure, formalizing and reasoning about these aspects can
be difficult and can become a major part of the verification effort,
thereby blurring the essential content of the proofs.

\section{The Treatment of Object-Level Binding Structure}
\label{sec:intro_binding}

The need to represent and analyze binding structure arises in a number
of contexts that involve the manipulation of formal objects such as
programs, formulas, types and proofs. A common core can be identified
to the issues that have to be treated in a formalization framework
that is well-suited to these varied contexts.
First, the framework must enable the representation of
the objects of interest in a way that makes the binding constructs
within them explicit. Second, it must provide a simple and verifiably
correct encapsulation of logical notions relating to binding
structure; these notions include, at the very least, equivalence
modulo renaming of bound variables and logically correct
substitution. Third, it must provide a means for analyzing the
representations of the formal objects in a way that takes into account
the binding constructs within them. Finally, the framework must
provide a way to prove properties about the represented objects that
incorporates an understanding of the binding structure that is
manifest in them.

Recent research has highlighted the need to deal with issues of the
kind described above and a few different approaches have been
developed towards this end. We discuss some of the more prominent
approaches that have emerged from this work below with the goal of
putting the approach that will be the subject of this thesis in
context.

\subsection{Approaches based on first-order representations}
\label{subsec:first_order}

The basis for this class of approaches is a first-order representation
of syntax that is augmented with special devices for interpreting the
parts that deal with binding notions within such a representation.
In their simplest form, these special devices consist of library
functions that can be used to realize binding sensitive operations
over the representations of objects. The most common example of this
approach is one that uses a scheme devised by de Bruijn that
eliminates names for bound variables, using indices for their
occurrences which unambiguously indicate the abstractions binding
them\cspc\cite{debruijn72}. A virtue of such a ``nameless''
representation is that it renders
all object language expressions that differ only in the names of bound
variables into a unique form, thereby making it trivial to determine
equality modulo renaming. Substitution and other relevant operations
can be defined relative to such a representation and are typically
realized through auxiliary programs.
When it comes to reasoning about properties of the
represented objects in a way that takes into account the binding
constructs within them, it becomes necessary of course to incorporate
into the process the task of reasoning about the programs that realize
substitution and related operations.
It has been noted that considerable effort can be taken up in proving
``boilerplate'' properties related to binding in this context. For
example, it is observed in \cite{chargueraud11ln} that when the
nameless representation is used for programs in a typed language , it
becomes necessary to prove weakening lemmas for typing in the course
of proving type soundness for the language and that this requirement
significantly complicates the overall proof.

A few variants of the basic approach discussed above have been
developed to solve specific problems. For example, it is sometimes
useful to treat the free variables in an expression differently from
the bound variables and a representation called the \emph{locally
  nameless} representation has been developed towards this end
\cspc\cite{chargueraud11ln,aydemir08popl}. Although
these variants simplify the treatment of some aspects of binding
structure, they do not alter the first-order nature of the
representation and hence do not overcome its primary drawback.

\subsection{The nominal logic approach}
\label{subsec:nominal_logic}

A different approach that still uses a first-order representation of
syntax is that based on nominal logic\cspc\cite{pitts03ic}. The
defining characteristic of this approach is that it provides a logical
treatment of equivalence of expressions under a renaming of
bound variables; the technical device it uses to realize this
capability is that of equivalence of expressions under permutations of
names. This representation obviously subsumes the benefits of the
nameless representation discussed earlier. A further virtue of the
approach is that it provides a logical treatment of free and bound
variables that can be useful in reasoning about the correctness of
manipulations of syntactic structure. However, nominal logic
representations do not provide an intrinsic treatment of substitution
and hence do not also intrinsically support the analysis of 
expressions with binding structure under substitution. 
The realizations of these and related aspects
have to be encoded in user programs that must then also be explicitly
reasoned about.

\subsection{The functional higher-order approach}
\label{subsec:ncho_approach}

When the formalization framework is based on a functional language, it
is possible to use expressions of function type to represent objects
that need to be manipulated or reasoned about. In this case we can
potentially use abstraction in the meta-language to represent binding
structure in formal objects. Such an idea has in fact been
investigated and we refer to it here as the \emph{functional higher-order
approach}.

One of the benefits of using meta-language abstraction to encode
binding is that it leads to a simple treatment of substitution: the
need to avoid inadvertent capture of free variables in the expression
being substituted and to replace the right variable occurrences are
both built into the treatment of function evaluation at the
meta-level. Benefit has been derived from this observation in a
variety of tasks. We mention two that are closely related to the work
in this thesis. In \cspc\cite{guillemette07plpv}, Guillemette has used
this idea in implementing the Continuation Passing Style (CPS)
transformation---a common transformation in the compilation of
functional languages---in Haskell. Similarly, Hickey and Nogin have
exploited this idea in implementing a version of closure conversion in
their MetaPRL logical framework\cspc\cite{hickey06hosc}.

There is, however, a serious limitation to the functional higher-order
approach: it is not capable of supporting the examination of the
structure of objects that embed binding constructs. The reason for
this is that the notion of equality of expressions in the context of a
functional language includes a lot more than just bound variable
renaming and $\beta$-conversion. To support computation in an adequate
fashion, such a language must include at least the conditional and
fixed-point combinators together with their associated notions of
evaluations. Further, the programmer has the ability to add to the
equality theory by defining new functions or combinators. It is not
surprising, therefore, that the uses of the functional higher-order
approach have not included serious analyses of syntactic structure
modulo binding, such as the computation of free and bound variables in
expressions. As a concrete example, the closure conversion
implementation of Hickey and Nogin makes the simplifying assumption
that every variable bound by an external abstraction appears free in
an expression of function type that they are wanting to convert into a
closure.

%% Guillemette encodes a CPS transformation in Haskell by using
%% %GN In good writing, you should avoid using acronyms such as GADT
%% %without introducing them first.
%% the non-canonical HOAS approach and GADT~\cite{guillemette07plpv}. He argues that
%% the type checking in Haskell ensures typing is preserved by the
%% transformation. Because of the incapability of analyzing variables in
%% Haskell, closure conversion cannot be encoded by using the non-canonical
%% HOAS approach. Instead, he falls back to Debruijn indexes for encoding closure
%% conversion and proves that the transformation is type
%% preserving~\cite{guillemette07haskell}.  Hickey and Nogin proposed a
%% formalization of closure conversion in the MetaPRL logical
%% framework~\cite{hickey06hosc} by using the non-canonical HOAS approach. To avoid
%% analysis such as computing free variables and forming fresh
%% environments, they simply combine an abstraction and its enclosing
%% environment to form a closure without any analysis. Because of the inability of
%% analyzing higher-order objects in the non-canonical HOAS approach,
%% none of the work above proved the
%% semantics preservation of the encoded transformations.

\subsection{The higher-order abstract syntax approach}
\label{subsec:hoas_approach}

One way to derive the benefits of the functional higher-order approach
while still retaining the ability to examine the structure of
expressions is to use a $\lambda$-calculus for representation that has
very weak computational power. One way to do this is to use a typed
$\lambda$-calculus that does not include built-in combinators or the
provision to define them. This idea underlies what has been called
the \emph{higher-order abstract syntax} or
\HOAS approach\cspc\cite{miller87slp,pfenning88pldi}.  Similar to the
functional higher-order approach, binding constructs in object
language syntax can be encoded directly via abstraction in the
meta-language.
The critical difference from the functional higher-order approach is
that equality of expressions is governed solely by bound variable
renaming and $\beta$-conversion which, in this weak setting,
corresponds only to substitution.
The result of this is that the \HOAS approach is capable of supporting
an analysis of the binding structure of objects.
In contrast to the first-order approaches, these capabilities derive
from features of the meta-language and hence do not need to be
reasoned about explicitly.
Moreover, with a properly configured logic, it is possible also to
reason about objects that embody binding notions using rich principles
such as case analysis and induction.

The systems that have been developed for specifying and reasoning
about formal systems based on the \HOAS approach include
\Twelf\cspc\cite{pfenning99cade}, \Beluga\cspc\cite{pientka10ijcar},
Hybrid\cspc\cite{felty12jar},
\LProlog\cspc\cite{miller12proghol,nadathur88iclp} and
\Abella\cspc\cite{gacek08ijcar,baelde14jfr}. These specification and reasoning
systems have been used in many formalization efforts related to
objects that embody binding constructs and the benefits described
above have been shown to be real through these applications. A survey
of these efforts can be found in \cite{felty15p1jar,felty15p2jar}.

\section{Verified Compilation Using the \HOAS Approach}
\label{sec:intro_vcfl_hoas}

The focus of this thesis, as we have previously noted, is on the
implementation and verification of compilers for functional
languages. Our contention is that these tasks are considerably
simplified by the use of the \HOAS approach. We provide evidence for
this claim by carrying out the exercise of implementing and verifying
a compiler for a representative functional language within a framework
that provides support for the approach.

%GN Why do you repeat this discussion here? Section 1.2 has made these
%observation in more detail, Section 1.3 has provided a further
%discussion!
%% As we have discussed in Section\cspc\ref{sec:intro_vcfl_problems}, a
%% flexible treatment to binding structure is essential to effectively
%% describe and verify compiler transformations on functional
%% programs. The \HOAS approach provides such a treatment by using
%% meta-level $\lambda$-abstractions to represent the binding operators
%% of function objects. As a result, the substitution or renaming
%% operations are automatically captured by $\beta$- and
%% $\alpha$-conversions. Analysis of binding structure of functions such
%% as identification of free variables also becomes possible. Because of
%% the higher-order representation of syntax, reasoning about binding
%% structure can be carried out in an concise and effective manner.

The framework we use in this work comprises the executable
specification language \LProlog\cspc\cite{miller12proghol,nadathur88iclp} and the
theorem-proving system \Abella\cspc\cite{baelde14jfr,gacek09phd}.
A natural way to characterize a formal system is to describe it
through relations that are defined via inference rules based on the
syntactic structure of the objects of interest in the system.
The \LProlog language is a suitable vehicle for formalizing such
descriptions: the language is based on a fragment of intuitionistic
logic that allows for the transparent encoding of rule-based
relational descriptions.
A further virtue of \LProlog is that it supports the use of a version
of \HOAS, which has been labeled \emph{$\lambda$-tree syntax}
in\cspc\cite{miller00cl}, in constructing these relational
specifications; we discuss this variant of \HOAS in more detail in 
Chapter~\ref{ch:framework}.
The \LProlog language has an executable interpretation.
As a consequence, specifications written in it can also serve as
implementations.
The \Abella theorem-prover is also based on an intuitionistic logic
that supports relational specifications exploiting the
$\lambda$-tree syntax approach.
However, this logic is more richly configured than the one underlying
\LProlog: in particular, it provides mechanisms for case analysis
and induction based reasoning.
While \Abella can be used to reason directly about relational
specifications, there is also the intriguing possibility that it could
be used to do this indirectly by reasoning about programs written in
\LProlog.
In fact, features have been built into \Abella to make this
possibility a reality: the logic underlying \LProlog has been embedded
into \Abella via a definition and particular specifications in \LProlog
can then be reasoned about via this encoding.
In summary, the combination of \LProlog and \Abella gives us a
framework for both implementing relational specifications and
reasoning about such implementations.

The methodology that we will use to realize verified compilation is
the following.
We will first articulate each transformation that underlies the
compilation of our functional language in the form of a rule-based
relational specification.
We will then render this specification into a \LProlog program.
The executability of \LProlog specifications means that our program
serves directly as an implementation of the transformation.
We then use the \Abella system to reason about the properties of the
\LProlog specification, thereby proving the correctness of the
compiler transformation it embodies.
In both the specification/programming and the reasoning phases, we
will show how the support the framework provides for the
$\lambda$-tree syntax approach can be used to advantage.

To keep the exposition manageable, the exercise we describe above will
be applied to a functional language that is restricted but still
contains all the features needed to make the results convincing.
The particular language that we will treat will be a superset of the
language commonly known as PCF\cspc\cite{plotkin77} that includes
recursion.
The sequence of transformations that we will consider will render
source language programs into ones in a language similar to Cminor,
which is the back-end language for the CompCert
project\cspc\cite{leroy09jar}.
There are still some transformation steps that need to be carried out
in order to obtain actual machine-executable code from programs in our
Cminor-like language.
However, we do not investigate these steps in this thesis for two
reasons.
First, the verified implementation of these steps has been studied
elsewhere in the literature and we do not have new ideas to add to that
discussion; in particular, we do not feel that the \HOAS approach
holds significant benefits after code in the Cminor-like language has
been produced.
Second, narrowing the scope of our work in this way allows us to focus
more sharply on the part of compilation that is novel to functional
languages and where we believe the \HOAS approach really shines.

Part of the work we describe also has the objective of enhancing
the framework we use to make it more suitable for the relevant
application domain.
These enhancements are essentially to the \Abella theorem-prover.
The version of the system that we started with had limited the
collection of \LProlog specifications that could be reasoned about.
We have removed this limitation so that the full logical structure of
\LProlog specifications can now be used to encode compiler
transformations.
The second extension incorporates polymorphism into \Abella.
A concrete result of this extension is that the \LProlog programs
about which we reason can also be polymorphic.
This actually has an important practical benefit:
Polymorphism allows us to share code that pertain, for example, to
manipulation of generic data structures.
Perhaps even more importantly, the sharing of code also allows us to
prove its properties just once and to then use these properties in
whichever place we eventually use the code.

\section{The Contributions of The Thesis}
\label{sec:contributions}

In summary, the work underlying this thesis contributes to the
state-of-the-art in three ways:

\begin{itemize}

\item It leads to extensions of the \Abella theorem prover that make
  it a more versatile tool for applications such as compiler
  verification. At a conceptual level, those extensions
  provide a means for reasoning about the
  full range of \LProlog specifications and they lead to logically sound
  support for polymorphism in specifications (which also double up as
  implementations in the context of \LProlog) as well as in the theorems
  that are proved about the specifications. We have also incorporated 
  these extensions into the implementation of the \Abella system.

\item It demonstrates the benefits of the \HOAS approach in implementing
  compilers for functional languages. This goal is achieved by
  developing a methodology for implementing compiler transformations
  in \LProlog that exploits and also show-cases the facilities the
  language possesses for supporting the \HOAS approach. An auxiliary
  effect of this exercise is that it yields a compiler that produces
  intermediate code similar to that in a popular low-level language for a
  representative functional language that includes recursion.

\item It demonstrates the benefits of the \HOAS approach in verifying
  compilers for functional languages. To achieve this objective, it
  develops a methodology for verifying compiler transformations
  expressed in \LProlog using \Abella. It also applies this
  methodology to verify the compiler for the representative functional
  language developed in this work, in the process illuminating the way
  in which the logical structure of \LProlog implementations and the
  \HOAS techniques can conspire to simplify correctness proofs.
\end{itemize}

We note that the work in this thesis is not the first one to undertake
the verification of properties related to compilation-oriented
transformations using the \HOAS approach. In\cspc\cite{hannan92lics},
Hannan and Pfenning have exploited this approach in specifying some
simple transformations---such as the translation of conventional
$\lambda$-terms to their de Bruijn forms---using the dependently typed
$\lambda$-calculus LF and in verifying these specifications
using the Elf meta-language (later renamed to
\Twelf). In\cspc\cite{tian06cats}, Tian has mechanized a CPS
transformation for the simply typed $\lambda$-calculus in LF and
proved its correctness in \Twelf. Finally, in
\cite{belanger13cpp}, Belanger \etal have developed a collection
of compiler transformations that includes the CPS transformation,
closure conversion and code hoisting in Beluga; their specification of
these transformations is such that the fact that it passes type
checking ensures that types are preserved between the source and
target language versions of the program. What distinguishes our work
is that it is a systematic study of the implementation and
verification of compilers for functional programs that examines the
use of the \HOAS approach in verifying deep properties such as meaning
preservation. We also note that while we conduct our work in the
context of a framework defined by \LProlog and \Abella, the ideas we
develop should be applicable to compiler verification using systems
like Twelf and Beluga, to the extent that these systems support
reasoning principles that are strong enough to carry out the relevant
tasks.

\section{An Overview of the Thesis}
\label{sec:thesis_overview}

The rest of this thesis develops the ideas that we have discussed in
this chapter.
In Chapter\cspc\ref{ch:framework}, we introduce the
\LProlog language and the \Abella theorem prover. The discussion in
this chapter also brings out the $\lambda$-tree syntax
approach and its use in simplifying both the implementation and
the verification of rule-based relational specifications. We then
present the two extensions to the \Abella theorem prover in
Chapter\cspc\ref{ch:extensions}. With the enriched framework in place,
we are ready to start the discussion of our verified compilation
work.
Chapter\cspc\ref{ch:vericfl}, provides an overview of this work.
We begin this chapter by presenting
the compilation model that we will use and the approaches to
describing compilers in the rule-base and relational fashion and to
verifying the correctness of compilers described in this manner. We
then show how our framework can be used to formalize such
implementation and verification and how the $\lambda$-tree syntax
approach can be used to simplify these tasks. Although in this thesis we will
characterize the correctness of compiler
transformations using \emph{logical
  relations}\cspc\cite{statman85ic}, we discuss in
Chapter\cspc\ref{ch:vericfl} some other notions of meaning
preservation that could have been used and also how the \HOAS approach
can simplify the verification task in their context as well. We
conclude this chapter with an overview of the exercise that we will
carry out towards demonstrating the effectiveness of
our approach to verified compilation of functional languages. That is,
we outline the structure of a compiler for the source language that we
have chosen and we present an overview of the steps in its verified
implementation. In Chapters\cspc\ref{ch:cps}, \ref{ch:closure},
\ref{ch:codehoist} and \ref{ch:complete}, we describe how the compiler
transformations constituting this compiler are implemented and
verified using our approach. We describe the related work in
Chapter\cspc\ref{ch:related}. Finally, we conclude the thesis and
discuss future work in Chapter\cspc\ref{ch:conclusion}.

% The framework
%%%%%%%%%%%%%%%%%%%%%%%%%%%%%%%%%%%%%%%%%%%%%%%%%%%%%%%%%%%%%%%%%%%%%%%%%%%%%%%
% framework.tex: The framework
%%%%%%%%%%%%%%%%%%%%%%%%%%%%%%%%%%%%%%%%%%%%%%%%%%%%%%%%%%%%%%%%%%%%%%%%%%%%%%%%
\chapter{A Framework for Verified Implementation}
\label{ch:framework}

The formal systems that are of interest in the thesis can be naturally
described via rules for deriving relations on syntactic objects.  To
specify and to reason about such \emph{rule-based relational
  specifications}, we use the framework consisting of the
specification language \LProlog\cspc\cite{miller12proghol,nadathur88iclp} and the
theorem-proving system \Abella\cspc\cite{baelde14jfr,gacek09phd}.
\LProlog is a language suitable for encoding rule-based relational
specifications. A characteristic of the \LProlog language is that
specifications written in it are executable and therefore they serve
also as implementations; such specifications can, for example, be
executed using the \Teyjus system that implements
\LProlog\cspc\cite{teyjus.website}.
\Abella is an interactive theorem prover for reasoning about
relational specifications.  It is possible to encode derivability in
\LProlog as a relation in \Abella. In fact, \Abella builds in such an
encoding of \LProlog. Further, it allows us to reason about
specifications written in \LProlog through this encoding using what is
referred to as the two-level logic
approach\cspc\cite{gacek12jar,mcdowell02tocl}. We will show in later
chapters how this structure can be exploited to realize the goal of
verified compilation: we will implement compiler
transformations in \LProlog and we will prove these implementations
correct in \Abella using the two-level logic approach. Another important
feature of both \LProlog and \Abella is that they support a
realization of the \HOAS approach that has been called the \emph{$\lambda$-tree
  syntax approach}\cspc\cite{miller00cl}. A key part of this thesis is to
show that this approach can help simplify the task of verified
compilation for functional programs.

We devote this chapter to exposing the various aspects of the
framework that we will make use of in the rest of the thesis. We start
by describing \LProlog and the logic it is based on and by introducing
the methodology they support for implementing rule-based
relational specifications in Section\cspc\ref{sec:spec_lang}. We then
discuss the logic underlying \Abella for reasoning about relational
specifications in Section\cspc\ref{sec:gee}. In
Section\cspc\ref{sec:abella}, we introduce Abella and the two-level logic
style of reasoning. The idea of $\lambda$-tree syntax will be implicit
throughout the chapter. In the concluding section of the chapter, we
discuss explicitly the benefits that can be derived in
implementation and in reasoning by using $\lambda$-tree syntax.

\section{The Specification Language}
\label{sec:spec_lang}
The \LProlog language is based on a fragment of a first-order
intuitionistic logic known as the logic of Hereditary Harrop
formulas\cspc\cite{miller91apal}. We will call this logic \HHw
here. \HHw is suitable for encoding rule-based description of
relations: rules for deriving relations translate naturally into
logical formulas that yield the desired kind of proof-theoretic
behavior in the logic. The formal systems that are of interest to us
usually concern objects that embody binding structure. \HHw supports
the representation of such objects by using simply typed
$\lambda$-terms, rather than the more commonly used first-order terms,
as ``data structures,'' \ie, as the arguments of predicates. Further,
\HHw possesses logical capabilities for manipulating these terms in a
way that respects and understands the abstraction operation they
contain. \HHw specifications can be given an operational interpretation
and this is, in fact, what \LProlog realizes as a programming
language.

We expand on the remarks above in the subsections that follow. In the
first subsection, we describe \HHw. We then explain the manner in
which \HHw can be used to encode relational specifications. Finally,
we outline how the \LProlog language allows us to execute such
specifications.

\subsection{The specification logic \HHw}
\label{subsec:spec_logic}

The logic \HHw is a fragment of Church's Simple Theory of
Types\cspc\cite{church40}. The expressions in this logic are those of
the simply typed $\lambda$-calculus or the \STLC. These expressions are
actually split into two categories: the types and the terms.

The type expressions are generated from \emph{atomic types} using the
\emph{function} or \emph{arrow} type constructor. Their syntax is
given as follows, assuming that $\tau$ stands for types and $a$ stands
for atomic types:
\begin{gather*}
  \tau \;::=\; a \sep (\tau \to \tau)
\end{gather*}
%
%GN What does this mean and why is it relevant? In a syntax rule,
%non-terminals may be used on the right and that just means that they
%have to be replaced by expressions that are also generated by the
%rule.
%% We may add subscripts to $\tau$ to distinguish between different
%% types. This convention is implied for the symbols to be introduced in
%% the rest of the thesis.
%
%GN This is not true of the ``simple theory of types'' it is true of
%how you mean to use them!
%% In the context of the simple theory of types, the atomic types consist
%% of a user-defined collection plus the distinguished type $\omic$ that
%% is used for formulas as explained below.
In this thesis, we shall take the atomic types to correspond to
a user-defined collection plus the distinguished type $\omic$ that
is used for formulas as explained below. We assume there to be at
least one user-defined atomic type. We drop parentheses in writing
arrow types, i.e. types of the second form, by using the convention
that the arrow operator $\to$ associates to the right. For instance,
$\tau_1 \to \tau_2 \to \tau_3$ stands for $(\tau_1 \to (\tau_2 \to
\tau_3))$. Using this associativity convention, every type can be
written in the form $\tau_1 \to \ldots \to \tau_{n} \to \tau_0$ where
$\tau_0$ is an atomic type. When a type is written in this fashion,
$\tau_i$ for $1\leq i \leq n$ are called its \emph{argument types} and
$\tau_0$ is called its \emph{target type}.

In building terms, we assume a vocabulary of constants and variables
where each constant and variable has a type associated with it. Terms
are then specified together with their types by the following
inductive rules, assuming that $t$ stands for terms:
\begin{itemize}
\item A constant or a variable of type $\tau$ is a term of
  type $\tau$.

\item If $t$ is a term of type $\tau'$ and $x$ is a variable of type
  $\tau$, then the expression $(\typedabs x \tau t)$ in is a term of type
  $\tau \to \tau'$. Such a term is referred to as an \emph{abstraction}
  that has the variable $x$ as its binder and $t$ as its
  body or scope. Further, all occurrences of $x$ in $t$ that are not
  in the scope of any abstraction in $t$ with $x$ as its binder 
  are considered bound by the abstraction.

\item If $t_1$ is a term of type $\tau_1 \to \tau_2$ and $t_2$ is a
  term of type $\tau_1$, then the expression $(t_1 \app t_2)$ is a
  term of type of type $\tau_2$. Such a term is called an
  \emph{application} that has   $t_1$ as its function part and $t_2$
  as its argument.
\end{itemize}
We drop parentheses in terms by assuming applications associate to the
left and applications bind more tightly than abstractions. For
instance, $(t_1 \app t_2 \app t_3)$ represents $((t_1 \app t_2) \app
t_3)$ and $(\typedabs x \tau {t_1 \app t_2})$ represents $(\typedabs x \tau
{(t_1 \app t_2)})$. We will often drop the type $\tau$ in an abstraction
$(\typedabs x \tau t)$, abbreviating it as $(\abs x t)$, when this type
is not important to our understanding or can be inferred uniquely from
the context. We will also need to refer below to the \emph{free variables} of
a term. These are the variables in the term that are not bound
by any abstraction occurring in it. 
We will often need to indicate a variable $x$ or a constant $c$
together with its type $\tau$. We do this by by writing $x:\tau$ or
$c:\tau$.
%
%% We will often need to indicate a term $t$, such
%% as a variable or constant, together with its type $\tau$. We will do
%% this by writing $t:\tau$.
%
We will also need to talk about a term $t$ with its type $\tau$ in a
context where the constants and free variables in $t$ and their types
are given. For this we will use the following notation
\begin{gather*}
  \{c_1:\tau_1,\ldots,c_m:\tau_m,x_1:\tau_1',\ldots,x_n:\tau_n'\} \tseq t:\tau
\end{gather*}
where $\{c_1,\ldots,c_m\}$ and $\{x_1,\ldots,x_n\}$ respectively
contain all the constants and free variables in $t$.

The underlying logic assumes an equality relation between terms that
is explained as follows:
%
%GN The way alpha step was described earlier is actually incorrect!
%you need the variable y not to appear in t, bound or free, for this
%style of presentation to work.
\begin{itemize}
\item
  An $\alpha$-step allows us to rename the variables bound by
  abstractions. Specifically, two terms are related by an
  $\alpha$-step if the second can be obtained from the first by replacing
  a subterm of the form $(\abs x t)$ by $(\abs y t')$   where $y$ is a
  variable that does not occur free in $t$ and $t'$ is the result of
  replacing the free occurrences of $x$ in $t$ by $y$. Two terms are
  related by $\alpha$-conversion if one can be obtained from the other
  by repeated applications of $\alpha$-steps.

\item The $\beta$-conversion relation captures the idea of equivalence
  under ``function evaluation.'' A term of the form $((\abs x {t_1}) \app
  t_2)$ is referred to as a $\beta$-redex. A term $u$
  $\beta$-contracts to a term $v$ if $v$ can be obtained by replacing
  such a $\beta$-redex in $u$ by the result of substituting $t_2$ for
  the free occurrences of $x$ in $t_1$ provided that the free
  variables in $t_2$ do not occur bound in $t_1$. Conversely, if $v$
  results from $u$ by $\beta$-contraction, then $u$ results from $v$
  by $\beta$-expansion. Finally, $\beta$-conversion is the reflexive
  and transitive closure of the union of the $\beta$-contraction,
  $\beta$-expansion and $\alpha$-step relations.

\item The $\eta$-rule reflects the idea of extensional equality for
  functions. We refer to a term of the form $(\abs x {t \app x})$
  where $x$ does not occur free in $t$ as an $\eta$-redex. A term $u$
  $\eta$-contracts to $v$ if $v$ can be obtained from $u$ by replacing such
  a $\eta$-redex $(\abs x {t \app x})$ in $u$ by $t$. Conversely, $v$ $\eta$-expands to
  $u$ if $u$ $\eta$-contracts to $v$. Finally, $\lambda$-conversion is
  the reflexive and transitive closure of the $\beta$-conversion,
  $\eta$-contraction and $\eta$-expansion relations.
\end{itemize}
Equality is given by the strongest of these relations, i.e., by
$\lambda$-conversion. It can be seen that this is an equivalence
relation, thereby meeting the basic criterion for an equality
relation.

We say that a variable or constant has arity $n$ if its type has $n$
argument types. We also say that its occurrence in a term is
\emph{fully applied} if it is applied to as many arguments as its
arity. A term is said to be in $\beta\eta$-long normal form if it does
not contain a $\beta$-redex and, further, every variable or constant
in it is fully applied. It is known that every term in the \STLC
$\lambda$-converts to a $\beta\eta$-long normal form that is unique up
to $\alpha$-conversion. We refer to such a form as a $\beta\eta$-long
normal form \emph{for} the term. It is further known that any term in
the \STLC can be transformed into (one of) its $\beta\eta$-long normal
form through a terminating process.
A further point to note is that the different operations such as
substitution that we consider on $\lambda$-terms commute with the
conversion rules. This allows us to always work with the
$\beta\eta$-long normal forms of terms, a fact that we will use
implicitly in the following discussions.

We will need to consider substitutions into terms. A substitution,
denoted by $\theta$, is a type-preserving mapping from variables to
terms that is the identity at all but a finite number of variables. We
write $(t_1/x_1,\ldots,t_n/x_n)$ for the substitution that maps
$x_1,\ldots,x_n$ to $t_1,\ldots,t_n$, respectively and each of the remaining
variables to itself. Given such a substitution, the set of variables
$\{x_1,\ldots,x_n\}$ is called its domain and each of the terms in
$t_1,\ldots,t_n$ is called the value of the mapping on the variable it
corresponds to. We write $\dom{\theta}$ for the domain of
$\theta$.
%GN There was a jump here! Substitutions up to this point are just
%mappings on variables. What does it mean then to require them to be
%``capture avoiding?'' This notion becomes meaningful only when you
%think of lifting substitutions to apply to terms!
Substitutions can be lifted to be mappings on terms rather than just
on variables. For this lifting to be logically correct, it must be
defined in such a way that it is ``capture-avoiding.'' One way to
formalize the idea to ensure this is as follows. Given a term
$t$ and a substitution $\theta$ such that $\theta =
(t_1/x_1,\ldots,t_n/x_n)$, the expression $t[\theta]$ denotes the term
%GN This stuff is really bizarre, like throwing away all the work you
%have done up to now and starting from scratch!
%% that is obtained by simultaneously replacing the free occurrences of
%% the variables $x_i$ in $t$ by the terms $t_i$ for $1 \leq i \leq
%% n$. Note, however, that in doing this replacement we must rename the
%% binders of abstractions in $t$ where this is necessary to prevent them
%% from inadvertently binding the free variables in $t_1,\ldots,t_n$. Such a
%% renaming has a logical justification: it corresponds to the use of
%% $\alpha$-steps that preserve equality between terms.
$((\abs {x_1} {\ldots \abs {x_n} {t}}) \app t_1 \app \ldots \app
t_n)$; observe that this term is equal under $\lambda$-conversion to
$t$ with each variable $x_i$ replaced by $t_i$ with relevant renamings
done within $t$ to avoid inadvertent capture of free variables in $t_i$.

%Sometimes we need to express two substitutions as a composed one.
The composition of two substitutions $\theta$ and $\rho$, denoted by
$\substcomp{\theta}{\rho}$, is defined as follows:
\begin{tabbing}
  \qquad\=$(\substcomp{\theta}{\rho})(x)$ \quad
    \=$:=$\quad \=$(\theta(x))[\rho]$ \qquad\=\kill
  \>$(\substcomp{\theta}{\rho})(x)$ \>$:=$ \>$(\theta(x))[\rho]$
    \>$x \in \dom{\theta}$\\
  \>$(\substcomp{\theta}{\rho})(x)$ \>$:=$ \>$\rho(x)$ \>otherwise.
\end{tabbing}
It is easy to check that $t[\theta][\rho] =
t[\substcomp{\theta}{\rho}]$ for any $t$. Thus, the application of a
composition of substitutions to a term corresponds, as one might
expect, to their composition as mappings on terms.

The definition of $\lambda$-terms relies on a collection of
constants. In the context of \HHw, we distinguish between
\emph{logical} constants and \emph{non-logical} constants. The set of
logical constants is fixed and consists of $true$ of type $\omic$, $\sconj$ and $\simply$,
both of type $\omic \to
\omic \to \omic$, and, for each type $\tau$ that does not contain
$\omic$, $\sfall_\tau$ of type $(\tau \to \omic) \to \omic$. The
constants $\sconj$ and $\simply$, also called the logical connectives,
correspond to
conjunction and implication and are usually written as infix
operators. The family of constants $\sfall_\tau$ represents universal
quantifiers: the term $(\sfall_\tau \app (\typedabs x \tau t))$
corresponds to the universal quantification of $x$ over $t$. In
writing this expression, we will often use the suggestive abbreviation
$\typedforall {\tau} x t$ and will also drop the type
annotation---i.e. we will simply write $\forallx x t$---when the type
can be inferred or when its knowledge is not essential to the
discussion. Furthermore, we will often abbreviate the formula
$\forallx {x_1\ldots} {\forallx {x_n} t}$ to $\forallx {x_1,\ldots,x_n} t$.
Note that given any term $\sfall_\tau\app t$, its subterm $t$ must have the
$\beta\eta$-long normal form $\abs x {t'}$ for some $t'$. As a result,
$\sfall_\tau\app t$ can always be represented
as $\typedforall {\tau} x {t'}$ for some $t'$.

The set of non-logical constants is also called a \emph{signature}. We
shall use the symbol $\Si$ to denote signatures. There is a
restriction on the constants allowed in a signature in \HHw: their
argument types must not contain the type $\omic$. A non-logical
constant whose target type is $\omic$ is called a predicate symbol or
predicate constant.  Such constants are used to form \emph{atomic
  formulas} that represent relations. Specifically, a term of the form
$(p \app t_1 \app \ldots \app t_n)$ in which $p$ is a predicate symbol
of arity $n$ constitutes an atomic formula; we shall call $p$ the
\emph{predicate head} of the atomic formula.\footnote{As previously
  mentioned, we work only with terms in $\beta\eta$-long normal
  form. Thus, we apply this and similar terminology to terms only
  after they have been transformed into their normal forms.}  We use
the symbol $A$ to denote atomic formulas.

The terms of type $\omic$, that are also called \emph{formulas}, have
a special status in the logic: they are the expressions to which the
derivation rules pertain. The \HHw logic is determined by two
particular kinds of formulas called \emph{goal formulas}, or simply
\emph{goals}, and \emph{program clauses}, or simply
\emph{clauses}. These formulas are denoted by the symbols $G$ and $D$,
respectively, and are given by the following syntax rules:
\begin{tabbing}
\qquad\=$G$\qquad\=::=\qquad\=\kill
\>$G$\>::=\>$true \sep A\sep G \sconj G\sep D\simply G \sep \typedforall{\tau}{x}{G}$\\
\>$D$\>::=\>$G\simply A\sep \typedforall{\tau}{x}{D}$
\end{tabbing}
Goal formulas of the form $D \simply G$ are called \emph{hypothetical
  goals}. Goal formulas of the form $\typedforall{\tau}{x}{G}$ are
called \emph{universal goals}. Note that a program clause has the form
$\typedforall{\tau_1}{x_1}{\ldots\typedforall{\tau_n}{x_n}{(G \simply
    A)}}$. We refer to $A$ as the head of such a clause. Further, we
call $G$ the body of the clause.

In the intended use of the \HHw logic, a collection of program clauses
and a signature constitutes a specification, also called a
\emph{program}. A user provides these collections towards defining
specific relations. The relations that are so defined are determined
by the atomic formulas that are derivable from a program. We formalize
the notion of derivability through a sequent
calculus\cspc\cite{gentzen35}. In the context of interest, a sequent
has the structure
\[
\Si; \G; \D \sseq G
\]
where $\Si$ is a signature, $\G$ is a multi-set of program clauses
called the \emph{static context} that contains user-defined
program clauses, $\D$ is a multi-set of program clauses called the
\emph{dynamic context} that contains clauses dynamically added during
the derivation of the sequent, and $G$ is a goal formula called the
\emph{goal} of the sequent that is to be shown derivable from the static and
dynamic contexts.
A program given by the clauses $\G$ and the signature $\Si$ then
specifies the relations represented by all the atomic formulas $A$
such that the sequent $\Si; \G; \emptyset \sseq A$ is derivable. If we
consider the sequents in a derivation from the perspective of how they
arise in the course of searching for a proof, the dynamic context is
initially empty. However, as we shall see presently, the process of
constructing derivations may add clauses to this context and may also
extend the signature.

The rules for deriving sequents of the form described are presented in
Figure\cspc\ref{fig:hhw-rules}. These rules are of three kinds: the
rule $\trueR$ is used to finish the proof, the rules $\andR$, $\impR$
and $\forallR$ are used to simplify non-atomic goals and the rule
$\backchain$ applies once the goal has been reduced to atomic form. In
\forallR we use $(\Si,c:\tau)$ to denote $\Si \cup \{c:\tau\}$ and in
\impR we write $(\D, D)$ for $\D \cup \{D\}$.
The rules \impR and \forallR are the ones that impart a dynamic
character to sequents. In particular, the \forallR rule causes the
signature to grow through the addition of a previously unused constant
in the course of searching for a derivation and the \impR rule
similarly causes additions to the dynamic context. These two rules are
crucial for specifying systems involving binding operators, as we
shall see in Section\cspc\ref{subsec:ecd_bindings}.  The $\backchain$
rule captures the following intuition for solving atomic goals: we look
for a clause in the dynamic or static context whose head matches the
goal we want to solve and then reduce the task to solving the relevant
instance of the body of the clause.

\begin{figure}[ht!]
    \begin{gather*}
      \infer[\trueR]{
        \Si; \G; \D \sseq true
      }{}
      \qquad
      \infer[\andR]{
        \Si; \G; \D \sseq G_1 \sconj G_2
      }{
        \Si; \G; \D \sseq {G_1}
        & \Si; \G; \D \sseq G_2
      }
    \end{gather*}
    \begin{gather*}
      \infer[\impR]{
        \Si; \G; \D \sseq {D \simply G}
      }{
        \Si; \G; \D, D \sseq G
      }
      \qquad
      \begin{array}{c}
        \infer[\forallR]{
          \Si; \G; \D \sseq {\typedforall{\tau}{x}{G}}
        }{
          \Si, c:\tau; \G; \D \sseq {G[c/x]}
        }
        \\
        \mbox{\small (where $c \notin \Si$)}
      \end{array}
    \end{gather*}
    %
    %% \begin{gather*}
    %%   \infer[\succeed]{
    %%       \Si; \G; \D \sseq A
    %%     }{
    %%       \typedforall{\tau_1}{x_1}{
    %%         \ldots \typedforall{\tau_n}{x_n}{A'}} \in \Gamma \cup \Delta
    %%     }
    %% \end{gather*}
    %
    \begin{gather*}
      \infer[\backchain]
        {
          \Si; \G; \D \sseq A
        }{
          \typedforall{\tau_1}{x_1}{
            \ldots \typedforall{\tau_n}{x_n}{G \simply A'}} \in \Gamma \cup \Delta
          &
          \Si; \G; \D \sseq G[t_1/x_1,\ldots,t_n/x_n]
        }
        \\
        \mbox{\small
          (where $\Si \tseq t_i:\tau_i$ for $1 \leq i \leq n$ and
          $A'[t_1/x_1,\ldots,t_n/x_n] = A$ in \backchain)
        }
    \end{gather*}

    \caption{Derivation Rules of \HHw}
    \label{fig:hhw-rules}
\end{figure}

An important observation is that a derivation in \HHw is guided by
the syntactic form of its goal $G$. If $G$ is $true$ then only $\trueR$ is applicable. 
If $G$ is not an atomic formula or $true$,
then exactly one of the $\impR$, $\andR$ and $\forallR$ rule is
applicable. Those rules simplify the goal until it becomes atomic. At
that point only $\backchain$ is applicable. The
derivation continues by
applying $\backchain$ which generates a subgoal. This process repeats
until all subgoals are proved.

The \LProlog language, which is a realization of \HHw, provides a
concrete syntax in which a user can present \HHw programs. In this
syntax, a multiset of clauses is written as a sequence, with each
clause being terminated by a period. Further, universal quantifiers at
the outermost level in a clause may be omitted by using names starting
with capital letters for the occurrences of variables that they bind.
Abstraction is written as an infix operator. More specifically, the
term $(\typedabs x \tau M)$ is written as $(x{:}\tau \mlam M)$ in
\LProlog. When the type of the bound variable can be inferred uniquely
from the context, this expression can also be simplified to $(x \mlam
M)$. Conjunction in goals is denoted by a comma: a goal of the form
$(G_1 \sconj G_2)$ is written as $(G_1 , G_2)$. The clause $(G \simply
D)$ is written in \LProlog as $(D \limply G)$. Finally, when $G$ is
$true$ the clause $G \simply D$ is further simplified to just $D$.

\subsection{Encoding rule-based relational specifications}
\label{subsec:ecd_rel_specs}

A natural way to present formal systems is to describe them via rules
deriving relations on syntactic objects. As an example, we consider
the rule-based descriptions of the append relation on lists of natural
numbers. Let $\mnil$ denote the empty list in our object language and
let $\mcons$ denote the constructor for lists such that $x \mcons l$
stands for a list where $x$ is the first element of the list, called
its head, and $l$ is a list containing the rest of the elements, called its
tail. If we limit our attention to lists that are constructed using
$\mnil$ and $\mcons$ as the only constructors of list type,
then we can define a ternary relation $\mappend$ given by the
following rules:
\begin{gather*}
  \infer[\appdnil]{
    \mappend \app \mnil \app l \app l
  }{}
  \qquad
  \infer[\appdcons]{
    \mappend \app (x \mcons l_1) \app l_2 \app (x \mcons l_3)
  }{
    \mappend \app l_1 \app l_2 \app l_3
  }
\end{gather*}
The content of this definition is that $\mappend \app l_1 \app l_2
\app l_3$ holds if and only if it can be derived using these rules.
It is then not difficult to see that $\mappend \app l_1 \app l_2 \app
l_3$ holds just in the case that $l_1$, $l_2$ and $l_3$ are lists
consisting of ground elements and $l_3$ contains the elements of $l_1$
followed by those of $l_2$.

Program clauses in \HHw provide a natural way to capture rule-based
specifications of relations: a relation can be represented by a
predicate constant and each rule of the relation translates naturally
to a program clause defining the predicate constant, with
the conclusion of the rule becoming the head of the clause and the
premises, if any, becoming its body.

We illustrate the above idea by considering the specification of the
append relation. In encoding this specification, let us assume
that we have designated the type $\knat$ to represent the type of natural numbers
and $\klist$ to represent the type of lists of natural numbers. Further, let us
assume that we have introduced into the signature the constants
$0,1,2,3,\ldots$ of type $\knat$ to represent the natural numbers and the
following constants to represent the list constructors:
\begin{align*}
  \knil &: \klist &
  \cons &: \knat \to \klist \to \klist
\end{align*}
Mirroring the convention in the object language, we will write
$\cons$ in infix form also in the \LProlog presentation and treat it
as right-associative; the
\LProlog language provides users a way to present such conventions at
the time when they identify a signature, but the details of this
process are orthogonal to the present discussion. We then use the
predicate
symbol
\[
\kappend : \klist \to \klist \to \klist \to \omic
\]
to represent the append relation and we encode the rules defining the
relation in the following clauses where the first clause encodes the
\appdnil rule and the second clause encodes the \appdcons rule:
\begin{tabbing}
  \qquad\=$\kappend \app (X \cons L_1) \app L_2 \app (X \cons L_3)$
  \quad\=$\limply$\quad\=\kill
  \>$\kappend \app \knil \app L \app L.$ \\
  \>$\kappend \app (X \cons L_1) \app L_2 \app (X \cons L_3)$\=
    $\limply$\=$\kappend \app L_1 \app L_2 \app L_3.$
\end{tabbing}

Let $\G$ be a static context consisting of the clauses for $\kappend$
shown above, let $l_1$, $l_2$
and $l_3$ be three lists, and let $L_1$, $L_2$ and $L_3$ be their
encodings in \LProlog. Then, showing that $(\mappend\app l_1\app l_2\app
l_3)$ holds given the specification of $\mappend$ is
equivalent to showing the following sequent is derivable, where $\Si$
% GN This is WRONG, L_i are not variables, they are terms.
%contains the variables $L_1$, $L_2$ and $L_3$ and the constants we
contains the constants we have introduced for encoding lists and the
append relation:
\[
\Si; \G; \emptyset \sseq \kappend\app L_1\app L_2\app L_3
\]
Note that the way the program clauses of \kappend would be used
in constructing
derivations for this sequent has a transparent relationship to the way
in which the
rules specifying $\mappend$ would be used in establishing that the
relation $(\mappend\app l_1\app l_2\app
l_3)$ holds: for instance, the attempt to construct a derivation would
begin by matching $(\kappend \app L_1 \app L_2 \app L_3)$
with the head of one of the clauses for $\kappend$ (that
corresponds to the conclusions of the rules in the object-language
specification) and by reducing the task to
deriving the corresponding body (that corresponds to the premise of
the relevant rule). In this sense, the \LProlog
encoding also reflects the \emph{derivation behavior} of the rule
based specification of the append
relation.
%GN This example is not really useful or necessary, the point should
%be amply clear at this stage.
%%  Figure\cspc\ref{fig:appd_spec_exm} shows such an example in
%% which an \HHw sequent is derived by first applying \backchain on the
%% second clause of \kappend (which corresponds to applying \appdcons)
%% and then applying \succeed on the first clause of \kappend (which
%% corresponds to applying \appdnil).

%% \begin{figure}[ht!]
%%   \centering
%%   \begin{subfigure}[t]{\textwidth}
%%     \begin{gather*}
%%       \infer[\appdcons]{
%%         \mappend\app (x \mcons \mnil)\app l\app (x \mcons l)
%%       }{
%%         \infer[\appdnil]{
%%           \mappend\app \mnil\app l\app l
%%         }{
%%         }
%%       }
%%     \end{gather*}
%%     \caption{Deriving a $\mappend$ relation via rules}
%%   \end{subfigure}
%%   \\
%%   \begin{subfigure}[t]{\textwidth}
%%     \begin{gather*}
%%       \infer[\backchain]{
%%         \Si; \G; \emptyset \sseq
%%         \kappend\app
%%           (X \cons \knil)\app
%%           L\app
%%           (X \cons L)
%%       }{
%%         \infer[\succeed]{
%%           \Si; \G; \emptyset \sseq
%%           \kappend\app \knil\app L\app L
%%         }{}
%%       }
%%     \end{gather*}
%%     \caption{The corresponding derivation in \HHw}
%%   \end{subfigure}
%%   \caption{An Example of Reflecting Derivation Behaviors in \HHw}
%%   \label{fig:appd_spec_exm}
%% \end{figure}

\subsection{Encoding specifications over objects with binding structure}
\label{subsec:ecd_bindings}

Our interest in this work is in treating specifications of formal
systems such as logics, programming languages and compilers.  An
important characteristic of such systems is that the expressions they
treat contain variable binding operators. The traditional approach to
encoding such operators is to use a first-order representation and to
let the user build in properties of binding through additional
specifications or programs. A defining aspect of the \LProlog language
is that it supports a different, more abstract, approach to treating
such binding operators. This language provides us with
$\lambda$-terms, rather than just first-order terms, to represent
expressions in an object language. The abstraction operator present in
these terms gives us a meta-language level mechanism to capture the
binding notions present in the expressions over which we want to
specify and carry out computations. In addition to representing the
syntax of objects, we also often need to support the ability to
specify properties by recursion over their structure. The \LProlog
language has mechanisms that are specially geared towards realizing
such recursion over binding operators. These mechanisms arise from the
presence of universal and hypothetical goals in the language.

To illustrate the different mechanisms mentioned above, we consider
the task of encoding the typing relation for the
simply-typed $\lambda$-calculus. Note that we are thinking of the
\STLC as an \emph{object system} in this example.
We will take the syntax of types and terms in this object system to be
given by the following rules:
\begin{gather*}
  \begin{array}{l@{\ }c@{\ }l}
    T & ::= & a \sep T_1 \to T_2 \\
    M & ::= & x \sep \typedabs x T M \sep M_1 \app M_2
  \end{array}
\end{gather*}
The typing relation that we want to encode is written as
\[
  \Gamma \stseq M:T
\]
where $M$ is a term, $T$ is a type and $\Gamma$ is a \emph{typing
  context} that has the form
\[
x_1:T_1, \ldots, x_n:T_n
\]
where each $x_i$ is a distinct variable.  Intuitively, such a judgment
represents the fact that $M$ is a well-formed term of type $T$,
assuming that $\Gamma$ provides the types for its free variables.  The
rules for deriving a judgment of this kind are the following:
\begin{gather*}
  \infer[\stlctvar]{
    \Gamma \stseq x:T
  }{
    x:T \in \Gamma
  }
  \qquad
  \infer[\stlctapp]{
    \Gamma \stseq M_1  \app  M_2 : T_2
  }{
    \Gamma \stseq M_1 : T_1 \to T_2
    &
    \Gamma \stseq M_2 : T_1
  }
  \\
  \infer[\stlctabs]{
    \Gamma \stseq \typedabs x {T_1} M : (T_1 \to T_2)
  }{
    \Gamma, x:T_1 \stseq M : T_2
  }
  \\
  \mbox{\small where $x$ is not already in $\Gamma$}
\end{gather*}
The only rule in this collection that is sensitive to binding
structure is \stlctabs, the rule for typing abstractions. This rule
asserts that the abstraction $\lambda x:T_1. M$ has the type
$T_1 \to T_2$ in a typing context $\Gamma$ if its body $M$ can be
given the type $T_2$ in a typing context that extends $\Gamma$ with
the type $T_1$ assigned to $x$, under the assumption, of course, that
$x$ is a variable that is not already assigned a type by
$\Gamma$. This rule exhibits many features that are typical to rules
that treat binding constructs that appear in formal objects. To encode
rules such as this one, we need a mechanism for capturing recursion
that is based on descending into the scope or body of the binding
construct, we need to be able to enforce conditions on the variables
we introduce to facilitate the recursion (in the case of \stlctabs,
this variable, which is named $x$, must be fresh to $\G$), and we need
to be able to augment the context with assumptions about the freshly
introduced variables (in the case of \stlctabs, $x$ must be assumed to
have type $T_1$). As we shall see below, \LProlog provides devices for
realizing all these aspects in a logically supported way.

To encode typing in the \STLC in \LProlog, we first need to represent the
expressions it pertains to. Towards this end, we identify the two
types \kty and \ktm.
Expressions in \LProlog of these types will correspond, respectively, to
types and terms in the object language. We then identify the following
constants for encoding types and terms:
\begin{gather*}
  \begin{array}{l@{\qquad}l}
  \kbase : \kty &  \karr : \kty \to \kty \to \kty\\
  \kabs : \kty \to (\ktm \to \ktm) \to \ktm
  &
  \kapp : \ktm \to \ktm \to \ktm
  \end{array}
\end{gather*}
The only constant in this signature that requires special mention is
\kabs. This constant is used to represent abstractions in the object
language. Note that the ``term'' component that this constant takes is
itself an abstraction in \LProlog. The idea underlying this
representation is that we isolate the binding aspect of
abstractions in the \STLC and allow these to be treated through the
understanding of abstraction in \LProlog. As a concrete example, the
\STLC term $(\lambda x: a\to a. \lambda y:a. x \app y)$ will be
encoded by
\[
\kabs \app
(\karr \app \kbase \app \kbase) \app (x\mlam \kabs \app {\kbase} \app
(y\mlam \kapp \app x \app y)).
\]
The virtue of this kind of encoding is that it allows us to use the
understanding of abstraction over $\lambda$-terms
present in \LProlog to transparently realize binding related
properties such as scoping, irrelevance of bound variable names and
(capture-avoiding) substitution over object language expressions.

We can now proceed to encoding the typing rules for the object
system. Towards this end, we first introduce the predicate symbol
\[
  \kof : \ktm \to \kty \to \omic
\]
to represent the typing relation. Note that the typing context is not
treated explicitly in this representation. Instead, it will be
realized implicitly via the dynamic context of the \HHw sequents we
try to derive. Specifically, the typing context
\[
  x_1:T_1, \ldots, x_n:T_n
\]
will be encoded by the dynamic context
\[
  \kof\app x_1\app T_1',\ldots,\kof\app x_n\app T_n'
\]
where, for $1 \leq i \leq n$, $T_i'$ is the encoding of $T_i$.

The typing rules $\stlctapp$ and $\stlctabs$ translate into the
following \HHw clauses defining the $\kof$ predicate:
\begin{tabbing}
  \qquad\=$\kof \app (\kabs \app T_1 \app M) \app (\karr \app T_1 \app T_2)$
  \;\=$\limply$\quad\=\kill
  \>$\kof \app (\kapp \app M_1 \app M_2) \app T_2$\>$\limply$
    \>$\kof \app  M_1 \app (\karr \app T_1 \app T_2) \scomma \kof \app M_2 \app T_1.$\\
  \>$\kof \app (\kabs \app T_1 \app M) \app (\karr \app T_1 \app T_2)$\>$\limply$
    \>$\forallx y {\kof \app y \app T_1 \simply \kof \app (M \app y) \app T_2}.$
\end{tabbing}

Assuming $\G$ is the static context consisting of these clauses, then
showing that the relation $\D \stseq M:T$ holds in the object system
is equivalent to showing that the sequent
\[
\Si; \G, [\D] \sseq \kof\app [M]\app [T].
\]
is derivable in
\HHw. We write $[E]$ here to denote the encoding in \LProlog of the object
language expression $E$, which might be a typing context, a term or a
% GN So what does the signature contain? Constants as you claim later? or
% (free) variables as you seem to be asserting here? You have to fix
% on a consistent terminology.
% type and we also assume that $\Si$ contains the constants we have
% introduced to encode terms and types and all the variables
% that are free in $[\D]$, $[M]$ and $[T]$.
type. We also assume that $\Si$ contains the constants we have
used to encode terms and types and all the constants we would have
introduced in the course of an \HHw derivation to represent the
variables that are assigned types by $[\D]$ as we explain below.

It is interesting to note that the way the typing rules are used
to establish $\D \stseq M:T$ in the object system is transparently
related to the way the sequent encoding this judgment is derived in
\HHw.
% $(\kof\app [M]\app [T])$ is derived from clauses in $[\D]$ and
% $\G$.
To see this,
first observe that application of the \stlctvar rule is mirrored in
the use of the \backchain and \trueR rules to close off a particular path in the
\HHw derivation: if $M$ is a variable $x$ that is given the type $T$
by the context $\D$, there there must be a formula
$(\kof\app x\app [T])$ in $[\D]$.
It is also easy to see that application of the \stlctapp rule is
naturally captured by an application in \HHw of \backchain based on
the clause that encodes \stlctapp.
Finally, the clause encoding $\stlctabs$ shows
how recursion over binding structure is realized by using universal and
hypothetical goals and how substitution is modeled by
$\beta$-conversion in \LProlog. When $[M]$ is $(\kabs \app
[t_1] \app (\abs x {[m]}))$ and $[T]$ is $(\karr \app [t_1] \app [t_2])$
for object language expressions $t_1$, $t_2$ and $m$ of the relevant
kinds, we would have to derive a sequent which has as its goal the
formula
\[
\kof \app (\kabs \app [t_1] \app (\abs x {[m]})) \app (\karr \app [t_1]
\app [t_2]).
\]
The only way this sequent can be derived is by backchaining on the
clause encoding $\stlctabs$. This would yield a sequent that has as
its goal the formula
\[
\forallx y {\kof \app y \app [t_1] \simply \kof \app ((\abs x {[m]}) \app y) \app [t_2]}.
\]
To derive this sequent in \HHw, we would have to introduce a new
constant $c$ and add the clause $(\kof \app c \app [t_1])$ to the
dynamic context before trying to derive a sequent that has as its goal
the formula $(\kof \app ((\abs x {[m]}) \app c) \app
[t_2])$. Observe here that the constant $c$ has been introduced to
represent the object language bound variable and that the ``newness''
of $c$ captures the freshness condition for this variable in the rule
$\stlctabs$. The addition of $(\kof \app c \app [t_1])$ to the dynamic
context of the sequent encodes the extension of the typing context in
the premise of the $\stlctabs$ rule and the \LProlog term $((\abs x
{[m]})\app c)$ is,
modulo $\lambda$-conversion, a representation of the object language
term $m$ in which $c$ is used to represent the variable bound by the
abstraction we have descended under. It is easy to see from all this
that the sequent we are left to prove corresponds precisely to the
premise of the $\stlctabs$ rule.

As an example, the following is a derivation of a typing relation
using the typing rules
\begin{gather*}
  \infer[\stlctabs]{
    \emptyset \stseq (\typedabs x {a \to a} {\typedabs y a {x \app y}) : (a \to a) \to a \to a}
  }{
    \infer[\stlctabs]{
      x: a \to a \stseq {(\typedabs y a {x \app y}) : a \to a}
    }{
      \infer[\stlctapp]{
        x: a \to a, y:a \stseq {x \app y} : a
      }{
        \infer[\stlctvar]{
          x: a \to a, y:a \stseq x : a \to a
        }{}
        &
        \infer[\stlctvar]{
          x: a \to a, y:a \stseq y:a
        }{}
      }
    }
  }
\end{gather*}
The corresponding derivation in \HHw is as follows:
\begin{gather*}
  \infer={
    \Si; \G; \emptyset \sseq
      \kof\app
      (\kabs \app
       (\karr \app \kbase \app \kbase) \app (x\mlam \kabs \app {\kbase} \app
       (y\mlam \kapp \app x \app y)))\app
      (\karr\app (\karr\app \kbase\app \kbase)\app (\karr\app \kbase\app \kbase))
  }{
    \infer={
      \Si, x:\ktm; \G; \kof\app x\app (\karr\app \kbase\app \kbase) \sseq
         {\kof\app (\kabs \app {\kbase} \app (y\mlam \kapp \app x \app y))\app
           (\karr\app \kbase\app \kbase)}
    }{
      \infer[\backchain]{
        \Si, x:\ktm, y:\ktm; \G;
        \kof\app x\app (\karr\app \kbase\app \kbase),\kof\app y\app \kbase
        \sseq
        \kof\app (\kapp \app x \app y)\app \kbase
      }{
        \infer[\backchain]{
          \Si'; \G; \D' \sseq \kof\app x\app (\karr\app \kbase\app \kbase)
        }{
          \infer[\trueR]{
            \Si'; \G; \D' \sseq true
          }{}
        }
        &
        \infer[\backchain]{
          \Si'; \G; \D' \sseq \kof\app y\app \kbase
        }{
          \infer[\trueR]{
            \Si'; \G; \D' \sseq true
          }{}
        }
      }
    }
  }
\end{gather*}
Here the double line stands for the successive application of
$\backchain$ on the clause encoding $\stlctabs$, $\forallR$ and
$\impR$, and $\Si'$ and $\D'$ are abbreviations of $(\Si, x:\ktm,
y:\ktm)$ and $(\kof\app x\app (\karr\app \kbase\app \kbase),\kof\app
y\app \kbase)$, respectively. Notice how backchaining on program
clauses followed by rules for simplifying the goal formula
represents the application of the typing rules for non-variable terms
and how the application of \backchain followed by \trueR represents
the application of the rule for typing variables. In general, it
should be clear from this discussion that, under the encoding that we
have used for the typing rules in the \STLC, the derivations we would
construct in \HHw closely follow the structure of those in the \STLC.

\subsection{\LProlog specifications as implementation}
\label{subsec:spec_impl}

It should be clear from the discussion up to this point that the
derivation rules for \HHw that are shown in Figure~\ref{fig:hhw-rules}
have an executable character: we proceed to find a derivation for a
sequent with a complex goal formula by simplifying the formula and
when we arrive at an atomic goal we look for a clause in the
static or dynamic context whose head matches the goal.
In the cases we have considered previously, the goals have all been
closed; in this mode our objective has been limited to checking if a
relation holds. However, we can think of extending this mechanism to
``compute answers'' by
including variables in a goal that we expect the derivation mechanism
to fill in with an actual expression for which the relation holds.
As a concrete example, letting $T$ be a variable of the kind just
described, we might submit the following query
\[
\kof \app (\kabs \app (\karr \app \kbase \app \kbase) \app (x\mlam
\kabs \app {\kbase} \app (y\mlam \kapp \app x \app y))) \app T
\]
in a context where the program consists of the clauses encoding the
typing rules in the \STLC. The only solution to this ``schematic'' query
is one where $T$ is instantiated with the expression $(\karr \app
(\karr \app \kbase \app \kbase) \app (\karr \app \kbase \app
\kbase))$.

The \LProlog language realizes this kind of an executable
interpretation. Many useful applications have been shown to exist for
this kind of ``logic programming'' interpretation for the
language\cspc\cite{miller12proghol}
and the \Teyjus system\cspc\cite{teyjus.website} has been developed to
provide efficient support for such applications. We will make use of
these facts in this dissertation. In particular, we will use \LProlog
to specify compiler transformations that we will then execute as
programs to effect compilation using the \Teyjus system.

\section{A Logic of Fixed-Point Definitions}
\label{sec:gee}

%GN This intro is not very accurate at least because G is more than a
%logic for reasoning about specifications in HHw. It is best not to
%introduce it as a logic for reasoning about specifications till you
%describe the two-level logic approach in a later section.
%GN You also need to be careful about references. To describe a
%``series of work'' leading to G that does not include McDowell and
%Miller that started it all is not good scholarship. These things
%matter!
%Finally, please pay attention to language. ``Series of work'' is
%incorrect. Also you cannot use ``leads to'' for something that
%happened in the past. ``Leads'' is present tense.
%% Having described a specification language for encoding relational
%% specifications, a natural question to ask next is how to reason about
%% such specifications. In the past decade, there has been a series of
%% work on this
%% topic\cspc\cite{miller05tocl,tiu06lfmtp,gacek08lics,gacek08lfmtp,gacek09phd}
%% which eventually leads to a logic named \Gee with rich capabilities
%% for reasoning about relational specifications of formal systems
%% involving binding constructs. The \Abella theorem prover that
%% implements \Gee is our choice of tool for verification of compiler
%% transformations in this thesis. We therefore give an exposition of
%% \Gee in this section.

The logic \HHw allows us to encode rule-based relational
specifications in such a way that we can reason about what should hold
in their context. Thus, using the formalization of typing for the \STLC,
we could show that the term
$(\typedabs x {a \to a} {\typedabs y a {x \app y}})$ has the type $(a \to a)
\to a \to a$. However, \HHw does not
  provide a means for capturing the fact that these encodings are
  complete in the sense that if a relation is not derivable from a
  program then it does not in fact hold. Rule-based specifications are
  usually intended to be interpreted in this way. For example, based
  on the specification of typing for the \STLC, we might want to conclude
  that every term in it has a unique type. Implicit to proving this is
  the fact that a typing judgment of the form
\[\emptyset \stseq (\typedabs x {a \to a} {\typedabs y a {x \app y}) : T}\]
is \emph{not} derivable for any type $T$ other than $(a \to a) \to a
\to a$.

The logic \Gee has been designed to provide the kind of complete
characterization of relational specifications that is discussed
above. This logic is the end-point of a sequence of developments that
started with work by McDowell and Miller about two decades
ago\cspc\cite{gacek11ic,mcdowell00tcs,miller05tocl,tiu06lfmtp}.
A defining characteristic of \Gee is that it interprets atomic
predicates as fixed-point definitions. These
fixed-point definitions provide a means not only for showing that a
relation holds, but
also for \emph{analyzing} why it holds. More specifically, they allow us
to carry out reasoning based on case analysis that occurs often in proving
properties of formal systems. Another important feature of \Gee is a
generic quantifier $\nabla$ (pronounced as ``nabla'') that, amongst
other things, requires us to provide a proof for the formula that it
scopes over that is independent of the instance chosen for the
quantified variable\cspc\cite{miller05tocl}.
Coupled with fixed-point definitions, this quantifier enables us to
use case analysis arguments over binding structure that is necessary
in many tasks of reasoning about formal systems. Definitions of
%definitions in \Gee can also be treated inductively,
% GN: ``Principal'' means ``head'' or ``chief,'' like in ``principal
% formula.'' I don't think you mean that here. ``Principle'' means
% ``rule'' or ``idea'' which is what you mean here.
%GN You are wrong to claim that induction allows you to reason about
%``infinite behaviors'' so PLEASE DO NOT make this mistake
%again. Induction has to be well-founded to be meaningful, which
%means that the behavior has to terminate, i.e. be
%finite. Co-induction is what allows you to reason about infinite
%behavior. What induction allows you to do is talk about finite
%behavior for an infinite collection of cases in a finite
%argument. This is quite a different thing from infinite behavior.
%% leading to the induction principal for reasoning about infinite
%% behaviors. These strong reasoning mechanisms combined together
%% allow
atomic predicates can also be given a \emph{least fixed-point}
interpretation, leading to the ability to reason about these
predicates in an inductive fashion.
These different capabilities give \Gee the ability to encode many
different forms of arguments that we might want to carry out over
relational specifications.

% GN: You cannot say ``flesh out'' and then immediately claim that
% your description is going to omit many details, these are
% contradictory!
%% We shall flesh out the aspects of \Gee outlined above in this section.
%% Our discussion pays particular attention to the features of \Gee that
%% make it a suitable logic for reasoning about relational specifications
%% that treat binding in object languages. Our representation of \Gee is
%% not a comprehensive one since we focus on explaining the features
%% relevant to our later applications.  A reader interested in a precise
%% description of \Gee may consult \cite{gacek09phd}.

In the rest of this section, we expose the features of \Gee
that are alluded to above. Our presentation is intended only to make
our use of the logic in explaining and constructing proofs in Abella
understandable. A reader interested in a more complete description of
\Gee may consult\cspc\cite{gacek11ic}.

\subsection{The syntax of \Gee}
\label{subsec:gee_syntax}

The logic \Gee is also based on an intuitionistic version of Church's
Simple Theory of Types. The
expressions of \Gee are similar to that of \HHw, \ie, the \STLC terms. One
difference is that \omic is replaced by \prop as the type of
formulas. Another difference is that different names are used for the
logical constants and the set of these constants is also larger. In
particular, the logical constants of \Gee consist of $\rtrue,\rfalse :
\prop$ that represent true and false, $\rand,\ror,\rimp : \prop \to
\prop \to \prop$ that represent conjunction, disjunction and
implication, and, for each type $\tau$ not containing \prop, the
constants $\rfall_\tau,\rexst_\tau : (\tau \to \prop) \to \prop$ that
correspond to (the family of) universal and existential
quantifiers. Following the style of \HHw, we write $\rand$, $\ror$ and
$\rimp$ as infix operators. Similarly, we abbreviate the expressions
$(\rfall_\tau\app (\typedabs x \tau B))$ and $(\rexst_\tau\app (\typedabs
x \tau B))$ by $(\typedrforall {\tau} x B)$ and $(\typedrexists
{\tau} x B)$. We also drop the type annotations in abstractions and
quantified formulas when they are irrelevant to the discussion or can
be inferred uniquely. Furthermore, we will often abbreviate the
formula $\rforallx {x_1\ldots} {\rforallx {x_n} t}$ ($\rexistsx {x_1\ldots}
{\rforallx {x_n} t}$) to $\rforallx {x_1,\ldots,x_n} t$ ($\rexistsx
{x_1,\ldots,x_n} t$) or $\rforallx {\vec{x}} t$ ($\rexistsx {\vec{x}} t$)
where $\vec{x}$ represents the sequence of variables $x_1,\ldots,x_n$.

%% When stating formulas that represent theorems we would like to prove
%% in \Abella, we shall use the concrete syntax of \Abella for universal
%% and existential quantified formulas in them. In this concrete syntax,
%% an universal formula of the form $(\rfall {x_1:\tau_1}. \ldots \rfall
%% {x_n:\tau_n}. M)$ is written as $(\rfall (x_1:\tau_1)\app \ldots\app
%% (x_n:\tau_n), M)$, or $(\rfall x_1\app \ldots\app x_n, M)$ when the
%% types of bound variables can be inferred uniquely from the
%% context. The concrete syntax for expressing existential formulas is
%% similar to universal ones.

\subsection{The generic quantifier $\nabla$}
\label{subsec:nabla}

Universal quantification in \Gee has an \emph{extensional}
reading. That is, to prove $\rforallx x B$ in \Gee we have to prove
$B[t/x]$ for each possible term $t$ but these proofs could be different
ones for different values of $t$.
%% For example, consider
%% proving a formula such as $\rforallx x {(q\app x \rimp p\app x)}$ where
%% $q$ and $p$ are predicate constants: one way to do this is to consider
%% each term $t$ that $q$ is true of and to construct a proof of $(p\app
%% t)$ that possibly treats different cases of $t$ differently.
However, when describing
specifications containing binding constructs, it is often desirable to
interpret a statement such as ``$B(x)$ holds for $x$'' as $B(t)$ holds
for every $t$ for the same \emph{structural} reasons that are
independent of the choice of $t$.

To provide a means for capturing this alternative style of reasoning,
\Gee includes a special \emph{generic}
quantifier\cspc\cite{miller05tocl}. Specifically, it includes a
constant $\nabla_\tau : (\tau \to \prop) \to \prop$ for every type
%GN: Do you use the fact that sometypes are designated as nominal
%types anywhere in the thesis? I am assuming not and therefore
%removing this qualification to avoid unnecessary noise. If you do use
%it, add it back.
%$\tau$ that does not contain \prop and that belongs to a designated
%set of \emph{nominal} types. As with other quantifiers, $\typednabla
$\tau$ that does not contain \prop.
As with other quantifiers, $\typednabla
{\tau} x B$ abbreviates $\nabla_\tau\app (\typedabs x \tau B)$. The
indexing type $\tau$ is omitted if its identity is not relevant to the
discussion or can be inferred uniquely from the context. The logical
meaning of the formula $\nabla_\tau x. B$ can be understood as
follows: to construct a proof for it, we pick a new constant of type
$\tau$ that does not appear in $B$ and then prove the formula that
results from instantiating $x$ in $B$ with this constant. The
constants that are to be used in this way are called \emph{nominal
  constants}. The treatment of $\nabla_\tau x. F$ as an assumption is
similar: we get to use $F$ with $x$ replaced by a fresh nominal
constant as an assumption instead. An important property of the
$\nabla$ quantifier is that each quantifier over the same formula
refers to a \emph{distinct} constant. Thus, $(\nabla x.\nabla y. x = y
\rimp \rfalse)$ is a theorem of \Gee. This property carries over to
nominal constants: two distinct nominal constants in a given formula
are treated as being different and, thus, the formula $(a_1 = a_2
\rimp \rfalse)$ where $a_1$ and $a_2$ are two different nominal
constants is a theorem. Another important property of nominal
constants is that their names have significance only in distinguishing
between different nominal constants in a single formula. For example,
given the predicate symbol $p$ and distinct nominal constants $a_1, a_2,
a_3,$ and $a_4$, the formula $(p\app a_1\app a_2)$ is considered to be
logically equivalent to $(p\app a_3\app a_4)$.

\subsection{Formalizing provability in \Gee}
\label{sec:gee_seq_cal}

The approach to treating $\nabla$ outlined in
Section\cspc\ref{subsec:nabla} was first described in the logic
\LGw\cspc\cite{tiu06lfmtp} and has been adopted by \Gee. Specifically,
we assume that there are an infinite number of nominal constants at
every %nominal
type. The collection of all nominal constants is denoted by
$\nominalcst$, which is disjoint from the collection of usual,
non-nominal constants denoted by $\normalcst$. We define the support
of a formula or a term $t$ as the set of nominal constants occurring
in it, denoted by $\support{t}$. We define a permutation $\pi$ of
nominal constants as a bijection from $\nominalcst$ to
$\nominalcst$. We write $\pi.B$ for the result of applying $\pi$ to
the formula $B$, which is defined as follows:
\begin{gather*}
\pi.c := c \quad \mbox{where $c \in \normalcst$}
\qquad\qquad
\pi.a := \pi(a) \quad \mbox{where $a \in \nominalcst$}\\
\pi.x := x
\qquad\qquad
\pi.(t_1\app t_2) = \pi.t_1\app \pi.t_2
\qquad\qquad
\pi.(\abs x t) = \abs x {(\pi.t)}
\end{gather*}
We use the proposition $B \feq B'$ to denote that the formulas $B$ and
$B'$ are equal modulo the permutation of nominal constants occurring
in them, that is, there exists some $\pi$ such that $\pi.B =
B'$. Given the fact that we treat the names of nominal constants
having scope only over individual terms, substitution must satisfy an
extra condition to be logically correct: it should avoid confusion
between the names of nominal constants in the terms being substituted
and in the terms being substituted into. Specifically, $B[\theta]$ now
stands for a formula obtained by first applying a permutation $\pi$
which maps the nominal constants in $B$ to nominal constants that do
not occur in the values of $\theta$ and then applying the substitution
$\theta$ on the resulting formula. Such substitution is ambiguous
since the permutation for nominal constants is not unique. However,
because formulas are considered to be logically equivalent modulo
permutation of nominal constants, the ambiguity turns out to be harmless.

Provability in \Gee is once again formalized via a sequent
calculus. Sequents in this logic have the form
\[
\Si: \G \rseq B
\]
where $\G$ is a multi-set of formulas called the context of the
sequent and the formulas in it are called the assumptions or
hypotheses of the sequent, $B$ is a formula called the conclusion of the
sequent, and $\Si$ is a signature containing the free variables
in $\G$ and $B$. The intuitive interpretation of the sequent is that
the conclusion $B$ is provable from the set of assumptions in $\G$.

The core rules defining provability in \Gee are shown in
Figure\cspc\ref{fig:gee_core_rules}.
The structure of these rules as well as the content of many of them
should be clear from a familiarity with sequent calculus style
formulations of intuitionistic logics.
We therefore limit ourselves here to elaborating only those aspects
that are peculiar to \Gee.
The \rinit and the \rcut rules in \Gee differ from the more familiar
versions in that they build in equality under the
permutation of nominal constants.
The $\nablaR$ rule formalizes the interpretation of the
$\nabla$ quantifier that was described informally in
Section\cspc\ref{subsec:nabla}; in this rule, $a$ is a ``new''
nominal constant that does not occur in $B$. The $\nablaL$ rule
encodes the reading of a $\nabla$-quantified formula as an
assumption that follows naturally from what it means to prove such a
formula.
In the usual reading of universal quantification, to prove that a
formula of the form $\typedrforall{\tau}{x}{B}$ holds, it suffices to
show that $B$ holds for all instantiations of $x$. The standard
formalization of this understanding is based on replacing $x$
with a new variable, called an {\it eigenvariable}, and then showing
that the resulting formula holds no matter what actual value is chosen
for that variable. In using this idea in the context of \Gee, we have
to be careful to respect the scope of $\nabla$ quantifiers:
in particular, we must consider instantiations for the eigenvariable
to include the nominal constants already appearing in $B$ but not the
ones that may be introduced for $\nabla$ quantifiers
appearing within $B$.
This requirement is encoded in $\rfallR$ by using a
technique called \emph{raising}\cspc\cite{miller92jsc}: an
eigenvariable $h$ is introduced, $x$ is replaced with the application
of $h$ to the nominal constants in the support of $B$, and we only consider
instantiations for eigenvariables that do not contain nominal
constants. Observe that instances of the quantified variable that use
the permitted nominal constants (but not the disallowed ones) can be
obtained by substituting a term for the eigenvariable that uses its
arguments in a suitable way.
The natural counterpart to this interpretation of universal
quantification is that to prove a sequent in which
$\typedrforall{\tau}{x}{B}$  appears as an assumption, it suffices to
show that the sequent is derivable when the quantified formula is
instantiated by a term that perhaps contains the nominal constants
appearing in $B$. The $\rfallL$ rule encapsulates this idea. In this
rule, the typing judgment $\Si, \normalcst, \nominalcst \tseq t :
\tau$ asserts that $t$ is a term of type $\tau$ that is constructed
using the variables, nominal constants and regular constants in $\Si$,
$\normalcst$ and $\nominalcst$; the nominal constants appearing in $t$
could be further restricted to being ones in the support of $B$ but,
as shown in\cspc\cite{gacek11ic}, the derivable sequents remain
unchanged even without this restriction. Similar explanations can be
given for the rules $\rexstL$ and $\rexstR$ which are, respectively,
the duals of $\rfallR$ and $\rfallL$.

\begin{figure}[ht!]
  \begin{gather*}
    \infer[\rinit]{
      \Si: \G, B \rseq B'
    }{
      B \feq B'
    }
    \quad
    \infer[\rcut]{
      \Si: \G \rseq C
    }{
      \Si: \G \rseq B
      &
      B \feq B'
      &
      \Si: \G, B' \rseq C
    }
  \end{gather*}
  \\
  \vspace{-1cm}
  \begin{gather*}
    \infer[\rcontra]{
      \Si: \G, C \rseq B
    }{
      \Si: \G, C, C \rseq B
    }
    \qquad
    \infer[\rfalseL]{
      \Si: \G, \rfalse \rseq B
    }{}
    \qquad
    \infer[\rtrueR]{
      \Si: \G \rseq \rtrue
    }{}
  \end{gather*}
  \\
  \vspace{-1cm}
  \begin{gather*}
    \infer[\randL, i\in\{1,2\}]{
      \Si: \G, B_1 \rand B_2 \rseq C
    }{
      \Si: \G, B_i \rseq C
    }
    \qquad
    \infer[\randR]{
      \Si: \G \rseq B \rand C
    }{
      \Si: \G \rseq B
      &
      \Si: \G \rseq C
    }
  \end{gather*}
  \\
  \vspace{-1cm}
  \begin{gather*}
    \infer[\rorL]{
      \Si: \G, B \ror C \rseq D
    }{
      \Si: \G, B \rseq D
      &
      \Si: \G, C \rseq D
    }
    \qquad
    \infer[\rorR, i\in\{1,2\}]{
      \Si: \G \rseq B_1 \ror B_2
    }{
      \Si: \G \rseq B_i
    }
  \end{gather*}
  \\
  \vspace{-1cm}
  \begin{gather*}
    \infer[\rimpL]{
      \Si: \G, B \rimp C \rseq D
    }{
      \Si: \G \rseq B
      &
      \Si: \G, C \rseq D
    }
    \qquad
    \infer[\rimpR]{
      \Si: \G \rseq B \rimp C
    }{
      \Si: \G, B \rseq C
    }
  \end{gather*}
  \\
  \vspace{-1cm}
  \begin{gather*}
    \infer[\rfallL]{
      \Si: \G, \typedrforall{\tau}{x}{B} \rseq C
    }{
      \Si, \normalcst, \nominalcst \tseq t : \tau
      &
      \Si: \G, B[t/x] \rseq C
    }
    \quad
    \infer[\rexstR]{
      \Si: \G \rseq \typedrexists{\tau}{x}{B}
    }{
      \Si, \normalcst, \nominalcst \tseq t : \tau
      &
      \Si: \G \rseq B[t/x]
    }
  \end{gather*}
  \\
  \vspace{-1cm}
  \begin{gather*}
    \begin{array}{c}
      \infer[\rexstL]{
        \Si: \G, \typedrexists{\tau}{x}{B} \rseq C
      }{
        \Si, h: \tau': \G ,B[(h \app a_1\app \ldots\app a_n)/x] \rseq C
      }
      \quad
      \infer[\rfallR]{
        \Si: \G \rseq \typedrforall{\tau}{x}{B}
      }{
        \Si, h:\tau': \G \rseq B[(h \app a_1\app \ldots\app a_n)/x]
      }
      \\
      \mbox{\small
        assuming that  $\support{B} = \{a_1,\ldots,a_n\}$, that, for $1 \leq i \leq n$,
        $a_i$ has type $\tau_i$, }
      \\
      \mbox{\small
        $h$ is variable of type $\tau_1 \to \ldots \to \tau_n \to \tau$
      and $h \not\in \dom{\Si}$ in \rfallR and \rexstL}
    \end{array}
  \end{gather*}
  \\
  \vspace{-1cm}
  \begin{gather*}
    \infer[\nablaL]{
      \Si: \G, \typednabla{\tau}{x}{B} \rseq C
    }{
      \Si: \G, B[a/x] \rseq C
    }
    \qquad
    \infer[\nablaR]{
      \Si: \G \rseq \typednabla{\tau}{x}{B}
    }{
      \Si: \G \rseq B[a/x]
    }
    \\
    \mbox{\small
      provided $a \not\in \support{B}$ in \nablaL and \nablaR}
  \end{gather*}
  \caption{The Core Rules of \Gee}
  \label{fig:gee_core_rules}
\end{figure}

\subsection{An informal understanding of fixed-point definitions}
\label{subsec:def}

The logic \Gee is actually better thought of as a \emph{family} of
logics, each parameterized by a \emph{definition} of the predicate
symbols in the vocabulary. Such a definition is given by a
possibly infinite collection of \emph{definitional
  clauses}. In the simplest form, each such clause has the
structure\footnote{The full form of a definitional clause permits the
  head of the clause to have $\nabla$ quantifiers over an atomic
  formula, as we shall explain later in this subsection.}
\[
\rfall \vec{x}. A \rdef B
\]
where $A$ is an atomic formula and $B$ is an arbitrary formula. We
call $A$ the head of such a clause and we call $B$ its body. Further, if $p$
is the predicate head of $A$ we say that the clause is
\emph{for} $p$. There are actually some provisos on the form of $A$
and $B$ for such a clause to be considered acceptable. First,
neither of them should contain nominal constants. Second, every
variable that has a free occurrence in $B$ must also occur in $A$ and
all of the free variables of $A$ should appear in $\vec{x}$. We shall
think of a definition as consisting of a sequence of blocks of
definitional clauses with the requirement that all the clauses for any
given predicate symbol be confined to one block. In this context, a
further requirement is that all the predicate symbols appearing in the
body of a definitional clause must have have their own definitional
clauses in the current or preceding blocks. Actually, occurrences of
predicates whose definitional clauses appear in the present block must
be further constrained to guarantee consistency of the logic. These
constraints can be stated in a few different ways and are also
somewhat complicated to describe. We therefore do not do this here but
refer the interested reader to\cspc\cite{baelde12lics}
or\cspc\cite{tiu12ijcar} for two alternatives. We note that
all the
definitions that we will use in the thesis will satisfy both forms of
restrictions.

The informal understanding of a definition is that it assigns a meaning
to each predicate symbol through the clauses it contains. This meaning
is obtained intuitively by collecting all the clauses for a predicate
and then thinking of any closed atomic formula that has that predicate
as its head being true exactly when it is an instance of the head of
one of the clauses and the corresponding instance of the body is
true. This intuition can be formalized by describing rules for
deriving a sequent in which an atomic formula appears as an assumption
or as a conclusion in a sequent. In the latter case, it suffices to
show that the body of an instance of any clause whose head is
identical to the atomic formula follows from the same assumptions. In
the former case, we consider all the possible ways in which the atomic
predicate could be the head of an instance of a
definitional clause and we show that the corresponding instance of the
sequent with the atomic formula replaced by the body of the clause
instance has a proof. Note that this treatment of an atomic formula
that appears on the left of a sequent corresponds to a case analysis
style of reasoning: we consider all the possible ways in which the
atomic formula could be true based on the definition and we show that
the sequent must be derivable in each case.

A simple example of a definition is one that encodes the equality
relation between terms. For any given type $\tau$ not containing
\prop, we identify a predicate $\keq_\tau : \tau \to \tau \to \prop$
and add to the definition the following sole clause for
it:\footnote{As mentioned earlier, definitional clauses must be
  provided in blocks. That they are presented in this way will be
  implicit in most of this thesis with one exception: when we discuss
  the addition of a form of polymorphism to Abella in
  Chapter~\ref{ch:extensions}, we will need to make explicit use of
  the idea of a block of definitional clauses.}
\[
\qquad \keq_\tau\app M\app M \rdef \rtrue
\]
In showing this and other clauses, we use the convention of making the
outermost universal quantifiers in the clause implicit by
choosing tokens that begin with uppercase letters for the occurrences
of the variables they bind. Thus, in a fully explicit form, this
clause would be written as
\[ \rfall M. \keq_\tau\app M\app M \rdef \rtrue. \]
By this definition, $\keq_\tau\app M_1\app M_2$ is provable if $M_1$
and $M_2$ are equal modulo $\lambda$-conversion. Conversely, if
$\keq_\tau\app M_1\app M_2$ occurs as an assumption, then it must be
the case that $M_1$ and $M_2$ are equal terms modulo
$\lambda$-conversion. In the following discussion, we shall write
$\keq_\tau\app M_1\app M_2$ as $M_1 =_\tau M_2$, as usual dropping the
type annotation if it does not add to the discussion.

Fixed-point definitions provide a natural way to encode rule-based
relational specifications in \Gee: predicate symbols are used to name
relations and each rule translates into a definitional clause such
that its conclusion becomes the head of the clause and its premises
(if any) become the body of the clause.  We use the append relation on
lists of natural numbers again to illustrate this idea. As in the case
of the encoding in \HHw, we
use the atomic type $\klist$ for representations of lists, the
constants $\knil$ and $\cons$ to construct such representations and
the predicate symbol
\[\kappend : \klist \to \klist \to \klist \to \prop\]
to encode the append relation. The rules defining the append
relation then translate into the following clauses:
\begin{tabbing}
\qquad\=$\kappend \app (X \cons L_1) \app L_2 \app (X \cons L_3)$
  \quad \=$\rdef$ \quad\=\kill
\>$\kappend \app \knil \app L \app L$ \>$\rdef$ \>$\rtrue$\\
\>$\kappend \app (X \cons L_1) \app L_2 \app (X \cons L_3)$
  \>$\rdef$ \>$\kappend \app L_1 \app L_2 \app L_3$
\end{tabbing}

To understand the way definitions are meant to be treated in \Gee, let us
consider using the definition of $\kappend$ in derivations.  First,
suppose that we want to show that the following is a theorem, \ie,
that it is provable in an empty context:
\begin{gather*}
  \rforallx L {\kappend \app \knil \app L \app L}.
\end{gather*}
To do this, we would have to show that the following holds, regardless
of what actual term of type $\klist$ we put in for $l$:
\begin{gather*}
  \kappend \app \knil \app l \app l
\end{gather*}
This atomic formula is an instance of the head of the first clause for
\kappend and so the task reduces to proving $\rtrue$, something that is
immediate in the logic.

The reasoning example above shows similarities between clauses in \HHw
and definitional clauses in \Gee when the latter are used to prove
atomic formulas; the transformation is effectively what we would
obtain through backchaining in \HHw. The difference between the two
logics is brought out by considering the formula
\begin{gather*}
\rforallx L {\kappend \app (1 \cons 2 \cons \knil) \app L \app (1 \cons 3 \cons L) \rimp \rfalse.}
\end{gather*}
%
%GN I hope you have been more careful than this with your other
%``fixes.'' What is L here? Also is it only saying that it is not
%provable or something stronger: MUST NOT hold? The latter
%seems to be the right interpretation of negation, the former is too
%encompassing in that many things are not provable without leading to
%contradictions. Also, what is the referrent of ``it'' here, it is not
%obvious in English.
%% It states that the assumption $\kappend \app (1 \cons 2
%% \cons \knil) \app L \app (1 \cons 3 \cons L)$ is false, \ie not provable, in \Gee.
%
This formula states that $\kappend \app (1 \cons 2
\cons \knil) \app L \app (1 \cons 3 \cons L)$ is false in the sense
that it {\it must not} hold for any value of $L$.
Such formulas that show the falsity of particular assumptions cannot be
proved by using the ``positive'' interpretation of fixed-point
definitions, \ie, by considering how the assumption can be derived
from the definitional clauses. However, they are provable in \Gee
because this logic also encodes the closed-world nature of fixed-point
definitions.
More specifically, the attempt to prove the particular formula at hand
will reduce in \Gee to showing that
$\rfalse$ holds whenever we have $\kappend \app (1 \cons 2 \cons
\knil) \app l \app (1 \cons 3 \cons l)$ for any list $l$.
%
%GN: What does it mean to be a prefix of a list? Also why is it
%necessary for the first list to be a prefix of $l$? Also, what are
%the first and third arguments of the assumption? This is different
%from talking about the the first and third arguments of a formula.
%% This must the case for the following reason: the assumption cannot be
%% true for \emph{any} value of $l$ because its first argument cannot be
%% a prefix of its third argument.  This form of reasoning is realized
This must the case for the following reason: the assumption formula
cannot be true for \emph{any} value of $l$ because the elements of the
list that is its first argument cannot be the initial elements of a
list that is its third argument. This form of reasoning is realized
in \Gee by considering the different ways the clauses for \kappend
might apply to
the assumption; we are, of course, permitted to specialize $l$ in
different ways so as to consider different instances of the assumption
in the process.  In this particular situation, only the second clause
for \kappend is applicable and so the ``case analysis'' yields a
single case where the assumption has been transformed to $\kappend
\app (2 \cons \knil) \app l \app (3 \cons l)$. We now try a further
case analysis and easily realize that no clause applies, i.e. the
assumption must not in fact be true. The desired theorem therefore
follows.

The above style of reasoning works well in proving theorems when the
``unfolding'' of an assumption via definitions terminates in a finite
number of steps. However, this property
does not hold in many reasoning contexts that are of interest. For
example, consider the formula
\begin{gather*}
  \rforallx {L_1 , L_2 , L_3 , L_3'}
    {\kappend \app L_1 \app L_2 \app L_3 \rimp \kappend \app L_1 \app L_2 \app L_3'
      \rimp L_3 = L_3'}
\end{gather*}
that states that \kappend is deterministic in its third argument. Case
analysis of either the first or the second assumption in this case
exhibits a looping structure, reflecting the fact that the lists that
we have to consider may be of \emph{a priori} undetermined
lengths. To prove this formula,
we must actually know that the \kappend predicate holds only by virtue
of a \emph{finite} number of unfoldings using the clauses and we must
have a means for using this knowledge in an argument. It is possible
to do this in \Gee by giving a \emph{least-fixed point} or inductive
interpretation to the definition of \kappend.
The logic \Gee allows definitions to be treated in this
way.
More specifically, we can mark the definition of \kappend as inductive
in \Gee, and this will give us the ability to apply an induction
principle to the first or second assumption in the determinacy formula
above.
We will give a formal account of how this is done in
Section\cspc\ref{subsec:def_rules} and will introduce an effective way to
construct inductive proofs in
Section\cspc\ref{subsec:annot_ind}.\footnote{There is also the dual
  possibility of giving the definition of particular predicates a
  \emph{greatest-fixed point} or co-inductive interpretation. We do
  not use co-induction in this thesis and hence do not discuss it
  further here.}

Like \HHw, the logic \Gee provides a set of mechanisms for
encoding relational specifications over syntactic objects that embody
binding constructs. Meta-level abstraction can be used as before to
represent object-level binding operators and $\alpha$- and
$\beta$-conversions capture the binding related notions such as
renaming and substitution. Further, the $\nabla$ quantifier enables
recursion over binding structure by providing a means for moving
binding in term structure to formula level and, eventually to proof
level binding. We use the formalization of
the typing rules for the \STLC in \Gee as an example to illustrate these
ideas. We use a representation for the terms in this calculus that is
identical to the one described in
Section\cspc\ref{subsec:ecd_bindings}. Since implication has a
different reading in \Gee from that in \HHw, we cannot use it to treat
typing contexts implicitly in the encoding of typing judgments. We
therefore represent these judgments via the three-place
relation
\[
\kof : \kclist \to \ktm \to \kty \to \prop
\]
in which the first argument is intended to be an encoding of the
typing context. We
use the type $\kclelem$ for classifying the type assignments for
variables in typing contexts and the constant $\kvty : \ktm \to \kty
\to \kclelem$ for encoding such assignments. We then identify the
following constants to represent the empty list and the cons operator
for $\kclist$, respectively.
\begin{gather*}
\kclnil : \kclist
\qquad
\kclcons : \kclelem \to \kclist \to \kclist
\end{gather*}
To encode type checking of a variable in a typing context, we identify
a predicate constant $\kvof : \kclist \to \ktm \to \kty \to
\prop$. The clauses defining this checking is given as follows:
\begin{tabbing}
\qquad\=$\kvof\app (\kclcons\app (\kvty\app X\app T)\app L)\app X\app T$
  \quad \=$\rdef$\quad \=\kill
\>$\kvof\app (\kclcons\app (\kvty\app X\app T)\app L)\app X\app T$
   \>$\rdef$ \>$\rtrue$\\
\>$\kvof\app (\kclcons\app E\app L)\app X\app T$ \>$\rdef$ \>$\kvof\app L\app X\app T$
\end{tabbing}
Then the typing rules in the \STLC can be encoded as the following clauses
for $\kof$:
\begin{tabbing}
\qquad\=$\kof\app L\app (\kabs\app R)\app (\karr\app T_1\app T_2)$
  \quad \=$\rdef$ \quad \=\kill
  \>$\kof\app L\app X\app T$ \>$\rdef$ \>$\kvof\app L\app X\app T$\\
  \>$\kof\app L\app (\kapp\app M_1\app M_2)\app T_1$ \>$\rdef$
    \>$\rexistsx {T_2} {\kof\app L\app M_1\app (\karr\app T_2\app T_1)
    \rand \kof\app L\app M_2\app T_2}$\\
  \>$\kof\app L\app (\kabs\app R)\app (\karr\app T_1\app T_2)$ \>$\rdef$
    \>$\nablax {x} {\kof\app (\kclcons\app (\kvty\app x\app T_1)\app L)\app
                       (R\app x)\app T_2}$
\end{tabbing}
It is easy to see that the first and second clauses capture the
typing rules for variables and applications. To see that the last rule
captures the typing rule for abstractions, assume that we would like
to prove the atomic formula $(\kof\app l\app (\kabs\app r)\app
(\karr\app t_1\app t_2))$ given particular encoded expressions $l$,
$r$, $t_1$ and $t_2$. Then, by the definition of $\kof$, we must prove
the following formula:
\[
\nablax {x} {\kof\app (\kclcons\app (\kvty\app x\app t_1)\app l)\app
                       (r\app x)\app t_2}
\]
The only way to do this is to introduce a new nominal constant $a$ for
$x$ and prove the following formula:
\[
{\kof\app (\kclcons\app (\kvty\app a\app t_1)\app l)\app
                       (r\app a)\app t_2}
\]
Like in \HHw, the $\beta$-redex $(r\app a)$ represents the result of
substituting $a$ (which represents the bound variable of the
abstraction) for the occurrence of $x$ in the body of the abstraction.
The only way to prove $(\kof\app a\app t)$ is to match $t$ with $t_1$
by using the first clause since the nominal constant $a$ is different
from both $\kapp$ and $\kabs$ and it is therefore impossible to match
$(\kof\app a\app t)$ with the second or the third clause. As a result,
deriving the formula above is equivalent to showing that the body of
the abstraction has type $t_2$ in the extended typing context, which
matches exactly the behavior of the typing rule for abstractions.

%% An interesting point to note about the manner in which the
%% specification above is constructed is the following: in a judgment
%% $(\kof\app L\app M\app T)$ that we want to prove, $M$ represents a
%% possibly open term whose free variables are represented by nominal
%% constants and $L$ identifies the types associated with these
%% constants. This kind of treatment, where nominal constants play a role
%% in representing open terms, is typical of formalizations in \Gee that
%% are based on the $\lambda$-tree syntax approach.

As we have seen in the \kappend example, we can prove properties of
the relations through their encoding as fixed-point definitions in
\Gee. The situation becomes more complicated when dealing with
relational specifications with binding structure. Proving properties
of such specifications often requires the ability to characterize the
binding structure in them. To see this, consider proving the property
that the typing rules for the \STLC assign unique types to
terms. Using the encoding of typing rules as described above, we may
express this property through the formula
\[
\rforallx {L, T_1, T_2, M}
  {\kof\app L\app M\app T_1 \rimp \kof\app L\app M\app T_2 \rimp
  T_1 = T_2}.
\]
To prove this formula, though, we need some restrictions on the typing
context $L$: it should assign types only to nominal constants and each
of these assignments should be unique.  These properties are satisfied
by any typing context that arises in deriving a typing judgment with
an initially empty typing context. However, to prove the unique type
assignment theorem it does not suffice that these properties are
true. We also need to make their truth explicit so that we can exploit
that knowledge in the argument.

The properties described above are ones about the structures of terms
and, more specifically, about the occurrences of nominal constants in
them. Definitions in \Gee include a mechanism for making such aspects
explicit. Specifically, \Gee allows the head of a definitional clause
to include $\nabla$ quantifiers over atomic formulas. Thus, the full form of
definitional clauses is in fact
\[
\rforallx {\vec{x}} {(\nablax {\vec{z}} A) \rdef B}
\]
%where the scope of $\vec{z}$ is $A$ and both ${(\nablax {\vec{z}} A)}$
%and $B$ should not contain any nominal constant.
In generating
instances of such clauses, the $\nabla$ quantifiers at the head must
be instantiated by distinct nominal constants. A further point to note
is the scope of the universal quantifiers: since the $\nabla$
quantifiers appear within their scope, the nominal constants that
instantiate them cannot appear in the instantiations of the universal
quantifiers.

As an example of the use of this extended form of definitional
clauses, consider the following clause defining the
predicate constant $\kname: \ktm \to \prop$:
\begin{tabbing}
\qquad\=\kill
\>$\nablax {x:\ktm} {\kname\app x} \rdef \rtrue$.
\end{tabbing}
An atomic formula $\kname\app M$ is derivable if and only if it
matches with this clause. For this to be possible, $M$ must be a
nominal constant of type $\ktm$. As another example, consider the
following clauses defining the predicate constant $\kfresh: \ktm \to
\ktm \to \prop$:
\begin{tabbing}
\qquad\=\kill
\>$\nablax {x:\ktm} {\kfresh\app x\app M} \rdef \rtrue$.
\end{tabbing}
The formula $\kfresh\app X\app M$ holds true if and only if $X$ is a
nominal constant and $M$ is a term that does not contain this nominal
constant. Thus, this second clause encodes the property of a nominal
constant not occurring in, or being fresh to, a term.

Using clauses of this general form, we can characterize
legitimate typing contexts using the following clauses for the
predicate $\kctx : \kclist \to \prop$:
\begin{tabbing}
  \qquad\=$\nablax {x} {\kctx \app (\kclcons\app (\kvty\app x\app T)\app L)}$
    \quad \=$\rdef$ \quad\=\kill
  \>$\kctx \app \kclnil$ \>$\rdef$ \>$\rtrue$\\
  \>$\nablax x {\kctx \app (\kclcons\app (\kvty\app x\app T)\app L)}$
    \>$\rdef$ \>$\kctx \app  L$
\end{tabbing}
The first clause asserts that an empty list encodes a valid typing
context. In any instance of the second clause, $x$ must be a nominal
constant and the list $L$ cannot contain the nominal constant for $x$
since it is bound outside of $x$. Furthermore, $L$ itself must encode
a valid typing context. These clauses thus define a relation $\kctx$
such that $\kctx\app L$ holds exactly when $L$ assigns unique types to
a collection of distinct nominal constants. Based on this definition,
the uniqueness of type assignment can be restated in the formula
\[
\rforallx {L, T_1, T_2, M}
  \kctx\app L \rimp \kof\app L\app M\app T_1
  \rimp \kof\app L\app M\app T_2 \rimp
  T_1 = T_2.
\]
that is in fact provable by induction on the second or third
assumption in \Gee. The detailed proof can be found in
\cite{baelde14jfr}.

\subsection{Formalizing fixed-point definitions}
\label{subsec:def_rules}

The informal exposure that we have provided to fixed-point definitions
should suffice for most of the discussions in this thesis. However, we
will need a more formal understanding of how these definitions are
realized within \Gee when we introduce a schematic polymorphism
capability in reasoning into Abella in Chapter~\ref{ch:extensions}. To
facilitate that discussion, we now present the proof rules that formalize
the treatment of definitions in \Gee.

As might be expected, the treatment of definitions is characterized by
rules for introducing atoms on the left and the right sides of
sequents. The \emph{definition left} or $\defL$ rule captures the case
analysis style reasoning based on the clauses defining the atom. The
definition right rule or \defR captures backchaining on definitional
clauses. To describe these rules formally, we need the notion of an
instance of a definitional clause. Given a definitional
clause $\rforallx{\vec{x}} {(\nablax {\vec{z}} A) \rdef B}$ and a
substitution $\theta$ that assigns distinct nominal constants to
$\vec{z}$ and terms not containing such constants to $\vec{x}$, we say
that $A[\theta] \rdef B[\theta]$ is an instance of the original
clause. Then the rules for definitions are shown in
Figure\cspc\ref{fig:def_rules}. In \defL, $\Si\theta'$ stands for the
signature obtained from $\Si$ by removing variables in the domain of
$\theta'$ and adding free variables in the values of $\theta'$; $\G[\theta']$
stands for the context $\{B[\theta'] \sep B \in \G\}$.
%It is easy to see
%these rules conform to the our more informal presentation of
%definitions described in Section\cspc\ref{subsec:def}.

\begin{figure}[ht!]
  \begin{gather*}
    \infer[\defR]{
      \Si: \G \rseq p\app \vec{t}
    }{
      \Si: \G \rseq B
    }
    \\
    \mbox{\small where $p\app \vec{t} \rdef B$ is an instance of a definitional clause for $p$}
  \end{gather*}
  \vspace{-1cm}
  \begin{gather*}
    \infer[\defL]{
      \Si: \G, p\app \vec{t} \rseq C
    }{
      \{\Si\theta' : \G[\theta'], B \rseq C[\theta'] \sep
        \mbox {$(p\app \vec{t})[\theta'] \feq A$ and $A \rdef B \in \mathcal{D}_p$}\}
    }
    \\
    \mbox{\small where
        $\mathcal{D}_p = \{A \rdef B \sep A \rdef B \mbox{ is an instance of a definitional clause for } p\}$}
  \end{gather*}
  \caption{Rules for Definitions}
  \label{fig:def_rules}
\end{figure}

Note that the set of premises in any given instance of the \defL rule
could be infinite; this would be the case if there are
an infinite number of instances of definitional clauses that match
with the atomic formula $p\app \vec{t}$ in the manner indicated. In
determining provability in practice, it would be useful to be able to
cover this kind of infinite branching possibility in a finitary way.
Towards this end, we can provide an alternative formulation of the
\defL rule that makes use of the idea of a \emph{complete set of
  unifiers} or \emph{CSU} between the heads of clauses and the atomic formula $p\app
\vec{t}$.

To present this alternative formulation, we first make precise what is meant
by a complete set of unifiers.
\begin{mydef}\label{def:csu}
  Given two terms or formulas $A$ and $B$, a complete set of unifiers
  for $A$ and $B$, denoted by $\csu{A}{B}$, is a set of substitutions
  such that
  \begin{itemize}
  \item
    for any $\theta \in \csu{A}{B}$, $A[\theta] = B[\theta]$;
  \item
    for any substitution $\rho$ such that $A[\rho] = B[\rho]$ there is
    a substitution $\theta \in \csu{A}{B}$ and a substitution
    $\gamma$ such that $\rho = \theta \circ \gamma$.
  \end{itemize}
\end{mydef}
\noindent Intuitively, a CSU for $A$ and $B$ consists of a set of substitutions
that unify $A$ and $B$ and that ``covers'' all the unifiers of $A$ and
$B$ in the sense that any such unifier can be obtained by further
instantiating a substitution in the CSU. Note that there need not
exist a unique set of substitutions satisfying this requirement and,
in this sense, the notation $\csu{A}{B}$ is ambiguous. However, this
ambiguity will be harmless in the discussions below in the following
sense: we will use the notation to select some complete set of
unifiers for $A$ and $B$ and there will be no sensitivity to the choice
that is actually made.

Looking at the \defL rule in Figure~\ref{fig:def_rules}, we see that
when matching an assumption $(p\app \vec{t})$ of the sequent with the
clause $\rforallx {\vec{x}} {(\nablax {\vec{z}} A) \rdef
  B}$, we must allow the variables in ${\vec{x}}$ to be instantiated
with nominal constants in $(p\app \vec{t})$.
Similarly, we must allow the variables in the sequent to be
instantiated with nominal constants that we choose for $\vec{z}$. We
use raising to build in both possibilities. First, given a clause
$\rforallx
{x_1,\ldots,x_n} {(\nablax {\vec{z}} A) \rdef B}$, we define a version of
it raised over the sequence of nominal constants $\vec{a}$ away from a
signature $\Si$ to be a clause of the form
\[
\rforallx {h_1,\ldots,h_n}
  {\nablax {\vec{z}}
           {A[(h_1\app \vec{a})/x_1,\ldots,(h_n\app \vec{a})/x_n]}
  \rdef B[(h_1\app \vec{a})/x_1,\ldots,(h_n\app \vec{a})/x_n],}
\]
where $h_1,\ldots,h_n$ are variables
that do not appear in $\Si$. Next, we define a version of the sequent $\Si : \G \rseq B$,
where $\Si = \{y_1:\tau_1,\ldots,y_m:\tau_m\}$, raised over the sequence of nominal
constants $\vec{c}$ to be a sequent of the form
\[
  \Si' : \G[(y_1'\app \vec{c})/y_1,\ldots,(y_m'\app \vec{c})/y_m] \rseq
  B[(y_1'\app \vec{c})/y_1,\ldots,(y_m'\app \vec{c})/y_m]
\]
where, for $1 \leq i \leq m$, $y_i'$ is an variable of suitable type
and $\Si' = \{y_1',\ldots,y_m'\}$. Finally, we combine these notions
together with the idea of complete sets of unifiers to identify a set
of premises arising from a definitional definitional clause that is
useful in formulating an alternative version of the \defL rule.
\begin{mydef}\label{def:defl_csu_premises}
  Let $H$ be the sequent $\Si: \G, p\app \vec{t} \rseq F$ and let $C$
  be the definitional clause $\rforallx {\vec{x}} {(\nablax {\vec{z}}
    A)\rdef B}$. Further, let $\support{p\app \vec{t}}$ be $\{\vec{a}\}$
  and let $\vec{c}$ be a sequence of nominal constants that is of the
  same length as $\vec{z}$ and such that each constant in the sequence
  has a type identical to that of the corresponding variable in
  $\vec{z}$ and is also distinct from the constants in
  $\vec{a}$. Finally, let $\rforallx {\vec{h}} {(\nablax {\vec{z}}
    A')\rdef B'}$ be a version of the clause $C$ raised over $\vec{a}$ away from $\Si$
  and let $\Si': \G', p\app \vec{t'} \rseq F'$ be a version of $H$
  raised over $\vec{c}$. Then
\begin{tabbing}
\quad\=$\deflcsuprem{H}{p\app \vec{t}}{C}$\ \=$=$\ \=\qquad\=\kill
\>$\deflcsuprem{H}{p\app \vec{t}}{C}$\>$=$\>$\{\Si'[\theta]: \G'[\theta],
    \pi.B'[\theta] \rseq F'[\theta] \sep$\\
\>\>\>\>$\pi$ is a permutation of the nominal constants in\\
\>\>\>\>$\{\vec{c},\vec{a}\}$ and $\theta \in \csu{p\app \vec{t'}}{\pi.A'[\vec{c}/\vec{z}]} \}$.
\end{tabbing}
\end{mydef}
\noindent Intuitively, $\deflcsuprem{H}{p\app
  \vec{t}}{C}$ corresponds to the premise sequents we would get from
unfolding the assumption formula $(p \app \vec{t})$ in the sequent $H$
based on the clause $C$ and using only the substitutions in the complete
set of unifiers for the head of $C$ and $(p\app \vec{t})$. In the process, we have
to instantiate the $\nabla$ quantifiers in the head of $C$ by nominal
constants. Further, we have to take care to allow all possible
substitutions for the universally quantified variables in $C$ and
the variables in $\Si$, an aspect that is treated by raising.

The definition of $\kdeflcsuprem$ that we have presented is actually
ambiguous: the set of sequents that it identifies is dependent on the
the variables we select when raising the sequent and the clause, the
names we use for the nominal constants that instantiate the $\nabla$
quantifiers in the head of the clause and the particular complete set
of unifiers we choose. The ambiguity arising from the last aspect will
be harmless for the reasons already noted. The ambiguity arising from
how we make the other choices is also
inconsequential. It is easy to see that differences in these choices
give rise to sets of sequents that are identical under a ``renaming''
of the variables and nominal constants. The inconsequentiality of
how these choices are made then follows from an easily proved property
of \Gee that two sequents that differ only in the names of the
variables and the nominal constants appearing in them are
equi-derivable in a strong sense: their proofs have an identical
structure and can in fact be obtained one from the other by using the
same renaming.

  %% Given a sequent $H = (\Si: \G, p\app \vec{t} \rseq C)$ where
  %% $\support{p\app \vec{t}} = \{\vec{a}\}$, the relation
  %% $\deflcsuprem{p\app \vec{t}}{H}{H'}$ holds if $H' = ({\Si'[\theta]: \G'[\theta],
  %%   \pi.B[\theta] \rseq C'[\theta]})$ and
  %% \begin{itemize}
  %% \item $\rforallx {\vec{x}} {(\nablax {\vec{z}} A)\rdef B}$ is a
  %%   version of a clause for $p$ raised over $\vec{a}$ away from $\Si$;
  %% \item $\Si': \G', p\app \vec{t'} \rseq C'$ is a version of the
  %%   original sequent raised over a sequence of distinct nominal
  %%   constants $\vec{c}$ with the same length as $\vec{z}$;
  %% \item $\theta$ is a substitution not containing nominal constants
  %%   and $\theta \in \csu{p\app \vec{t'}}{\pi.A[\vec{c}/\vec{z}]}$
  %%   where $\pi$ is a permutation of the nominal constants
  %%   $\{\vec{c},\vec{a}\}$.
  %% \end{itemize}
%\end{mydef}
%
%% \noindent Intuitively, given a sequent $H$ that contains the atomic assumption
%% $(p\app \vec{t})$
%% \[
%% \deflcsuprem{p\app \vec{t}}{H}{H'}
%% \]
%% holds if there exists a clause for $p$ such that, after raising
%% $(p\app \vec{t})$ and this clause over relevant nominal constants, the
%% clause head matches $(p\app \vec{t})$ under a substitution in the CSU
%% for them modulo the permutation of nominal constants, and the sequent
%% $H'$ is a premise of \defL generated from this matching.

We can finally present an alternative to the \defL rule that uses the
idea of complete sets of unifiers. This rule is shown in
Figure\cspc\ref{fig:defl_csu}. Using an approach similar to that in
\cspc\cite{mcdowell00tcs}, this rule can be shown to be inter-admissible
with the \defL rule in the context of the other rules defining \Gee. As a
result, we can replace \defL with \defLCSU without affecting the
provability of sequents in \Gee. If we limit ourselves to using
\defLCSU only when the CSU is finite, then the proofs constructed by
using the rule have a finite branching character. It turns out that we
can construct proofs in many interesting situations under this
limitation. In fact, all the examples that we consider in this thesis
will use the \defLCSU with finite CSUs.
\begin{figure}[ht!]
  \begin{gather*}
    \infer[\defLCSU]{
      \Si: \G, p\app \vec{t} \rseq F
    }{
      \{H \in \deflcsuprem{\Si: \G, p\app \vec{t} \rseq F}{p\app
        \vec{t}}{C} \sep C \in {\cal D}\}
    }\\
    \mbox{\rm where}\ {\cal D}\ \mbox{\rm is the definition parameterizing \Gee}
  \end{gather*}
  \caption{The Definition Left Rule using CSU}
  \label{fig:defl_csu}
\end{figure}

The last rule for definitions we consider is induction. This rule is
shown in Figure\cspc\ref{fig:induction}; this rule is applicable only
to those predicate that are specifically marked as being inductively
defined, as described in the previous subsection.
%GN I am not sure I understand what you say below. Specifically, what
%do you mean by an ``inductive invariant of p?'' I am changing just
%this part, but please check that you are comfortable with the rest of
%what you have written.
%% Figure\cspc\ref{fig:induction} to such a definition. The intuition of
%% this rule is given as follows. Given an inductively defined predicate
%% $p$, if $p$ holds of the arguments $\vec{t}$ then any inductive
%% invariant of $p$ must also hold of $\vec{t}$. In the rule \indL, the
%% provability of the set of premises on the left ensures that $S$ is an
%% inductive invariant of $p$. The invariant $S$ is thus made available
%% as an assumption in the right premise for deriving the goal formula.
The intuition underlying this rule is the following. If some property
$S$ satisfies the clauses that define $p$ then $S\app t$ must hold
whenever $p\app t$ holds; this follows by virtue of $p$ being the
least fixed-point of its defining clauses. But then if some formula
$C$ follows from assuming $S\app t$, it must also follow any time
$p\app t$ holds.

Since the property $S$ can be very verbose, it is not very
convenient from a user's perspective to use \indL directly in proof
construction. We will introduce a more natural approach to do
induction in Section\cspc\ref{subsec:annot_ind}. This approach is what
we use for constructing proofs via induction in this thesis.

\begin{figure}[ht!]
  \begin{gather*}
    \infer[\indL]{
      \Si: \G, p\app \vec{t} \rseq C
    }{
      \{
      \vec{x} : B[S/p] \rseq \nablax{\vec{z}}{S\app \vec{t}_i}
       \sep
        \rforallx{\vec{x}}{\nablax{\vec{z}}{p\app \vec{t}} \rdef
          B} \in {\cal D}\}
      &
      \Si: \G, S\app \vec{t} \rseq C
    }
    \\
    \mbox{ provided $p$ is inductively defined by the set of clauses
      $\cal D$}
    \\
    \mbox{ and $S$ is a term with no nominal constants and of the same type as $p$}
  \end{gather*}
  \caption{The Induction Rule}
  \label{fig:induction}
\end{figure}

\section{The Abella Theorem Proving System}
\label{sec:abella}

%% In this section, we provide an introduction to the theorem prover
%% \Abella that implements \Gee. The purpose of this discussion is to
%% expose the ideas and terminologies for presenting proof developments in
%% the following chapters.

\Abella is a tactics-based interactive theorem prover like
Coq\cspc\cite{coq02manual} or Isabelle\cspc\cite{paulson94book} that
helps in the construction of proofs in \Gee. A user types in
commands known as \emph{tactics} to incrementally build up proofs for
the theorems she/he want to prove. Tactics are designed to
correspond to reasoning steps that are more natural to mathematical
arguments but that, at the same time, can translate into a combination
of proof rules of \Gee. Thus, with the help of tactics,
developing a proof in \Abella can be made to have a flavor similar to
developing proofs on paper.

We would like to use \Abella to prove properties of specifications
written in \LProlog. This is realized via the \emph{two-level logic
  approach} to reasoning\cspc\cite{gacek12jar,mcdowell02tocl}. In this
approach, the logic of \LProlog is itself encoded as a fixed-point
definition in \Gee and the \LProlog specifications are then reasoned
about through this encoding.  With the two-level logic approach, we
can use \LProlog programs both as implementations and as inputs to
\Abella in which their properties are stated and proved.

We will introduce these features of \Abella in the rest of this
section. We successively describe the interactive approach to proof
construction in \Abella, an effective way to construct induction
proofs by using tactics and the two-level logic approach to reasoning
about \LProlog specifications.

\subsection{Interactive proof construction}
\label{subsec:interactive}

In \Abella, a user initiates a proof construction process by stating a
formula, called a theorem or a lemma, he/she wants to prove. At any
moment, the proof state is represented by a set of \emph{subgoals} all
of which must be proved to prove the original theorem. A subgoal
essentially represents a sequent whose proof is still to be found. It
consists of a multi-set of formulas called its \emph{hypotheses} or
\emph{assumptions}, which constitute the context of the corresponding
sequent, and a formula called its \emph{conclusion},
which corresponds to the conclusion of the sequent. Initially, there is only
one subgoal with no hypotheses and whose conclusion is the theorem to be
proved.

At any moment, a user applies some tactic to one of the subgoals to
make progress to the proof state.\footnote{In practice, the subgoals
  are listed in sequence. The first subgoal in the sequence must be
  proved before a user can move on to the next one. This continues
  until there is no subgoal left.} A tactic corresponds to a scheme
for applying a collection of rules in \Gee. The changes to the proof
state depend on the applied tactic. The following are some examples:
the subgoal may be proved by the tactic and disappear, new hypotheses
may be added to the subgoal, the subgoal may be replaced by several
other subgoals as a result of case analysis, or a warning that the
tactic is not applicable may be issued and the subgoal remains
unchanged. These changes reflect the effects of applying the proof
rules represented by the tactic on the sequent corresponding to the
subgoal. The proof construction process ends when there are no
subgoals left. At that point, we have essentially constructed a proof
for the original theorem in \Gee.

As an example, consider the following theorem whose proof we
have informally described in Section\cspc\ref{subsec:def}
\begin{gather*}
\rforallx L {\kappend \app (1 \cons 2 \cons \knil) \app L \app (1 \cons 3 \cons L) \rimp \rfalse}.
\end{gather*}
This theorem is proved by using tactics as follows. Initially, we have
a single subgoal with the formula above as its conclusion and with no
hypothesis. We first apply an ``introduction'' tactic to this subgoal
which introduces the variable $L$ and the following hypothesis
\[
\kappend \app (1 \cons 2 \cons \knil) \app L \app (1 \cons 3 \cons L)
\]
and makes $\rfalse$ as the new conclusion; application of this tactic
mirrors the application of the \rfallR and \rimpR rules to the corresponding
sequent. We then apply the ``case analysis'' tactic to this
hypothesis, which corresponds to applying the $\defL$ rule to the corresponding
sequent. Since the hypothesis can only be derived from the second
clause for \kappend, case analysis replaces the subgoal with a new one
whose hypothesis is reduced to
\[
\kappend \app (2 \cons \knil) \app L \app (3 \cons L)
\]
and whose conclusion is $\rfalse$. Applying another case analysis
tactic to this reduced hypothesis results in no more subgoals because
there are no clauses whose heads match the hypothesis. Since there are
no subgoals to be solved, the proof is concluded.

A common approach to constructing a large proof is to break it down
into smaller lemmas and build towards a final result. \Abella provides
this capability by allowing for proving theorems separately and using
them in proof construction as lemmas.
An established theorem can be freely used as a hypothesis at any
point of proof construction. Such usage of a theorem corresponds to
applying the \rcut rule to introduce the theorem as a new
hypothesis.

Theorems in \Abella often have the form
\[
\rforallx{x_1, \ldots, x_n} {
  H_1 \rimp \ldots \rimp H_m \rimp B
}
\]
Suppose we have such a theorem, we can then use the ``apply'' tactic
to match $H_i$ with some hypotheses to generate a new hypothesis $B$
under proper instantiation of $x_i$. This corresponds to application
of a collection of \rfallR, \rimpL and \rinit
rules\cspc\cite{gacek09phd}.  For example, consider the following
theorem:
\begin{gather*}
{\kappend \app (1 \cons 2 \cons \knil) \app (4 \cons \knil) \app (1 \cons 3 \cons 4 \cons \knil) \rimp \rfalse}.
\end{gather*}
It is proved as follows. We first introduce the following hypothesis:
\[
\kappend \app (1 \cons 2 \cons \knil) \app (4 \cons \knil) \app (1 \cons 3 \cons 4 \cons \knil)
\]
and try to prove the conclusion $\rfalse$. We then apply the following
theorem
\begin{gather*}
\rforallx L {\kappend \app (1 \cons 2 \cons \knil) \app L \app (1 \cons 3 \cons L) \rimp \rfalse}.
\end{gather*}
which has been proved above, with $L$ substituted by $(4 \cons nil)$ to
match the above hypothesis and generate $\rfalse$ as a new
hypothesis. The proof is concluded by matching the conclusion $\rfalse$ with
the generated hypothesis, which corresponds to applying the \rinit
rule.

Later we will see theorems of the following form in which the
top-level universal quantifiers embed some $\nabla$ quantifiers which
further embed an implication formula:
\[
\rforallx{x_1, \ldots, x_n} {\nablax{z_1, \ldots, z_m} {
  H_1 \rimp \ldots \rimp H_k \rimp B
}}.
\]
We can apply such theorems in a way similar to what we have described
above. The only important point to note is that $z_1,\ldots,z_m$ must be
instantiated with distinct nominal constants and $x_1,\ldots,x_n$ must be
instantiated with terms that do not contain the nominal constants for
$z_1,\ldots,z_m$.

\subsection{An annotated style of induction}
\label{subsec:annot_ind}

Up to now we have only discussed interactive proofs with case analysis
that has a finite structure. When case analysis has a looping
structure, we need induction to complete the proof. As we have
described in Sections\cspc\ref{subsec:def} and \ref{subsec:def_rules},
\Gee supports inductive reasoning by interpreting definitions as least
fixed-points and by providing a rule to perform induction on
inductively defined predicate symbols. However, the induction rule is
not easy to use since it requires explicitly identifying and using an
inductive invariant, which can be rather complicated, in proof
construction.

To solve this problem, \Abella implements an annotated style of
induction that, on the one hand, mimics how we perform inductive reasoning
in practice and, on the other hand, translates into a sequence of rule
applications in \Gee that includes the induction
rule\cspc\cite{gacek09phd}. The annotated style of
induction works as follows. Suppose that we are interested in proving
a theorem of the form
\[
\rforallx {x_1,\ldots,x_m} {F_1 \rimp \ldots \rimp A \rimp \ldots \rimp F_n \rimp B}
\]
where $A$ is an atomic formula whose head is an inductively defined
predicate.  We can choose to prove this formula inductively by
applying the ``induction'' tactic with respect to $A$. Doing so adds
the formula
\[
\rforallx {x_1,\ldots,x_m} {F_1 \rimp \ldots \rimp A^* \rimp \ldots \rimp F_n \rimp B}
\]
as an \emph{induction hypothesis} to the assumption set and changes
the formula to be proved to the following
\[
\rforallx {x_1,\ldots,x_m} {F_1 \rimp \ldots \rimp A^@ \rimp \ldots \rimp F_n \rimp B}.
\]
 We can now
advance proof search by introducing the variables $x_1,\ldots,x_m$
and adding $F_1,\ldots,A^@,\ldots,F_n$ as hypotheses, leaving $B$ as
the conclusion to be shown. Note the annotations on $A^*$ and
$A^@$. Their meaning is the following: $A^*$ will match only with
another formula that has a similar annotation and the only way to
produce a formula with that annotation is to use a definitional clause
to unfold $A^@$. In other words, we get to use the induction
hypothesis only on an atomic formula that is smaller in the unfolding
sequence than the one in the original formula to be proved.

As an example, consider proving the theorem that the inductive
predicate \kappend is deterministic, something we considered in
Section\cspc\ref{subsec:def}:
\begin{gather*}
  \rforallx {L_1, L_2, L_3, L_3'}
            {\kappend \app L_1 \app L_2 \app L_3 \rimp \kappend \app L_1 \app L_2 \app L_3'
              \rimp L_3 = L_3'}.
\end{gather*}
The proof is by induction on the first assumption. More
specifically, we apply the induction tactic to add
\[
  \rforallx {L_1, L_2, L_3 , L_3'}
            {\kappend \app L_1 \app L_2 \app L_3^* \rimp \kappend \app L_1 \app L_2 \app L_3'
              \rimp L_3 = L_3'}
\]
as an induction hypothesis and we transform the original subgoal into
one with the following hypotheses respectively named $\kwd{H1}$ and
$\kwd{H2}$:
\begin{tabbing}
\qquad\=\kill
\>$\hyp{\kwd{H1}}{\kappend \app L_1 \app L_2 \app L_3^@}$\\
\>$\hyp{\kwd{H2}}{\kappend \app L_1 \app L_2 \app L_3'}$
\end{tabbing}
and with $L_3 = L_3'$ as the conclusion we want to show;
$L_1$, $L_2$ and $L_3$ are variables here. We then analyze the
possible cases of \kwd{H1} as follows:
\begin{itemize}
\item It holds by virtue of the first clause for $\kappend$. In
  this case $L_1$ must be $\knil$ and $L_2 = L_3$. A case analysis on \kwd{H2}
  reveals $L_2 = L_3'$ and hence the conclusion follows;

\item
It holds because of the second clause for $\kappend$.
  Here, there must be some $L_1'$ and $L_3''$ such that $L_1
  = X \cons L_1'$ and $L_3 = X \cons L_3''$ and such that
\begin{tabbing}
\qquad\=\kill
\>$\hyp{\kwd{H3}}{\kappend \app L_1' \app L_2 \app L_3''^*}$
\end{tabbing}
  holds. Note the changed annotation on \kwd{H3},
  corresponding to the fact that it is obtained by an unfolding of the
  definition. Case analysis on the corresponding instance of \kwd{H2} leads
  us to assume that there is some $L_3'''$ such that $L_3' = X \cons
  L_3'''$ and we add
\begin{tabbing}
\qquad\=\kill
\>$\hyp{\kwd{H4}}{\kappend \app L_1' \app L_2 \app L_3'''}$
\end{tabbing}
  to the hypothesis set. At this point, we can apply the inductive
  hypotheses to \kwd{H3} and \kwd{H4} to get $L_3'' = L_3'''$. The
  desired conclusion now easily follows.
\end{itemize}
As we can see from this example, inductive reasoning using the
annotated style is very similar to what we would have done on paper.

\Abella also supports the annotated style of induction for theorems
of the following form:
\[
\rforallx {x_1,\ldots,x_m}
 {\nablax{z_1,\ldots,z_k} {F_1 \rimp \ldots \rimp A \rimp \ldots \rimp F_n \rimp B}}
\]
in a way similar to what we have described above.

\subsection{The two-level logic approach to reasoning}
\label{subsec:twolevel}

We are eventually interested in reasoning about the specifications of
compiler transformations. One possible approach might be to encode
these specifications via fixed-point definitions and to reason based
on this encoding. The problem with this approach is that it would not
then yield a correctness proof of an \emph{implementation} of the
transformations.

%% As we have just explained, \Abella is an effective tool for reasoning
%% about rule-based relational specifications. We would like to use it
%% for reasoning about \LProlog programs. One obvious approach would be
%% to encode \LProlog programs as fixed-points definitions and to reason
%% about the properties of these definitions in \Abella; this is the
%% approach we used with the specification of typing for the \STLC in
%% Section\cspc\ref{subsec:def}. However, this requires that we construct
%% two versions of the specifications, one in \LProlog and one in
%% \Abella. Moreover, the reasoning in \Abella does not directly apply to
%% the \LProlog program. Ideally, we would like to have only \emph{one}
%% specification that serves directly as an implementation and that is
%% also shown to be correct.

To obtain a proof of a specification that is also an implementation,
we use the so-called \emph{two-level logic approach} to
reasoning\cspc\cite{gacek12jar,mcdowell02tocl}.  In the setting of
\Abella, this approach translates into the following. We continue to
write specifications in \HHw. What we also do is encode \HHw into a
fixed-point definition in \Gee that captures the derivability relation
of \HHw. We then lift specifications in \HHw
into \Gee through the encoding. Finally, we reason about the \HHw
specifications using \Abella via the translation. This style of
reasoning yields the desired goal: we end up proving properties
about actual, executable programs in \LProlog. An auxiliary benefit to
the approach comes from the fact that \HHw is itself a logic with special
meta-theoretic properties that can be useful in reasoning about
derivations in it. Since it has been encoded into \Gee, such
properties can be proved as theorems in \Gee. Once they have been
proved, they become available for use in other reasoning tasks.

The Abella system is in fact specifically structured to support the
two-level logic approach described above. Prior to the work in this
thesis, \Abella actually encoded a
weaker logic than \HHw called \HHtwo. The difference between \HHtwo
and \HHw is that the syntax of goal formulas is limited in the
former: specifically, the antecedents of implications in such formulas
are required to be atomic. In other words, the syntax of goal formulas
in \HHtwo is
given by the following rule:
%program clauses. Its syntax is given as follows where $G$ denotes the
%goal formulas and $D$ denotes the program clauses:
%% \begin{tabbing}
%% \qquad\=$G$\qquad\=::=\qquad\=\kill
%% \>$G$\>::=\>$A\sep G \sconj G\sep D\simply G \sep \typedforall{\tau}{x}{G}$\\
%% \>$D$\>::=\>$A\sep A\simply A\sep \typedforall{\tau}{x}{D}$
%% \end{tabbing}
\begin{tabbing}
\qquad\=$G$\qquad\=::=\qquad\=\kill
\>$G$\>::=\>$true \sep A\sep G \sconj G\sep A\simply G \sep \typedforall{\tau}{x}{G}$
\end{tabbing}
As a result of this restriction, the dynamic program context is
limited to being a collection of atomic formulas.
The initial reason for restricting attention to the \HHtwo
specification logic was that it is simpler to realize the two-level
logic approach with this kind of a dynamic program context.
This restriction also turned out to be acceptable because a large
class of rule-based relational specifications naturally fall within
the structure of \HHtwo.
%% The interest in \HHtwo specifications is two-fold: First, a large
%% class of rule-based relational specifications can be described in
%% \HHtwo.
%% %GN It is rather odd to say ``a large class'' and then offer exactly
%% %one example!
%% %% (\eg typing for the \STLC as described in
%% %% Section\cspc\ref{subsec:ecd_bindings});
%% Second, it is straightforward
%% to realize the two-level logic approach to reasoning about \HHtwo
%% specifications.
We shall describe the realization of two-level logic approach with
respect to \HHtwo in this section. A contribution of this thesis is
that it extends the two-level logic approach to deal with the full
\HHw specifications. We discuss this extension in
Chapter\cspc\ref{ch:extensions}.

The embedding \HHtwo within \Abella is achieved by encoding the
derivability relation of \HHtwo in a fixed-point definition. In order
to exploit the meta-logical properties of the \HHtwo logic in
reasoning, we want to be able to prove theorems about the derivability
%GN Calling the measure the "height" is not quite accurate.
relation in \Abella. These proofs typically rely on a measure of the
size of a derivation. To 
facilitate their construction, we parameterize the encoding of the
derivability relation by a natural
number. Concretely, we identify \knat as the type of natural numbers
that we encode using the constants $\kz: \knat$ for representing $0$
and $\ks:\knat \to \knat$ for representing the successor function. The
inductive nature of this representation of natural numbers is
formalized by introducing the predicate  $\knat : \knat \to \prop$
that is defined as the least fixed point of the following clauses:
\begin{tabbing}
\qquad\=$\knat\app (\ks\app N)$ \quad\=$\rdef$ \quad\=\kill
\>$\knat\app \kz$ \>$\rdef$ \>$\rtrue$\\
\>$\knat\app (\ks\app N)$ \>$\rdef$ \>$\knat\app N$
\end{tabbing}
Finally, we use the predicate symbol $\kseq : \knat \to \olist \to
\omic \to \prop$ to encode the derivability of \HHtwo. Here, $\olist$ is a
type that is reserved for lists of \HHtwo formulas and such lists are
constructed using the constants $\knil$ and $\cons$. An \HHtwo
sequent $\Sigma;\Gamma;\Delta \sseq G$ is encoded as $(\kseq \app
N\app \Delta \app G)$; intuitively, this predicate holds if the
corresponding sequent has a derivation whose size is at most $N$.

An aspect to be noted about the encoding of \HHtwo sequents is that it
treats the static context and the signature implicitly. How exactly
this implicit treatment is realized will become clear when we present
an actual fixed-point definition for \kseq in
Chapter\cspc\ref{ch:extensions}. However, we sketch the general idea
that is used here. A development in \Abella that is 
about an \HHtwo specification begins with loading in that specification.
This loading action fixes the static context for the development and
Abella manages the use of this context separately. With regard to the
signature, one part of it arises from the \HHtwo specification that is
loaded. Abella treats this part by adding it directly to is own
vocabulary of non-logical constants. The other, dynamic part of the
signature arises from the treatment of universal goals in proof
search. The encoding of derivability in \HHtwo handles universally
quantified goals in \HHtwo by transforming them into formula-level
$\nabla$ quantifiers in \Gee. The ``constants'' that are added 
to the signature when these $\nabla$ quantifiers are processed
therefore  end up being represented by nominal constants and so
there is already a means for identifying them in the proof
development.  It has been shown in \cite{gacek09phd} that
$(\kseq\app N \app \Delta \app G)$ is derivable for some $N$ in \Gee
if and only if $\Sigma;\Gamma;\Delta \sseq G$ is derivable in
\HHtwo. This result provides the basis for our thinking of properties
of our encoding directly as properties of derivability in \HHtwo.

We often want to prove properties of \HHtwo derivations based on an
induction on their sizes. Towards this end, we use the notation
$\oseq{\Delta \sseq G}$ introduced by \cite{gacek09phd} to represent
the formula $\rexistsx N {\knat\app N \rand \kseq\app N\app \Delta\app
  G}$ in the propositions we want to prove in \Abella. 
% As should be
% evident from the
% following discussion, in constructing proofs interactively for the
% \Abella theorems, this notation will also be used to represent a
% packet of formulas $\knat\app N$ and $\kseq\app N\app \Delta\app G$ for some $N$
% which is logically equivalent to $\rexistsx N {\knat\app N \rand
%   \kseq\app N\app \Delta\app G}$ in the context where the packet
% occurs. 
We abbreviate $\oseq{ \sseq G}$ as $\oseq{G}$. Given a
formula $T$ of the form
\[
\rforallx {x_1,\ldots,x_m} {F_1 \rimp \ldots \rimp \oseq{\Delta \sseq G} \rimp \ldots \rimp F_n \rimp B}.
\]
applying the induction tactic on $\oseq{\Delta \sseq G}$ is equivalent
to first transforming $T$ into the following logically equivalent
formula:
\[
\rforallx {x_1,\ldots,x_m,N} {F_1 \rimp \ldots \rimp
  \knat\app N \rimp \kseq\app N\app \Delta\app G \rimp \ldots \rimp F_n \rimp B}
\]
and then applying the induction tactic with respect to $(\knat\app N)$. The
annotated style of induction is extended to represent this process in a
user-friendly way. Specifically, induction on $\oseq{\Delta \sseq G}$
in $T$ introduces the following inductive hypothesis
\[
\rforallx {x_1,\ldots,x_m} {F_1 \rimp \ldots \rimp \oseq{\Delta \sseq G}^* \rimp \ldots \rimp F_n \rimp B}
\]
which represents
\[
\rforallx {x_1,\ldots,x_m,N} {F_1 \rimp \ldots \rimp
  (\knat\app N)^* \rimp \kseq\app N\app \Delta\app G \rimp \ldots \rimp F_n \rimp B}
\]
and changes the conclusion to 
\[
\rforallx {x_1,\ldots,x_m} {F_1 \rimp \ldots \rimp \oseq{\Delta \sseq G}^@ \rimp \ldots \rimp F_n \rimp B}
\]
which represents
\[
\rforallx {x_1,\ldots,x_m,N} {F_1 \rimp \ldots \rimp
  (\knat\app N)^@ \rimp \kseq\app N\app \Delta\app G \rimp \ldots \rimp F_n \rimp B}.
\]
When the notation $\oseq{\Delta
  \sseq G}$ is used as a hypothesis, it represents the packet of
formulas $\knat\app N$ and $\kseq\app N\app \Delta\app G$ for some
variable $N$. This representation coincides with the fact that
applying the proof rules to eliminate the logical constants in the
hypothesis $\rexistsx N {\knat\app N \rand \kseq\app N\app \Delta\app
  G}$ will result in the packet we just described.
When this notation is used with an annotation in a hypothesis, \ie,
when the hypothesis has the form $\oseq{\Delta \sseq G}^{\square}$
with $\square$ being either $@$ or $*$, then it represents
$(\knat\app N)^{\square}$ and $\kseq\app N\app \Delta\app G$ for some
$N$.  Case analysis on $\oseq{\Delta \sseq G}$ treats the packet as a
whole. Specifically, it has the effect of first analyzing the cases of
$\kseq\app N\app \Delta\app G$ and then analyzing $\knat\app N$ if $N$
has been instantiated with $(\ks\app N')$ for some $N'$. As a result,
case analysis on $\oseq{\Delta \sseq G}^@$ might produce branches that
have hypotheses
of the form $\oseq{\Delta' \sseq G'}^*$ for some $\Delta'$ and $G'$.
The inductive hypothesis then becomes applicable to such hypotheses.

To illustrate how we can state and prove properties about \HHtwo
specifications in \Abella, suppose that we have included into an
\Abella development the %\HHw
program from
Section\cspc\ref{subsec:ecd_bindings} that specifies the typing rules
for the \STLC.\footnote{This program, which was
  presented as one in \HHw, is also one in \HHtwo. We use this fact
  systematically to refer to the development here as one related to \HHtwo
  rather than in \HHw.}
Assuming that $L$ is a list of formulas that encode the
dynamic typing context that arises in the relevant \HHtwo derivation, it
is the case that $(\kof\app M\app T)$ is derivable in \HHtwo if and only
if $\oseq{L \sseq \kof\app M\app T}$ is provable in \Gee.\footnote{We
  use a systematic abuse of notation here and elsewhere, confusing the
  items introduced into the signature by universal goals in \HHtwo
  derivations with the nominal constants that arise in corresponding
  \Gee derivations.}  We can actually characterize legitimate
representations for typing contexts in \Gee by using the predicate
symbol $\kctx : \kolist \to \prop$ that is defined by the following
clauses:
\begin{tabbing}
\qquad\=$\nablax x {\kctx \app (\kof \app x \app T \cons L)}$
 \quad\=$\rdef$ \quad\=\kill
\>$\kctx \app \knil$ \>$\rdef$ \>$\rtrue$\\
\>$\nablax x {\kctx \app (\kof \app x \app T \cons L)}$
  \>$\rdef$ \>$\kctx \app  L$
\end{tabbing}
In this setting, the uniqueness of type assignment in the \STLC is
captured by the following formula:
\begin{gather*}
  \rforallx {L, M, T, T'} {\kctx \app L \rimp \oseq{L \sseq \kof \app M \app T} \rimp \oseq{L \sseq \kof \app M \app T'} \rimp T = T'}.
\end{gather*}
It is provable by induction on the assumption $\oseq{L \sseq \kof \app
  M \app T}$. A proof of this formula can be found, for instance, in
the \Abella tutorial\cspc\cite{baelde14jfr}.

In the thesis work, we will use the two-level logic approach to verify
compiler transformations implemented in \LProlog. As we shall see in
Chapters\cspc\ref{ch:cps}, \ref{ch:closure}, \ref{ch:codehoist} and
\ref{ch:complete}, the two-level logic approach is essential for
reasoning about such implementations. Combining it with the other rich
reasoning devices in \Abella, we will be able to prove deep properties
such as preservation of meaning by compiler transformations in an
effective manner.

\section{A Higher-Order Abstract Syntax Approach}
\label{sec:hoas}

In the previous sections of this chapter, we have exposed the support
that both \LProlog and \Abella provide for the $\lambda$-tree syntax
version of the \HOAS approach. We have also provided some examples
that should have motivated the benefits of this approach to
representing, manipulating and reasoning about objects that embody
binding constructs. In this section, we provide a few more directed
examples that illustrate particular paradigms related to the use of
the $\lambda$-tree syntax approach in programming and reasoning within
the framework of interest. These paradigms will turn out to have
repeated applications when we consider the implementation and
verification of compilers for functional languages in later chapters
in the thesis.

%% The primary goal of this thesis is to show that these benefits of the
%% $\lambda$-tree syntax approach can be exploited to simplify the
%% implementation and verification of compilers for functional
%% languages. We therefore expose the $\lambda$-tree syntax approach in
%% this section to build up the terminologies for the later discussions.

\subsection{Analyzing object structure in specifications}
\label{subsec:lts_spec}

In Section\cspc\ref{subsec:ecd_bindings} we have explained the key
idea underlying the representation of formal objects that underlies
the use of the $\lambda$-tree approach in specifications within our
framework: object-level binding translates into abstraction in the
\LProlog representation. We have also discussed several benefits that
flow for this approach, including the fact that the programmer derives
help from the meta-language in realizing operations such as bound
variable renaming, the correct implementation of substitution and
recursion over binding structure. One aspect that we have not
explicitly touched upon that we will use repeatedly later is the
capabilities of structure analysis that we get from unification in the
meta-language.

We illustrate this capability by considering the encoding of
$\eta$-conversion over the object-language of $\lambda$-terms.
Suppose that we have designated the \LProlog type $\ktm$ for the
representations of these terms and that we use the constants $\kapp:
\ktm \to \ktm \to \ktm$ and $\kabs: (\ktm \to \ktm) \to \ktm$ to
encode the object language applications and abstractions. Now, let
$\keta$ be a \LProlog predicate symbol of type $\ktm \to \ktm \to
\omic$ and consider the following clause:
\begin{tabbing}
\qquad\=\kill
\>$\keta\app (\kabs\app (x \mlam \kapp\app M\app x))\app M.$
\end{tabbing}
This clause succinctly captures the basic component of the object
language $\eta$-conversion relation: the atomic formula $\keta\app
t\app t'$ is derivable from it exactly when $t$ encodes an
$\eta$-redex and $t'$ encodes its contracted form. To see this,
observe that the binding order requires that in generating an instance
of the clause $M$ must not be instantiated with a term that contains
$x$. To derive $\keta\app t\app
t'$, we need to unify it with the clause. This unification must respect
the constraint just described and therefore it will succeeds only if $t =
(\kabs\app (x \mlam \kapp\app M\app x))$ for some $M$ not
containing $x$ and $t$ is equal to $M$.

A closer examination of the example shows that we have used the idea
of unification to tell us when a particular part of an object language
expression does not depend on a bound variable. This and related forms
of structure analysis will be a recurring theme in the specification
of compiler transformations  that we provide in later chapters. For
example, this particular kind of structure analysis will play an
important role in specifying the code hoisting transformation in
Chapter\cspc\ref{ch:codehoist}.

\subsection{Exploiting $\lambda$-tree syntax in reasoning}
\label{subsec:lts_reason}

%% \Abella also supports the $\lambda$-tree syntax approach to encoding
%% rule-based relational specifications. The only difference between the
%% $\lambda$-tree syntax approach in \Abella and that in \LProlog is that
%% the former realizes recursion over binding operators via
%% $\nabla$-quantified formulas while the latter via universal and
%% hypothetical goals. We have already discussed these aspects of the
%% $\lambda$-tree syntax approach in \Abella via the example of encoding
%% typing rules for the \STLC in Section\cspc\ref{subsec:def}.

The ability to analyze the structure of terms via unification also
plays an essential role in reasoning, specifically, in case
analysis. In effect, the case analysis style of reasoning in \Abella
examines the possible ways an atomic formula can unify with the
head of clauses that define it. This was manifest in the examples
that make use of case analysis in Section\cspc\ref{subsec:def}.

By exploiting the logical structure of the $\lambda$-tree syntax
representations and the rich reasoning capabilities provided by
\Abella, we are able to prove properties related to binding structure
in an effective manner. We illustrate this point via the example of
defining substitution as an explicit relation and proving properties
%GN STLC is not a proper noun, it expands into a sequence of
%words. Read the sentence to check if you can actually drop the
%article.
%% about this relation. Assume that we are working with \STLC and the
%% encoding
about this relation. Assume that we are working with the \STLC and the
encoding
of its syntax that we have already described. We now use the type \kmap
and the constant
$\kmap:\ktm \to \ktm \to \kmap$ to represent mappings from variables
(encoded as nominal constants) to terms and a list of such mappings to
represent a substitution. We use the \Abella type $\kmaplist$ to
represent the type of lists of mappings and the constants $\kmlnil :
\kmaplist$ and $\kmlcons : \kmap \to \kmaplist \to \kmaplist$ as the
constructors of such lists.
Then the predicate $\kappsubst: \kmaplist \to \ktm \to \ktm \to \prop$
such that $\kappsubst\app S\app M\app M'$ holds exactly when $M'$ is
the result of applying the substitution $S$ to $M$ can be defined by
the following clauses:
\begin{tabbing}
\qquad\=$\nablax x {\kappsubst\app (\kmlcons\app (\kmap\app x\app V)\app S)\app (R\app x)\app M}$
\quad\=$\rdef$ \quad\=\kill
\>$\kappsubst\app \kmlnil\app M\app M$ \>$\rdef$ \>$\rtrue$\\
\>$\nablax x {\kappsubst\app (\kmlcons\app (\kmap\app x\app V)\app S)\app (R\app x)\app M}$
  \>$\rdef$ \>$\kappsubst\app S\app (R\app V)\app M$
\end{tabbing}
For obvious reasons, we shall take $\kappsubst$ to be inductively
defined. The first clause encodes the fact that an empty substitution
has no effect on a term. The second clause
applies to a non-empty substitution that maps $x$ to $V$ and that has
additional mappings given by $S$. This clause uses the pattern $(R\app x)$
to match the term to be substituted into. The quantification
ordering ensures that $R$ cannot contain $x$. For this reason, $R$
must be bound to the result of abstracting over all the free occurrences of
$x$ in the ``source term'' for the substitution. The $(R\app V)$ then
is equal modulo $\lambda$-conversion to the result of
replacing all the free occurrences of $x$ in the source term with
$V$. The body of the clause represents the application of the
remainder of the substitution to the resulting term.

This style of encoding of the substitution relation that takes
advantage of the meta-level understanding of binding structure makes
it extremely easy to prove structural properties of the relation.
For example, the facts that substitution distributes over applications
and abstractions can be stated as follows:
\begin{tabbing}
\qquad\=\qquad\=\kill
\>$\rfall {S, M_1, M_2, M'}.\kappsubst\app S\app (\kapp\app M_1\app M_2)\app M' \rimp$\\
\>\>$\rexistsx {M_1', M_2'} M' = \kapp\app M_1'\app M_2' \rand
               \kappsubst\app S\app M_1\app M_1' \rand
               \kappsubst\app S\app M_2\app M_2'$.\\
\>$\rfall {S, R, T, M'}.\kappsubst\app S\app (\kabs\app T\app R)\app M' \rimp$\\
\>\>$\rexistsx {R'} {M' = \kabs\app T\app R' \rand
                \nablax x {\kappsubst\app S\app (R\app x)\app (R'\app x)}}$.
\end{tabbing}
These properties are easily proved by induction on the only
assumption, as follows. The base cases are immediately proved. In the
inductive cases, we apply the induction hypothesis to the result of
unfolding the first assumption to generate new hypotheses, from which
the conclusion is immediately derived. The simplicity of these proofs
comes from the fact that the actual substitution in the definition of
\kappsubst is carried out via meta-level $\beta$-reduction. As a
result, the properties above are just explicit statements of the
corresponding properties of $\beta$-reduction which are immediately
true given the meta-language.

Further benefits in reasoning about binding structure can be derived
by combining definitions in \Abella with the two-level logic approach.
As an example, we may want to characterize relationships between
closed terms and substitutions. For this, we can first define
the well-formed \STLC terms through the following \HHw clauses:
\begin{tabbing}
\qquad\=$\ktm\app (\kabs\app T\app R)$ \quad\=$\limply$ \quad\=\kill
\>$\ktm\app (\kapp\app M\app N)$ \>$\limply$ \>$\ktm\app M \scomma \ktm\app N.$\\
\>$\ktm\app (\kabs\app T\app R)$ \>$\limply$ \>$\forallx x {\ktm\app x \simply \ktm\app (R\app x)}.$
\end{tabbing}
This definition is similar to the definition of typing rules for the \STLC
except that $\ktm$ does not record the type information of terms. We
characterize the context used in \ktm derivations in \Abella as
follows:
\begin{tabbing}
\qquad\=$\nablax x {\ktmctx\app (\ktm\app x \cons L)}$ \quad\=$\rdef$ \quad\=\kill
\>$\ktmctx\app \knil$\\
\>$\nablax x {\ktmctx\app (\ktm\app x \cons L)}$ \>$\rdef$ \>$\ktmctx\app L.$
\end{tabbing}
Intuitively, if $\ktmctx\app L$ and $\oseq{L \sseq \ktm\app M}$ hold,
then $M$ is a well-formed term whose free variables are given by $L$.
Clearly, if $\oseq{\ktm\app M}$ holds, then $M$ is closed. We can also
state a ``pruning'' property about a term that says that if a term is
well-formed in a context that does not include a particular variable,
then the variable does not occur in that term:
\begin{gather*}
\rforallx {M, L} {\nablax {x:\ktm}
  {\ktmctx\app L \rimp \oseq{L \sseq \ktm\app (M\app x)} \rimp
    \rexistsx {M'} {M = y \mlam M'}}}.
\end{gather*}
This formulation of the property of interest makes critical use of
$\lambda$-tree syntax. The pattern $(M\app x)$ expresses the
\emph{possible} dependence of a term on the variable $x$ but with $M$
being the result of abstracting over \emph{all} free occurrences of $x$ in
the term; this is the case because of the order of the quantification
over $M$ and $x$. Also because of the order of the quantifiers, $L$
cannot contain $x$. The conclusion $\rexistsx {M'} {M = y \mlam M'}$
states that $M$ is a vacuous abstraction since  $M'$ cannot contain
any occurrences of the bound variable $y$ because of the order of the
binders.
The reason why the conclusion must be true is that if $\ktm\app (M\app
x)$ is derivable from $L$ in the specification language, it must be
the case that $x$ does not appear free in $(M\app x)$. As might be
expected from this observation, the pruning property can be proved
by an induction on the judgment $\oseq{L \sseq \ktm\app (M\app
  x)}$. The details of the argument are considerably simplified by
using the meta-language treatment of binding.

As a special case of the pruning property, we derive that a closed
term do not depend on any free variable.
\begin{gather*}
\rforallx {M} {\nablax {x:\ktm}
  \oseq{\ktm\app (M\app x)} \rimp \rexistsx {M'} {M = y \mlam M'}}.
\end{gather*}
Now we can state the fact that substitution has no effect on closed
terms as follows:
\begin{gather*}
\rforallx {S, M, M'} {\oseq{\ktm\app M} \rimp
  \kappsubst\app S\app M\app M' \rimp M = M'}.
\end{gather*}
This property is proved by an induction on the judgment
$\kappsubst\app S\app M\app M'$. In the inductive case, we make use of
the pruning property that we have proved to discharge the dependence
of the closed term $M$ on variables in $S$, thereby discharging the
effect of the substitution.

% Extensions to Abella
%%%%%%%%%%%%%%%%%%%%%%%%%%%%%%%%%%%%%%%%%%%%%%%%%%%%%%%%%%%%%%%%%%%%%%%%%%%%%%%
% extensions.tex: Extensions to Abella
%%%%%%%%%%%%%%%%%%%%%%%%%%%%%%%%%%%%%%%%%%%%%%%%%%%%%%%%%%%%%%%%%%%%%%%%%%%%%%%%
\chapter{Extensions to the Abella System}
\label{ch:extensions}

The framework consisting of \LProlog and \Abella is a powerful tool
for specifying and reasoning about rule-based relational descriptions
of formal systems. The specification language \LProlog has been worked
on for many years and is at a reasonably mature state of development
to be used unchanged in our work on verified compilation. However, the
\Abella system is newer and is still evolving. In this chapter we
describe some extensions to \Abella that make it a more flexible and
powerful reasoning tool. These extensions make it easier for \Abella
to carry out large-scale proof developments such as the verification
compilation work we will describe in later chapters.

The extensions we have developed consist of two parts. The first part
is a treatment of \LProlog specifications in the full
\HHw. Previously, the two-level logic approach in \Abella only
supported reasoning over a subset of \HHw, \ie, \HHtwo. We have given
a full encoding of \HHw in \Gee and developed a methodology to reason
about \HHw specifications through this encoding. This methodology
allows us to reason about rule-based relational specifications that
involve higher-order contexts, \ie, contexts with not only atomic but
also universal and implicational assumptions. We demonstrate the
usefulness of this extension by proving non-trivial properties of one
such specification that encodes a common compiler transformation on
functional programs.

The second extension we have developed is to provide support of
polymorphism in \Abella. The current proof theory of \Gee is based on
the simply typed $\lambda$-calculus which does not include polymorphic
types. To encode general data structures such as lists in \Abella, we
have to define a separate version of them at every concrete type where
the data structure is needed. Furthermore, we have to prove their
properties individually at each such concrete type, even through these
properties and their proofs are mostly replications of a general
pattern under different type instantiations. This is a recurring theme
that we have witnessed when working on the implementation and
verification of compiler transformations. To solve these problems, we
have developed support for an approach to describing data structures,
to writing definitions and statements we want to prove and to
providing proofs that is schematic at the type level. Using this kind
of ``schematic polymorphism,'' we are able to avoid a lot of the
duplication of code,
definitions and proofs that we have alluded to above.\footnote{A
  related approach has been developed by Kaustuv Chaudhuri and has
  already been incorporated into the \Abella system. However, that
  approach does not support a type schematic treatment of data
  structures or of \HHw specifications which is the main focus of the
  work here.}

We describe the two extensions in detail in the rest of this
chapter. We first describe the extension to the two-level logic
approach that enables reasoning about the full \HHw specifications in
Section\cspc\ref{sec:embed_hhw}. We then describe the extension to
support polymorphism in \Abella in Section\cspc\ref{sec:poly}.

\section{Embedding \HHw in \Abella}
\label{sec:embed_hhw}

We have described a two-level logic approach to reasoning about
\LProlog specifications in Section\cspc\ref{subsec:twolevel}. The idea
is to embed the logic of \LProlog as a fixed-point definition in \Gee
and to reason about \LProlog specifications through the embedding. The
two-level logic approach adopted by \Abella previously, as described
in Section\cspc\ref{subsec:twolevel}, embeds a subclass of \HHw called
\HHtwo that allows for only dynamic addition of atomic formulas during
derivations. It is good enough for proving properties of rule-based
relational specifications that use contexts that contain only atomic
assumptions, as demonstrated in the proof of determinacy of typing for the
\STLC in Section\cspc\ref{subsec:twolevel}. On the other hand, a lot
of formal systems are naturally described via rules that operate with
contexts containing non-atomic formulas, or \emph{higher-order
  contexts}. For example, a common compiler transformation that
converts $\lambda$-terms into their de Bruijn forms can be elegantly
described via rules that record the rule for transforming variables
into de Bruijn indexes as implicational formulas in the contexts. Such
rule-based specifications can be elegantly encoded in \HHw. However,
they do not have a direct representation in \HHtwo and thereby cannot
be reasoned about using the two-level logic approach described
previously.

One reason for the previous two-level logic approach to embed \HHtwo
instead of \HHw was for simplifying the case analysis based reasoning
about derivability in \LProlog. To analyze a derivation in \LProlog,
we need to take into account of the possibility that the goal formula
is derived from the dynamically added program clauses, or \emph{dynamic
clauses} in short. In \HHtwo, this case is possible only if the goal
formula is a member of the dynamic context because the dynamic context
contains only atomic formulas and such an atomic formula can be used
to prove the goal only if it matches exactly with the goal. Analyzing
this case is thus equivalent to checking membership. However, in \HHw,
since the dynamic clauses can have the form of arbitrary program
clauses, the derivation may proceed by backchaining on some dynamic
clause and further yielding subgoals. Analysis of such cases becomes
much more complicated and may not even be possible if no proper
constraint is placed on the form of the dynamic context and the
structure of derivations.

In this section, we describe an extension of the two-level logic
approach to support reasoning about the full class of \HHw
specifications\cspc\cite{wang13ppdp}. The critical observation that
enables this extension is that clauses added dynamically during a
derivation must take the form of some subformulas in the original
specification. Since we always fix the \LProlog specification at the
beginning of proof developments, the dynamic clauses must have finite
forms. We can therefore give the dynamic context an inductive
definition in \Gee. To support reasoning over backchaining steps that
result from using dynamic clauses, we embed a version of \HHw based on
the technique called \emph{focusing}\cspc\cite{andreoli92jlc} as a
fixed-point definition in \Gee. With this extension, we are able to
reason about rule-based relational specifications that involve
higher-order contexts in \Abella.

We elaborate on the above ideas in the following subsections. We first
introduce the transformation of $\lambda$-terms into the de Bruijn
form as an example to motivate the extension in
Section\cspc\ref{subsec:hhw_mtv_exm}. We then explain the difficulties
with case analysis on higher-order contexts in
Section\cspc\ref{subsec:dyn_ctx_problem} and present our solution in
Section\cspc\ref{subsec:dyn_ctx_solution}. We conclude the discussion
by demonstrating the power of the extended two-level logic approach
through proving non-trivial properties about the motivating example in
Section\cspc\ref{subsec:hhw_exm}.

\subsection{A motivating example}
\label{subsec:hhw_mtv_exm}

To understand the issues with reasoning about rule-based descriptions
of relations involving higher-order contexts, consider the
transformation of $\lambda$-terms into their de Bruijn forms as an
example. The de Bruijn form is an alternative notion of
$\lambda$-terms due to de Bruijn in which bound variables are not
named and their occurrences are represented instead by indexes that
count the abstractions up to the one binding them
\cite{debruijn72}. Its syntax is given as follows where $t$ denotes a
term in this form and $i$ is a natural number denoting the index of a
variable occurrence:
\begin{gather*}
  t \;::=\; i \sep t\app t \sep \dabs t
\end{gather*}
Note that the binding structure is already implicitly captured by the
representation of variables as de Bruijn indexes. As a result, there
is no need to represent binding variables in abstractions.

There is a natural mapping between closed $\lambda$-terms and their de Bruijn
representation. The $\lambda$-terms can be translated into the
nameless form by converting every variable occurrences into de Bruijn
indexes and dropping the binding variables, and vice versa. For example,
the term $(\abs {x} {(\abs{y} {x\app y})\app x})$ is mapped to the de
Bruijn term $(\dabs {(\dabs {2\app 1})\app 1})$.
We give a rule-based description of such mapping. Writing $\G \vdash
\dbrel m h d$ to denote the correspondence between the $\lambda$-term
$m$ that occurs at \emph{depth} $h$ (\ie, under $h$
$\lambda$-abstractions) and the de Bruijn term $d$ where $\G$
determines the mapping between free variables in the two
representations, we can define this relation via the rules in
Figure\cspc\ref{fig:dbrel}.
The rule for relating applications is straightforward. To relate
$(\abs x m)$ to a De Bruijn term at depth $h$, we must relate each
occurrence of $x$ in $m$, which must be at a depth $h + k$ for some $k
> 0$, to the de Bruijn index $k$. To encode this correspondence, the
context is extended in the premise of rule \kdbabs with a (universally
quantified) implicational formula. Note also that this rule carries
with it the implicit assumption that the name $x$ used for the bound
variable is fresh to $\G$, the context for the concluding
judgment. Eventually, when the $\lambda$-term on the right of the
turnstile is a variable, the rule \kdbvar provides the means to
complete the derivation by using the relevant assumption from the
context $\G$.

\begin{figure}[!ht]
\begin{gather*}
  \infer[\kdbapp]{
    \G \vdash \dbrel {m\app n} h {d\app e}
  }{
    \G \vdash \dbrel m h d
    &
    \G \vdash \dbrel n h e
  }
  \\
  \infer[\kdbabs]{
    \G \vdash \dbrel {\abs x m} h {\dabs d}
  }{
    \G, \rforallx {i, k} {((h + k = i) \rimp \dbrel x i k)}
    \vdash \dbrel m {h + 1} d
  }
  \\
  \infer[\kdbvar]{
    \G \vdash \dbrel x i k
  }{
    \rforallx {i, k} {((h + k = i) \rimp \dbrel x i k) \in \G}
    &
    h + k = i
  }
\end{gather*}
\caption{The Rules Relating $\lambda$-Terms and their de Bruijn Forms}
\label{fig:dbrel}
\end{figure}

The above rule-based description can be elegantly encoded in
\LProlog. We first encode the syntax of the $\lambda$-terms and the de
Bruijn terms in \LProlog as follows. We use the \LProlog type $\ktm$
to represent the type of $\lambda$-terms and build their encoding
around the two constructors $\kapp: \ktm \to \ktm \to \ktm$ and
$\kabs: (\ktm \to \ktm) \to \ktm$.  Using the type \kdtm for the
representation of $\lambda$-terms in the de Bruijn form, we can encode
them via the constructors $\kdvar : \knat \to \kdtm$, $\kdapp : \kdtm
\to \kdtm \to \kdtm$ and $\kdabs : \kdtm \to \kdtm$. Compared the
encoding of $\lambda$-terms, variables in this form are formed
explicitly by applying $\kdvar$ to de Bruijn indexes. Abstractions are
formed by applying $\kdabs$ to terms representing their body.

The rules in Figure\cspc\ref{fig:dbrel} make use of addition of
natural numbers. We use \knat to represent the type of natural numbers
and encode natural numbers by using the constants $\kz: \knat$ for
representing $0$ and $\ks:\knat \to \knat$ for constructing the
successor of a natural number. We then encode addition as a relation
$\kadd : \knat \to \knat \to \knat \to \omic$ such that given natural
numbers $n_1$, $n_2$, $n_3$ and their encoding $N_1$, $N_2$ and $N_3$,
$n_1 + n_2 = n_3$ if and only if $\kadd\app N_1\app N_2\app N_3$
holds. This relation is defined through the following clauses:
\begin{tabbing}
\qquad\=$\kadd\app (\ks\app N_1)\app N_2\app (\ks\app N_1)$ \quad
  \=$\limply$ \quad\=\kill
\>$\kadd\app \kz\app N\app N$.\\
\>$\kadd\app (\ks\app N_1)\app N_2\app (\ks\app N_3)$
  \>$\limply$ \>$\kadd\app N_1\app N_2\app N_3$.
\end{tabbing}
We then designate the predicate constant $\khodb: \ktm \to \knat \to
\kdtm \to \omic$ to represent the ternary relation $\equiv$. The rules
in Figure\cspc\ref{fig:dbrel} translate into the following clauses:
\begin{tabbing}
\qquad\=\qquad\=\kill
\>$\khodb\app (\kapp\app M_1\app M_2)\app H\app (\kdapp\app E_1\app E_2)$
  \quad $\limply$ \quad
  $\khodb\app M_1\app H\app E_1 \scomma \khodb\app M_2\app H\app E_2$.\\
\>$\khodb\app (\kabs\app M)\app H\app (\kdabs\app D)$ \quad $\limply$ \quad \\
\>\>$\forallx x
      {(\forallx {i, k} {\kadd\app H\app k\app i \simply
           \khodb\app x\app i\app (\kdvar\app k)})
       \simply {\khodb\app (M\app x)\app (\ks\app H)\app D}}$.
\end{tabbing}
The first and second clause encode the \kdbapp and \kdbabs rules,
respectively. Assume we use the dynamic contexts in \HHw sequents to
represent the contexts in the original rules. Then backchaining
$\khodb\app M_1\app H\app M_2$ on any one of the clauses followed by
simplification of the goal formula corresponds to applying the
relevant rule to the relation encoded by $\khodb\app M_1\app H\app
M_2$. This is easy to see for the first clause. In the case of the
second clause, backchaining succeeds if $M_1 = \kabs\app M$ and $M_2 =
\kdabs\app D$ and results in the following new goal:
\[
\forallx x
{(\forallx {i, k} {\kadd\app H\app k\app i \simply
           \khodb\app x\app i\app (\kdvar\app k)})
       \simply {\khodb\app (M\app x)\app (\ks\app H)\app D}}
\]
To derive this goal we need to introduce a new constant $c$ for $x$
and derive the goal
\[
{(\forallx {i, k} {\kadd\app H\app k\app i \simply
           \khodb\app c\app i\app (\kdvar\app k)})
       \simply {\khodb\app (M\app c)\app (\ks\app H)\app D}}
\]
The only way to derive this goal is to extend the dynamic context with
the assumption $(\forallx {i, k} {\kadd\app H\app k\app i \simply
  \khodb\app c\app i\app (\kdvar\app k)})$ and then derive the goal
${\khodb\app (M\app c)\app (\ks\app H)\app D}$. This corresponds to
deriving the premise of \kdbabs.
Like the encoding of typing rules for the \STLC described in
Section\cspc\ref{subsec:ecd_bindings}, this example uses universal
goals to realize recursion over abstractions and to capture the side
condition and it uses hypothetical goals to introduce assumptions
about binding variables. The only difference is that the dynamically
introduced assumptions here are not atomic formulas but have the full
form of program clauses. Note that an explicit encoding of the \kdbvar
rule is not necessary as the application of this rule is implicitly
captured by backchaining on the relevant dynamic clause.

We use a concrete example to further demonstrate how the above
specification works. Let $\Si$ be the signature containing constants
we have defined so far and $\G$ the static context containing the
clauses defining $\kadd$ and $\khodb$. Consider showing that the
$\lambda$-term $\abs x {\abs y {(y\app x)}}$ corresponds to the De
Bruijn term $\dabs \dabs (1\app 2)$. This amounts to proving the
following \HHw sequent:
\begin{tabbing}
\qquad\=$\Si ; \G ; \emptyset \sseq \khodb\app $\=\kill
\>$\Si ; \G ; \emptyset \sseq \khodb\app
     (\kabs\app (x\mlam {\kabs\app (y\mlam {\kapp\app y\app x})}))\app \kz$\\
\>\>$(\kdabs\app
     (\kdabs\app
       (\kdapp\app (\kdvar\app (\ks\app \kz))\app
                   (\kdvar\app (\ks\app (\ks\app \kz))))))$.
\end{tabbing}
The only choice is backchaining on the second clause for $\khodb$,
which changes the proof obligation to:
\begin{tabbing}
\qquad\=$\Si, x:\knat ; \G ; $\=$\khodb\app $\=\kill
\>$\Si, x:\knat ; \G ;
     (\forallx {i, k} {\kadd\app \kz\app k\app i \simply
                       \khodb\app x\app i\app (\kdvar\app k)})
     \sseq$  \\
\>\>$\khodb\app (\kabs\app (y\mlam {\kapp\app y\app x})) \app (\ks\app \kz)$ \\
\>\>\>$(\kdabs\app (\kdapp\app (\kdvar\app (\ks\app \kz))\app
                                (\kdvar\app (\ks\app (\ks\app \kz)))))$.
\end{tabbing}
Attempting to backchain the new dynamic clause will fail because the
new signature constant $x$ does not unify with \kabs. Hence, the sole
possibility that remains is backchaining on the second clause for
$\khodb$ again, yielding:
\begin{tabbing}
\qquad\=$\Si, x:\knat, $\=$y:\knat ; \G ;$ \=\kill
\>$\Si, x:\knat, y:\knat ; \G ;
     (\forallx {i, k} {\kadd\app \kz\app k\app i \simply
                       \khodb\app x\app i\app (\kdvar\app k)}),$\\
\>\>\>$(\forallx {i, k} {\kadd\app (\ks\app \kz)\app k\app i \simply
                       \khodb\app y\app i\app (\kdvar\app k)})
      \sseq$ \\
\>\>$\khodb\app (\kapp\app y\app x)\app (\ks\app (\ks\app \kz))\app
                (\kdapp\app (\kdvar\app (\ks\app \kz))\app (\kdvar\app (\ks\app (\ks\app \kz))))$.
\end{tabbing}
Now we can only backchain on the first clause for $\khodb$ to yield
two new proof obligations, the first of which is:
\begin{tabbing}
\qquad\=$\Si, x:\knat, $\=$y:\knat ; \G ;$ \=\kill
\>$\Si, x:\knat, y:\knat ; \G ;
     (\forallx {i, k} {\kadd\app \kz\app k\app i \simply
                       \khodb\app x\app i\app (\kdvar\app k)}),$\\
\>\>\>$(\forallx {i, k} {\kadd\app (\ks\app \kz)\app k\app i \simply
                       \khodb\app y\app i\app (\kdvar\app k)})
      \sseq$ \\
\>\>$\khodb\app y\app (\ks\app (\ks\app \kz))\app (\kdvar\app (\ks\app\kz))$.
\end{tabbing}
The only clause that we can select for backchaining is the second
dynamic clause for $y$; none of the other clauses have a matching
head. This modifies the goal to:
\begin{tabbing}
\qquad\=$\Si, x:\knat, $\=$y:\knat ; \G ;$ \=\kill
\>$\Si, x:\knat, y:\knat ; \G ;
     (\forallx {i, k} {\kadd\app \kz\app k\app i \simply
                       \khodb\app x\app i\app (\kdvar\app k)}),$\\
\>\>\>$(\forallx {i, k} {\kadd\app (\ks\app \kz)\app k\app i \simply
                       \khodb\app y\app i\app (\kdvar\app k)})
      \sseq$ \\
\>\>$\kadd\app (\ks\app \kz)\app (\ks\app \kz)\app (\ks\app (\ks\app \kz))$.
\end{tabbing}
This sequent is then proved by backchaining on clauses for \kadd. The
other proof obligation is handled similarly.

\subsection{Difficulty with dynamic contexts}
\label{subsec:dyn_ctx_problem}

We have shown how relational specifications involving higher-order
contexts can be faithfully encoded in \LProlog. We are interested in
reasoning about these specifications through their encoding by using
\Abella. For example, we might be interested in showing that the
relation that we have defined above identifies a bijective mapping
between the two representations of $\lambda$-terms. One part of
establishing this fact is proving that the relation is deterministic
from left to right, \ie, that every term in the named notation is
related to at most one term in the nameless notation.  Writing
$\oseq{\G \vdash \dbrel m h d}$ to denote derivability of the judgment
$\G \vdash \dbrel m h d$ by virtue of the rules \kdbapp, \kdbabs and
\kdbvar, this property is stated as follows:
\begin{gather*}
  \rforallx {\G, m, h, d, e}
  {\oseq{\G \vdash \dbrel m h d} \rimp
  \oseq{\G \vdash \dbrel m h e} \rimp
  d = e}.
\end{gather*}
Adopting the two-level logic notation provided in
Section\cspc\ref{subsec:twolevel}, we denote derivability of the \HHw
sequent $\Si; \Th; \D \vdash F$ by $\oseq{\D \vdash F}$ where $\Th$
contains the clauses for $\kadd$ and $\khodb$ and $\Si$ is the
signature for this encoding. Then the property above translates into
the following theorem in \Abella:
\begin{gather*}
  \rforallx {\G, m, h, d, e}
  {\oseq{\G \sseq \khodb\app m\app h\app d} \rimp
  \oseq{\G \sseq \khodb\app m\app h\app e} \rimp
  d = e}.  \tag{\kdbdet}
\end{gather*}
A proof of the property \kdbdet must obviously be based on an
induction of derivability of the \HHw sequents. A particular
difficulty in articulating such inductive arguments relative to \HHw
specifications is that they need to analyze the cases of derivations
that rely on assumptions in changing dynamic contexts. For example, a
proof of \kdbdet must accommodate the fact that $\G$ can be
dynamically extended in a derivation of $\oseq{\G \sseq \khodb\app
  m\app h\app d}$ and that the particular content of $\G$ influences
the derivation in the variable case via backchaining. Without
well-defined constraints on $\G$, it is difficult to predict how the
dynamical added clauses might be used and indeed the above property
may not even be true.

The difficulties described above can be summarized in two parts:
First, it is necessary to finitely characterize the structure of the
dynamic contexts; Second, it is necessary to have an approach to
structuring reasoning about backchaining steps on the clauses in the
dynamic contexts. These problems have straightforward solutions when
\LProlog specifications only introduce atomic formulas into the
dynamic contexts, \ie, when they are \HHtwo specifications. In this
case, we observe that the dynamic clauses must come from atomic
formulas on the left of the implication symbols in program
clauses. Since there are only finitely many such formulas in a
\LProlog specification, the dynamic contexts can be characterized as
fixed-point definitions. Moreover, backchaining the goal formula on a
dynamic clause succeeds if and only if it matches exactly with that
clause. Thus analysis of such backchaining steps is equivalent to
analysis of membership of the dynamic context. The two-level logic
approach described in Section\cspc\ref{subsec:twolevel} is based on
those observations and provides an elegant way to reason about \HHtwo
specifications. The proof of determinacy of typing in that section
illustrated this approach.

The above two problems are much more difficulty to solve when dealing
with full \HHw specifications. The \HHw logic allows for arbitrarily
nested universal and implicational formulas in program clauses. As a
result, derivations may dynamically introduce implicational or
universal clauses. Backchaining on such clauses may yield subgoals
that require sub-derivations to be constructed. These sub-derivations
may further add clauses into the dynamic context. Since this process
can be repeated in derivations, it is not clear how to characterize
the structure of the dynamically added clauses and how to effectively
perform case analysis over backchaining steps on these clauses.

However, in the example under consideration, there is an easy
resolution to these problems. We observe that initially clauses can be
introduced dynamically only by backchaining on the clause encoding
\kdbabs. These clauses all have the form
\[
  \forallx {i, k} {\kadd\app h\app k\app i \simply
    \khodb\app x\app i\app (\kdvar\app k)}
\]
where $h$ is some natural number and $x$ is some fresh
variable. Moreover, backchaining further on such clauses does not add
new clauses. The elements of the dynamic context $\G$ must therefore
all be of the above form. Thus, the structure of $\G$ can be encoded
into an inductive definition in $\Gee$ and treated in a finitary
fashion by the machinery that $\Gee$ already provides for reasoning
about backchaining steps.

\subsection{The solution to the problem}
\label{subsec:dyn_ctx_solution}

We provide a solution to the problems of reasoning about the full \HHw
specifications by extending the two-level logic approach described in
Section\cspc\ref{subsec:twolevel}. The key insight underlying our
solution is that the observation we made for the example in the last
section generalizes cleanly to other reasoning situations that involve
dynamic contexts with nested universal and implicational
formulas. Concretely, the dynamic contexts that need to be considered
in these situations are completely determined by the additions that
can be made to them. Further, the structure of such additions must
already be manifest in the original specification because the way
derivations are constructed determines only subformulas that have the
form of program clauses in the original specification can be added
dynamically. Since there are only a finite number of such forms, the
dynamic context can always be encapsulated in an inductive
definition. To take advantage of this observation we encode a version
of \HHw that is based on the technique called focusing as a
fixed-point definition in Abella to support reasoning also over the
backchaining steps that result from using dynamically added
assumptions.

We shall elaborate on this extension of the two-level logic approach
in this section. We first present a focused version of \HHw. We then
provide an encoding of it in \Gee. We lastly describe how to use the
inductive definitions of dynamic contexts and the \HHw encoding to
analyze backchaining steps on dynamic contexts. We will illustrate the
power of this approach through a complete example in the next section.

\subsubsection{A focused version of \HHw}

Focusing is a technique proposed in \cite{andreoli92jlc} for
formulating sequent calculi in a way that non-determinism in searching
for proofs is reduced while the completeness of the focused systems
with respect to the original calculi is retained. Here we shall use
focusing to characterize backchaining in a fine-grained fashion to
facilitate the analysis of backchaining on dynamic contexts. The
presentation of \HHw in Section\cspc\ref{subsec:spec_logic} already
implicitly built in the notion of focusing. To derive a \HHw sequent
$\Si; \G; \D \sseq A$ where $A$ is an atomic goal, we pick some
formula $D$ from $\G$ or $\D$ and backchain $A$ on $D$. If $D$ exactly
matches the atomic formula $A$, then the proof is completed. If $D$
has an universal or implicational structure, then we match the head of
$D$ with $A$ and try to derive its body (if any). We observe that
backchaining is completed in one big step without any
divergence. Using the terminology of focusing, the formula $D$ is
``focused'' on throughout the backchaining step. After that, we
simplify the body to atomic formulas and repeat the ``focusing'' phase
(backchaining) again. This process continues until there are no more
subgoals left to prove.

Based on the above observations, we give a focused version of \HHw
that makes the steps of ``focusing'' on formulas explicit and breaks
up the ``big-step'' backchaining into ``small-step'' rules. Besides
sequents of the form $\Si ; \G ; \D \sseq G$ which we call
\emph{goal-reduction sequents} and which we have already seen in
Section\cspc\ref{subsec:spec_logic}, we also introduce \emph{focused
  sequents} of the form
\[
\Si ; \G ; \D, [D] \sseq A
\]
where $A$ is an atomic formula and $D$ is the formula being focused
on. The rules deriving such sequents are shown in
Figure\cspc\ref{fig:focused_hhw}. The rules \andR, \impR and \forallR
reduce the goal formula when it is not atomic. The rules \prog and
\dyn respectively focus on a clause in the static and dynamic contexts
when the goal formula is atomic. The rules \match, \forallL and \impL
reduce the focused formula; together they realize the backchaining
steps on it. If we compare the system in
Figure\cspc\ref{fig:focused_hhw} and the one in
Figure\cspc\ref{fig:hhw-rules}, we will see that they have the same
reasoning power. Specifically, notice that the right rules are the
same in the two systems; successive application of $\prog$,
$\dyn$, $\forallL$, $\impL$, $\trueR$ and $\match$ for an appropriate number of
times is equivalent to applying $\backchain$.

\begin{figure}[ht!]
    \begin{gather*}
      \infer[\trueR]{
        \Si; \G; \D \sseq true
      }{}
      \qquad
      \infer[\andR]{
        \Si; \G; \D \sseq G_1 \sconj G_2
      }{
        \Si; \G; \D \sseq {G_1}
        & \Si; \G; \D \sseq G_2
      }
      \\\vspace{0.1cm}
      \infer[\impR]{
        \Si; \G; \D \sseq {D \simply G}
      }{
        \Si; \G; \D, D \sseq G
      }
      \qquad
      \begin{array}{c}
      \infer[\forallR]{
        \Si; \G; \D \sseq {\typedforall{\tau}{x}{G}}
      }{
        \Si, c:\tau; \G; \D \sseq {G[c/x]}
      }
      \\\vspace{0.1cm}
      \mbox{\small (where $c \notin \Si$)}
      \end{array}
      \\\vspace{1cm}
      \infer[prog]{
          \Si; \G; \D \sseq A
        }{
          \Si; \G; \D, [D] \sseq A
          &
          (D \in \G)
        }
      \qquad
      \infer[dyn]{
          \Si; \G; \D \sseq A
        }{
          \Si; \G; \D, [D] \sseq A
          &
          (D \in \D)
        }
      \\\vspace{1cm}
      \infer[\match]{
          \Si; \G; \D, [A] \sseq A
        }{}
      \\\vspace{0.1cm}
      \infer[\forallL]{
        \Si; \G; \D, [\forallx {x:\tau} D] \sseq A
      }{
        \Si \tseq t:\tau
        &
        \Si; \G; \D, [D[t/x]] \sseq A
      }
      \qquad
      \infer[\impL]{
        \Si; \G; \D, [G \simply A'] \sseq A
      }{
        \Si; \G; \D \sseq G
        &
        \Si; \G; \D, [A'] \sseq A
      }
    \end{gather*}
    \caption{A Focused Version of \HHw}
    \label{fig:focused_hhw}
\end{figure}

\subsubsection{Encoding of \HHw in \Gee}

To encode \HHw sequents in \Gee, we first note that \Gee and \HHw
share the same type system. The \HHw signatures can therefore be
imported transparently into \Gee, so the signatures of \HHw sequents
will not be explicitly encoded. The contexts of \HHw are represented
in \Gee as lists of \HHw formulas (\ie, lists of terms of type \omic).
The type \olist with constructors $\knil : \olist$ and $\cons :
\omic \to \olist \to \olist$ is used for these lists, and, per
tradition, the $\cons$ constructor is written infix. Membership in a
context is defined inductively as a predicate $\kmember : \omic \to
\olist \to \prop$ with these clauses:
\begin{tabbing}
\qquad\=$\kmember\app E\app (F \cons L)$ \quad \=$\rdef$ \quad\=\kill
\>$\kmember\app E\app (E \cons L)$ \>$\rdef$ \>$\rtrue$ \\
\>$\kmember\app E\app (F \cons L)$ \>$\rdef$ \>$\kmember\app E\app L$.
\end{tabbing}
Observe that the two clauses have overlapping heads; there will be as
many ways to show $\kmember\app E\app L$ as there are occurrences of
$E$ in $L$. This validates the view of \HHw contexts as multisets.

The encoding of \HHw is parameterized by the \LProlog specification
that we would like to reason about. We identify the predicate symbol
$\kprog : \omic \to \omic \to \prop$ for representing this \LProlog
specification and translate every program clause $\forallx {x_1}{
  \ldots \forallx {x_n}{G \simply A}}$ in the \LProlog specification
to the definitional clause $\kprog\app G\app A$.

The sequents of \HHw are then encoded in \Gee using the predicates
$\kseq: \knat \to \olist \to \omic \to \prop$ and $\kbc: \knat \to
\olist \to \omic \to \omic \to \prop$. A goal-reduction sequent $\Si;
\G; \D \sseq G$ is encoded as $\kseq\app N\app \D\app G$ for some $N$
and a focused sequent $\Si; \G; \D, [F] \sseq A$ is encoded as
$\kbc\app N\app \D\app F\app A$ for some $N$, where $N$ encodes
the size of a derivation for the sequent. Here we omit the static context $\G$ which is the
parameterizing \LProlog specification and encoded explicitly in \kprog
as described above. We also omit the signature $\Si$ which is absorbed
into the signature of \Gee.

The rules of the \HHw proof system in Figure\cspc\ref{fig:focused_hhw}
are used to build mutually recursive definitions of the $\kseq$ and
$\kbc$ predicates. This definition is depicted in
Figure\cspc\ref{fig:hhw_encoding}. The goal reduction rules are
systematically translated into the clauses for $\kseq$, the only
novelty being that universally quantified variables of the
specification logic are represented as nominal constants in \Gee using
the $\nabla$ quantifier. This use of $\nabla$ accurately captures the
rule for reducing universally quantified goal formulas in \HHw: to
derive such a goal, a new (nominal) constant $c$ that is different
from any other constructs or constants must be introduce for its
binding variable $x$ and the formula obtained by replacing the
occurrences of $x$ in its body with $c$ must be derivable.
The backchaining rules of \HHw are encoded as clauses for \kbc in a
straightforward manner.
For the $\dyn$ and $\match$ rules of \HHw, we have to enforce the
invariant that the goal of the sequent is atomic. This is achieved by
means of a predicate $\katomic : \omic \to \prop$ defined by the
following clause:
\begin{tabbing}
\qquad\=$\katomic\app (G_1 \simply G_2)$ \quad \=$\rdef$ \quad\=\kill
\>$\katomic\app (\sfall_\tau\app G)$ \>$\rdef$ \>$\rfalse$\\
\>$\katomic\app (G_1 \sconj G_2)$ \>$\rdef$ \>$\rfalse$\\
\>$\katomic\app (G_1 \simply G_2)$ \>$\rdef$ \>$\rfalse$\\
\>$\katomic\app true$ \>$\rdef$ \>$\rfalse$
\end{tabbing}
Effectively, \katomic characterizes atomic formulas negatively by
saying that an atomic formula cannot be constructed with an \HHw
connective or $true$. The \dyn rule transparently translates into a clause for
\kseq. The \prog rule does not have a corresponding clause. Instead,
it is absorbed into a direct encoding of backchaining on program
clauses.  Specifically, because the exact forms of the program clauses
in the \LProlog specification is known, backchaining on such clauses
%GN This is incorrect English: you cannot ``went through'' you can
%perhaps ``go through.''
%can be went through in one big step. Such backchaining is encoded in
can be analyzed in one big step. Such backchaining is encoded in
the following clause which combines the application of $\prog$ to
focus on a clause in the static context and the application of
$\forallL$, $\impL$ and $\match$ for a number of times to realize
backchaining on the clause:
\def\progbc{\kwd{prog-bc}}
\[
\kseq\app (\ks\app N)\app L\app A \quad \rdef \quad
\katomic\app A \rand
     \rexistsx {G} {\kprog\app G \app A \rand \kseq\app N\app L\app G}
\tag{\progbc}
\]

\begin{figure}[!ht]
\begin{tabbing}
\qquad\=$\kbc\app (\ks\app N)\app L\app (D_1 \sconj D_2)\app A$ \quad\=$\rdef$ \quad\=\kill
\>$\kseq\app N\app L\app true$ \>$\rdef$ \>$\rtrue$\\
\>$\kseq\app (\ks\app N)\app L\app (G_1 \sconj G_2)$ \>$\rdef$
  \>$\kseq\app N\app L\app G_1 \rand \kseq\app N\app L\app G_2$
\\
\>$\kseq\app (\ks\app N)\app L\app (D \simply G)$ \> $\rdef$
  \>$\kseq\app N\app (D \cons L)\app G$
\\
\>$\kseq\app (\ks\app N)\app L\app (\sfall_\tau\app G)$ \>$\rdef$
  \>$\nablax {x:\tau} {\kseq\app N\app L\app (G\app x)}$
\\[1ex]
\>$\kseq\app (\ks\app N)\app L\app A$ \>$\rdef$
  \>$\katomic\app A \rand
     \rexistsx {D} {\kmember\app D\app L \rand \kbc\app N\app L\app D\app A}$
\\
\>$\kseq\app (\ks\app N\app L\app A$ \>$\rdef$
  \>$\katomic\app A \rand
     \rexistsx {G} {\kprog\app G \app A \rand \kseq\app N\app L\app G}$
\\[2ex]
\>$\kbc\app (\ks\app N)\app L\app (D_1 \sconj D_2)\app A$ \>$\rdef$
  \>$\kbc\app N\app L\app D_1\app A \ror \kbc\app N\app L\app D_2\app A$
\\
\>$\kbc\app (\ks\app N)\app L\app (G \simply A')\app A$ \>$\rdef$
  \>$\kseq\app N\app L\app G \rand \kbc\app N\app L\app A'\app A$
\\
\>$\kbc\app (\ks\app N)\app L\app (\sfall_\tau\app D)\app A$ \>$\rdef$
  \>$\rexistsx {t:\tau} {\kbc\app N\app L\app (D\app t)\app A}$
\\
\>$\kbc\app N\app L\app A\app A$ \>$\rdef$ \>$\rtrue$
\end{tabbing}

  \caption{Encoding of \HHw Rules as Inductive Definitions in \Gee}
  \label{fig:hhw_encoding}
\end{figure}

%GN What does it mean to be a ``small issue''? It is an issue or is
%not. You can use the word small in front of issue if you end up
%showing that it is in the end a non-issue but you do not do that
%here.
%GN I don't actually like this discussion at all because the
%polymorphism solution is not one to the finiteness of definitions, it
%is a solution to being able to write the clauses at all.
%% It is important to note that there is a small issue with the
%% definition of \kseq, \kbc, and \katomic: they treat $(\sfall_\tau\app G)$
%% as if it were a single object, but, since the reasoning and
%% specification logics share the \STLC as the type system, it actually
%% stands for all instances for the type $\tau$. To keep these
%% definitions finite, we would require polymorphism, which we will
%% discuss later in Section\cspc\ref{sec:poly}.
We note that while the definitions of \kseq, \kbc, and \katomic that
we have presented above are sensible at a schematic level, they are
not actually ones that can be written explicitly in \Gee. The reason
for this is that some of the clauses that we have shown---such as the
one for $\kseq$ that pertains to an \HHw goal of the form
$(\sfall_\tau\app G)$---are parameterized by a type. In the present
context, an explicit rendition of such a ``clause'' would involve
writing down a separate clause for exact distinct type and there may
be an infinite collection of such types. The schematic form of
polymorphism that we introduce in Section\cspc\ref{sec:poly} will
overcome this difficulty.

%% Note that the meta-theory
%% of \Gee does not require that inductive definitions have finitely many
%% clauses, so even an infinitary interpretation of the clauses of
%% Figure\cspc\ref{fig:hhw_encoding}, as was done in~\cite{gacek12jar},
%% is compatible with our approach.

The faithfulness of our encoding allows us to state and prove known
properties of \HHw in \Gee. For example, the cut-elimination property
of \HHw can be proved as a theorem in \Gee and freely used in
reasoning. Employing such meta-properties can greatly simplify the
reasoning about \LProlog specifications in many
circumstances\cspc\cite{gacek09phd,wang13ppdp}.

\subsubsection{Reasoning about \HHw specifications through their encoding}

We are now in a position to talk about the approach to reasoning about
\HHw specifications through the above encoding of \HHw. We shall use
the curly brace notations described in Section~\ref{subsec:twolevel}
for encoding derivability in \HHw. That is, we use $\oseq{L \sseq G}$
and $\oseq{L, [D] \sseq G}$ to respectively represent the formula
$\rexistsx E {\knat\app N \rand \kseq\app N\app L\app G}$ and
$\rexistsx E {\knat\app N \rand \kbc\app N\app L\app D\app A}$ in
describing theorems. When they occur as hypotheses of a proof state,
they respectively represent a packet of formulas $\knat\app N$ and
$\kseq\app N\app L\app G$ and a packet of formulas $\knat\app N$ and
$\kbc\app N\app L\app D\app A$. The annotated style of induction and
case analysis on such notations work in a way similar to what we have
described in Section~\ref{subsec:twolevel}. That is, induction on
$\oseq{L \sseq G}$ or $\oseq{L, [D] \sseq G}$ is actually induction on
the size measures associated with them and case analysis of them is
actually case analysis on the packets they represent.

The major difficulty in reasoning about \HHw derivations is in
performing case analysis on backchaining steps, \ie, on derivations of
$\oseq{L \sseq A}$ where $A$ can either backchain on a clause in the
\LProlog specification or in $L$. Analysis of the former case is easy
because the form of program clauses in the original specification is
already known and the way backchaining proceeds is fixed; this is also
reflected in the clause \progbc that encodes backchaining on such
clauses. The difficulty part is to analyze backchaining on clauses in
the dynamic context $L$. By the encoding of \HHw, $\oseq{L \sseq A}$
is derivable only if there exists some $D$ such that $\kmember\app
D\app L$ holds---\ie, $D$ is a member of $L$---and $\oseq{L,[D] \sseq
  A}$ is derivable. Further analysis is not possible without knowing
the form of $D$. Fortunately, as we have observed previously, clauses
in $L$ must come from the original specification, thereby must have a
fixed number of forms. As a result, we can give $L$ a fixed-point
definition and prove this structural property as a theorem about the
definition. By applying the structural property of $L$, we can reveal
the possible forms of $D$. Once this is done, the way $A$ can
backchain on $D$ is fixed and we can reduce the formula $\oseq{L,[D]
  \sseq A}$ to one representing a goal-reduction sequent. Now we have
finished analysis of the backchaining step and can go on with
subsequent reasoning.

\subsection{An illustration of the extended system}
\label{subsec:hhw_exm}

We demonstrate the power of the extension we described above through
its use in explicitly proving the bijectivity property of the
transformation between $\lambda$-terms and de Bruijn terms presented
in Section\cspc\ref{subsec:hhw_mtv_exm}. We do this by showing that
\khodb is deterministic in both its first and third arguments.
As expected, we work within \Abella with the encoding of \HHw
described in the previous section. We also assume that the program
clauses defining \khodb and \kadd have been reflected into the
definition of \kprog in this context.

In the rest of this section, we describe the proof of determinacy of
\khodb. The discussion is organized as follows: We first present the
inductive definition of the dynamic contexts for \khodb derivations
and prove their structural properties through the definition, we then
present the determinacy theorems and their proofs; the structural
properties of dynamic contexts play a critical role in constructing
these proofs.

\subsubsection{Formalizing the dynamic contexts and their structural properties}

As mentioned in the previous section, we will need to finitely
characterize the dynamic contexts during the derivation of
\khodb. They are given through the following clauses that inductively
define the predicate $\kctx : \olist \to \prop$:
\begin{tabbing}
\qquad\=$\nablax {x}
   {\kctx\app ((\forallx {i, k}
                         {\kadd\app H\app k\app i \simply
                          \khodb\app x\app i\app (\kdvar\app k)})\app
               \cons L)}$
  \quad\= $\rdef$ \quad\=\kill
\>$\kctx\app \knil$ \>$\rdef$ \>$\rtrue$\\
\>$\nablax {x}
   {\kctx\app ((\forallx {i, k}
                         {\kadd\app H\app k\app i \simply
                          \khodb\app x\app i\app (\kdvar\app k)})\app
               \cons L)}$
  \>$\rdef$ \>$\kctx\app L$.
\end{tabbing}
Note again that capitalized symbols are implicitly universally
quantified over the entire clause. In the second clause, the $\nabla$
quantification guarantees that $x$ must match a nominal constant and
the quantification ordering guarantees that $x$ does not occur in
$L$. Therefore, in any $L$ for which $\kctx\app L$ holds, it must be
the case that there is exactly one clause $(\forallx {i, k} {\kadd\app
  H\app k\app i \simply \khodb\app x\app i\app (\kdvar\app k)})$ for
each such $x \in \support{L}$. It is easy to establish this fact in
terms of a pair of lemmas.

The first of these lemmas, called \kctxinv, characterizes the form of
clauses in $L$.
\begin{tabbing}
\qquad\=\qquad\=\kill
\>$\rfall L, E. \kctx\app L \rimp \kmember\app E\app L \rimp$ \\
\>\>$\rexistsx {x, H}
        {E = (\forallx {i, k} {\kadd\app H\app k\app i \simply
                               \khodb\app x\app i\app (\kdvar\app k)})
        \rand \kname\app x}$.
\end{tabbing}
Here, $\kname$ is defined by the single clause $\nablax {x}
{\kname\app x} \rdef \rtrue$. Therefore, $\kname\app x$ is a predicate
that asserts that $x$ is a nominal constant.
To prove \kctxinv, we proceed by induction on the first hypothesis,
$\kctx~L$. As mentioned in Section\cspc\ref{subsec:annot_ind}, this is
achieved by assuming a new \emph{inductive hypothesis} \IH:
\begin{tabbing}
\qquad\=\qquad\=\kill
\>$\rfall L, E. (\kctx\app L)^* \rimp \kmember\app E\app L \rimp$ \\
\>\>$\rexistsx {x, H}
        {E = (\forallx {i, k} {\kadd\app H\app k\app i \simply
                               \khodb\app x\app i\app (\kdvar\app k)})
        \rand \kname\app x}$.
\end{tabbing}
Moreover, the proof state is transformed to have the following
hypotheses converted from the assumptions of the lemma where $L$ and
$E$ are promoted to variables at the proof level:
\begin{tabbing}
\qquad\=\kill
\>$\hyp{\kwd{H1}}{\kctx\app L}^@$\\
\>$\hyp{\kwd{H2}}{\kmember\app E\app L}$
\end{tabbing}
and have
\begin{tabbing}
\qquad\=\kill
\>$\rexistsx {x, H} {E = (\forallx {i, k} {\kadd\app H\app k\app i \simply
                               \khodb\app x\app i\app (\kdvar\app k)})
        \rand \kname\app x}$
\end{tabbing}
as the new conclusion.

The \IH cannot be immediately used because the annotation of \kwd{H1}
does not match. To make progress, we need to unfold \kwd{H1} to get
$\kctx\app L'$ for some $L'$ whose annotation matches with \IH. This
amounts to finding all ways of unifying $\kctx\app L$ with the heads
of the clauses in the definition of $\kctx$. There are two cases to
consider here: when $L = \knil$ and when $L = ((\forallx {i, k}
{\kadd\app H\app k\app i \simply \khodb\app \kn\app i\app (\kdvar\app
  k)}) \cons L')$ for some new variables $H$ and $L'$ and a
nominal constant $\kn$. In the latter case we also have a new
hypothesis, $(\kctx~L')^*$, that comes from the body of the second
clause for $\kctx$. There are two things to note: first, the $\nabla$
quantified $x$ at the head of the second clause of $\kctx$ is turned
into a nominal constant in the proof obligation, and the $*$
annotation indicates that $(\kctx~L')$ is derivable in fewer steps and
hence suits \IH.

In each case for $L$, the argument proceeds by analyzing the second
hypothesis, $\kmember\app E\app L$. The case of $L = \knil$ is
vacuous, because there is no way to infer $\kmember\app E\app \knil$,
making that hypothesis equivalent to false.
In the case of
\begin{tabbing}
\qquad\=\kill
\>$L = ((\forallx {i, k} {\kadd\app H\app k\app i \simply
  \khodb\app \kn\app i\app (\kdvar\app k)}) \cons L')$,
\end{tabbing}
we have two possibilities for $\kmember\app E\app L$: either
\begin{tabbing}
\qquad\=\kill
\>$E = (\forallx {i, k} {\kadd\app H\app k\app i \simply
  \khodb\app \kn\app i\app (\kdvar\app k)})$
\end{tabbing}
or $\kmember\app E\app L'$.
The former possibility is exactly the conclusion that we seek, so this
branch of the proof finishes. The latter possibility lets us apply
\IH to the hypotheses $(\kctx\app L')^*$ and $\kmember\app E\app
L'$, which also yields the desired conclusion.

The second necessary lemma, called \kctxdet asserts that there is at
most a single clause for each variable in the dynamic context.
\begin{tabbing}
\qquad\=\qquad\=\kill
\>$\rfall L, x, H_1, H_2. \kctx\app L \rimp$ \\
\>\>$\kmember\app (\forallx {i, k}
                         {\kadd\app H_1\app k\app i \simply
                          \khodb\app x\app i\app (\kdvar\app k)})
                  L \rimp$\\
\>\>$\kmember\app (\forallx {i, k}
                         {\kadd\app H_2\app k\app i \simply
                          \khodb\app x\app i\app (\kdvar\app k)})
                  L \rimp$\\
\>\>$H_1 = H_2$.
\end{tabbing}
Note that from $H_1 = H_2$, we are able to conclude that the two
dynamic clauses relating $x$ to a de Bruijn index must be the
same. Like the previous lemma, it is proved by induction on the
hypothesis $\kctx\app L$.

\subsubsection{Proving the determinacy theorems for \khodb}

We can now show both directions of determinacy for \khodb by using the
proved lemmas of \kctx. In the forward direction the statement is as
follows.
\begin{tabbing}
\qquad\=\kill
\>$\rforallx {L, M, H, D, E}
             {\kctx\app L \rimp
              \oseq{L \sseq \khodb\app M\app H\app D} \rimp
              \oseq{L \sseq \khodb\app M\app H\app E} \rimp D = E}$.
\end{tabbing}
We prove this by induction on $\oseq{L \sseq \khodb\app M\app H\app
  D}$. The induction introduces the \IH below:
\begin{tabbing}
\qquad\=\kill
\>$\rforallx {L, M, H, D, E}
             {\kctx\app L \rimp
              \oseq{L \sseq \khodb\app M\app H\app D}^* \rimp
              \oseq{L \sseq \khodb\app M\app H\app E} \rimp D = E}$.
\end{tabbing}
Moreover, the proof state is changed to have the following hypotheses
\begin{tabbing}
\qquad\=\kill
\>$\hyp{\kwd{H1}}{\kctx\app L}$\\
\>$\hyp{\kwd{H2}}{\oseq{L \sseq \khodb\app M\app H\app D}^@}$\\
\>$\hyp{\kwd{H3}}{\oseq{L \sseq \khodb\app M\app H\app E}}$
\end{tabbing}
where ${L, M, H, D, E}$ are new variables and the conclusion to be
proved becomes $D = E$.

Now, $\oseq{L \sseq \khodb\app M\app H\app D}^@$ is just a notation
for the packet of formulas $(\knat\app N)^@$ and $\kseq\app L\app
(\khodb\app M\app H\app D)$ whose definition is given by the clauses
in Figure\cspc\ref{fig:hhw_encoding}. Unfolding the definition amounts
to finding all the clauses in Figure\cspc\ref{fig:hhw_encoding} whose
heads match $\kseq\app L\app (\khodb\app M\app H\app D)$. Only the
final two clauses of \kseq, corresponding to focusing on a dynamic
clause and backchaining on a static program clause, are therefore
relevant.

%\subsubsection{Backchaining on static clauses}

Let us consider backchaining the static clauses first. Unfolding
$\oseq{L \sseq \khodb\app M\app H\app D}^@$ introduces the following
new hypotheses for some new variable $G$:
\begin{tabbing}
\qquad\=\kill
\>$\hyp{\kwd{H4}}{\kprog\app G\app (\khodb\app M\app H\app D)}$\\
\>$\hyp{\kwd{H5}}{\oseq{L \sseq G}^*}$
\end{tabbing}
There are only a finite number of static clauses, so the assumption
$\kprog\app G\app (\khodb\app M\app H\app D)$ can be immediately
turned into a branched tree with one case for every static program
clause. The cases for the static clauses of $\kadd$ are immediately
dismissed because their heads do not match with $(\khodb\app M\app
H\app D)$. We are left with cases for the static clauses of
$\khodb$.

For the first static clause of \khodb, analysis of $\kprog\app G\app
(\khodb\app M\app H\app D)$ instantiates $M$ with $(\kapp\app M'\app
N')$, $D$ with $(\kdapp\app D'\app E')$ and $G$ with $(\khodb\app
M'\app H\app D' \sconj \khodb\app N'\app H\app E')$ for fresh
variables $M', N', D', E'$. The goal reduction sequent $\oseq{L
  \sseq G}^*$ becomes:
\begin{tabbing}
\qquad\=\kill
\>$\hyp{\kwd{H5}}{\oseq{L \sseq (\khodb\app M'\app H\app D' \sconj
                                 \khodb\app N'\app H\app E')}^*}$
\end{tabbing}
which is reduced by the second clause for \kseq to:
\begin{tabbing}
\qquad\=\kill
\>$\hyp{\kwd{H6}}{\oseq{L \sseq (\khodb\app M'\app H\app D')}^*}$\\
\>$\hyp{\kwd{H7}}{\oseq{L \sseq (\khodb\app N'\app H\app E')}^*}$
\end{tabbing}

We can almost apply the induction hypothesis \IH---we know $\kctx~L$
and \kwd{H6} already---but we still must find the third argument.  To
get this argument we need to case analyze the other hypothesis,
$\oseq{L \sseq \khodb\app M\app H\app E}$, which becomes $\oseq{L
  \sseq \khodb\app (\kapp\app M'\app N')\app H\app E}$ as a result of
the previous instantiation. It has no size annotations because the
induction was on the first hypothesis.  Nevertheless, we can perform a
case analysis of its structure by unfolding its definition. Once
again, we have a choice of using a static program clause or a dynamic
clause from $L$. If we use a static clause, then by a similar argument
to the above we will get the following fresh hypotheses, for new
variables $D''$ and $E''$ such that $E = \kdapp\app D''\app E''$:
\begin{tabbing}
\qquad\=\kill
\>$\hyp{\kwd{H8}}{\oseq{L \sseq (\khodb\app M'\app H\app D'')}}$\\
\>$\hyp{\kwd{H9}}{\oseq{L \sseq (\khodb\app N'\app H\app E'')}}$
\end{tabbing}
We can now apply the \IH twice, one to $\kctx\app L$, \kwd{H6},
\kwd{H8} and one to $\kctx\app L$, \kwd{H7}, \kwd{H9}, yielding $D' =
D''$ and $E' = E''$, so $D = \kdapp\app D'\app E' = \kdapp\app D''\app
E'' = E$.

If, on the other hand, we use a dynamic clause in $L$, then the two fresh
hypotheses we get are:
\begin{tabbing}
\qquad\=\kill
\>$\hyp{\kwd{H8}}{\kmember\app D\app L}$\\
\>$\hyp{\kwd{H9}}{\oseq{L, [D] \sseq \khodb\app (\kapp\app M'\app N')\app H\app E}}$
\end{tabbing}
for some new variable $D$. This is the first place where the
context characterization hypothesis $\kctx\app L$ becomes useful. By
applying the lemma \kctxinv above to $\kctx\app L$ and \kwd{H8}, we
should be able to conclude that $D$ is of the following form
\begin{tabbing}
\qquad\=\kill
\>$(\forallx {i, k}
     {\kadd\app H'\app k\app i \simply \khodb\app \kn\app i\app (\kdvar\app  k)})$
\end{tabbing}
for some term $H'$ and some nominal constant $\kn$. By looking at the
clauses for \kbc in Figure\cspc\ref{fig:hhw_encoding}, it is clear
that there is no way to prove the sequent \kwd{H9}, because the term
$\kn$ will never unify with $\kapp\app M'\app N'$. Hence this
hypothesis is vacuous, which closes this branch.
We have now accounted for the cases of backchaining the first static
clause of \khodb for the inductive assumption $\oseq{L \sseq
  \khodb\app M\app H\app D}^@$.

For backchaining the second static clause of \khodb, analysis of
$\kprog\app G\app (\khodb\app M\app H\app D)$ instantiates $M$ with
$(\kabs\app M')$, $D$ with $(\kdabs\app D')$ and $G$ with
\begin{tabbing}
\qquad\=\kill
\>$\forallx x
      {(\forallx {i, k} {\kadd\app H\app k\app i \simply
           \khodb\app x\app i\app (\kdvar\app k)})
       \simply {\khodb\app (M'\app x)\app (\ks\app H)\app D'}}$
\end{tabbing}
for fresh variables $M'$ and $D'$. The goal reduction sequent
$\oseq{L \sseq G}^*$ is then reduced first by the fourth and then by
the third clause for \kseq to:
\begin{tabbing}
\qquad\=\kill
\>$\hyp{\kwd{H6}}
       {\oseq{L,
          (\forallx {i, k} {\kadd\app H\app k\app i \simply
             \khodb\app \kn\app i\app (\kdvar\app k)})
          \sseq
          (\khodb\app (M'\app \kn)\app (\ks\app H)\app D')}^*}$
\end{tabbing}
where $\kn$ is a nominal constant. To apply \IH, we need another
assumption that accompanies \kwd{H6}. For this, we do a case analysis
of \kwd{H3}, which has become $\oseq{L \sseq \khodb\app (\kabs\app
  M')\app H\app E}$ as a result of the previous instantiation, by
unfolding its definition. Again, there are two cases to consider here:
either \kwd{H3} is proved by backchaining on the second static clause
of \khodb or on the dynamic context. In the former case, $E$ is
instantiated with $(\kdabs\app E')$ for some $E'$ and we got the
following assumption from the case analysis followed by goal reduction
steps:
\begin{tabbing}
\qquad\=\kill
\>$\hyp{\kwd{H7}}
       {\oseq{L,
          (\forallx {i, k} {\kadd\app H\app k\app i \simply
             \khodb\app \kn\app i\app (\kdvar\app k)})
          \sseq
          (\khodb\app (M'\app \kn)\app (\ks\app H)\app E')}^*}$
\end{tabbing}
Because $\kctx\app L$ holds, by the definition of \kctx, we know
the following hypothesis holds:
\begin{tabbing}
\qquad\=\kill
\>$\hyp{\kwd{H8}}
    {\kctx\app
        ((\forallx {i, k} {\kadd\app H\app k\app i \simply
             \khodb\app \kn\app i\app (\kdvar\app k)}) \cons L)}$
\end{tabbing}
We now apply \IH to \kwd{H8}, \kwd{H6} and \kwd{H7}, yielding $D' =
E'$, so $D = (\kdabs\app D') = (\kdabs\app E') = E$.
The case that \kwd{H3} is proved by backchaining on the dynamic
context $L$ is dismissed by following the same argument described
previously when $M$ is an application: by applying \kctxinv to
$\kctx\app L$ we observe that $(\kabs\app M')$ needs to unify with a
nominal constant, which is impossible.
We have now accounted for all the cases of backchaining a static
clause for the inductive assumption $\oseq{L \sseq \khodb\app M\app
  H\app D}^@$.

%\subsubsection{Backchaining on dynamic clauses}

We are left with only backchaining on dynamic clauses in $L$ for
$\oseq{L \sseq \khodb\app M\app H\app D}^@$. This corresponds to
reducing $\oseq{L \sseq \khodb\app M\app H\app D}^@$ by the second to
last clause of \kseq, yielding the following pair of new hypotheses:
\begin{tabbing}
\qquad\=\kill
\>$\hyp{\kwd{H4}}{(\kmember\app D\app L)^*}$\\
\>$\hyp{\kwd{H5}}{\oseq{L, [D] \sseq \khodb\app M\app H\app D}^*}$
\end{tabbing}
As these hypotheses come from unfolding an inductive assumption, they
are $^*$-annotated. Once again, we can apply the \kctxinv lemma to
conclude that
\begin{tabbing}
\qquad\=\kill
\>$D =
    (\forallx {i, k}
      {\kadd\app H'\app k\app i \simply
       \khodb\app \kn\app i\app (\kdvar\app k)})$
\end{tabbing}
for some variable $H'$ and nominal constant $\kn$. We then
perform a case analysis on \kwd{H5}. The only possibility here is to
use the definitional clauses for \kbc to unify $M$ with $\kn$ and $D$
with $(\kdvar\app k)$ for some variable $k$. Moreover, the case
analysis generates the following fresh hypothesis:
\begin{tabbing}
\qquad\=\kill
\>$\hyp{\kwd{H6}}{\oseq{L \sseq \kadd\app H'\app k\app H}^*}$
\end{tabbing}
We now analyze how the other hypothesis related to \khodb, \ie
${\oseq{L \sseq \khodb\app \kn\app H\app E}}$, can be derived. Since
\kn cannot unify with the normal constants $\kabs$ or $\kapp$, the
only way to prove ${\oseq{L \sseq \khodb\app \kn\app H\app E}}$ would
be to use a dynamic clause $D'$ in $L$. From this analysis we get the
following hypotheses:
\begin{tabbing}
\qquad\=\kill
\>$\hyp{\kwd{H7}}{\kmember\app D'\app L}$\\
\>$\hyp{\kwd{H8}}{\oseq{L, [D'] \sseq \khodb\app \kn\app H\app E}}$
\end{tabbing}
Once again, by the lemma \kctxinv and unfolding the definition of \kbc
as above, we can derive that
\begin{tabbing}
\qquad\=\kill
\>$D' =
    (\forallx {i, k}
      {\kadd\app H''\app k\app i \simply
       \khodb\app \kn\app i\app (\kdvar\app k)})$\\
\>$E = (\kdvar\app k')$
\end{tabbing}
for some variable $H''$ and $k'$ and obtain the following new
hypothesis:
\begin{tabbing}
\qquad\=\kill
\>$\hyp{\kwd{H9}}{\oseq{L \sseq \kadd\app H''\app k'\app H}}$
\end{tabbing}
We can now apply the lemma \kctxdet to \kwd{H4} and \kwd{H7} to show
that $H' = H''$.
To finish this case, we need the following lemma that shows \kadd is
deterministic in its second argument:
\begin{tabbing}
\qquad\=\qquad\=\kill
\>$\rfall {L,N_1,N_2,N_3,N_2'}.$\\
\>\>${\kctx\app L \rimp
       \oseq{L \sseq \kadd\app N_1\app N_2\app N_3} \rimp
       \oseq{L \sseq \kadd\app N_1\app N_2'\app N_3}
      \rimp N_2 = N_2'}$.
\end{tabbing}
This lemma is proved by induction on the second assumption. Note that
in the case that $\kadd\app N_1\app N_2\app N_3$ backchains on $L$, we
derive a contradiction by applying the lemma \kctxinv to show that
$\kadd$ must match with $\khodb$, which is not possible. We apply the
above lemma to \kwd{H6} and \kwd{H9}, yielding $k = k'$, so $D =
(\kdvar\app k) = (\kdvar\app k') = E$.
We now have finished proving the theorem that $\khodb$ is
deterministic in its third arguments.

The \khodb relation is also deterministic in its first argument, \ie,
given a de Bruijn indexed term, there is at most a single \HOAS term
it corresponds to. This is stated as the following theorem which is
proved in a similar fashion:
\begin{tabbing}
\qquad\=\kill
\>$\rforallx {L, M, N, H, D}
             {\kctx\app L \rimp
              \oseq{L \sseq \khodb\app M\app H\app D} \rimp
              \oseq{L \sseq \khodb\app N\app H\app D} \rimp M = N}$.
\end{tabbing}
Thus, the \khodb relation is manifestly an isomorphism between the two
representations of $\lambda$-terms.

The example we have considered here has dynamic clauses with
implicational formulas only at the top level. It is possible for \HHw
specifications to dynamically introduce clauses that have nested
implications. The same approach described in this section applies to
reasoning abut such specifications. Such an example can be found in
\cite{wang13ppdp}.

\section{Schematic Polymorphism in \Abella}
\label{sec:poly}

The logic \Gee is based on the simply typed $\lambda$-calculus. As a
consequence, we cannot identify data structures
such as lists and sets within it that will work at any type. Instead
we have to
define such data structures separately for each type at which they are
needed and prove properties about each of these versions
independently. This drawback also affects the way in which data
structures must be defined and used in the specification logic if we
are to be able to reason about them in \Abella: to be embedded into
\Abella, the specifications must also be provided separately for each
concrete type. All this has two adverse consequences for a large
development such as that involved in verified compilation. First,
it leads to code duplication and loss of modularity at the
implementation level. Second, it requires proofs that have an
identical structure to be repeated at different types, thereby
expanding the proof development effort.

These problems can be alleviated by utilizing a schematic form of
polymorphism. Such an approach is, in fact, already used in
\LProlog. Constructors for data structures such as lists can be
parameterized by types in \LProlog. This kind of type parameterization
is extended to predicates and thereby also to the \HHw clauses that
define them. However, in the end, the underlying logic and language
remain simply typed and the parameterization functions simply as a
notational shorthand for identifying separate constructors, predicate
symbols and clauses at each of their respective type instances.

Our goal in this section is to extend this kind of schematic
polymorphism to \Abella so as to allow for the kind of programming in
\LProlog that is described above and also to modularize the construction
of proofs in \Abella. We achieve this goal through the following
steps:
\begin{itemize}
\item as in \LProlog, we allow constants---which could be
  constructors or predicate names---to have types that are
  parameterized by other types, with the interpretation that they each
  actually stand for a family of constants that are obtained by
  instantiating the type parameters with concrete types;

\item we allow definitional clauses to be parameterized by types, with
  the interpretation that each such clause stands for a collection
  of clauses in \Gee, each member of which is obtained by instantiating the
  type parameters with concrete types;

\item we allow blocks of definitional clauses to be parameterized by
  types, with the interpretation that a block of this kind
  actually stands for a collection of blocks in \Gee that are obtained
  by instantiating the type parameters with concrete types;

\item we permit theorems to be parameterized by types, with the
  interpretation that each such ``theorem'' actually stands for the
  collection of theorems in \Gee that are obtained by instantiating
  the type parameters with concrete types; and

\item we lift the proof rules for \Gee to permit the construction of
  proofs for type parameterized theorems that are such that
  instantiating the proof with concrete types will yield a proof in
  \Gee for the ``concrete'' theorem.
\end{itemize}
We refer to the proofs that are constructed in this context as
\emph{schematic proofs}.

We elaborate on the steps described above in the rest of this section.
We begin by providing motivating examples for the extension. We then
describe a ``polymorphic'' form for definitions and theorems in
Section\cspc\ref{subsec:schm_poly}. In
Section~\cspc\ref{subsec:schm_prf_thy} we present a lifted form of
the proof rules for \Gee and show that using them indeed leads to
schematic proofs for polymorphic theorems under the intended
interpretation of polymorphic definitions.
We conclude the section by showing how the resulting system can be
used to construct proofs for the motivating examples.

\subsection{Some motivating examples}
\label{subsec:poly_mtv_exms}

%% When implementing a compiler, we may need to represent lists of terms
%% in the source, intermediate and target languages. Moreover, we may
%% need to prove properties about these lists in the course of verifying
%% the compiler. Towards this end, we would reflect the representations
%% of lists into \Abella and then establish the properties we need.

When implementing a compiler, we may need to represent lists of terms
in the source language. Assume that the \Abella type $\ktm$ is the
type of terms in this language, that $\ktmlist$ is the type of lists of such
terms and that the constructors for such lists are
\[
\knil : \ktmlist
\qquad
\qquad
\cons : \ktm \to \ktmlist \to \ktmlist
\]
As usual, we will write $\cons$ in infix form. In the course of
reasoning about these lists, we may need to describe a membership
relation based on them. Such a relation may be represented by the
predicate $\kmember: \ktm \to \ktmlist \to \prop$ that is defined by
the following clauses:
\begin{tabbing}
\qquad\=$\kmember\app X\app (X \cons L)$
  \quad \=$\rdef$ \quad\=\kill
\>$\kmember\app X\app (X \cons L)$
  \>$\rdef$ \>$\rtrue$\\
\>$\kmember\app X\app (Y \cons L)$
  \>$\rdef$ \>$\kmember\app X\app L$
\end{tabbing}
Now, in the reasoning process, we may need to articulate a property of
the following kind concerning list membership: If a term $M$ is a
member of the list $L$ and we know that a nominal constant $c$ does
not appear in $L$, then $c$ could not possibly appear in $M$.
This property is an easy consequence of the following statement that
we refer to as a \emph{pruning property} for lists of terms:
\[
\rforallx {M, L}
  {\nablax {x} {\kmember\app (M\app x)\app L \rimp \rexistsx {M'} {M = y \mlam M'}}}.
\]
Note that $x$ is of type $\ktm$, $M$ is of type $\ktm \to \ktm$ and
$L$ is of type $\ktmlist$ in this formula.
%
%GN I do not see that this is a beta redex
%The $\beta$-redex $(M\app x)$ in this formula captures the idea that
The expression $(M\app x)$ in this formula captures the idea that
the term that is a member of $L$ might contain the variable
$x$ that represents the nominal constant. Since $L$ is bound outside
the scope of $x$, the term instantiating it cannot contain this
nominal constant. As should
be clear from previous discussions, $(\rexistsx {M'} {M
  = y \mlam M'})$, the conclusion of the formula, asserts that $M$
represents a vacuous abstraction under the assumed circumstances.

The pruning property presented above can be proved by induction on the
definition of the membership predicate that is used in the antecedent
of the 
property. In the base case, in an assumption of the form $\kmember\app
(M\app n)\app L$, $L = (E \cons L')$ and $E = (M\app \kn)$
for some variables $E$ and $L'$ and some nominal constant
$\kn$. Since $E$ cannot depend on $\kn$, the only solution to $E =
(M\app \kn)$ is $M = y \mlam E$. Thus, this case is concluded. In the
inductive case, $L = (E \cons L')$ and $\kmember\app (M\app \kn)\app
L'$ holds. Applying the inductive hypothesis to $\kmember\app (M\app
\kn)\app L'$ concludes this case.

Suppose that the source and target languages of the compiler are
different. We would then have to use a different type, say
$\ktm'$, to represent target language expressions. To
represent lists of such expressions, we would need a new
type, say $\ktmlist'$, and also new constructors, say $\knil' :
\ktmlist'$ and $\cons' : \ktm' \to \ktmlist' \to \ktmlist'$.
%We would
%then need to implement all the list related operations afresh at this
%type in \HHw.
To reason about the properties of such lists, we would
again need to reflect this type into \Gee. As we shall see in later
chapters, it is also not unusual that
we would want to define properties such as membership for lists of
target language terms and to prove facts about it such as the
pruning property. All this work would be identical in structure to
what we have described earlier for source language terms except that
it is done now using a different type for terms.

If this kind of repetitive work looks bad, it can get worse.
As we shall see in Chapter\cspc\ref{ch:vericfl}, we are interested in
verifying compilers that transform source
programs into target code through a sequence of compilation passes. It
is often the case that these passes will take code in one intermediate
language as input and produce code in a different language as
output. We may need to repeat the sequence of steps described above
for every distinct intermediate language. This can get to be quite tedious.

The schematic polymorphism we want to introduce into  \Abella will
solve the kind of problem that we have described above.
The key is to observe that the constructors for lists differ only in
the type of the elements in the list. Similarly, the definition of
membership in lists and the pruning property can be parameterized by
this type. Finally, the proof of the pruning property does not
actually refer to the type, \ie, it has the same structure regardless
of the type of the list elements. Thus, if we could parameterize the
types of constants, definitions and theorems by types with the
interpretation that they stand for the family of their type instances
and if we could provide a means for constructing ``proofs'' that are both
independent of this type and will yield a correct proof when the type
is instantiated, then we would be able to cover all the cases in one
go.

The above discussion motivates the parameterization of a block of
definitional clauses by types. Thus, if the clauses defining
$\kmember$ constitute a block and we parameterize it by the type of
the elements of the list, what we would be doing is providing a
concise description of a collection of blocks of clauses, one for each
possible concrete type. We also want to provide for the
parameterization of individual clauses within a given block by types,
with the interpretation that such a clause abbreviates a collection of
clauses \emph{within} the block, each obtained by replacing the type
parameters by concrete types. The primary motivation for this addition
is that we want to be able to support a schematic form of polymorphism
in the writing of specifications in \HHw. As mentioned earlier, this
kind of polymorphism is already realized in \LProlog and it is
essential to the convenient development of programs in the language.

To understand why this second kind of polymorphism is needed and also
some of the issues involved in supporting it, let us suppose that we
need a ``member'' predicate for lists not just for reasoning about
compiler transformations but also for specifying or implementing
them. Following the approach we have been discussing---which is
actually implemented in \LProlog---we might identify
the predicate $\kmemb : A \to \klist\app A \to \omic$ in the
specification logic and define it using the \HHw-style clauses shown
below.
\begin{tabbing}
\qquad\=$\kmemb\app X\app (X \cons L)$
  \; \=$\limply$ \; \=\kill
\>$\kmemb\app X\app (X \cons L).$ \\
\>$\kmemb\app X\app (Y \cons L)$
  \>$\limply$ \>$\kmemb\app X\app L$.
\end{tabbing}
The token $A$ in the type shown for $\kmemb$ is to be interpreted as a
type variable that schematizes this constant in the sense that we
should think of
$\kmemb$ as representing a collection of constants, each indexed by a
concrete type. This schematization extends to the clauses that define
$\kmemb$: each of these clauses also abbreviates a collection of
clauses obtained by using $\kmemb$ and the list constructor $\cons$ at
particular concrete types. Note that if the collection of types is
infinite, then what results is in fact an infinite collection of
clauses.

Now consider what happens when we think of reflecting specifications
provided in such an abbreviated form into \Gee. Following the approach
we have described in Section~\ref{sec:embed_hhw}, we would have to
provide a definitional clause for $\kprog$ for each clause in the
specification. It can be quite tedious if we have to do this for
each type instance of the clauses for $\kmemb$. In fact, it is
something that is impossible to do when the collection of types is
infinite. Our proposal to permit type parameterization at the level of a
definitional clause is intended to solve this problem. Using such a
parameterization over suitable definitional clauses for $\kprog$, we
can provide a concise, finite description of what might actually be an
infinite collection of clauses in \Gee.

We would, of course, want to reason about the specifications that we
allow to be written in the abbreviated way that we have described. For
example, we may want to prove a pruning property that is based this
time on the membership predicate in the specification logic:
\[
\rforallx {M, L}
  {\nablax {x} {\oseq{\kmemb\app (M\app x)\app L} \rimp \rexistsx {M'} {M = y \mlam M'}}}.
\]
In proving this statement, we would eventually have to carry out a
case analysis on $\oseq{\kmemb\app (M\app x)\app L}$. This would
require us to consider the $\kprog$ clauses that encode the definition
of $\kmemb$ in the specification logic. It is not too difficult to see
how this would proceed if the statement we are trying to prove uses
$\kmemb$ at a particular type such as $\ktm \to \klist\app \ktm \to
\omic$. However, the situation changes if the statement that we are
interested in
proving is actually parameterized by types. In this case, the case
analysis would, in principle, need to consider all the different type
instantiations. Each of these could match the definitional clauses in
different ways, potentially leading to different proofs for each
case.

The problem we have described above can show up even if we confine
ourselves to reasoning about relations defined within \Gee, albeit in
a polymorphic way. Our solution to the issue in both cases is
identical. We will consider constructing only those proofs for
``polymorphic'' theorems that have an identical structure regardless
of the type instance that is chosen. This design principle translates
into two specific constraints in proof construction. First, we do not
permit the instantiations of types in goal sequents in the course of
constructing proofs. Second, we permit the use of case analysis in the
reasoning process only when the way the atomic assumption matches the
head of a definitional clause is independent of the type
instantiation; this means, in particular, that the unifiers in each
case will have an identical structure, regardless of the type
instance. The technical machinery for realizing these constraints is
what underlies our lifting of the proof rules of \Gee to the context
of schematic polymorphism.

\subsection{The schematization of definitions and theorems}
\label{subsec:schm_poly}

%% In this section we describe the light-weight polymorphism extension
%% of \Abella, which we call \emph{schematic polymorphism}.
%% %
%% The first step is to extend the term language of \Abella to allow for
%% polymorphic expressions. For this, we allow type constants to take
%% other types as arguments. We now call type constants also as
%% \emph{type constructors}. A type constructor is associated with an
%% \emph{arity} $n$ such that it can be applied to $n$ type expressions
%% (whose construction we shall talk about shortly) to form an
%% \emph{atomic type}. For example, given \klist of arity $1$ and \knat
%% of arity $0$, we can construct the atomic type $(\klist \app \knat)$.
%% %
%% The syntax of type expressions is given as follows, assuming $\tau$
%% stands for type expressions, $A$ stands for type variables, $a$ stands
%% for type constructors and $(a \app \tau\app \ldots\app \tau)$ stands for
%% atomic types:

Our interest in polymorphism arose from a desire to parameterize data
structures by types. The starting point in building in such a facility
is to generalize type constants to \emph{type constructors} that could
take other types as arguments to form atomic types. With each type
constructor must be associated an arity that indicates how many
argument types it needs. As examples, we might identify \klist as a
constructor of arity $1$ that is can be used to form the type of lists
of different types of elements. In this terminology, \knat, which
represents the types of natural numbers, would also be a type
constructor whose arity is $0$.

In addition to type constructors, type expressions will now also
include type variables. The syntax of such expressions is given by the
rule below, in which we denote type variables by $A$, a type
constructor of arity $n$ by $a_n$ and we use $\tau$ with subscripts to
denote types.
\begin{gather*}
\tau \;::=\; A \sep (\tau \to \tau) \sep (a_n \app \tau_1\app \ldots\app \tau_n)
\end{gather*}
We call type expressions not containing type variables \emph{concrete
  types} or \emph{ground types}.  We think of type variables and
expressions of the form $(a_n \app \tau_1\app \ldots\app \tau_n)$ as
atomic types and we then extend the notions of argument types and target
type in the obvious way to the type expressions defined here. We
shall refer to a type that has \prop as its target type as a predicate
type.

We will now associate \emph{type schemata} rather than types with
term-level constants. A type schema has the form $([A_1,\ldots,A_n]\tau)$
where $\tau$ is a type expression all of whose type variables are
contained in the sequence of distinct type variables
$A_1,\ldots,A_n$. We shall write
$(c:[A_1,\ldots,A_n]\tau)$ to associate such a schema with $c$.
A declaration of this kind identifies $c$ as representing a family of
constants each of which has a type obtained by instantiating
$(A_1,\ldots,A_n)$ in $\tau$ with ground types. We write
$c_{\tau_1,\ldots,\tau_n}$ to represent the constant of type
$\tau[\tau_1/A_1,\ldots,\tau_n/A_n]$ in the family represented by $c$.
For example, we can declare the following constants for constructing
lists where $\cons$ is written in infix form:
\[
\knil : [A] \klist \app A \qquad
\cons : [A] A \to \klist \app A \to \klist \app A
\]
In this context,  $\cons_{\knat}$ has the type $\knat \to \klist \app \knat \to
\klist \app \knat$ and similarly $\cons_{\ktm}$ has the type $\ktm \to \klist
\app \ktm \to \klist \app \ktm$.
%
%GN Changing this because the switch from object level to meta level
%can be a bit confusing. We can change this back if needed.
%% As another example, we can now represent the quantifiers
%% in \Gee as the following constants:
%% \[
%%   \rfall, \rexst, \nabla : [A] (A \to \prop) \to \prop
%% \]
As another example, the universal quantifier in \HHw might be
identified by the declaration $(\sfall : [A] (A \to \omic) \to
\omic)$ in \Gee. We shall use $\normalcst$ to represent the set of
constant declarations that are operative in a given context.

%% %
%% Terms are constructed in the same way as described in
%% Section\cspc\ref{subsec:spec_logic}, except that now they are build up
%% from constants and variables that may have polymorphic types.
%% %
%% For example, the following is a list of natural numbers
%% \[
%% 1 \;\cons_\knat\; 2 \;\cons_\knat\; 3 \;\cons_\knat\; \knil_\knat
%% \]
%% %
%% and the following is a list of elements of type $A$ where the
%% variables $X$, $Y$ and $Z$ are of type $A$.
%% \[
%% X \;\cons_A\; Y \;\cons_A\; Z \;\cons_A\; \knil_A
%% \]
%% We omit the type indexes of constants that occur in expressions when
%% they can be uniquely inferred from the context. For instance, the
%% example lists above can be written as $(1 \;\cons\; 2 \;\cons\; 3
%% \;\cons\; \knil)$ and $(X \;\cons\; Y \;\cons\; Z \;\cons\;
%% \knil)$. As a matter of fact, we never actually explicit index
%% constants with types in practice. Instead, we provide enough type
%% annotations so that the types of constants can always be inferred from
%% the context. Under this convention, type quantifications in constant
%% declarations are not necessary and can be omitted. For example, we can
%% write the constructors of lists as
%% \[
%% \knil : \klist \app A \qquad
%% (\cons) : A \to \klist \app A \to \klist \app A
%% \]
%% This convention will be assume in the rest of the thesis.

Like constants, terms can also be parameterized by types such
that by instantiating the type parameters with ground types they
become terms in \Gee. Of course, to function this way they must
satisfy some constraints even with variables in type expressions. We
formalize this idea via a \emph{schematic typing judgment}. Towards
this end, we identify \emph{schematic typing
  contexts for variables}, denoted by $\Si$, of the form
$(x_1:\tau_1,\ldots,x_n:\tau_n)$ where $x_i$ are variables and $\tau_i$
are type expressions.\footnote{We will extend $\Si$ to include type
  associations for predicate names being defined in a declaration
  block later in this subsection.}
   We also identify \emph{schematic typing contexts
  for nominal constants}, denoted by $\nominalcst$, of the form
$(a_1:\tau_1,\ldots,a_n:\tau_n)$ where $a_i$ are nominal constants and
$\tau_i$ are type expressions. A schematic typing judgment then
   has
the form $\Psi; \nominalcst, \Si \tseq_{\normalcst} t:\tau$ where the
free variables and nominal constants in $t$ are respectively bound in
$\Si$ and $\nominalcst$, and $\Psi$ is a set of type variables that
contains all the type variables appearing in $\Si$, $\nominalcst$, $t$
and $\tau$. We refer to the variables in $\Psi$ as the type parameters
of the judgment.

In describing the rules for deriving schematic typing judgments we
will make use of the notion of a \emph{type substitution} that is
written as $(\tau_1/A_1,\ldots,\tau_n/A_n)$. We will use the notation
$\Psi' \stseq \Phi : \Psi$ to denote the fact that $\Phi$ is
substitution $(\tau_1/A_1,\ldots,\tau_n/A_n)$ where $\Psi$ is the set of
variables $\{A_1,\ldots,A_n\}$ and, for $1 \leq i \leq n$, all the type
variables in the type expression $\tau_i$ are contained in $\Psi'$.
The rules for deriving schematic typing judgments are then those shown
in Figure~\ref{fig:schm_typing_rules}. All the rules except $\tcst$
are similar to the typing rules in the \STLC. The rule $\tcst$ assigns the
constant $c_{\tau_1,\ldots,\tau_n}$ a type expression obtained by
instantiating the type parameters of $c$ with
$(\tau_1,\ldots,\tau_n)$. Note that the right premise ensures that each
$\tau_i$ in this collection must be a well-formed type expression that
uses only the variables in $\Psi$.

\begin{figure}[ht!]
  \begin{gather*}
  \infer[\tvarnorm]{
    \Psi; \nominalcst, \Si \tseq_{\normalcst} d : \tau
  }{
    d:\tau \in \nominalcst \cup \Si
  }
  \\\vspace{0.1cm}
  \infer[\tcst]{
    \Psi; \nominalcst, \Si \tseq_{\normalcst} c_{\tau_1,\ldots,\tau_n} : \tau[\tau_1/A_1,\ldots,\tau_n/A_n]
  }{
    c:[A_1,\ldots,A_n]\tau \in \normalcst
    &
    \Psi \stseq (\tau_1/A_1,\ldots,\tau_n/A_n) : \{A_1,\ldots,A_n\}
  }
  \\\vspace{0.1cm}
  \infer[\tapp]{
    \Psi; \nominalcst, \Si \tseq_{\normalcst} t_1\app t_2 : \tau
  }{
    \Psi; \nominalcst, \Si \tseq_{\normalcst} t_1 : \tau_1 \to \tau
    &
    \Psi; \nominalcst, \Si \tseq_{\normalcst} t_2 : \tau_1
  }
  \quad
  \infer[\tabs]{
    \Psi; \nominalcst, \Si \tseq_{\normalcst} \typedabs x {\tau_1} t : \tau_1 \to \tau
  }{
    \Psi; \nominalcst, \Si, x:\tau_1 \tseq_{\normalcst} t:\tau
  }
  \end{gather*}
  \caption{The Schematic Typing Rules}
  \label{fig:schm_typing_rules}
\end{figure}

Given the rules in Figure~\ref{fig:schm_typing_rules} we have the
following lemma which states that schematic typing judgments are
preserved under the instantiation of type parameters:
\begin{mylemma}\label{thm:tyinst}
If $\Psi; \nominalcst, \Si \tseq_{\normalcst} t:\tau$ and ${\Psi'}
\stseq \Phi : \Psi$, then $\Psi'; \nominalcst[\Phi], \Si[\Phi] \tseq_{\normalcst}
       {t[\Phi]} : {\tau[\Phi]}$.
\end{mylemma}
\noindent This lemma is proved by an easy induction on $\Psi;
\nominalcst, \Si \tseq_{\normalcst} t:\tau$. When $\Psi'$ is empty,
${t[\Phi]}$ is a well-formed term with a ground type. Consequently, if
$\Psi; \nominalcst, \Si \tseq_{\normalcst} t:\tau$ holds, we can think
if $t$ as a schematic term of type $\tau$ parameterized by
$\Psi$. When $\tau$ is $\prop$, we call $t$ a schematic formula.

We now move on to schematizing definition blocks.
%% We have all the devices to talk about definitional clauses and
%% blocks of definitional clauses parameterized by types. We shall first
%% present their formalization and then provide their interpretation in
%% \Gee.
As a first step, we identify a \emph{pre-definitional clause} as an
expression of the form
\[
\rforallx {\vec{x}:\vec{\tau_x}} {(\nablax {\vec{z}:\vec{\tau_z}} A) \rdef B}
\]
in which $A$ must have the structure of an atomic formula that
contains no nominal constants and all of whose free variables appear
in $\vec{x}$ or $\vec{z}$ and $B$ has the structure of a formula that,
once again, does not contain any nominal constants and all of whose
free variables also occur free in $(\nablax {\vec{z}:\vec{\tau_z}}
A)$. If the constant at the head of $A$ is $c$, then the clause is
said to be \emph{for} $c$.\footnote{The expressions ``atomic
  formula,'' ``formula'' and the ``head'' of an atomic formula are
  being used loosely here because we have not yet enforced typing
  constraints on $A$ and $B$. However, the structural properties being
  imposed should be clear and will become even more
  apparent after the definition of a well-defined schematic definition
  block that is provided below.}
Let $C$ denote a pre-definitional clause. A \emph{schematic
  definitional clause} has the form $[\Psi]C$ where $\Psi$
is a set of type variables; $\Psi$ is said to be the set of types
parameterizing the definitional clause.
%GN this statement does not make sense here because we have not talked
%about types and typing for a schematic definitional clause
%yet. Perhaps a better idea to bring this issue up in the examples.
% Note that $\Psi$ may contain
% only some some of the type variables appearing in the type expressions
% in $C$.
Finally, a \emph{block of definitional clauses parameterized by types}
or a \emph{schematic definition  block} is constituted by a finite set of
type variables $\Psi'$, a finite set of predicate constants
$\{c_1:\tau_1,\ldots,c_n:\tau_n\}$ and a collection of schematic
definitional clauses each of which is for one of the constants in
$c_1,\ldots,c_n$. Such a definitional block is said to be
parameterized by the type variables in $\Psi'$.

Not all schematic definition blocks are considered to be
well-formed. To have that property, they must satisfy the typing
constraints that are described below.
\begin{mydef}\label{def:schm_block}
Given a schematic definitional block parameterized by $\Psi'$ and its
associated predicate constants $\{c_1:\tau_1,\ldots,c_n:\tau_n\}$. Let
$\Si = (c_1:\tau_1,\ldots,c_n:\tau_n)$. Then this block is
well-defined if for 
every clause $[\Psi]\rforallx {\vec{x}:\vec{\tau_x}} {(\nablax
  {\vec{z}:\vec{\tau_z}} A) \rdef B}$ in it, $\Psi$ is disjoint from
$\Psi'$, the typing judgments $
\Psi', \Psi; \Si, \vec{x}:\vec{\tau_x},\vec{z}:\vec{\tau_z}
\tseq_{\normalcst} A:\prop$ and $\Psi', \Psi; \Si,
\vec{x}:\vec{\tau_x} \tseq_{\normalcst} B:\prop$ hold,
and all the type variables that occur in $B$ also
occur in $\nablax {\vec{z}:\vec{\tau_z}} A$.
\end{mydef}
\noindent Observe that the way we have phrased the definition above,
the constants $c_1,\ldots,c_n$ identified by a schematic definitional
block are required to be used at their ``defined types'' at every
occurrence in the block. Thus, these constants are schematic at the
block level. When these constants are added to the signature for
defining other blocks or in writing formulas to be proved, then each
$c_i$ is to be thought of as having the type expression $[\Psi']\tau_i$
associated with it, \ie, it can be used at instances of its schematic
type.
Note also that we require that all the type variables in the body of a
schematic definitional clause occur in its head. This requirement
ensures that whenever the type variables in the head are fixed the
type variables in the body are also fixed. As we shall see in
Section~\ref{subsec:schm_prf_thy}, this is a desired property for case
analysis on the schematic definitional clauses in the construction of
schematic proofs.

A schematic definition block is intended to be an abbreviated
representation of a collection of definition blocks in \Gee that are
obtained via type instantiations as follows. First, we instantiate the
type variables that parameterize the block with concrete types. Then,
within the structure of each ``block'' generated in this fashion, we
generate all the versions of each schematic definitional clause it
contains by instantiating the type variables that parameterize it with
all available concrete types. Using Lemma~\ref{thm:tyinst}, it can be
seen that each block that results from this process is a well-formed
definition block in \Gee. A point to be noted is that both the
collection of definition blocks and the collection of definitional
clauses within each block is sensitive to the vocabulary of types in
existence at a particular point. However, the schematic proofs whose
construction we will support will be such that they will allow us to
prove only those statements whose instances have derivations in \Gee
independently of the available type signature.

We consider some examples to illustrate the definition we have
presented. Each of the following expressions is a schematic
definitional clause:
\begin{tabbing}
\qquad\=$\kmember\app X\app (X \cons L)$
  \quad \=$\rdef$ \quad\=\kill
\>$\kmember\app X\app (X \cons L)$
  \>$\rdef$ \>$\rtrue$\\
\>$\kmember\app X\app (Y \cons L)$
  \>$\rdef$ \>$\kmember\app X\app L$
\end{tabbing}
The set of variables that parameterize each of the clauses above is
empty. Assuming the type schema we have provided earlier for $\cons$,
combining these clauses with the predicate signature $\kmember : A \to
\klist\app A \to \prop$ and parameterizing the result by $A$
produces a well-formed schematic definition block; in ensuring that
schematic definitional clauses within the block ``type-check,'' we
will have to use $\cons$ in them at the type $A \to \klist\app A \to
\klist\app A$. To get definition blocks in \Gee from this schematic
definition block, we would have to instantiate $A$ with concrete
types. If we instantiate it with $\ktm$, we would get the definition
block
\begin{tabbing}
\qquad\=$\kmember_\ktm \app X\app (X \cons_{\ktm} L)$
  \quad \=$\rdef$ \quad\=\kill
\>$\kmember_\ktm \app X\app (X \cons_{\ktm} L)$
  \>$\rdef$ \>$\rtrue$\\
\>$\kmember_\ktm \app X\app (Y \cons_{\ktm} L)$
  \>$\rdef$ \>$\kmember_\ktm \app X\app L$
\end{tabbing}
We can similarly generate a definition block that, for instance,
provides us clauses for $\kmember_\knat$.

As another example, suppose that our signature contains the following constant
\[\kmemb : [B] B \to \klist\app B \to \omic\]
and then consider the following
as a schematic definitional clause:
\begin{tabbing}
\qquad\=\kill
\>$[A] \kprog\app \ktrue\app (\kmemb_{A}\app X\app (X \cons L)) 
\rdef \top$
\end{tabbing} 
It represents a number of clauses
for \kprog, one in fact for each concrete type that can be used to
instantiate $A$.

Given the schematic definitions, we would like to state theorems about
them. We adopt a schematic view of such theorems as well. A
\emph{schematic theorem} has the form $[A_1,\ldots,A_n] F$ that is such
that $A_1,\ldots,A_n; \emptyset \tseq F : \prop$ holds; intuitively, $F$
must be a closed formula such that all the type variables occurring in
it are contained in $\{A_1,\ldots,A_n\}$. We say that such a theorem is
parameterized by the type variables $A_1,\ldots,A_n$. Given the ground
types $\tau_1,\ldots,\tau_n$, we can
generate the theorem $F[\tau_1/A_1,\ldots,\tau_n/A_n]$ in \Gee. In
effect, a schematic theorem parameterized by a non-empty set of type
variables stands for an infinite collection of theorems in \Gee under
the instantiation of its parameterizing type variables with ground
types. In other words, a theorem of this kind should be provable with
our extended machinery only if every type instance is provable in
\Gee. A schematic theorem not parameterized by any type variable
coincides with a theorem in \Gee.

As an example, consider the pruning property of \kmember we mentioned
in Section\cspc\ref{subsec:poly_mtv_exms}. Given variables $M$ of type
$(B \to A)$, $L$ of type $\klist\app A$, $x$ of type $B$ and $M'$ of
type $A$, that theorem can be formulated as the following schematic
theorem where $\kmember$ is an instance of the membership predicate and
has the type
$A \to \klist\app A \to \prop$:
\[
[A,B] \rforallx {M, L}
  {\nablax {x} {\kmember\app (M\app x)\app L \rimp \rexistsx {M'} {M = y \mlam M'}}}.
\]
By instantiating $B$ with $\ktm$ and A with $\ktm$, we obtain the
pruning property for lists of $\ktm$, where $M$ is of type $(\ktm \to
\ktm)$, $L$ is of type $\klist\app \ktm$, $x$ and $M'$ is of type
$\ktm$:
\[
\rforallx {M, L}
  {\nablax {x} {\kmember_\ktm\app (M\app x)\app L \rimp \rexistsx {M'} {M = y \mlam M'}}}.
\]
Similarly, by instantiating $B$ with $\ktm'$ and A with $\ktm'$, we
obtain the pruning property for lists of $\ktm'$. We shall discuss an
approach to proving such schematic theorems in the next section.

The ability to parameterize each definitional clause by type variables
allows us to encode \HHw in a situation where specifications utilize
the schematic polymorphism supported by \LProlog. Crucial to the
embedding of \HHw in \Gee are definitional clauses for \kprog that
encode \HHw clauses. However, we have already seen how this encoding
can be realized. For example, suppose that the \HHw program has the
clause
\begin{tabbing}
\qquad\=$\kmemb\app X\app (X \cons L)$
  \; \=$\limply$ \; \=\kill
\>$\kmemb_{A}\app X\app (X \cons L).$ \\
\>$\kmemb_{A}\app X\app (Y \cons L)$
  \>$\limply$ \>$\kmemb_{A}\app X\app L$.
\end{tabbing}
where \kmemb has been defined to have the type $A \to \klist\app A \to
\omic$ that is polymorphic in the \LProlog sense. We would first
translate $\kmemb$ into a constant of the same name that has a type
scheme associated with it and we would add the clauses
\begin{tabbing}
\qquad\=\kill
\>$[A] \kprog\app \ktrue\app (\kmemb_{A}\app X\app (X \cons L))
\rdef \top$\\
\>$[A] \kprog\app (\kmemb_{A}\app X\app L)\app (\kmemb_{A}\app X\app (Y \cons L))
\rdef \top$
\end{tabbing}
to the appropriate schematic definition block.

In Section\cspc\ref{subsec:dyn_ctx_solution}, we pointed out a problem
in the encoding of \HHw that manifest itself in the following clauses
that treat quantifiers:
\begin{tabbing}
\qquad\=$\kbc\app L\app (D_1 \sconj D_2)\app A$ \quad\=$\rdef$ \quad\=\kill
\>$\kseq\app L\app (\sfall_\tau\app G)$ \>$\rdef$
  \>$\nablax {x:\tau} {\kseq\app L\app (G\app x)}$\\
\>$\kbc\app L\app (\sfall_\tau\app D)\app A$ \>$\rdef$
  \>$\rexistsx {t:\tau} {\kbc\app L\app (D\app t)\app A}$\\
\>$\katomic\app (\sfall_\tau\app G)$ \>$\rdef$ \>$\rfalse$
\end{tabbing}
This problem is easily solved with our extended syntax. In particular,
our encoding can use the following schematic definitional clauses:
\begin{tabbing}
\qquad\=$[\tau]\kbc\app L\app (D_1 \sconj D_2)\app A$ \quad\=$\rdef$ \quad\=\kill
\>$[\tau]\kseq\app L\app (\sfall_\tau\app G)$ \>$\rdef$
  \>$\nablax {x:\tau} {\kseq\app L\app (G\app x)}$\\
\>$[\tau]\kbc\app L\app (\sfall_\tau\app D)\app A$ \>$\rdef$
  \>$\rexistsx {t:\tau} {\kbc\app L\app (D\app t)\app A}$\\
\>$[\tau]\katomic\app (\sfall_\tau\app G)$ \>$\rdef$ \>$\rfalse$
\end{tabbing}
%
% Note that because of this treatment, the encoding of \HHw contains an
% infinite number of clauses and therefore cannot be treated as an
% inductive definition. This means we cannot construct proofs by direct
% induction on judgments of the form $\oseq{L \sseq G}$ or $\oseq{L, [D]
%   \sseq G}$. This problem can be easily solved by adding an extra
% argument that encodes the height of the \HHw derivation to \kseq and
% \kbc and perform induction on this argument instead. Such an encoding
% of \HHw has been developed and its uses in the two-level logic
% approach has been investigated before
% (see\cspc\cite{gacek08lfmtp,gacek09phd}). We therefore omit a
% discussion of it.

Our extension also allows us to formulate theorems about polymorphic
specifications in \HHw.
% Continuing the above example, the polymorphic program clauses
% for \kmemb translate into the following clauses for \kprog:
% %
% \begin{tabbing}
% \qquad\=$[A''] \kprog\app (\kmemb_{A''}\app X\app (Y \cons L))\app
%         (\kmemb_{A''}\app X\app L)$ \quad\=$\rdef$ \quad\=\kill
% \>$[A'] \kprog\app (\kmemb_{A'}\app X\app (X \cons L))\app \ktrue$
%   \>$\rdef$ \>$\rtrue$\\
% \>$[A''] \kprog\app (\kmemb_{A''}\app X\app (Y \cons L))\app
%    (\kmemb_{A''}\app X\app L)$
%   \>$\rdef$ \>$\rtrue$.
% \end{tabbing}
% %
For example, we can state the pruning property of \kmemb that we
discussed in the previous subsection as the following
schematic theorem, where $M$ is of type $(B \to A)$, $L$ is of type
$\klist\app A$, $x$ is of type $B$ and $M'$ is of type $A$:
\[
[A,B] \rforallx {M, L}
  {\nablax {x} {\oseq{\kmemb\app (M\app x)\app L}
      \rimp \rexistsx {M'} {M = y \mlam M'}}}.
\]
% We discuss an approach to constructing proofs for such schematic
% theorem in the next section.

\subsection{Proving schematic theorems}
\label{subsec:schm_prf_thy}

A schematic theorem stands for a possibly infinite collection of
theorems obtained by instantiating its parameterizing type
variables with concrete types. One possible way to try to prove such a
theorem is by providing a collection of possibly different proofs that
cover all the type instances. However, there are two drawbacks with
this approach. First, it could involve the kind of duplication of
effort that we are wanting to avoid by introducing schematic
polymorphism. Second, and perhaps more importantly, we do not intend
the schematic theorem to be true for only those concrete type
instances that are known in a particular context. In fact, we intend
these theorems to hold for any vocabulary of types; the ``closed world
assumption'' applies only to the collection of defining
clauses available for a predicate.

% Every theorem in this collection must be proved in order to
% prove the original schematic theorem. One way to achieve this is to
% provide an individual proof for every theorem in the collection. This
% is not practical because the collection of theorems may be
% infinite. We observe that there exists a large class of schematic
% theorems such that for that every theorem in it the proofs of its
% instances have the same structure if we ignore their type
% information. The pruning property for the membership relation as
% described in Section\cspc\ref{subsec:poly_mtv_exms} is such an
% example.

To avoid these pitfalls, we propose to support the construction of
only those proofs for schematic theorems that work \emph{the same way} at
\emph{all} types, \ie, our proofs must be structures that yield
concrete proofs for concrete types simply through the process of type
instantiation. We refer to such proofs as \emph{schematic proofs}.
The basis for realizing our approach is to lift the proof rules for
\Gee to a set of \emph{schematic proof rules}; each of these schematic
proof rules must yield actual proof rules in \Gee under type
instantiation. Most of the proof rules for \Gee are easily made
schematic. The one complicated case is that of the \defL rule that
supports case analysis. Much of the discussion in this subsection is
devoted to a consideration of this rule.

To begin the formalization process, we will assume that our schematic
proof rules derive \emph{schematic sequents} that have the following
form
\[
  \Psi; \Si : \G \rseq B.
\]
These sequents augment those in \Gee with a set $\Psi$ of type
variables that binds the type variables in $\Si$, $\G$ and
$B$. This set will remain unchanged throughout the derivation of the
sequent. Thus, the variables in this set will function as placeholders
for arbitrary types but will be like ``black boxes'' in that they will
not allow us to look at or use the particular structures of the types
that fill them. To prove a schematic theorem $[A_1,\ldots,A_n]F$, we need
to derive the sequent $A_1,\ldots,A_n; \emptyset : \emptyset \rseq F$.

The schematic versions of the core rules in
Figure\cspc\ref{fig:gee_core_rules} are shown in
Figure~\ref{fig:schm_core_rules}. They are obtained from the rules in
Figure~\ref{fig:gee_core_rules} by abstracting their conclusions and
premises over the set $\Psi$ of type variables. It is easy to see the
schematic nature of these rules: by instantiating the premises and
conclusions of each rule with any substitution of ground types for
variables in $\Psi$, we get a rule in \Gee.

\begin{figure}[ht!]
  \begin{gather*}
    \infer[\srinit]{
      \Psi; \Si: \G, B \rseq B'
    }{
      B \feq B'
    }
    \quad
    \infer[\srcut]{
      \Psi; \Si: \G \rseq C
    }{
      \Psi; \Si: \G \rseq B
      &
      B \feq B'
      &
      \Psi; \Si: \G, B' \rseq C
    }
  \end{gather*}
  \\
  \vspace{-1cm}
  \begin{gather*}
    \infer[\srcontra]{
      \Psi; \Si: \G, C \rseq B
    }{
      \Psi; \Si: \G, C, C \rseq B
    }
    \quad
    \infer[\srfalseL]{
      \Psi; \Si: \G, \rfalse \rseq B
    }{}
    \quad
    \infer[\srtrueR]{
      \Psi; \Si: \G \rseq \rtrue
    }{}
  \end{gather*}
  \\
  \vspace{-1cm}
  \begin{gather*}
    \infer[\srandL, i\in\{1,2\}]{
      \Psi; \Si: \G, B_1 \rand B_2 \rseq C
    }{
      \Psi; \Si: \G, B_i \rseq C
    }
    \quad
    \infer[\srandR]{
      \Psi; \Si: \G \rseq B \rand C
    }{
      \Psi; \Si: \G \rseq B
      &
      \Psi; \Si: \G \rseq C
    }
  \end{gather*}
  \\
  \vspace{-1cm}
  \begin{gather*}
    \infer[\srorL]{
      \Psi; \Si: \G, B \ror C \rseq D
    }{
      \Psi; \Si: \G, B \rseq D
      &
      \Psi; \Si: \G, C \rseq D
    }
  \end{gather*}
  \\
  \vspace{-1cm}
  \begin{gather*}
    \infer[\srorR, i\in\{1,2\}]{
      \Psi; \Si: \G \rseq B_1 \ror B_2
    }{
      \Psi; \Si: \G \rseq B_i
    }
  \end{gather*}
  \\
  \vspace{-1cm}
  \begin{gather*}
    \infer[\srimpL]{
      \Psi; \Si: \G, B \rimp C \rseq D
    }{
      \Psi; \Si: \G \rseq B
      &
      \Psi; \Si: \G, C \rseq D
    }
    \qquad
    \infer[\srimpR]{
      \Psi; \Si: \G \rseq B \rimp C
    }{
      \Psi; \Si: \G, B \rseq C
    }
  \end{gather*}
  \\
  \vspace{-1cm}
  \begin{gather*}
    \infer[\srfallL]{
      \Psi; \Si: \G, \typedrforall{\tau}{x}{B} \rseq C
    }{
      \Psi; \nominalcst, \Si \tseq_{\normalcst} t : \tau
      &
      \Psi; \Si: \G, B[t/x] \rseq C
    }
  \end{gather*}
  \\
  \vspace{-1cm}
  \begin{gather*}
    \infer[\srexstR]{
      \Psi; \Si: \G \rseq \typedrexists{\tau}{x}{B}
    }{
      \Psi; \nominalcst, \Si \tseq_{\normalcst} t : \tau
      &
      \Psi; \Si: \G \rseq B[t/x]
    }
  \end{gather*}
  \\
  \vspace{-1cm}
  \begin{gather*}
    \begin{array}{c}
      \infer[\srexstL]{
        \Psi; \Si: \G \typedrexists{\tau}{x}{B} \rseq C
      }{
        \Psi; \Si, h: \tau': \G ,B[(h\app a_1\app \ldots\app a_n)/x] \rseq C
      }
      \\[5pt]
      \infer[\srfallR]{
        \Psi; \Si: \G \rseq \typedrforall{\tau}{x}{B}
      }{
        \Psi; \Si, h:\tau': \G \rseq B[(h\app a_1\app \ldots\app a_n)/x]
      }
      \\[3pt]
      \mbox{\small
       assuming that  $\support{B} = \{a_1,\ldots,a_n\}$, that, for $1 \leq i \leq n$,
        $a_i$ has type $\tau_i$} \\
      \mbox{\small
       $h$ is variable of type $\tau_1 \to \ldots \to \tau_n \to \tau$
      and $h \not\in \dom{\Si}$ in \srfallR and \srexstL}
     \end{array}
  \end{gather*}
  \\
  \vspace{-1cm}
  \begin{gather*}
    \infer[\snablaL]{
      \Psi; \Si: \G, \typednabla{\tau}{x}{B} \rseq C
    }{
      \Psi; \Si: \G, B[a/x] \rseq C
    }
    \qquad
    \infer[\snablaR]{
      \Psi; \Si: \G \rseq \typednabla{\tau}{x}{B}
    }{
      \Psi; \Si: \G \rseq B[a/x]
    }
    \\
    \mbox{\small
      where $a \not\in \support{B}$ in \nablaL and \nablaR}
  \end{gather*}
  \caption{The Schematic Core Rules}
  \label{fig:schm_core_rules}
\end{figure}

To describe schematic forms of the \defL and \defR rules, we introduce
the notion of \emph{the reduced form} of a schematic definitional
clause.
\begin{mydef}\label{def:red_schm_clause}
Let $[\Psi]C$ be a schematic definitional clause in a schematic
definition block parameterized by $\Psi'$. Then the reduced form of
$[\Psi]C$ is $[\Psi'']C$ where $\Psi''$ is the subset of
$\Psi \cup \Psi'$ such that $A \in \Psi''$ if and only if $A$ occurs in
$C$.
\end{mydef}
%
% \begin{mydef}\label{def:red_schm_clause}
% Let $[\Psi]C$ be a schematic definitional clause in a schematic
% definition block
% parameterized by $\Psi'$ and associated with predicate constants
% $c_1:\tau_1,\ldots,c_n:\tau_n$. Let $\vec{A_i}$ be the free variables in
% $\tau_i$ and $c_i' = {c_i}_{\vec{A_i}}$ for $1 \leq i \leq n$. Let
% $\theta$ be the substitution $(c_1'/c_1,\ldots,c_n'/c_n)$. Then the
% schematic definitional clause derived from $[\Psi]C$ is
% $[\Psi''](C[\theta])$ where $\Psi''$ is the subset of $\Psi \cup
% \Psi'$ such that $A \in \Psi''$ if and only if $A$ occurs in
% $(C[\theta])$.
% \end{mydef}
%
\noindent Intuitively, the reduced form of a schematic definitional
clause is obtained by ``pulling down'' the type parameters at the
block level to the clause level and by further removing type variables
that do not occur in the clause. Note that because the type
variables in the body of the clause must occur in its head, $\Psi''$
contains exactly the variables that occur in the clause head.

The schematic version of \defR rule can be easily derived by
abstracting it over a set $\Psi$ of type variables as before. To
present this rule we need to formalize the notion of instances of
schematic definitional clauses, as follows. Given a schematic
definitional clause $[A_1,\ldots,A_n]\rforallx{\vec{x}} {(\nablax
  {\vec{z}} A) \rdef B}$, a type substitution $\Phi =
[\tau_1/A_1,\ldots,\tau_n/A_n]$ and a substitution $\theta$ that assigns
distinct nominal constants to $\vec{z}$ and terms not containing such
constants to $\vec{x}$, we say that $A[\theta][\Phi] \rdef
B[\theta][\Phi]$ is an instance of the original clause. Then the
schematic definition right rule is depicted in
Figure\cspc\ref{fig:schm_defr}.

\begin{figure}[ht!]
  \begin{gather*}
    \infer[\sdefR]{
      \Psi; \Si: \G \rseq p\app \vec{t}
    }{
      \Psi; \Si: \G \rseq B
    }
    \\
    \mbox{\small where $p\app \vec{t} \rdef B$ is an instance
      of the reduced form of some schematic definitional clause}
  \end{gather*}
  \caption{The Schematic Definition Right Rule}
  \label{fig:schm_defr}
\end{figure}

% The definition left rule cannot be easily made schematic. The major
% difficulty lies in the fact that, in general, the solutions to
% unification problems involving polymorphic expressions may vary under
% instantiation of types. However, since we are only interested in
% developing schematic proofs that are sound, we can derive the
% schematic definition left rule from \defLCSU by restricting it to work
% only for unification problems with solutions that do not vary under
% the instantiation of types. We characterize such solutions as follows:

We will consider treating atomic formulas on the assumptions side of a
sequent only by a schematic version of the \defLCSU rule. In
describing a lifted form of this rule, we have to pay
attention to the fact that the structure of unifiers in the context of
the simply typed $\lambda$-calculus can depend on the particular types
assigned to constants\cspc\cite{nadathur92types}. To get around this
issue, we will limit the schematic version of the rule to apply only
in those cases where the complete set of unifiers can be calculated
without paying attention to types. The following definition provides
the basis for doing so by lifting the notion of a CSU to schematic
terms.
% The definition left rule cannot be easily made schematic. The major
% difficulty lies in the fact that, in general, the solutions to
% unification problems involving polymorphic expressions may vary under
% instantiation of types. However, since we are only interested in
% developing schematic proofs that are sound, we can derive the
% schematic definition left rule from \defLCSU by restricting it to work
% only for unification problems with solutions that do not vary under
% the instantiation of types. We characterize such solutions as follows:
%
\begin{mydef}\label{def:gencsu}
  Given two schematic terms $B$ and $C$ of the same type parameterized
  by $A_1,\ldots,A_n$, a type generic
  complete set of unifiers (type generic CSU) for $B$ and $C$, denoted
  by
  $\gencsu{B}{C}$, is a set of substitutions for variables in $B$ and
  $C$ such that for any ground types $\tau_1,\ldots,\tau_n$ the set
  $\{\theta[\tau_1/A_1,\ldots,\tau_n/A_n] \sep \theta \in
  \gencsu{B}{C}\}$ is a complete set of unifiers for
  $B[\tau_1/A_1,\ldots,\tau_n/A_n]$ and $C[\tau_1/A_1,\ldots,\tau_n/A_n]$.
\end{mydef}
One approach to calculating type generic CSUs for two terms is to use
the unification algorithm due to Miller that is known as \emph{pattern
  unification}\cspc\cite{miller91jlc}. This approach in fact suffices
for all the theorems we have considered in the work in this thesis.

A schematic version of the \defLCSU rule will require us to analyze
all the ways in which an atomic formula can match with the head of a
schematic definitional clause. In doing so, we need to consider the
possibility that there are nominal constants in the formula and,
conversely, that the $\nabla$ quantifiers in the head will be
instantiated with nominal constants. To deal with this issue, we once
again use the idea of raising.
In Section\cspc\ref{subsec:def_rules}, we have defined the notions of
raising a definitional clause over a sequence of nominal constants and
away from the variables in a sequent, and of raising a sequent
over a sequence of nominal constants.
These notions are, in a sense, orthogonal to typing structure and
therefore adapt in an obvious way to the situation of schematized
sequents and definitional clauses.
We shall assume such an adaptation here.
% Given a schematic definitional clause
% $
% [A_1,\ldots,A_n]\rforallx {x_1,\ldots,x_n} {(\nablax {\vec{z}} A) \rdef
%   B},
% $
% a version of it raised over the sequence of nominal constants
% $\vec{a}$ away from a signature $\Si$ is identified as the formula
% \[
% [A_1,\ldots,A_n]\rforallx {h_1,\ldots,h_n} {\nablax {\vec{z}} {A[(h_1\app
%       \vec{a})/x_1,\ldots,(h_n\app \vec{a})/x_n]} \rdef B[(h_1\app
%     \vec{a})/x_1,\ldots,(h_n\app \vec{a})/x_n]}
% \]
% where $h_1,\ldots,h_n$ are variables that do not occur in $\Si$.
% %
% We similarly define the raising of a sequent over a sequence of
% nominal constants. This definition is identical to the one provided in
% Section\cspc\ref{subsec:def_rules} and we do not repeat it here.

For us to be able to use case analysis over a particular atomic
formula in proving a schematic sequent in a schematic way, it is
necessary for the analysis to be generic with respect to \emph{all}
available schematic definitional clauses. We identify a set of
requirements that ensure this to be the case below.
\begin{mydef}\label{def:amenable}
Let $S = (\Psi; \Si:
\G, p\app \vec{t} \rseq D)$ be a schematic sequent and let
$\support{p\app \vec{t}}$ be $\{\vec{a}\}$. Let $C$
be a schematic definitional clause and $[A_1,\ldots,A_n]\rforallx {\vec{x}} {(\nablax {\vec{z}}
  A)\rdef B}$ be $C$
raised over $\vec{a}$ and away from $\Si$. Also let $\Psi \cup \{A_1,\ldots,A_n\}; \Si': \G',
p\app \vec{t'} \rseq D'$ be a version of the original sequent raised
over a sequence of distinct nominal constants $\vec{c}$ for
$\vec{z}$.
Then $S$ is said to be analyzable in a generic way with respect to the
clause $C$ and the atomic formula $(p\app t)$ if one of the following
conditions hold for every permutation $\pi$ of the nominal constants
$\{\vec{c},\vec{a}\}$:
\begin{enumerate}
\item $(p\app \vec{t'})$ and $(\pi.A[\vec{c}/\vec{z}])$ are not
  unifiable under any instantiation of type variables.

\item For some type expressions $\tau_1,\ldots,\tau_n$ whose type
  variables are bound in $\Psi$, these is a type generic CSU for
  $(p\app \vec{t'})[\tau_1/A_1,\ldots,\tau_n/A_n]$\footnote{These type
    substitutions must be applied to $(p\app t')$ as well because the
    type variables $A_1,\ldots,A_n$ may appear in the types of the
    formula via the nominal constants in $\vec{c}$.}and
  $(\pi.A[\vec{c}/\vec{z}])[\tau_1/A_1,\ldots,\tau_n/A_n]$ and for any
  other type substitution $\Phi$ for $A_1,\ldots,A_n$, it is the case that
  the formulas $(p\app \vec{t'})[\Phi]$ and
  $(\pi.A[\vec{c}/\vec{z}])[\Phi]$ are not unifiable.
\end{enumerate}
The sequent $S$ is said to be amenable to case analysis with respect
to the atomic formula $(p\app t)$ if it is analyzable with respect to
the reduced version of every available definitional clause.
\end{mydef}
%
% Suppose that we want to prove the schematic sequent $S = (\Psi; \Si:
% \G, p\app \vec{t} \rseq C)$ where $\support{p\app \vec{t}} =
% \{\vec{a}\}$. Let $[A_1,\ldots,A_n]\rforallx {\vec{x}} {(\nablax {\vec{z}}
%   A)\rdef B}$ be a reduced version of a schematic definitional clause
% raised over $\vec{a}$ and away from $\Si$. Also let $\Psi; \Si': \G',
% p\app \vec{t'} \rseq C'$ be a version of the original sequent raised
% over a sequence of distinct nominal constants $\vec{c}$ for
% $\vec{z}$. For the schematic form of the \defLCSU rule to be
% applicable, we require one of the following conditions to hold for
% every permutation $\pi$ of the nominal constants $\{\vec{c},\vec{a}\}$:
% \begin{itemize}
% \item $(p\app \vec{t'})$ and $(\pi.A[\vec{c}/\vec{z}])$ are not
%   unifiable under any instantiation of type variables or

% \item for some type expressions $\tau_1,\ldots,\tau_n$ whose type
%   variables are bound in $\Psi$, these is a type generic CSU for
%   $(p\app \vec{t'})[\tau_1/A_1,\ldots,\tau_n/A_n]$\footnote{These type
%     substitutions must be applied to $(p\app t')$ as well because the
%     type variables $A_1,\ldots,A_n$ may appear in the types of the
%     term via the nominal constants in $\vec{c}$.}and
%   $(\pi.A[\vec{c}/\vec{z}])[\tau_1/A_1,\ldots,\tau_n/A_n]$ and for any
%   other type substitutions for $A_1,\ldots,A_n$ it is the case that
%   the formulas $(p\app \vec{t'})[\Phi]$ and
%   $(\pi.A[\vec{c}/\vec{z}])[\Phi]$ are not unifiable.
% \end{itemize}
%
\noindent The condition that the definition above requires to hold
with respect to each schematic definitional clause may be understood
as follows.
The first possibility applies when the formula being analyzed
cannot match with the head of the clause no matter how the type variables
are instantiated.
The second possibility applies when the formula being analyzed can
match with the head of the schematic definitional clause. The
requirement in this case has two parts. First, case analysis should
uniquely fix the type variables parameterizing the (reduced) schematic
definitional clause\footnote{Since these variables are exactly the
  type variables in the clause head and is a superset of the variables
  in the clause body, case analysis only needs to fix the type
  variables in the clause head through unification.}; without this, we
would have to explore different possible instantiations for these type
variables for the rule to be sound. Second, case analysis via the
schematic definition left rule should be type generic.
If one of these possibilities hold, then the effect of the
schematic definitional clause on the case analysis is guaranteed to be
independent of the types involved.

To actually articulate the schematic version of the \defLCSU rule, we
have to identify the set of premises arising from any given
definitional clause.
Towards this end, we define the relation $\kdeflgenprem$ which is
similar to $\kdeflcsuprem$ used for \defLCSU in
Section~\ref{subsec:def_rules}.
\begin{mydef}\label{def:defl_csu_premises}
  Let $H$ be the schematic sequent $\Psi; \Si: \G, p\app \vec{t} \rseq
  F$ and let $C$ be the schematic definitional clause $[A_1,\ldots,A_n]\rforallx {\vec{x}}
  {(\nablax {\vec{z}} A)\rdef B}$. Further, let $\support{p\app
    \vec{t}}$ be $\{\vec{a}\}$ and let $\vec{c}$ be a sequence of
  nominal constants that is of the same length as $\vec{z}$ and such
  that each constant in the sequence has a type identical to that of
  the corresponding variable in $\vec{z}$ and is also distinct from
  the constants in $\vec{a}$. Finally, let $[A_1,\ldots,A_n]\rforallx {\vec{h}}
  {(\nablax {\vec{z}} A')\rdef B'}$ be a version of the clause $C$
  raised over $\vec{a}$ and let $\Si' \cup \{A_1,\ldots,A_n\}: \G', p\app \vec{t'} \rseq F'$
  be a version of $H$ raised over $\vec{c}$. Then
\begin{tabbing}
\ \=$\deflgenprem{H}{p\app \vec{t}}{C}$\ \=$=$\ \=\ \=\kill
\>$\deflgenprem{H}{p\app \vec{t}}{C}$\>$=$\>$\{\Si'[\theta]: \G'[\theta],
    \pi.B'[\theta] \rseq F'[\theta] \sep$\\
\>\>\>\>$\pi$ is a permutation of the nominal constants in $\{\vec{c},\vec{a}\}$\\
\>\>\>\>and $\Phi = (\tau_1/A_1,\ldots,\tau_n/A_n)$ is a type substitution\\
\>\>\>\>such that $\gencsu{(p\app \vec{t'})[\Phi]}{(\pi.A'[\vec{c}/\vec{z}])[\Phi]}$\\
\>\>\>\>exists and $\theta \in \gencsu{(p\app \vec{t'})[\Phi]}{(\pi.A'[\vec{c}/\vec{z}])[\Phi]}$\}.
\end{tabbing}
\end{mydef}

%% \begin{mydef}\label{def:defl_csu_premises}
%%   Given a schematic sequent $H = (\Psi; \Si: \G, p\app \vec{t} \rseq
%%   C)$ where $\support{p\app \vec{t}} = \{\vec{a}\}$, the relation
%%   $\deflgenprem{p\app \vec{t}}{H}{H'}$ holds if $H'$ is
%% \[
%% \Psi; {\Si'[\theta]: \G'[\theta], \pi.B[\theta] \rseq C'[\theta]}
%% \]
%%   and
%%   \begin{itemize}
%%   \item $[A_1,\ldots,A_n]\rforallx {\vec{x}} {(\nablax {\vec{z}} A)\rdef B}$ is a
%%     version of a schematic definitional clause for $p$ raised over $\vec{a}$;
%%   \item $\Psi; \Si': \G', p\app \vec{t'} \rseq C'$ is a version of the
%%     original sequent raised over a sequence of distinct nominal
%%     constants $\vec{c}$ for $\vec{z}$;
%%   \item There exists a permutation $\pi$ of nominal constants
%%     $\{\vec{c},\vec{a}\}$ and a type substitution $\Phi =
%%     (\tau_1/A_1,\ldots,\tau_n/A_n)$ such that a type generic CSU exists
%%     for $(p\app \vec{t'})[\Phi]$ and $(\pi.A[\vec{c}/\vec{z}])[\Phi]$.
%%     Then $\theta$ is a substitution not containing nominal constants
%%     such that $\theta \in \gencsu{(p\app \vec{t'})[\Phi]}{(\pi.A[\vec{c}/\vec{z}])[\Phi]}$.
%%   \end{itemize}
%% \end{mydef}
%
\noindent The schematic definition left rule \sdefL is shown in
Figure\cspc\ref{fig:schm_defl}. This rule has a proviso associated
with it: it is only applicable if the lower sequent of the rule is
amenable to case analysis with respect to $(p\app \vec{t})$ as
formally described in Definition~\ref{def:amenable}.

\begin{figure}[ht!]
  \begin{gather*}
    \infer[\sdefL]{
      \Psi; \Si: \G, p\app \vec{t} \rseq F
    }{
      \{H \in \deflgenprem{\Psi; \Si: \G, p\app \vec{t} \rseq F}{p\app
        \vec{t}}{C} \sep C \in {\cal D_S}\}
    }\\
    \mbox{\rm where}\ {\cal D_S}\ \mbox{\rm is the set of all the
      reduced forms of the schematic}\\
    \mbox{ definitional clauses obtained from all the schematic
      definition blocks}
  \end{gather*}
  \caption{The Schematic Definition Left Rule}
  \label{fig:schm_defl}
\end{figure}

%% \begin{figure}[ht!]
%%   \begin{gather*}
%%     \infer[\sdefL]{
%%       \Psi; \Si: \G, p\app \vec{t} \rseq C
%%     }{
%%       \{H \sep \deflgenprem{(p\app \vec{t})}{(\Psi; \Si: \G, p\app \vec{t} \rseq C)}H\}
%%     }
%%   \end{gather*}
%%   \caption{The Schematic Definition Left Rule}
%%   \label{fig:schm_defl}
%% \end{figure}

Up to this point, we have allowed schematic definition blocks only to
be interpreted as abbreviations for collections of fixed-point
definitions in \Gee. We actually allow such blocks also to be treated
as generators of inductive definitions. For a block to be designated
in this way, it must be the case that the schematic definitional
clauses within it are each parameterized by an empty set of type
variables; when this constraint is met, each schematic definitional
clause gives rise to exactly one clause in the definition block in
\Gee that is generated by instantiating the type variables
parameterizing the block with concrete types. Furthermore, for each
clause $\rforallx {\vec{x}} {\nablax {\vec{z}} {p\app \vec{t}} \rdef
  B}$ in the block, the type variables in the type of $p$ must contain
all the type variables in $\nablax {\vec{z}} {p\app \vec{t}}$. As a
  result, the type of the entire clause is fixed when the type of $p$
  is fixed.

We provide the following auxiliary definition for formalizing a schematic induction rule.
%% At last, we need to design the schematic version of the induction
%% rule. For this, we allow for designation of some schematic
%% definitional blocks as inductive, with the intention that every
%% instance of the block consists of a set of clauses that inductively
%% define the instantiated predicate constants associated with the
%% block. A schematic definitional block can be designated as inductive
%% only if every schematic clause in it has the form $[\Psi]C$ where
%% $\Psi$ is empty. In the thesis we are only concerned with induction
%% over schematic definition blocks that contain clauses for exactly one
%% predicate constant.  We define the notion of the instance of such a
%% block relative to a type instance of that predicate constant as follows:
%
%% \begin{mydef}
%%   Let $B$ be an inductive schematic definition block that has
%%   associated with it only the predicate constant $p$ and that is
%%   parameterized by $\{A_1,\ldots,A_n\}$.  Then the instance of $B$
%%   relative to
%%   $p_{\vec{\tau}}$ where $\tau = (\tau_1,\ldots,\tau_n)$ consists of the
%%   following set of clauses
%%   \[
%%   \{C[\tau_1/A_1,\ldots,\tau_n/A_n]\sep C \in B\}
%%   \]
%% \end{mydef}
%
\begin{mydef}
  Let $B$ be an inductive schematic definition block that has
  associated with it only the predicate constant $p$.  Then the
  clauses in $B$ for a type instance $p_{\vec{\tau}}$ of $p$ are the
  instances of clauses in $B$ obtained by instantiating their
  parameterizing type variables with $\vec{\tau}$.
\end{mydef}
\noindent The schematic induction rule is then given in
Figure~\ref{fig:schm_ind}, which is very similar to the induction rule
in \Gee.

\begin{figure}[ht!]
  \begin{gather*}
    \infer[\sindL]{
      \Psi; \Si: \G, p_{\vec{\tau}}\app \vec{t} \rseq C
    }{
      \{
      \Psi; \vec{x} : D[S/p_{\vec{\tau}}] \rseq
      \nablax{\vec{z}}{S\app \vec{t}}\sep
      \rforallx{\vec{x}}{\nablax{\vec{z}}{p_{\vec{\tau}}\app
          \vec{t}} \rdef D} \in \mathcal{C}
      \}
      &
      \Psi; \Si: \G, S\app \vec{t} \rseq C
    }
    \\
    \mbox{if $B$, the schematic block for $p$, has only $p$ associated
      with it}
    \\
    \mbox{and $\mathcal{C}$ is the set of all the clauses in $B$ for $p_{\vec{\tau}}$}
    \\
    \mbox{ and $S$ is a term with no nominal constants and of the same type as $p_{\vec{\tau}}$}
  \end{gather*}
  \caption{The Schematic Induction Rule}
  \label{fig:schm_ind}
\end{figure}

%% As we have discussed before, induction works only for predicate
%% constants associated with schematic blocks that do not contain any
%% clause of the form $[\Psi]C$ where $\Psi$ is empty. The induction rule
%% of \Gee can therefore be easily lifted to be schematic by abstracting
%% its premises and conclusion over a set $\Psi$ of type variables, as
%% shown in Figure\cspc\ref{fig:schm_ind}.

%% \begin{figure}[ht!]
%%   \begin{gather*}
%%     \infer[\indL]{
%%       \Psi; \Si: \G, p\app \vec{t} \rseq C
%%     }{
%%       \{
%%       \Psi; \vec{x_i} : B_i[S/p] \rseq \nablax{\vec{z}_i}{S\app \vec{t}_i}
%%       \}
%%       &
%%       \Psi; \Si: \G, S\app \vec{t} \rseq C
%%     }
%%     \\
%%     \mbox{ provided $p$ is inductively defined by the set of clauses
%%       $\{\rforallx{\vec{x}_i}{\nablax{\vec{z}_i}{p\app \vec{t}_i} \rdef B_i}\}$}
%%     \\
%%     \mbox{ and $S$ is a term with no nominal constants and of the same type as $p$}
%%   \end{gather*}
%%   \caption{The Schematic Induction Rule}
%%   \label{fig:schm_ind}
%% \end{figure}

The ``proofs'' constructed using the schematic proof rules are
schemata for generating actual proofs in \Gee under the instantiation
of type variables. This property is stated as follows:
\begin{mythm}\label{thm:schm_sound}
  If a schematic sequent $\Psi; \Si: \G \rseq B$ is derivable by using
  the schematic proof rules in Figures\cspc\ref{fig:schm_core_rules},
  \ref{fig:schm_defr}, \ref{fig:schm_defl} and \ref{fig:schm_ind},
  then given any type substitution $\Phi$ for variables in $\Psi$ such
  that $\emptyset \stseq \Phi : {\Psi}$ holds, there exists a proof
  for ${\Si[\Phi]} : {\G[\Phi]} \rseq {B[\Phi]}$ in \Gee.
\end{mythm}
\begin{proof}
  We proceed by induction on the derivation of $\Psi; \Si: \G \rseq
  B$, analyzing the last schematic rule it uses. In most cases, the
  argument follows a set pattern: we invoke the inductive hypothesis
  on the premises of the last rule, we then apply the non-schematic
  version of the rule to conclude. When the last rule is either
  $\srexstR$ or $\srfallL$, we apply Lemma\cspc\ref{thm:tyinst} to its
  left premise to get a well-typed term for substituting for the
  binding variable. The only case that needs further explanation is
  when the last rule is $\sdefL$. Given any schematic definitional
  clause, the provisos of this rule ensure that if the matching
  between the atomic formula being analyzed with the head of the
  clause fails, then it will also fail under the instantiation of type
  variables. The provisos also ensure that when the matching succeeds,
  it also succeeds under the type instantiation. Moreover, the type
  generic natural of the matching ensures the structure of the premise
  generated from the matching is preserved under the type
  instantiation. From these observations, we can follow the set
  pattern to finish the proof for this case.
\end{proof}
The proof of the above theorem is constructive. Its procedural
interpretation provides us a function for constructing proofs in \Gee
from the schematic proofs.
%GN Isn't this already quite clear and hasn't it also been said a few
%times already?
%% Moreover, all the proofs in \Gee computed
%% from a particular schematic proof have the same structure despite of
%% their differences in types. As such, the schematic proofs indeed
%% function as schemata for generating proofs in \Gee.
%
%% The following corollary states the property that schematic theorems
%% proved by using schematic rules are schemata for generating theorems
%% in \Gee; it is immediate from Theorem\cspc\ref{thm:schm_sound}:
An easy consequence of the theorem is also the following:
%% The following corollary states the property that schematic theorems
%% proved by using schematic rules are schemata for generating theorems
%% in \Gee; it is immediate from Theorem\cspc\ref{thm:schm_sound}:
%
\begin{mycoro}\label{coro:schm_thm}
  If a schematic theorem $[A_1,\ldots,A_n]F$ is provable by using
  schematic proof rules, then given any ground types
  $\tau_1,\ldots,\tau_n$, $F[\tau_1/A_1,\ldots,\tau_n/A_n]$ is provable is
  \Gee.
\end{mycoro}

We have implemented the schematic polymorphism extension in
\Abella.
%% Before our implementation, there is already some very
%% primitive support of polymorphism in \Abella implemented by Kaustuv
%% Chaudhuri. This primitive polymorphism is based on treating type
%% variables as constants with a global scope. Our implementation reuses
%% some of the code for implementing this polymorphism.
%
The interactive style of reasoning works mostly in the same
way as before with our polymorphism extension. One difference is that
\Abella now keeps track of the type variables parameterizing schematic
theorems during the proof developments, which are like constants in
that they cannot be instantiated and are also like unknown objects in
that they cannot be compared with other types for equality. Another
difference is that when performing case analysis, \Abella checks if
the unification problems can be solved in a way that satisfies the
provisos of the \sdefL rule. If not, for example, if unification needs
to compare two different parameterizing variables for equality, then
the application of the case analysis tactic fails.
%% the proof development is suspended as \Abella does not know how to
%% proceed with the proof.

\subsection{Some example interactive proof developments}
\label{subsec:schm_poly_exms}

To illustrate the usefulness of the schematic polymorphism extension,
let us consider proving the pruning property of the membership
relation using the extension. The schematic definition of the
membership relation has already been given in
Section\cspc\ref{subsec:schm_poly}. So does the schematic theorem that
states the pruning property. We repeat the theorem as follows where
$M$ is of type $(B \to A)$, $L$ is of type $\klist\app A$, $x$ is of
type $B$ and $M'$ is of type $A$:
\[
[A,B]\rforallx {M, L}
  {\nablax {x} {\kmember\app (M\app x)\app L \rimp \rexistsx {M'} {M = y \mlam M'}}}.
\]
In Section\cspc\ref{subsec:poly_mtv_exms}, we have discussed the
proofs of the two instances of this pruning property obtained by
instantiating $A$ and $B$ with $\ktm$ and with $\ktm'$,
respectively. The proofs for these instances have exactly the same
structure and do not rely on the type information. This implies a
schematic proof exists for this theorem. We show that this is indeed
the case by formally constructing a schematic proof for it.

At the beginning of the proof development, $A$ and $B$ are marked as
type variables that need to be tracked implicitly. The theorem to be
proved becomes
\[
\rforallx {M, L}
  {\nablax {x} {\kmember\app (M\app x)\app L \rimp \rexistsx {M'} {M = y \mlam M'}}}.
\]
We perform an induction on the only assumption, introducing the
following inductive hypothesis \IH:
\[
\rforallx {M, L}
  {\nablax {x} {(\kmember\app (M\app x)\app L)^* \rimp \rexistsx {M'} {M = y \mlam M'}}}
\]
The proof state is changed to have the following hypothesis
\begin{tabbing}
\qquad\=\kill
\>$\hyp{\kwd{H1}}{(\kmember\app (M\app \kn)\app L)^@}$
\end{tabbing}
where ${L, M}$ are new variables and $\kn$ is a nominal
constant. The conclusion to be proved becomes ${\rexistsx {M'} {M = y \mlam
    M'}}$. We then unfold \kwd{H1}, resulting in two cases.

In the first case, we have $L = E \cons L'$ and $(M\app \kn) = E$ for
some variables $E$ and $L'$. By solving the unification problem
$(M\app \kn) = E$, we get $M = y \mlam E$. We can now close this
branch by instantiating $M'$ with $E$.

In the second case, we have $L = E \cons L'$ and the following new
hypothesis
\begin{tabbing}
\qquad\=\kill
\>$\hyp{\kwd{H2}}{(\kmember\app (M\app \kn)\app L')^*}$
\end{tabbing}
Because \kwd{H2} comes from the unfolding of \kwd{H1}, it is
*-annotated, indicating it is derivable in fewer steps than
\kwd{H1}. We can therefore apply \kwd{IH} to \kwd{H1}, yielding
$\rexistsx {M'} {M = y \mlam M'}$ which is exactly the conclusion we want to
prove. This concludes the proof.

Notice that when we do case analysis in the above proof, we have not
inspected the types of terms because the solutions to the unification
problems are completely determined by the structure of terms and are
oblivious to any type information. Applying the case analysis tactic
here corresponds to applying the $\sdefL$ rule whose provisos are
satisfied because of the type generic nature of these unification
problems.

We have also given the encoding of the membership relation in \LProlog
and the theorem describing its pruning property in
Section\cspc\ref{subsec:schm_poly}. We repeat the theorem as follows
where $M$ is of type $(B \to A)$, $L$ is of type $\klist\app A$, $x$
is of type $B$ and $M'$ is of type $A$:
\[
[A,B] \rforallx {M, L}
  {\nablax {x} {\oseq{\kmemb\app (M\app x)\app L}
      \rimp \rexistsx {M'} {M = y \mlam M'}}}.
\]
We can prove this theorem by induction on the only assumption and by
following conceptually the same steps as in the previous
proof. However, there is a significant difference in terms of how case
analysis is actually carried out. After unfolding $\oseq{\kmemb\app
  (M\app \kn)\app L}$ (where $M$ and $L$ are variables and $\kn$
is a nominal constant), we need to consider unifying the formula
$\kprog\app G\app (\kmemb\app (M\app \kn)\app L)$ where $G$ is some
variable with the heads of the following schematic definitional clauses:
\begin{tabbing}
\qquad\=$[A''] \kprog\app (\kmemb_{A''}\app X\app L)\app (\kmemb_{A''}\app X\app (Y \cons L))$ \quad\=$\rdef$ \quad\=\kill
\>$[A'] \kprog\app \ktrue\app (\kmemb_{A'}\app X\app (X \cons L))$
  \>$\rdef$ \>$\rtrue$\\
\>$[A''] \kprog\app (\kmemb_{A''}\app X\app L)\app (\kmemb_{A''}\app X\app (Y \cons L))$
  \>$\rdef$ \>$\rtrue$.
\end{tabbing}
Note that the unification problems have solutions only when $A'$ and
$A''$ unify with the type $A$. That is, only the following two clauses
among the instances of the above schematic definitional clauses are relevant to our
case analysis; all the other cases end in unification failure:
\begin{tabbing}
\qquad\=$\kprog\app (\kmemb_{A}\app X\app L)\app (\kmemb_{A}\app X\app (Y \cons L))$ \quad\=$\rdef$ \quad\=\kill
\>$\kprog\app \ktrue\app (\kmemb_{A}\app X\app (X \cons L))$
  \>$\rdef$ \>$\rtrue$\\
\>$\kprog\app (\kmemb_{A}\app X\app L)\app (\kmemb_{A}\app X\app (Y \cons L))$
  \>$\rdef$ \>$\rtrue$.
\end{tabbing}
As in the previous proof, the case analysis on those two clauses
proceeds in a type generic fashion. As a result, the provisos of the
\sdefL rule are satisfied and a schematic proof can be constructed
using our polymorphism extension.

As another example, consider the definition of substitutions as
relations. We have given a definition of the substitution relation on the
\STLC terms in Section\cspc\ref{subsec:lts_reason}. It can be
generalized to substitutions at any type as follows. We first identify
the type $\kmap : \ktype \to \ktype \to \ktype$ for mappings and its
constructor $\kmap : A \to B \to \kmap\app A\app B$. A substitution is
represented as a list of mappings. We then identify the substitution
relation as the predicate symbol $\kappsubst: \klist\app (\kmap\app
A\app A) \to B \to B \to \prop$ defined through the following clauses:
\begin{tabbing}
\qquad\=$\nablax x {\kappsubst\app ((\kmap\app x\app V) \cons S)\app (R\app x)\app M}$
\quad\=$\rdef$ \quad\=\kill
\>$\kappsubst\app \knil\app M\app M$ \>$\rdef$ \>$\rtrue$\\
\>$\nablax x {\kappsubst\app ((\kmap\app x\app V) \cons S)\app (R\app x)\app M}$
  \>$\rdef$ \>$\kappsubst\app S\app (R\app V)\app M$
\end{tabbing}
By the definition, $\kappsubst\app S\app M\app M'$ holds exactly when
$M'$ is the result of applying the substitution $S$ to $M$. Given this
schematic definition, we can easily prove general properties about the
substitution relation in \Abella. The following are some examples:
\begin{tabbing}
\qquad\=\qquad\=\qquad\=\kill
\>$[A,B] \rfall {(S:\klist\app (\kmap\app A\app A)), (M:B), M', M''}.$\\
\>\> ${\kappsubst\app S\app M\app M' \rimp \kappsubst\app S\app M\app M'' \rimp M' = M''}.$\\
\>$[A,B,C] \rfall (S:\klist\app (\kmap\app A\app A)), (M:B), M'. \nabla (x:C).$\\
\>\> ${\kappsubst\app S\app M\app (M'\app x) \rimp \rexistsx {M''} {M' = y \mlam M''}}$.\\
\>$[A,B,C] \rfall (S:\klist\app (\kmap\app A\app A)), (M_1:C), (M_2:C \to B), M_1', M_2'. \nabla (x:C).$\\
\>\> $\kappsubst\app S\app M_1\app M_1' \rimp \kappsubst\app S\app (M_2\app x)\app (M_2'\app x)$\\
\>\>\> $\rimp \kappsubst\app S\app (M_2\app M_1)\app (M_2'\app M_1').$
\end{tabbing}
The first theorem states that substitution is deterministic. The
second theorem is a pruning property of the substitution relation. The
last theorem is an ``instantiation'' theorem for the substitution
relation. All of these theorems have schematic proofs because their
proof construction does not make use of any type information.

%%%%%%%%%%%%%%%%%%%%%%%%%%%%%%%%%%%%%%%%%%%%%%%%%%%%%%%%%%%%%%%%%%%%%%%%%%%%%%%%

% Verified compilation of functional languages
%%%%%%%%%%%%%%%%%%%%%%%%%%%%%%%%%%%%%%%%%%%%%%%%%%%%%%%%%%%%%%%%%%%%%%%%%%%%%%%
% vericfl.tex: Verified compilation of functional languages
%%%%%%%%%%%%%%%%%%%%%%%%%%%%%%%%%%%%%%%%%%%%%%%%%%%%%%%%%%%%%%%%%%%%%%%%%%%%%%%%
\chapter{A Structure for Verified Compilation}
\label{ch:vericfl}

A common way to structure the compilation of functional programs is as
a \emph{multi-pass} process: Given a particular program, the compiler
transforms it through a sequence of steps into successively lower-level
forms till eventually it has produced code in the desired target
language, which might be an assembler or a low-level intermediate
language. It is this kind of a multi-pass compiler that will be
considered in this thesis.

A natural way to specify a compiler transformation is to describe it
via a set of rules in the Structural Operational Semantics (SOS)
style\cspc\cite{plotkin81} that derive relations on the source and
target programs of the transformation. The specification of a
multi-pass compiler then comprises the rule-based relational
specifications of its transformations.
To guarantee the correctness of a compiler, we must prove
that the meaning or \emph{semantics} of the source language program is
preserved by the target language program that is eventually produced.
In this thesis, like in other compiler
verification
projects such as CompCert\cspc\cite{leroy09cacm} or
CakeML\cspc\cite{kumar14popl}, our interest is in showing
that the compiler preserves the meaning of closed programs
at atomic types.
One could try to achieve this goal by taking on the verification of
the entire compiler in one step (\eg, see\cspc\cite{benton09icfp}).
% However, because of the complexity in
% compiling functional languages this approach is only applicable to
% simple functional languages and is not scalable in general.
We adopt an alternative approach here.
In this approach we prove that each individual transformation
preserves the semantics of programs and we then compose these results
to obtain the correctness property for the full compiler.
We feel that this approach is more manageable because it takes
advantage of the fact that each individual transformation is designed
to achieve one specific objective. As such, semantics preservation for
it has a clear definition and its proof is modular.

An obvious way to prove semantics preservation for compiler
transformations described via rule-based relational specifications is
to define semantics preservation as a relation between programs in the
source and target languages and show that it subsumes the
transformation relation. Since a compiler transformation in a
multi-pass compiler may transform programs in one intermediate
language into programs in a different language, the relation denoting
semantics preservation must be able to relate programs across multiple
languages. We use the device of \emph{logical
  relations}\cspc\cite{statman85ic} that have this property
to denote semantics preservation. Because we are only interested in
verifying compilation of closed programs at atomic types and logical
relations are composible at atomic types, proofs of semantics
preservation for the individual transformations can be composed to
form the correctness proof for the full multi-pass compiler.

As we have discussed in the introduction chapter, a major difficulty
in implementing and reasoning about the compilation of functional
languages lies in modeling and reasoning about the binding structure
of functional objects. Most existing proof systems for formal
verification provide only very primitive support for dealing with
binding structure. As a result, verification of compilers for
functional languages in such systems tends to be unintuitive and
laborious; a large amount of effort needs to be expended just to prove
the ``boilerplate'' lemmas about notions related to binding structure.

The main goal of the thesis is to show that our extended framework is
suitable for implementing and verifying compiler transformations on
functional programs. In
particular, we would like to show that the $\lambda$-tree syntax
approach supported by the framework significantly simplifies the
representation, manipulation, analysis and reasoning about binding
structure in the implementation and verification of compiler transformations
on functional programs.
We achieve this goal by presenting a methodology that formalizes the
approach to verified compilation described above using our
framework. Under this methodology, we encode 
transformations on functional programs as \LProlog specifications by utilizing the approach
described in Section\cspc\ref{sec:spec_lang}. Since \LProlog
specifications are executable, they serve also as implementations of
the transformations and the compiler. We argue that the $\lambda$-tree
syntax approach that is supported by \LProlog provides a convenient
way to realize manipulation and analysis of binding structure in these
rules. Thanks to these features, the \LProlog specifications of
transformations on functional programs are concise and transparently
relate to the original rule-based relational descriptions. Moreover,
they have a logical structure that can be exploited in the process of
reasoning about their correctness. In the methodology that we are
proposing, showing the correctness of compiler transformations amounts
to formally proving properties of the \LProlog specifications of
transformations on functional programs using \Abella. More precisely, for
every compiler transformation, we formalize the relation describing
semantics preservation in \Abella and then prove a theorem that this
relation subsumes the relation that encodes the compiler
transformation in \LProlog. These theorems are then composed together
to form the correctness proof for the full compiler. In the
construction of the correctness proofs, we show that the
$\lambda$-tree syntax approach supported by \Abella provides a
convenient way to formalize and prove properties about binding
structure.
To illustrate this methodology, we use it to implement a verified
compiler for a representative functional programming language that is
an extension of the Programming Computable Functions (PCF)
language\cspc\cite{plotkin77}.

In the following sections, we elaborate on the above ideas. In
Section\cspc\ref{sec:cpl_model} we present the model for compiling
functional languages we work with in this thesis. In
Section\cspc\ref{sec:vfc_approach} we expand on the approach to
verified compilation described in this preamble. In
Section\cspc\ref{sec:fm_vfc} we elaborate on the methodology for
formalizing this approach using our framework. There exist many
choices for the notion of semantics preservation beside logical
relations. In Section\cspc\ref{sec:sem_pres_nuances} we shall compare
those different notions and explain why the benefits of our framework
in verified compilation on functional programs observed in the thesis
can also be derived when other notions of semantics preservation are
used. In Section\cspc\ref{sec:vfc_exercise} we give an overview of the
exercise we shall carry out to verify the usefulness of our
methodology for verified compilation of functional programs.

\section{The Compilation Model}
\label{sec:cpl_model}

There exist two common models for compiling functional languages in
multiple passes. One compiles functional programs into
abstract machine code through a sequence of
transformations\cspc\cite{landin64,cousineau87,wand89tr,pjones92}. The
compiled code is then executed on the run-time infrastructures of the
abstract machines. The other model compiles functional programs into
executable code for real
hardware\cspc\cite{cardelli84,morrisett99toplas,tarditi96pldi}. Compilation
in this model usually comprises two phases. In the first phase, the
higher-order functional programs go through several compiler
transformations by which their higher-order features are gradually
removed. The output of the first phase are programs that resemble
those in procedural languages such as C in which all functions are at
the top-level (\ie, there are no nested functions) and the control flow
is explicit. At this point, the conventional techniques for
compiling procedural languages become applicable. In fact, the second
phase of the compilation usually consists of a sequence of
transformations that resemble those in compilers for procedural
languages. It takes the output of the first phase as input and
eventually generates executable code in the desired target language
such as an assembly language or machine language.

In this thesis, we shall work with the latter model. Within this
context, we will focus on the first phase described above for two
reasons. First, the transformations in this phase involve
complicated manipulation and analysis of binding structure and it is
in encoding and verifying such transformations that the $\lambda$-tree
syntax approach proves most useful. Second,
there is already a large body of work devoted to
verifying the compiler transformations in the second phase, such as
the series of papers from the CompCert
project\cspc\cite{leroy09cacm,leroy09jar,leroy06popl}. Since the
transformations in the second phase do not involve significant
manipulation of binding structure, we do not have many new ideas to
add to the work that has already been done in relation to this phase.
Rather than repeating a development of existing ideas,
we provide a minimal but functional implementation of the second phase
and verify this implementation.

\section{The Approach to Verified Compilation}
\label{sec:vfc_approach}

In this section, we elaborate on our approach to specifying and
verifying the multi-pass compilers for functional languages on
paper. To specify a multi-pass compiler, we give every compiler
transformation a rule-based relational
description. Specifically, we characterize a transformation via a
relation between programs in its source and target languages and a set
of rules for deriving the relation. The characterization must satisfy
the following property: a source program is transformed to a target
program by the transformation if and only if the relation holds of
them. In the setting of compiling functional programs, the
transformation rules usually involve manipulation and analysis of
binding structure. An example of such a specification is the
transformation of $\lambda$-terms into their de Bruijn forms described
in Section\cspc\ref{sec:embed_hhw}. The specifications for the
sequence of transformations of a multi-pass compiler together
constitute the specification of the compiler.

%% To verify a compiler is to prove that the meaning or semantics of a
%% program is preserved by compilation.
%% %
%% We will do this here by proving
%% semantics preservation for each of the transformations used and then
%% composing these results to yield correctness for the full
%% compiler.
%% %
%% One way to characterize the semantics of programs is to examine how
%% they interact with the outside world. The interaction is characterized
%% by their \emph{behaviors} that have obvious and fixed meanings, such as
%% termination and I/O events. Semantics preservation can then be
%% described as the preservation of behaviors~\cite{leroy09cacm}. We write
%% $F \Rightarrow F'$ to denote that if $F$ holds then $F'$ holds and
%% $F \iff F'$ to denote that $F \Rightarrow F'$ and $F' \Rightarrow F$.
%% We write $\behave t B$ to represent the judgment that the program $t$ has the
%% behavior $B$.
%% If $t'$ is the program that results from transforming the program $t$,
%% $t'$ preserves the semantics of $t$ if the following
%% property holds:

As we have noted already, a compiler that is composed of a sequence of
transformations can be verified by showing that each transformation
preserves the meaning or semantics of the program it is applied to.
How exactly we do the latter will, of course, depend on how we
characterize the semantics of programs.
One way to do this is to examine how a program interacts with the
outside world.
This interaction is characterized by their \emph{behaviors} that have
obvious and fixed meanings, such as termination and I/O
events.
Let us write $\behave t B$ to represent the judgment that the program
$t$ has the behavior $B$.
We describe a development of the idea of semantics preservation
based on such a judgment that has been used in the CompCert project by
Xavier Leroy~\cite{leroy09cacm}.
Let $t'$ be the program that results from transforming the program
$t$.
Then we may say at the outset that $t'$ preserves the semantics of $t$
if the following property holds:
\begin{equation*}\label{eq:bhpres}
\rfall B. \behave t B \iff \behave {t'} B.
\end{equation*}
We write $F \iff F'$ here to denote that $F \Rightarrow F'$ and $F'
\Rightarrow F$ hold where $F \Rightarrow F'$ ($F' \Rightarrow F$)
itself denotes that $F'$ ($F$) holds if $F$ ($F'$) holds.
Now, the strict notion of behavior preservation presented above is
often replaced by the notion of \emph{behavior refinement} that
corresponds to the following property:
\begin{equation*}\label{eq:bhrefine}
\rfall B. \behave {t'} B \Rightarrow \behave t B.
\end{equation*}
This property allows the target program to choose to follow only
\emph{some} behaviors of the source program. For instance, if the
evaluation order of expressions $e_1$ and $e_2$ is not fixed in the
source program, then it may be acceptable if a particular order has
been picked in the target program. We are also often not interested in
following specific behaviors of {\it all} source programs but only
those that satisfy some criterion. For example, we may be concerned
only about source programs that do not ``go wrong.'' In this case,
behavioral refinement can be further refined to
\begin{equation*}\label{eq:wbhrefine}
\wellbehaved t \Rightarrow \rfall B. \behave {t'} B \Rightarrow \behave t B
\end{equation*}
where $\mwellbehaved$ is defined as
\begin{equation*}\label{eq:wbh}
\wellbehaved{t} \iff (\rfall B. \behave t B \Rightarrow B \not\in \wrong);
\end{equation*}
here $\wrong$ represents the set of ``going wrong'' behaviors.
This definition of behavioral refinement gives the compiler the
freedom to pick whatever target program it wants if the source program
is one that will go wrong during execution. This is a choice exercised
by many C compilers when they encounter source programs with undefined
behaviors.

Behavior refinement is not easy to prove in general. However, if
the source and target languages have deterministic semantics, then it
is easy to show that the refined form of behavioral refinement
described above is implied by the following property
\begin{equation*}\label{eq:bhfwdsim}
\forall B \not\in \wrong. \behave t B \Rightarrow \behave {t'} B.
\end{equation*}
This property is much easier to prove since we can induct on the
evaluation of source programs.

In this thesis, we are only concerned with languages that have a
deterministic semantics and that do not have constructs with side
effects. In this context, there are only three kinds of behaviors to
consider: either a program gets stuck, it diverges, or it evaluates to
some value. We will consider getting stuck and divergence as ``going
wrong''. The property above then reduces to the following statement,
commonly known as \emph{forward simulation}:
\begin{quotation}
\noindent {\it If a source program evaluates to some value $v$, then
  its transformed version evaluates to some value $v'$ that is
  equivalent to $v$ in a sense that is intuitively well-motivated.}
\end{quotation}
Forward simulation can have many different interpretations depending
on the notion of equivalence it uses. In the setting of verifying
multi-pass compilers for functional languages, this notion must
possess the following properties. First, equivalence needs to be
defined between not only atomic values but also function
values. Second, since a compiler transformation may transform programs
in one language into ones in a different language, the equivalence
notion must be definable across multiple languages. We use logical
relations that possess these properties as the notion of semantics
preservation.
We could have made a different choice as we discuss in
Section\cspc\ref{sec:sem_pres_nuances}, but this choice suffices to
bring out the main ideas that are of concern in this thesis.
%% There exist many other choices for the notion of
%% semantics preservation. We shall compare those choices in
%% Section\cspc\ref{sec:sem_pres_nuances}.

A \emph{logical relation} defines a family of relations on
$\lambda$-terms indexed by their types such that the definition of
each relation at a certain type only refers to relations at smaller
types. We can use logical relations to describe the forward simulation
property for program transformations as follows. Given a program
transformation $P$, we identify the logical relation $\sim$ to
describe a simulation relation between terms in the source and target
languages and the logical relation $\approx$ to describe an
equivalence relation between values in the source and target
languages. Both $\sim$ and $\approx$ are indexed by the types of the
source terms (written as subscripts). We define $\sim$ such that
$\simulate \tau {t} {t'}$ holds if and only if
\begin{quotation}
  {\noindent \it for any value $v$, if $t$ evaluates to $v$, then
  there exists  some value $v'$ such that $t'$ evaluates to
  $v'$ and $\equal \tau v {v'}$ holds.}
\end{quotation}
To complete the description of the simulation relation, we need to
define when values in the source and target language are considered
equivalent. If the source and target languages contain the same atomic
objects, then an obvious choice for the notion of equivalence at
atomic types is identity. Identity would of course not be a good
choice for equivalence at function types if a transformation is
intended to accomplish something substantial. A better idea might be
to base this equivalence on the idea of semantics preservation---which
is characterized by simulation---when the two expressions are applied
to equivalent arguments.
%% At function types, a natural interpretation
%% of equivalence might be that the two expressions have equivalent
%% behavior on equivalent arguments.
Specifically, if $\tau$ is an arrow
type $\tau_1 \to \tau_2$, $\equal \tau {v} {v'}$ holds if and only if
$v$ and $v'$ are functions such that the following holds
\begin{quotation}
  {\noindent \it for any $v_1$ and $v_1'$, if $\equal {\tau_1}
    {v_1} {v_1'}$, then $\simulate {\tau_2} {(v\app
      v_1)} {(v'\app v_1')}$.}
\end{quotation}
Note that $\sim$ and $\approx$ are mutually recursively defined. In
the statement above we are using $\approx$ negatively, i.e. we are
assuming it is already defined at a certain type in defining it at
another type. This works because the type at which we assume that we
already know the relation is smaller in an inductive ordering. In
other words, this is a recursive definition based on the inductively
defined collection of types. Note also that the relations $\sim$ and
$\approx$ are defined only for closed terms since evaluation semantics
does not exist for terms containing free variables.

In this thesis, we consider compiling functional languages with
general recursion. In such a language, functions have the form $(\fix
f x t)$ where $f$ is a binding variable for the function itself, $x$
is the argument of the function and $t$ is the function body that may
refer to $f$ and $x$. Applying $(\fix f x t)$ to an argument $t'$ has
the effect of replacing $f$ with the function itself and $x$ with $t'$
in $t$ and recursively evaluating the resulting term $(t[(\fix f x
  t)/f,t'/x])$. Since functions are arguments to themselves, the
logical relation describing equivalence between them must have itself
as an assumption. One may consider defining this equivalence relation
as follows: if $\tau$ is an arrow type $\tau_1 \to \tau_2$, $\equal
\tau {v} {v'}$ holds if and only if $v$ and $v'$ are functions such
that the following holds
\begin{quotation}
  {\noindent \it for any $f$, $f'$, $v_1$ and $v_1'$, if $\equal {\tau} {f}
    {f'}$ $\equal {\tau_1} {v_1} {v_1'}$, then $\simulate {\tau_2}
    {(v\app f\app v_1)} {(v'\app f'\app \app v_1')}$.}
\end{quotation}
However, this relation is not well-defined because $\equal {\tau} {f}
{f'}$ occurs as an assumption: we are assuming that the equivalence
relation $\approx_\tau$ is already known in its very definition.

We solve the above problem by using the idea of \emph{step indexing}
logical
relations\cspc\cite{appel01toplas,ahmed06esop,ahmed08icfp}. Specifically,
we further index a logical relation with a natural number which stands
for the maximum number of evaluation steps in which the two terms
related by it cannot be distinguished from each other. We mutually
recursively define the step-indexed simulation relation $\sim$ and the
equivalence relation $\approx$ as follows. The relation $\simindex
\tau i t {t'}$ where $i$ is the step index holds if and only if
\begin{quotation}
  {\noindent \it for any value $v$ and any $j$ such that $j \leq i$, if $t$
    evaluates to $v$ in $j$ steps, then there exists some value $v'$
    such that $t'$ evaluates to $v'$ and $\equalindex \tau {i-j} v
    {v'}$ holds.}
\end{quotation}
If $\tau$ is an atomic type, $\equalindex \tau i {v} {v'}$ holds if
and only if $v$ and $v'$ are identical. If $\tau$ is an arrow type
$\tau_1 \to \tau_2$, $\equalindex \tau i {v} {v'}$ holds if and only
if $v$ and $v'$ are functions such that
\begin{quotation}
  {\noindent \it for any $f$, $f'$, $v_1$ and $v_1'$, and any $j$ such
    that $j < i$, if $\equalindex {\tau} j {f} {f'}$ and $\equalindex
    {\tau_1} j {v_1} {v_1'}$ hold, then $\simindex {\tau_2} j {(v\app
      f\app v_1)} {(v'\app f'\app \app v_1')}$.}
\end{quotation}
Every assumption in this definition is either indexed by a smaller
type or by the same type and a smaller evaluation steps. Therefore, we
can view it as a recursive definition based on a collection of pairs
of types and evaluation steps inductively defined by lexicographical
ordering. We say $t$ simulates $t'$ in $i$ steps at type $\tau$ if
$\simindex \tau i t {t'}$ holds. For a term $t$ of type $\tau$ to
simulate $t'$, the simulation relation must hold of them at all steps,
\ie, $\simindex \tau i t {t'}$ holds for any $i$. We use $\simulate
\tau t {t'}$ to represent this relation, which coincides with the
notation for simulation when no step-indexing is involved.

Given the above definitions, semantics preservation for a
transformation $P$ can be stated as follows:
\begin{myproperty}\label{thm:sem_pres}
If a closed term $t$ of type $\tau$ is transformed
into $t'$ by $P$, then $\simulate \tau t {t'}$ holds.
\end{myproperty}
\noindent When the transformation $P$ is given as a rule-based description of
relations, the property above is equivalent to saying that the
simulation relation subsumes the transformation relation. When
step-indexing logical relations are used, this property is equivalent
to the following 
\begin{myproperty}\label{thm:sem_pres_index_closed}
For any $i$, if a closed term $t$ of type $\tau$ is transformed
into $t'$ by $P$, then $\simindex \tau i t {t'}$ holds.
\end{myproperty}
\noindent This is the property we would like to prove towards showing the
correctness of the transformation $P$.

We may consider proving Property\cspc\ref{thm:sem_pres_index_closed} by
induction on the derivation of the transformation relation. However,
this will not work because transformations on functional programs
often manipulate expressions within the scope of an abstraction and,
in this sense, work recursively on open terms. In this
case the inductive hypothesis is not applicable because it assumes
terms involved in the transformation are closed. We solve this problem
by generalizing Property\cspc\ref{thm:sem_pres_index_closed} to accommodate
open terms by relating them under closed substitutions. Specifically,
given a transformation $P$, assume $\theta$ is a substitution
$(v_1/x_1,\ldots,v_n/x_n)$ in its source language, $\theta'$ is a
substitution $(v_1'/x_1',\ldots,v_n'/x_n')$ in its target language,
where $v_1,\ldots,v_n,v_1',\ldots,v_n'$ are closed values and $x_i$ is mapped
to $x_i'$ by $P$ for $1 \leq i \leq n$. We use $\equalindex \G i \theta
{\theta'}$ where $\G$ is a type context $(x_1:\tau_1,\ldots,x_n:\tau_n)$
to denote that $\equalindex {\tau_k} i {v_k} {v_k'}$ holds for $1 \leq k
\leq n$. Given two terms $t$ and $t'$ such that the free variables of
$t$ are contained in $\{x_1,\ldots,x_n\}$ and the free variables of $t'$
are contained in $\{x_1',\ldots,x_n'\}$, the generalized semantics
preservation property is stated as follows:
\begin{myproperty}\label{thm:sem_pres_index_open}
For any $i$, $\theta$, $\theta'$ and $\G$ such that $\equalindex \G i
\theta {\theta'}$ holds, if the free variables of $t$ are bound in
$\G$ and $t$ is of type $\tau$, and if $t$ is transformed into $t'$ by
$P$, then $\simindex \tau i {t[\theta]} {t'[\theta']}$ holds.
\end{myproperty}
\noindent This property can be proved by induction on the derivation
of the transformation relation. In the case that the transformation
goes under binders, we extend the substitutions $\theta$ and $\theta'$
with equivalent values for the binders and apply the inductive
hypothesis using the extended substitutions. Since $\simindex \tau i
\theta {\theta'}$ holds vacuously when $\theta$ and $\theta'$ are
empty, Property\cspc\ref{thm:sem_pres_index_closed} is just a special
case of Property\cspc\ref{thm:sem_pres_index_open}.

As we have described at the beginning of this chapter, we are
interested in verifying compilation of closed programs at atomic
types. Since the equivalence relation at atomic types is identity, we
can show that Property\cspc\ref{thm:sem_pres} is equivalent to the
following property when $\tau$ is an atomic type:
\begin{myproperty}\label{thm:sem_pres_atom}
If a closed term $t$ of atomic type $\tau$ is transformed into $t'$ by
$P$ and $t$ evaluates to $v$, then $t'$ evaluates to $v$.
\end{myproperty}
\noindent This is the forward simulation property of the transformation
$P$ we are eventually interested in.

Finally, we would like to compose the forward simulation properties of
individual transformations to form the correctness proof of the full
compiler. Notice that to apply these properties, we need to show that
the input terms of the transformations are well-typed. For this we
prove that every transformation preserves the types of its source
terms. Thus, if the source program of the compiler is well-typed, then
by the type preservation properties, every intermediate result in the
compilation sequence is also well-typed. A type preservation property
looks like the following:
\begin{quotation}
  {\noindent \it If a term $t$ has type $\tau$ and is
    transformed into $t'$ by $P$, then $t'$ has type $\tau$.}
\end{quotation}
%
%GN Please make this kind of correction in all the other places where
%it occurs.
Such a property is proved by induction on the transformation
relation. This proof is
usually straightforward and much simpler than that of semantics
preservation.

The semantics preservation property for a compiler $C$ is stated as
follows:
\begin{myproperty}\label{thm:full_sem_pres}
If a closed term $t$ of atomic type $\tau$ is compiled into $t'$ by
$C$ and $t$ evaluates to $v$, then $t'$ evaluates to $v$.
\end{myproperty}
\noindent Assume $C$ consists of a sequence of transformations
$P_1,\ldots,P_n$, the property above is equivalent to the following:
\begin{myproperty}\label{thm:sem_pres_seq}
If a closed term $t$ of atomic type $\tau$ is transformed into $t'$ by
$P_1,\ldots,P_n$ in sequence and $t$ evaluates to $v$, then $t'$
evaluates to $v$.
\end{myproperty}
\noindent We prove it by applying the semantics and type preservation
theorems of $P_1,\ldots,P_n$ in sequence, as follows. We know $t$ is
transformed into $t''$ by $P_1$ for some $t''$ and $t''$ is
transformed by the rest of the transformations into $t'$. By semantics
preservation of $P_1$, we know $t''$ evaluates to $v$. By type
preservation of $P_1$, we know $t''$ has type $\tau$. At this point we
have gathered enough assumptions for applying the semantics and type
preservation theorems of $P_2$ with $t''$ as its input. This process
is repeated for the rest of the transformations. In the end, we have $t'$
evaluates to $v$.

Up to now, we have presented the essential ideas for characterizing
and proving semantics preservation properties using logical
relations. When we talk about compiler transformations concretely in
the later chapters, we will notice that they work with languages with
richer constructs more than just atomic values and functions. In those
situations, the definitions of logical relations, the semantics
preservation theorems and their proofs will have more complicated
forms than those presented in this section. Nevertheless, the
fundamental ideas in this section still apply in those situations.

\section{Using the Framework in Verified Compilation}
\label{sec:fm_vfc}

In this section, we elaborate on the methodology for formalizing the
approach to implementing and verifying compilation of functional
programs discussed in the previous section. Under this methodology, we
encode compiler transformations as \LProlog specifications and prove
semantics preservation of the transformations in \Abella through the
two-level logic approach.

We follow the approach to implementing rule-based relational
specifications described in Section\cspc\ref{sec:spec_lang} to
implement compiler transformations. Given a compiler transformation,
we first encode the source and target languages of the transformation,
including their syntax and typing rules in \LProlog. For instance, the
typing relation of the source language may be represented by the
predicate $\kof$ such that $\kof\app M\app T$ holds if and only if $M$
has type $T$; the type relation of the target language is represented
by $\kof'$ in a similar way.
We then identify a predicate constant to represent the relation for
the compiler transformation and translate the transformation rules
into program clauses defining this predicate constant. We have already
seen an example illustrating this approach in
Section\cspc\ref{sec:embed_hhw}, \ie, the encoding of the
transformation from $\lambda$-terms to their de Bruijn forms.

As we have discussed in Section\cspc\ref{subsec:spec_impl}, the
\LProlog specifications function directly as implementations under a
logic programming interpretation. We can therefore use the encoding of
compiler transformations as an implementation of the compiler. Suppose
that a multi-pass compiler consists of a sequence of transformations
$P_1,\ldots,P_n$ and they are represented as binary predicate symbols
$p_1,\ldots,p_n$ such that $(p_i\app s\app t)$ holds if and only if the
program $S$ is transformed by $P_i$ into $T$ and $s$ and $t$ are
respectively the encodings of $S$ and $T$. Then given a source program
$S$ and its encoding $m_0$, we can identify variables $m_1,\ldots,m_n$
and query the following goal
\[
(p_1\app m_0\app m_1) \sconj (p_2\app m_1\app m_2) \sconj \ldots \sconj (p_n\app m_{n-1}\app m_n).
\]
Proof search in \LProlog will instantiate $m_1,\ldots,m_n$ with terms
such that $(p\app m_{i-1}\app m_i)$ holds for $1 \leq i \leq n$. The
term for $m_n$ is then the encoding of the result of compiling $S$.

We verify the compiler by formalizing the verification development
described in Section\cspc\ref{sec:vfc_approach} in \Abella. That is,
we encode semantics preservation of every transformation as a theorem
in \Abella, prove these theorems and compose them to form the
correctness proof of the full compiler. The key is to formally prove
Property\cspc\ref{thm:sem_pres_index_open} for every compiler
transformation. To formalize this property in \Abella, we need to
formalize the logical relations and the substitution operations. The
encoding of substitution as a relation has already been given in
Section\cspc\ref{subsec:schm_poly_exms}. We are left with the problem
of encoding logic relations.

Before we can talk about encoding logical relations, we have to encode
the evaluation semantics of the source and target languages of the
transformation. Since we need to keep track of the number of
evaluation steps in step-indexing logical relations, we give
small-step evaluation semantics to these languages. We define the
evaluation semantics as \LProlog specifications using the approach
described in Section\cspc\ref{sec:spec_lang}. Specifically, for the
source language we designate a predicate constant \kstep to represent
a one-step evaluation relation. It is defined through a set of program
clauses such that $\kstep\app t\app t'$ holds if and only if $t$
evaluates $t'$ in one step. We also identify the predicate constant
$\knstep$ such that $\knstep\app n\app t\app t'$ holds if and only if
$t$ evaluates to $t'$ in $n$ steps where $n \geq 0$. It is defined by
transitively composing the $\kstep$ relation. We use the predicate
constant $\keval$ to denote the big-step evaluation relation such that
$\keval\app t\app v$ holds if $t$ evaluates to the value $v$ in a finite
number of steps, \ie, $\knstep\app n\app t\app v$ holds for some $n$.
Similarly, we define the predicates $\kstep'$, $\knstep'$ and
$\keval'$ that represent corresponding evaluations relations for the
target language.

Assume $P$ is the transformation under consideration and $\sim$ and
$\approx$ represent the logical relations denoting the semantics
preservation property of $P$. We designate the predicate $\ksim$ to
represent $\sim$ and $\kequiv$ to represent $\approx$. We then
translate the definitions of $\sim$ and $\approx$ into clauses for
$\ksim$ and $\kequiv$ such that
\begin{itemize}
\item $\simindex \tau i t {t'}$ holds if and only if $\ksim\app T\app
  I\app M\app M'$ holds and
\item $\equalindex \tau i t {t'}$ holds if and only if $\kequiv\app
  T\app I\app M\app M'$ holds
\end{itemize}
where $T$, $I$, $M$ and $M'$ are encodings of $\tau$, $i$, $t$ and
$t'$, respectively. The translation is straightforward and makes use
of the encoding of evaluation semantics. Note that we need to indicate
that $\ksim$ and $\kequiv$ are defined only for closed terms. We use
the technique presented in Section\cspc\ref{subsec:lts_reason} to
characterized this closedness property. That is, we identify a
predicate $\ktm$ defined by a set of program clauses in \LProlog such
that $\oseq{\ktm\app M}$ holds if and only if $M$ is a well-formed
closed term.
Notice that $\ksim$ and $\kequiv$ are not given as a fixed-point
definition. Instead, they form a \emph{recursive definition} that is
based on an inductively defined set of pairs of types and step
indexes. The theoretical justification of recursive definitions in
\Abella is given in \cite{baelde12lics}.

Given the encoding of logical relations on closed terms, it is easy to
extend it to relate closed substitutions. We designate the predicate
$\ksubstequiv$ to represent the logical relation on closed
substitutions such that $\ksubstequiv\app L\app I\app S\app S'$ holds
if and only if $\equalindex \G i \theta {\theta'}$ where $\G$, $I$,
$S$ and $S'$ are encodings of $\G$, $i$, $\theta$ and $\theta'$. It is
given a fixed-point definition by using $\ksim$ and $\kequiv$ through
the following clauses:
\begin{tabbing}
\qquad\=\quad\=\kill
\>$\ksubstequiv\app \knil\app I\app \knil\app \knil$ \quad$\rdef$ \quad$\rtrue$\\
\>$\nabla x. \ksubstequiv\app (\kof\app x\app T \cons L)\app I\app
     (\kmap\app x\app M \cons S)\app (\kmap\app x\app M' \cons S') \quad\rdef$\\
\>\>$\kequiv\app T\app I\app M \app M' \rand \ksubstequiv\app L\app I\app S\app S'$
\end{tabbing}
%% \begin{tabbing}
%% \qquad\=$\nabla x. \ksubstequiv\app $\=$(\kmap\app x\app M \cons S)\app
%%        (\kmap\app x\app M' \cons S')$
%%   \quad\=$\rdef$
%%   \quad\=\kill
%% \>$\ksubstequiv\app \knil\app I\app \knil\app \knil$ \>\>$\rdef$ \>$\rtrue$\\
%% \>$\nabla x. \ksubstequiv\app (\kof\app x\app T \cons L)\app I$\\
%% \>\>$(\kmap\app x\app M \cons S)\app (\kmap\app x\app M' \cons S')$
%%   \>$\rdef$ \>$\ksubstequiv\app L\app I\app S\app S'$
%% \end{tabbing}

Suppose the relation describing the transformation $P$ is encoded as a
predicate constant $p$ such that $(p\app M\app M')$ holds if and only
if $t$ is transformed into $t'$ by $P$ where $M$ and $M'$ are
respectively the encodings of $t$ and $t'$. Then we can state
Property\cspc\ref{thm:sem_pres_index_open} for $P$ as the following
theorem in \Abella where $\oseq{L \stseq \kof\app M\app T}$ asserts
that the source term $M$ has type $T$ in the typing context $L$:
\begin{tabbing}
\qquad\=\quad\=\quad\=\kill
\>$\rfall L, I, S, S', M, M', T, N, N'.$\\
\>\>$\ksubstequiv\app L\app I\app S\app S' \rimp
  \oseq{L \stseq \kof\app M\app T} \rimp
  \oseq{p\app M\app M'} \rimp$\\
\>\>\>$\kappsubst\app S\app M\app N \rimp
  \kappsubst\app S'\app M'\app N' \rimp
  \ksim\app T\app I\app N\app N'$.
\end{tabbing}
This theorem is then proved by induction on $\oseq{p\app M\app
  M'}$. When the terms $M$ and $M'$ are closed and the substitutions
$S$ and $S'$ are empty, we get the following special case of the above
theorem, corresponding to Property\cspc\ref{thm:sem_pres_index_closed}:
\begin{tabbing}
\qquad\=\quad\=\quad\=\kill
\>$\rfall I, M, M', T.
  \oseq{\kof\app M\app T} \rimp
  \oseq{p\app M\app M'} \rimp \ksim\app T\app I\app M\app M'$.
\end{tabbing}
When $T$ is an atomic type, this theorem further degenerates into
the following, corresponding to Property\cspc\ref{thm:sem_pres_atom}:
\begin{tabbing}
\qquad\=\quad\=\quad\=\kill
\>$\rfall M, M', T.
  \oseq{\kof\app M\app T} \rimp
  \oseq{p\app M\app M'} \rimp
  \oseq{\keval\app M\app V} \rimp
  \oseq{\keval'\app M'\app V}$.
\end{tabbing}

To compose the above theorem with that for other transformations, we
need to prove the following type preservation property for the
transformation:
\begin{tabbing}
\qquad\=\quad\=\quad\=\kill
\>$\rfall M, M', T.
  \oseq{\kof\app M\app T} \rimp
  \oseq{p\app M\app M'} \rimp
  \oseq{\kof'\app M'\app T}$.
\end{tabbing}
It is proved by induction on $\oseq{p\app M\app M'}$, usually in a
straightforward manner.

Now, assume the predicate constants $p_1,\ldots,p_n$ encode the
transformation relations for the sequence of transformations in the
compiler and the predicate constant $c$ represent the compilation
relation such that $(c\app t_0\app t_n)$ holds if and only if there
exists $t_1,\ldots,t_{n-1}$ such that $(p_i\app t_{i-1}\app t_i)$ holds
for $1 \leq i \leq n$. Then semantics preservation of the full
compiler is encoded as follows which corresponds to
Property\cspc\ref{thm:full_sem_pres}:
\begin{tabbing}
\qquad\=\quad\=\quad\=\kill
\>$\rfall M, M', T.
  \oseq{\kof\app M\app T} \rimp
  \oseq{c\app M\app M'} \rimp
  \oseq{\keval\app M\app V} \rimp
  \oseq{\keval'\app M'\app V}$.
\end{tabbing}
Suppose we have proved the semantics and type preservation properties
for $p_1,\ldots,p_n$. We can then apply them to prove the above
theorem. The proof closely follows the informal one we described in
Section\cspc\ref{sec:vfc_approach}.

\section{Using $\lambda$-Tree Syntax in Verified Compilation}
\label{sec:lts_vfc}

%GN This is indeed repetitive as Kaustuv says but it is easily solved
%by removing it completely.
%% As described in the introduction chapter, the functional compiler
%% transformations often perform non-trivial manipulation and analysis of
%% binding structure. To formally specify functional compiler
%% transformations, we need to precisely characterize the notions related
%% to binding structure such as scopes of binding variables, renaming of
%% bound variables and substitution. Reasoning about binding structure is
%% an even bigger obstacle because the properties of binding structure
%% that are implicit in specification must be proved explicitly in
%% reasoning. If the complexity of dealing with binding structure is not
%% handled properly, it can overwhelm the reasoning task.

%% We exploit the $\lambda$-tree syntax approach provided by the
%% framework consisting of \LProlog and \Abella to simplify the
%% implementation and reasoning about binding structure in verified
%% compilation. The uses and benefits of this approach described in
%% Section\cspc\ref{sec:hoas} carry over to verified compilation. We
%% elaborate a bit further on how they play out in the setting of
%% verified compilation in the rest of this section.

By using the framework consisting of \LProlog and \Abella in verified
compilation of functional programs, we expect to draw on the benefits
for the $\lambda$-tree syntax approach that we outlined in
Section\cspc\ref{sec:hoas}. In this section, we motivate the way in
which these benefits will play out in the concrete developments that
we undertake in the next few chapters in the thesis.

In the implementations of transformations on functional programs, we use
meta-level $\lambda$-abstraction to represent the binding operators in
functional objects. As a result, the notions related to binding
structure such as renaming and substitution are captured by $\alpha$-
and $\beta$-conversions in a logical and precise fashion. For
instance, $\beta$-conversion will be used to model the
``administrative'' substitution operations that are an inherent part
of the CPS transformation as described in Chapter\cspc\ref{ch:cps},
resulting in a very concise implementation of the transformation.
When working on functional objects, the
compiler transformations often need to go under their binding
operators and recursively transform the function bodies. Furthermore,
these transformations often have side conditions for the binding
variables introduced by recursion. As described in
Sections\cspc\ref{subsec:ecd_bindings} and \cspc\ref{sec:hoas} we use
the universal and hypothetical goals to perform recursion over binding
operators and to enforce these side conditions.
Compiler transformations often perform non-trivial analysis on the
binding structure of function objects. Such analysis can be captured
in the \LProlog specifications concisely and logically. For example,
in Chapter\cspc\ref{ch:codehoist} we will present a transformation
that extracts closed functions to the top-level. For the extraction to
work, it is necessary to show that such functions do not refer to any
bound variable. This independence relation can be statically
characterized via quantification ordering and dynamically realized via
unification. Complicated analysis of binding structure can even be
specified through \LProlog programs. As an example, consider the
closure conversion transformation that was briefly touched in the
introduction and that will be presented in detail in
Chapter\cspc\ref{ch:closure}. To implement closure conversion, we need
to compute free variables in functions. This computation will be
presented as a relation between function objects and their free
variables and defined through a set of \LProlog program clauses.

The above uses of the $\lambda$-tree syntax approach also apply when
we treat definitions involving binding structure in the verification
of compiler transformations on functional programs in \Abella. Furthermore, the
logical structure of such treatments can be exploited to significantly
simplify reasoning about binding structure. For example, we have given
a definition of substitution as an explicit relation in
Section\cspc\ref{subsec:schm_poly_exms} which is an essential
component for describing the semantics preservation property
characterized as logical relations. We have also shown that a lot of
properties of the substitution relation can be easily established by
observing that they are just manifestation of the properties of
$\beta$-conversion in the meta-level language.

Rich properties of binding structure can be proved by combining the
$\lambda$-tree syntax approach and the two-level logic approach. This
manifests in the treatment of closedness property of $\lambda$-terms.
In Section\cspc\ref{subsec:lts_reason} we have seen that we can
characterize closed terms through a \LProlog specification for the
predicate \ktm such that $\oseq{\ktm\app M}$ holds if and only if $M$
is a closed term. This predicate is used in the definition of logical
relations which should hold only for closed terms. We can then easily
prove the property that substitution has no effect on closed terms by
using the explicit definition of substitution and the \LProlog
specification for closed terms. This property is critical to proving
semantics preservation of closure conversion as we shall see in
Chapter\cspc\ref{ch:closure}.

Finally, the logical structure of binding related treatments in the
implementation of compiler transformations can be exploited to
simplify reasoning about binding structure through the two-level logic
approach. For instance, the usage of $\beta$-conversion to model
``administrative'' substitution operations in the CPS transformation
makes it extremely easy to reason about the effects of such
operations, as we shall see in Chapter\cspc\ref{ch:cps}. As another
example, in Chapter\cspc\ref{ch:closure} we will need to prove the
following ``strengthening'' property for typing: if a term $t$ has
type $\tau$ in a typing context $\G$ and $\G'$ is a restriction of
$\G$ that contained only the free variables in $t$, then $t$ also has
type $\tau$ in $\G'$. We shall see that this property can be easily
established by exploiting the logical structure of the \LProlog
program for computing free variables.

\section{Nuances in Formalizing Semantics Preservation}
\label{sec:sem_pres_nuances}

We will be using a logical relations style characterization of
semantics preservation in the work in this thesis. There are, however,
a few other ways in which this notion has been characterized in the
literature. We discuss and contrast these different approaches in this
section.

%% We use the following criteria proposed by Neis
%% \etal\cspc\cite{neis15icfp} to assess the different approaches to
%% semantics preservation:
A good starting point for the discussion is a set of criteria proposed
by Neis \etal\cspc\cite{neis15icfp} to assess the different approaches
to formalizing semantics preservation:
\begin{itemize}
\item
  \emph{Modularity}: This is a property that allows us to build the
  correctness proof for a large program in the way we build the
  program itself, \ie by composing the correctness proofs for the
  modules constituting the program. Formally, this means that if we
  have shown that the target programs $T_1$ and $T_2$ preserve the
  semantics of the source programs $S_1$ and $S_2$ from which they
  were generated, then the code that results from linking $T_1$ and
  $T_2$  should also preserve the semantics of the result of composing
  $S_1$ and $S_2$ at the source level. This property is also called
  ``horizontal composibility.''

\item
  \emph{Flexibility}: This criterion amounts to saying that the
  definition of semantic preservation is fixed solely by the semantics
  of the source and target languages and is oblivious to the
  transformations that are performed. Together with modularity, it
  allows for combination of correctness proofs of modules that use the
  same definition of semantics preservation but are generated by
  different compiler transformations.

\item
  \emph{Transitivity}: This criterion amounts to saying that semantics
  preservation proofs for individual transformations can be composed
  to derive the semantic preservation proof for the full sequence of
  transformations. Transitivity is necessary for the separate
  verification of compilation passes in multi-pass
  compilers. Transitivity is also called ``vertical composibility''.
\end{itemize}
%
%% Based on the three criteria, we compare the strength and weakness of
%% three notions of semantics preservation for proving the correctness
%% of compiler transformations based respectively on logical relations,
%% a weak equivalence relation between values and a relation called
%% \emph{parametric inter-language simulations} or
%% PILS\cspc\cite{neis15icfp}.

%% We first discuss the strength of logical relations as the notion of
%% semantics preservation based on the above criteria.
Correctness proofs
based on logical relations enjoy the modularity property because of
the extensional reading underlying equivalence at function types. If
we think of modules with external references as functions whose
arguments are these references, then we can reduce the linking of
modules to function applications. By the definition of the equivalence
relation for function values, we can easily combine the semantics
preservation proofs of individual modules to form the proof for the
linked program.

Correctness proofs based on logical relations are also flexible.  This
is because the definitions of logical relations only depend on the
evaluation semantics of the source and target programs they hold
of. As a result, the correctness proofs of different transformations
that use the same logical relation as the notion of semantics
preservation can be combined without any problem.

Establishing transitivity of correctness proofs when these use a
notion of semantics preservation based on logical relations is more
difficult. Perhaps the most obvious way to obtain this property is to
show that the underlying logical relations are composable. In
particular, suppose that the logical relations $\simone$ and $\simtwo$
underlie the proofs of semantics preservation for two 
transformations on functional programs and we want to show the correctness of the composition
of these two transformations with respect to a third logical relation
$\simthree$. This would trivially be the case if we can show the
following:
\[
\rfall \tau\app t_1\app t_2\app t_3\app,
\simulateone \tau {t_1} {t_2} \rimp \simulatetwo \tau {t_2} {t_3} \rimp
\simulatethree \tau {t_1} {t_3}.
\]
However, this kind of property is generally hard to prove. In fact, it
may not hold even if the source and target languages for the
transformations are identical and all three logical relations are the
same. In \cite{ahmed06esop}, Ahmed proposed a way to restrict the
permitted forms of logical relations using properties based on types
that overcomes this difficulty.  Regardless of the merits of this
approach, it is not directly usable in the typical setting for
compiler verification. The reason for this is that compiler
transformations typically modify and simplify programs in a one
language into programs in a \emph{different} language that is equipped
with specialized constructs motivated by the relevant
transformation. As a result, the logical relations that are used to
characterize semantics preservation at each stage of a multi-stage
transformation are usually different ones and they also relate
programs in different languages. For example, $\simone$ may be a
logical relation for the continuation-passing style transformation
that will be described in Chapter\cspc\ref{ch:cps}, $\simtwo$ may be a
logical relation for the closure conversion transformation that will
be described in Chapter\cspc\ref{ch:closure}, and $\simthree$ is
correspondingly the logical relation that captures the notion of
semantics preservation between the source and target languages for the
composition of the two transformations. In this case, the definition
of $\simone$ would have to take into account the way we intend the
devices called \emph{continuations} to function, while the definition
of $\simtwo$ makes no assumption about continuations. In this kind of
situation, it is not clear that we can derive a suitable logical
relation between expressions in the initial and final languages simply
by composing the sequence of intermediate logical relations that have
been used.

The above discussion shows that if we use a definition of semantics
preservation that is based on logical relations, then it becomes
difficult to prove the correctness of a multi-stage compiler simply by
composing correctness proofs for each stage. We may, however, restrict
our attention to those programs that produce a value of atomic type;
doing so effectively means that we are focusing on compilation of
complete programs and then modularity and flexibility become
irrelevant issues. If we narrow our focus in this way and if
equivalence at atomic types is based on a simple one-to-one mapping of
values between relevant domains, then correctness results for multiple
stages can be composed to obtain a correctness result for the entire
sequence of transformations. In fact, this is the perspective we take
in this thesis and we have discussed how the composition works in
Section\cspc\ref{sec:vfc_approach}.

The idea of limiting attention to programs that produce values of
atomic type has, in fact, been widely used in compiler verification;
to take two recent examples, it underlies the
CompCert~\cite{leroy09cacm} and the CakeML~\cite{kumar14popl}
projects.  In this more restricted context, the main requirement of
the notion of program equivalence that we use is that it accord with
intuitions at atomic types, \ie, equivalence between values is reduced
to identity at atomic types. At function types, we may use extensional
equality as is done with logical relations style definitions of
equivalence, but we may also use other notions that, for example, help
us construct correctness proofs more easily. One particular notion of
equivalence that we may use is the relation induced by the
transformation itself; this typically preserves values at atomic
types, thus satisfying the key property we need of program
equivalence.

If we use the above idea, the forward simulation property reduces to
the following statement for any given transformation $P$:
\begin{quotation}
\noindent {\it For any $t$ that is transformed by $P$ into $t'$, if
  $t$ evaluates to some value $v$, then there exists an value $v'$
  such that $t'$ evaluates to $v'$ and $v$ is transformed by $P$ into $v'$.}
\end{quotation}
This statement, which amounts to saying that the transformation and
evaluation permute with each other, is usually proved by induction on
evaluation sequences in the source language. In order to be able to
prove such a statement, we often have to tune the definitions of the
transformation and of evaluation to each other.

By the above statement it is easy to show that correctness proofs
based on the refined approach we have just described are transitively
composible; note that the justification for this approach is that we
are eventually interested only in whole programs that produce a value
of atomic type.  Proofs in this style are, as a rule, not flexible
because they are often dependent on tuning the definition of
evaluation to the transformation being proved correct. Finally, the
simulation based approach does not automatically guarantee modularity
because the notion of equivalence it uses for functions is not
sufficiently constrained. Determining ways to apply this approach that
yield modularity is an active research topic; \cite{stewart15popl}
presents a recent development along these lines.

There has been some recent work led, not surprisingly by Neis and
colleagues that is aimed at designing a notion of semantics
preservation that actually meets all three desired criteria put
forward for such a notion. Most specifically, Neis
\etal\cspc\cite{neis15icfp} have introduced a relation called
parametric inter-language simulation or PILS towards this end.
PILS can be thought of as
a refinement to logical relations. Similar to logical relations, PILS
enjoys modularity and flexibility. Moreover, PILS also enjoys
transitivity that logical relations fail to support. The key idea is
to break up a single definition of logical relation into two relations
called ``global knowledge'' and ``local knowledge'' to dismiss the
negative self-reference that occurs in the definition of equivalence
at function types so that semantics preservation can be characterized
as fixed-point definitions. As we have discussed before, the negative
self-reference is the main reason that logical relations are not
transitively composible. By finessing the need for such a reference,
PILS is able to support transitivity.

%% Recently, some researchers working on compiler verification have been
%% striving to design notions of semantics preservation that enjoy all
%% the three properties. The relation called parametric inter-language
%% simulations or PILS introduced by Neis \etal\cspc\cite{neis15icfp} is
%% the most recent result of this line of work. PILS can be thought of as
%% a refinement to logical relations. Similar to logical relations, PILS
%% enjoys modularity and flexibility. Moreover, PILS also enjoys
%% transitivity that logical relations fail to support. The key idea is
%% to break up a single definition of logical relation into two relations
%% called ``global knowledge'' and ``local knowledge'' to dismiss the
%% negative self-reference that occurs in the definition of equivalence
%% at function types so that semantics preservation can be characterized
%% as fixed-point definitions. As we have discussed before, the negative
%% self-reference is the main reason that logical relations are not
%% transitively composible. By getting rid of it, PILS is able to support
%% transitivity.

We have chosen to use a logical relations style definition of
semantics preservation rather than one based on PILS mainly because
the latter is still in at an evolutionary stage; our focus in this
thesis is not so much on developing new and useful ideas concerning
semantics preservation as it is on exposing useful approaches to
formalizing them.
%GN Have you really already shown this? Then what is the purpose of
%the chapters that follow?
%% We have already shown that the $\lambda$-tree syntax approach can
%% greatly simplify the formal proofs of semantic preservation based on
%% logical relations in Section\cspc\ref{sec:lts_vfc}. This also applies
%% in the settings where other notions of semantics preservation are
%% used.
%
In this regard, we believe many of the lessons to be drawn from this
work also carry over to situations in which other notions of semantics
preservation are used.
The key observation is that representation, manipulation and analysis
of binding structure are inherent in compiler transformations of
functional programs and properties about them must be explicitly
proved in reasoning no matter what notions of semantics preservation
are used. Thus, the $\lambda$-tree syntax approach can always be used
to simplify these reasoning tasks.
For example, suppose we prove the correctness of the closure
conversion transformation by showing that it permutes with
evaluations. For this we need to keep an explicit representation of
environments in the evaluation semantics. The evaluation environments
can be easily encoded in \LProlog by using meta-level abstractions to
represent the bindings of the environments and their properties can be
easily proved in \Abella by exploiting the logical structure of the
encoding of environments.
As another example, the definition of PILS makes essential use of
substitution like logical relations. As a result, reasoning about the
substitution operations and their interaction with other program
constructs is an inherent part of the correctness proofs based on
PILS. Such reasoning can be carried out effectively by using
$\lambda$-tree syntax just as when logical relations are used as the
correctness notion.

\section{Exercise of Verified Compilation in the Thesis}
\label{sec:vfc_exercise}

To demonstrate the effectiveness of our methodology for implementing
and verifying transformations on functional programs, we will use it to develop a
verified compiler for a representative functional language. This
language extends the simply typed $\lambda$-calculus with general
recursion, conditional and arithmetic expressions; it can be thought
of as an extension to the well-known PCF
language\cspc\cite{plotkin77}. Our compiler takes programs in the
$\lambda$-tree syntax representation as input; transforming actual
programs into this form can be accomplished by standard tools for
parsing and extracting an internal representation. Compilation will be
achieved through several passes that translate the source programs
into successive intermediate languages, and finally producing code in
a language that is similar to the Cminor language used in the
CompCert project\cspc\cite{leroy09jar}.
We have chosen this as our target language because many other compiler
verification projects have used Cminor as an intermediate language and
we can therefore benefit from their work in completing the compiler
verification process.
%GN Why is it useful if other projects also target Cminor? That still
%leaves the rest of the way to be verified!
%% We choose this target language for the
%% following reasons. First, Cminor is the source language of the
%% back-end of CompCert and a well-known intermediate language that many
%GN Haven't you said this already? How many more times do you want to
%repeat it?
%% compiler verification projects target. Second, there is a minimum
%% need to deal with binding structure for compiling Cminor. Therefore,
%% the benefits of HOAS cannot be observed in the development of a
%% verified compiler for it using our framework. In fact, the
%% implementation and verification of the compiler for Cminor using our
%% framework will most likely use only the first-order features of our
%% framework and will be similar to the abundant existing work on
%% verified compilation of Cminor-like languages. We would like to avoid
%% repeating such work in this thesis.

\begin{figure}[!ht]
  \center
  \begin{tikzpicture}
    \draw (0cm,0cm) node[rotate=90]{\emph{PCF-like Code}};
    \draw (1.2cm,0cm) node[draw]{\Textbox{1.3cm}{1.7cm}{CPS Transformation}};
    \draw (2.5cm,0cm) node{$\Rightarrow$};
    \draw (3.7cm,0cm) node[draw]{\Textbox{1.3cm}{1.5cm}{Closure Conversion}};
    \draw (4.9cm,0cm) node{$\Rightarrow$};
    \draw (6.1cm,0cm) node[draw]{\Textbox{1.3cm}{1.5cm}{Code Hoisting}};
    \draw (7.3cm,0cm) node{$\Rightarrow$};
    \draw (8.5cm,0cm) node[draw]{\Textbox{1.3cm}{1.6cm}{Code Generation}};
    \draw (9.7cm,0cm) node[rotate=90]{\emph{Cminor-like Code}};
  \end{tikzpicture}

  \caption{The Compiler for a PCF-style Language}
  \label{fig:pcf_compiler}
\end{figure}
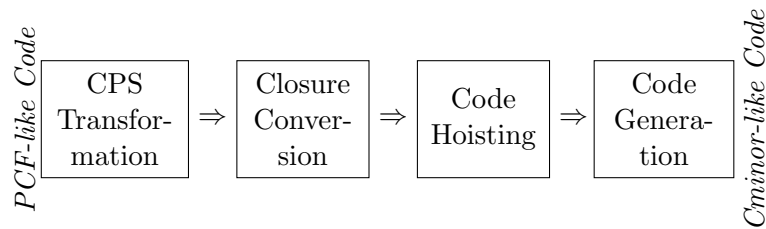

The structure of the compiler that we develop in this thesis and that
is shown in
Figure\cspc\ref{fig:pcf_compiler} is explained qualitatively as follows.
The CPS transformation makes the control flow such as evaluation
ordering explicit. The closure conversion transformation makes
(nested) functions independent of their context. It effectively makes
functions closed. The code hoisting transformation lifts these closed
functions to the top level. After this transformation, all the
higher-order features of the source programs are removed. The code
closely resembles that in procedural languages such as C. The code
generation phase makes the allocation of objects explicit and
generates Cminor-like code.

We use the methodology described in Section\cspc\ref{sec:fm_vfc} to
implement a formally verified version of this compiler.
Chapter\cspc\ref{ch:cps},\ref{ch:closure},\ref{ch:codehoist}
respectively describe the implementation and verification of the CPS
transformation, closure conversion and code
hoisting. Chapter\cspc\ref{ch:complete} describes the code generation
transformation and composition of semantics preservation of individual
transformations to form the correctness proof for the full
compiler.

\chapter{The Continuation Passing Style Transformation}
\label{ch:cps}

Continuation Passing Style (CPS) is a programming style in which
control flow is made explicit by using devices called
\emph{continuations} that record ``the remaining computation''. CPS has a
long history that dates back to
1960s\cspc\cite{reynolds93lsc}. Languages in this form are favored as
intermediate languages for a lot of compilers for functional
languages (\eg, see\cspc\cite{appel1992,sussman98hsc,kennedy07icfp})
because continuations provide an elegant representation of language
constructs such as function calls and pattern matching, and
expressions in the CPS form resemble statements in procedural languages
and thereby can be more easily related to executable code.

A CPS transformation translates functional programs in a direct style
of programming into a CPS
form\cspc\cite{fischer72sigplan,plotkin76,danvy92mscs}. It takes as
arguments a source term and a continuation representing an abstraction
of the remaining computation over the value of the source term. It
then transforms the source term into a more fine-grained form such
that control flow is made explicit---\ie, there is no ambiguity in the
order of evaluation---by recursively applying the transformation on
its subterms and accumulating the continuation in the process. A
common approach to describe the CPS transformation is to describe it
using $\lambda$-calculus~\cite{plotkin76,danvy92mscs}. In this approach
, it is important to distinguish
between $\beta$-redexes introduced by the transformation for realizing
substitution at the translation time and $\beta$-redexes that come
from the source term. The former kind of $\beta$-redexes are reduced
at the translation time as part of the transformation while the later
should not be reduced because that will count as partial evaluation of
the source term in the compilation phase. We adopt the terminology
in \cite{danvy92mscs} to denote the former as \emph{administrative
  $\beta$-redexes} and the later as \emph{dynamic $\beta$-redexes},
\ie, $\beta$-redexes that should only be reduced at run time.

The CPS transformation can be easily described in a rule-based and
relational style. However, the formalization of such a description can
be difficult. The major difficulty lies in correctly capturing the way
the administrative and dynamic $\beta$-redexes work. Moreover, this
difficulty is amplified when proving properties of the transformation
such as semantic preservation because the properties of binding
related notions for manipulating administrative and dynamic
$\beta$-redexes must be proved explicitly.

We solve the above problems by using the methodology for implementing
and verifying compiler transformations described in
Section\cspc\ref{sec:fm_vfc}. First, we encode the rule-based
relational description of the CPS transformation as a \LProlog
specification. We use meta-level $\beta$-redexes to encode the
administrative $\beta$-redexes. As such, the substitution operations
they represent are automatically captured by meta-level $\beta$-reduction. The
dynamic $\beta$-redexes are encoded as program constructs by using the
$\lambda$-tree syntax approach. Because \LProlog specifications are
executable, the encoding of the CPS transformation is also its
implementation. We then prove that the implementation preserves
semantics in \Abella by following the logical
relation based approach described in
Section\cspc\ref{sec:fm_vfc}. Because the logical structure of the
implementation is transparently reflected into \Abella via the
two-level logic approach, we are able to prove the properties of
administrative and dynamic $\beta$-redexes
easily by exploiting their $\lambda$-tree syntax representation. In the end
we get a concise and elegant proof of semantics
preservation for the implementation.

In the following sections, we illustrate the above ideas by
constructing a verified implementation of the CPS transformation in
our compiler described in Section\cspc\ref{sec:vfc_exercise}.
%% This
%% verified implementation follows the above approach and illustrates the
%% benefits of the $\lambda$-tree syntax in implementing and reasoning
%% about operations related to administrative and dynamic
%% $\beta$-redexes.
We first give an overview of the CPS transformation
in Section\cspc\ref{sec:cps_overview}. We then present the source and
target languages and give a rule-based relational description of the
transformation in Section\cspc\ref{sec:rule_based_cps}. We then encode
the rule-based description as a \LProlog specification in
Section\cspc\ref{sec:cps_impl}. The discussion here will focus on
using the $\lambda$-tree syntax approach to encode the
administrative and dynamic $\beta$-redexes. In
Section\cspc\ref{sec:informal_verify_cps}, we present the informal
semantics preservation proof of the CPS transformation. We lastly give
the formal semantic preservation proof of the implementation in
Section\cspc\ref{sec:formal_verify_cps}. The discussion will bring out
the benefits of $\lambda$-tree syntax approach in reasoning about the
administrative and dynamic $\beta$-redexes.

\section{An Overview of the Transformation}
\label{sec:cps_overview}

The CPS transformation we work with is based on the transformation
proposed by Danvy and Filinski in \cite{danvy92mscs} that reduces
administrative $\beta$-redexes on the fly. To describe the
transformation, we need to distinguish between the administrative
$\beta$-redexes which only manifest themselves during the
transformation and are
eliminated in the final result, and the dynamic $\beta$-redexes that
are constituting pieces of the result. We shall use $\cabs x t$ and
$(\capp {t_1} {t_2})$ to represent administrative abstractions and
applications and normal abstractions and applications to represent the
dynamic ones.

The transformation takes as input a term and a continuation that is an
administrative abstraction over the value of the term and that is
already in the CPS form. Intuitively, the source term can be thought
as a program fragment that represents the ``current computation'' and
eventually evaluates to a value. The continuation can be thought as a
program fragment that, when fed with the value of the source term,
%GN Periods must be inside the quotes. Check the convention with other
%punctuations on the web and make sure to follow it uniformly.
represents the ``remaining computation.'' The job of the CPS transformation
is to convert the source term into a form in which control flow
becomes explicit and to combine the result with the continuation to
form the complete program. The transformation proceeds by recursion
over the structure of the source term. At the beginning of this
process, a continuation representing the context in which the source
program will be used is given. The CPS transformation is then
recursively applied to the sub-expressions of the source term and the
results are accumulated into the continuation in an order that
reflects the control flow of the source program. The result of such
accumulation is the output of the CPS transformation.

We start by illustrating how the CPS transformation works on
basic expressions through some examples. Consider the base case
of the transformation, \ie, when the source term $t$ is a variable or
a constant. In this case, the control flow in $t$ is vacuously
fixed. Letting $\cabs v {t'}$ be the input continuation, we can
therefore apply $\cabs v {t'}$ to the source term to form the output
which is $(\capp {(\cabs v {t'})} t)$. The term $(\capp {(\cabs v
  {t'})} t)$ is an example of administrative $\beta$-redexes which
will be immediately reduced by the transformation. To illustrate the
transformation on compound expressions, consider the case when the
source term is an addition expression $t_1 + t_2$. Let $(\cabs v t)$
be the input
continuation and assume that the expression is evaluated from left to
right. Then we can break down the computation represented by $t_1 +
t_2$ into the following sequence: evaluating $t_1$ to $v_1$,
evaluating $t_2$ to $v_2$ and evaluating $v_1 + v_2$. By doing so the
control flow becomes explicit and unambiguous. We recursively perform
the CPS transformation on $t_2$ with the continuation
\[
c_2 = (\cabs {v_2} {\letexp v {v_1 + v_2} {\capp {(\cabs v t)} v}})
\]
to get a target program $t_2'$; note that $v_1$ is a free variable
in this expression that will be bound by an administrative abstraction
in the input continuation to the transformation of $t_1$ as we explain
presently. In the continuation $c_2$, we have used a let expression to
explicitly state that the addition of $v_1$ and $v_2$ must be computed
before we can use it for the remaining computation represented by $t$. This
is a common technique for enforcing evaluation ordering that will
arise often in the following discussion. Again notice the use of the
administrative $\beta$-redex $({\capp {(\cabs v t)} v})$ to represent
a substitution operation. Notice also that $c_2$ represents the
computation after evaluating $t_2$. As a result, $t_2'$ represents the
evaluation of $t_2$ followed by the remaining computation, which is
also the computation after evaluating $t_1$. We can therefore
recursively apply the transformation on $t_1$ with $\cabs {v_1}
{t_2'}$ as the input continuation to get the complete program in the
CPS form.

At this point we know how the basic expressions are transformed into
the CPS form. Another important part of the CPS transformation is to
convert functions and function applications into the CPS form. A
function in the CPS form takes a continuation as an extra
argument. This continuation is provided by the caller of the function
that represents the future computation after the function
returns. Instead of returning a value to the caller, evaluation of the
function body ends with calling the continuation with the return value
as its argument. As a result, in the CPS form function calls never
return and computation always goes forward by invoking other
continuations. By making the control flow for function calls and
returns explicit, we avoid having to treat a function return as a
program construct and enable optimizations that rely on control flow
analysis such as elimination of tail-calls.

We give a more concrete account of the transformation on functions and
function calls. Let $\abs x t$ be the source function and $k$ be the
input continuation. Letting $k'$ denote the continuation passed
over from the caller, we first recursively perform the transformation
on $t$ with the continuation $(\cabs a {k'\app a})$ to get the
function body $t'$ in the CPS form. Note that we cannot use $k'$
directly as an input to the recursive transformation because it is a
dynamic $\lambda$-abstraction. That is why we instead have used $(\cabs a {{k'}\app a})$
which is an administrative $\lambda$-abstraction equivalent to $k'$ as
the input continuation. Then the original function $\abs x t$ is
transformed into the following expression:
\begin{tabbing}
  \qquad\=\kill
  \>$\letexp f {\abs x {\abs {k'} {t'}}} {(\capp k f)}$
\end{tabbing}
Note that the original $\lambda$-abstraction is converted to the
dynamic abstraction ${\abs x {\abs {k'} {t'}}}$ and the
administrative $\beta$-redex $(\capp k f)$ represents the remaining
computation.

To transform a function application $(t_1\app t_2)$ given the input
continuation $k$, we need to pass $k$ as an extra argument to
$t_1$. We cannot use $k$ directly as an argument because it is an
administrative $\lambda$-abstraction while a dynamic $\lambda$-abstraction
is expected here. We therefore convert $k$ into a dynamic abstraction
$\abs a {(\capp k a)}$ and use it as the continuation argument. Let
$c_2$ be the continuation representing the computation after
evaluating $t_2$. This continuation has the form
\[
   \cabs {v_2} {v_1\app v_2\app (\abs a {(\capp k a)})}
\]
where $v_1$ is a free variable that, intuitively, represents the place
where the result of evaluating (the transformed version of) $t_1$ must
be filled in to complete the computation.
Now, to actually effect the transformation we first transform
$t_2$ with $c_2$ as the input continuation to get $t_2'$.
%Observe that
%the continuation $\cabs
%{v_1} {t_2'}$ represents the computation after evaluating $t_1$.
Then we transform $t_1$ with the
continuation $\abs {v_1} {t_2'}$ to get the complete result of
transforming $(t_1\app t_2)$ into the CPS form.

As an example, consider the following program expression:
\begin{tabbing}
  \qquad\=\kill
  \>$\letexp f {\abs x {x + 2}} {f\app 3}$
\end{tabbing}
The input continuation used for transforming this expression, as also
any expression at the top-level, is $\cabs x x$. Using this
continuation, the CPS transformation converts
the given expression into the following equivalent one in the CPS form:
\begin{tabbing}
\qquad\=\qquad\=\kill
\>$\letexp
    f
    {(\abs x {\abs {k'} {\letexp v {x + 2} {(k'\app v)}}})}
    {f\app 3\app (\abs {a} a)}$
\end{tabbing}
In producing this expression, we have contracted several
administrative redexes that arise, as we invite the reader to verify
by actually carrying out the steps of the transformation that we have
described.
%% The one ``administrative'' application that is left is
%% $(\capp k a)$. The $\beta$-redex $(\capp k a)$ can also be immediately
%% reduced because we known $k$ must be of the form $\cabs v t$ for some
%% $v$ and $t$, resulting in a term without any administrative redexes.

We have now covered all the important aspects of the CPS
transformation in the style of Danvy and Filinski. Realistic
functional programming languages may have richer program constructs. But the CPS
transformation for them follows the essential ideas exposed in this
section.

\section{A Rule-Based Description of the Transformation}
\label{sec:rule_based_cps}

The CPS transformation is the first pass of the compiler we described
in Section\cspc\ref{sec:vfc_exercise}. We give it a rule-based
relational description in this section based on the ideas presented in
the last section.

\subsection{The source and target languages}
\label{sec:cps_langs}

The source language of the CPS transformation, which is also the
source language of the whole compiler, is a slight variant of the PCF
language\cspc\cite{plotkin77}, a representative functional programming
language that is often studied in the literature on
functional programming languages and verification. The target language
of the transformation is the same as its source language, \ie, the CPS
transformation is a source-to-source transformation that does not
introduce any new program constructs. The syntax of this language is shown in
Figure\cspc\ref{fig:cps_src_lang}. In this figure, $T$, $M$ and $V$
stand respectively for the categories of types, terms and the terms
recognized as values.

\begin{figure}[ht!]
  \begin{tabbing}
    \qquad\qquad\qquad\qquad\=$T$ \quad\=$::=$ \quad\=\kill
    \> $T$ \>$::=$ \>$\tnat \sep T_1 \to T_2 \sep \tunit \sep  {T_1} \tprod {T_2}$
    \\[1ex]
    \> $M$ \>$::=$ \>$n \sep x \sep \pred M \sep M_1 + M_2 \sep$\\
      \>\>\>$\ifz {M_1} {M_2} {M_3} \sep$ \\
      \>\>\>$\unit \sep \pair {M_1} {M_2} \sep \fst M \sep \snd M \sep$\\
      \>\>\>$\letexp x {M_1} {M_2} \sep$ \\
      \>\>\>$\fix f x M \sep (M_1 \app M_2)$
    \\[1ex]
    \>$V$ \>$::=$ \>$n \sep \fix f x M \sep () \sep \pair {V_1} {V_2}$
  \end{tabbing}
  \caption{The Syntax of the Source/Target Language of the CPS transformation}
  \label{fig:cps_src_lang}
\end{figure}

We provide some intuition into the structure of the language whose
syntax is described by the rules in Figure\cspc\ref{fig:cps_src_lang};
this intuition will underlie the typing and evaluation judgments that
we will present later. The symbol $\tnat$ represents the type of
natural numbers and the symbol $\tunit$ represents a type whose sole
constructor is $\unit$ that is also pronounced as unit. Further, $T_1
\to T_2$ corresponds to the function type and ${T_1} \tprod {T_2}$ is
the type of pairs.
The collection of terms in the language is essentially an extension of
the terms constituting the simply typed $\lambda$-calculus.
More specifically, this collection can be understood as follows.
First, it includes the natural numbers; this collection is denoted by
$n$ in the syntax rules.
Second, it includes the arithmetic operators $\predsans$ and $+$ that
represent, respectively, the predecessor and addition functions on
natural numbers.
Third, it includes constructors and destructors for tuples:
$()$ is the unit constructor, $(M_1,M_2)$ is a pair whose first element
is $M_1$ and second element is $M_2$; $\fstsans$ and $\sndsans$ are
the projection operators on pairs to their first and second
elements.
Fourth, it includes the conditional expression $\ifz {M_1}
{M_2} {M_3}$. The behavior of such an expression is based on whether
or not the ``condition'' $M_1$ is zero: If so, it behaves the same as
$M_2$; Otherwise, it behaves the same as $M_3$.
Fifth, it includes \emph{let expressions} of the form $\letexp x {M_1}
{M_2}$ that are convenient for breaking up a program into more
manageable pieces; in any given instance of this expression, $x$
represents a local variable that is bound
to the value of $M_1$ and whose scope is limited to $M_2$.
Finally, the collection of terms includes the recursion or fixed-point
operator
$\fixsans$ which abstracts simultaneously the function $f$ and the
parameter $x$ over the function body $M$ to form the function $\fix f
x M$ and the usual function application expression $(M_1\app
M_2)$. The application of a function $\fix f x M$ to an argument $M'$
behaves by replacing $f$ with itself and $x$ with $M'$ in its body $M$
and evaluating the resulting term. We use $\abs x M$ to denote the
function $\fix f x M$ in which $f$ does not occur in $M$.

A typing judgment for the source/target language is written
as $\Gamma \stseq M : T$, where $\Gamma$ is a list of type assignments
for variables. They are derivable by using the rules in
Figure\cspc\ref{fig:cps_src_typing}. The typing rules are mostly
standard. The only interesting rule is $\offix$ for typing functions:
to give the function $\fix f x M$ an arrow type $T_1 \to T_2$, we need
to show its body $M$ has type $T_2$ in an extended context that
assigns $f$ with the type of the function itself and $x$ with the type
of its argument. This coincides with the intuitive interpretation of such
expressions that we have presented above.

\begin{figure}[ht!]
  \begin{gather*}
    \infer[\ofnat]{
      \Gamma \stseq n : \tnat
    }{}
    \quad
    \infer[\ofvar]{
      \Gamma \stseq x : T
    }{
      x:T \in \Gamma
    }
    \\
    \infer[\ofpred]{
      \Gamma \stseq \pred M : \tnat
    }{
      \Gamma \stseq M : \tnat
    }
    \quad
    \infer[\ofplus]{
      \Gamma \stseq M_1 + M_2 : \tnat
    }{
      \Gamma \stseq M_1 : \tnat
      &
      \Gamma \stseq M_2 : \tnat
    }
    \\
    \infer[\ofif]{
      \Gamma \stseq \ifz {M_1} {M_2} {M_3} : T
    }{
      \Gamma \stseq M_1 : \tnat
      &
      \Gamma \stseq M_2 : T
      &
      \Gamma \stseq M_3 : T
    }
    \\
    \infer[\ofunit]{
      \Gamma \stseq \unit : \tunit
    }{}
    \quad
    \infer[\ofpair]{
      \Gamma \stseq (M_1, M_2) : T_1 \tprod T_2
    }{
      \Gamma \stseq M_1 : T_1
      &
      \Gamma \stseq M_2 : T_2
    }
    \\
    \infer[\offst]{
      \Gamma \stseq \fst M : T_1
    }{
      \Gamma \stseq M : T_1 \tprod T_2
    }
    \quad
    \infer[\ofsnd]{
      \Gamma \stseq \snd M : T_2
    }{
      \Gamma \stseq M : T_1 \tprod T_2
    }
    \\
    \infer[\oflet]{
      \Gamma \stseq \letexp x {M_1} {M_2} : T
    }{
      \Gamma \stseq M_1 : T_1
      &
      \Gamma, x : T_1 \stseq M_2 : T
    }
    \\
    \mbox{\small (provided $x$ does not occur in $\G$)}
    \\
    \begin{array}{c}
    \infer[\offix]{
      \Gamma \stseq {\fix f x M} : {T_1 \to T_2}
    }{
      {\Gamma, f: T_1 \to T_2, x: T_1} \stseq M : {T_2}
    }
    \\
    \mbox{\small (provided $f$ and $x$ do not occur in $\G$)}
    \end{array}
    \quad
    \infer[\ofapp]{
      \Gamma \stseq {{M_1}\app {M_2}} : T
    }{
      \Gamma \stseq {M_1} : {T_1 \to T}
      &
      \Gamma \stseq {M_2} : {T_1}
    }
  \end{gather*}
  \caption{Typing Rules for the Source Language of the CPS
    transformation}
  \label{fig:cps_src_typing}
\end{figure}

\subsection{The transformation rules}
\label{subsec:cps_rules}

We give the rules for the CPS transformation based on the informal
ideas described in Section\cspc\ref{sec:cps_overview}. In general, we
must transform terms containing free variables. These free variables
must be tracked throughout the transformation. Thus, we specify the
transformation as a 4-place relation written as $\cps \rho M K {M'}$,
where $M$ and $M'$ are the input and output terms, $K$ is the input
continuation and $\rho$ is a set of variables that contains all the
free variables in $M$. We write $(\rho, x)$ to denote the extension of
$\rho$ with a variable $x$.
Figure\cspc\ref{fig:cps_rules} defines the $\cps \rho M K {M'}$
relation in a rule-based fashion. Note again that we use $\cabs x M$
and $(\capp {M_1} {M_2})$ to represent administrative abstractions and
applications. An administrative $\beta$-redex should be considered as
representing its $\lambda$-normal form, \ie, it is immediately reduced
by the transformation. Most of the rules follow directly from the
informal description of the CPS transformation given in
Section\cspc\ref{sec:cps_overview}. The rules that need further
explanation are $\cpsif$ for transforming the conditional expressions,
$\cpsfix$ for transforming functions and $\cpsapp$ for transforming
function applications.

\begin{figure}
\begin{gather*}
  \infer[\cpsnat]{
    \cps \rho n K {\capp K n}
  }{}
  \quad
  \infer[\cpsvar]{
    \cps \rho x K {\capp K x}
  }{
    x \in \rho
  }
  \\
  \infer[\cpsunit]{
    \cps \rho \unit K {\capp K \unit}
  }{}
  \\
  \infer[\cpspred]{
    \cps \rho {\pred M} K {M'}
  }{
    \cps \rho M {\cabs x {\letexp v {\pred x} {(\capp K v)}}} {M'}
  }
  \\
  \infer[\cpsplus]{
    \cps \rho {M_1 + M_2} K {M'}
  }{
    \cps \rho {M_2} {\cabs {x_2} {\letexp v {x_1+x_2} {(\capp K v)}}} {M_2'}
    &
    \cps \rho {M_1} {\cabs {x_1} {M_2'}} {M'}
  }
  \\
  \mbox{\small (provided $x_1$ does not occur in $\rho$, $M_2$ and $K$)}
  \\
  \infer[\cpspair]{
    \cps \rho {(M_1,M_2)} K {M'}
  }{
    \cps \rho {M_2} {\cabs {x_2} {\letexp v {(x_1,x_2)} {(\capp K v)}}} {M_2'}
    &
    \cps \rho {M_1} {\cabs {x_1} {M_2'}} {M'}
  }
  \\
  \mbox{\small (provided $x_1$ does not occur in $\rho$, $M_2$ and $K$)}
  \\
  \infer[\cpsfst]{
    \cps \rho {\fst M} K {M'}
  }{
    \cps \rho M {\cabs x {\letexp v {\fst x} {(\capp K v)}}} {M'}
  }
  \\
  \infer[\cpssnd]{
    \cps \rho {\snd M} K {M'}
  }{
    \cps \rho M {\cabs x {\letexp v {\snd x} {(\capp K v)}}} {M'}
  }
  \\
  \infer[\cpsif]{
    \cps \rho {\ifz {M_1} {M_2} {M_3}} K {M'}
  }{
    \begin{array}{c}
      \cps \rho {M_2} {\cabs x {k\app x}} {M_2'}
      \\
      \cps \rho {M_3} {\cabs x {k\app x}} {M_3'}
      \\
      \cps \rho {M_1}
           {\cabs
             {x_1}
             {\letexp k {\abs a {(\capp {K} a)}}
               {(\ifz {x_1} {M_2'} {M_3'})}}}
           {M'}
    \end{array}
  }
  \\
  \mbox{\small (provided $k$ does not occur in $\rho$, $M_2$ and $M_3$)}
  \\
  \infer[\cpslet]{
    \cps \rho {\letexp x {M_1} {M_2}} K {M'}
  }{
    \cps {\rho,x} {M_2} K {M_2'}
    &
    \cps \rho {M_1} {\cabs x {M_2'}} {M'}
  }
  \\
  \mbox{\small (provided $x$ does not occur in $\rho$ and $K$)}
  \\
  \infer[\cpsfix]{
    \begin{align*}
      & \rho \triangleright {(\fix f x M)}; K \leadsto_{cps}
      \\
      & \qquad
        {\letexp v {(\fix f p
          {\letexp k {\fst p}
            {\letexp x {\snd p} {M'}}})}
          {(\capp K v)}}
    \end{align*}
  }{
    \cps {\rho,f,x} M {\cabs y {k\app y}} {M'}
  }
  \\
  \mbox{\small (provided $f,x$ do not occur in $\rho$ and $k$ does not occur in $\rho$ and $M$)}
  \\[1ex]
  \infer[\cpsapp]{
    \cps \rho {{M_1}\app {M_2}} K {M'}
  }{
    \begin{array}{c}
      \cps \rho {M_2}
           {\lambda x_2.
             \letexp k {\abs a {(\capp {K} a)}}
                     {\letexp p {(k,x_2)} {(x_1\app p)}}}
           {M_2'}
      \\
      \cps \rho {M_1} {\cabs {x_1} {M_2'}} {M'}
    \end{array}
  }
  \\
  \mbox{\small (provided $x_1$ do not occur in $\rho$, $M_2$ and $K$)}
\end{gather*}
\caption{The Rules for the CPS Transformation}
\label{fig:cps_rules}
\end{figure}

Intuitively, a rule for transforming the conditional expressions can
be given as follows:
\[
  \infer[]{
    \cps \rho {\ifz {M_1} {M_2} {M_3}} K {M'}
  }{
    \cps \rho {M_2} K {M_2'}
    &
    \cps \rho {M_3} K {M_3'}
    &
    \cps \rho {M_1} {\cabs {x_1} {\ifz {x_1} {M_2'} {M_3'}}} {M'}
  }
\]
That is, we recursively transform both branches of the expression with
the input continuation $K$ since $K$ must be the remaining computation no
matter which branch is taken in the evaluation. We then recursively
transform the condition part $M_1$ with the continuation
\[
  {\cabs {x_1} {\ifz {x_1} {M_2'} {M_3'}}}
\]
that is formed from the branches in the CPS form and represents the
remaining computation after evaluating $M_1$ to get the CPS form of the
original expression. However, there is a problem with this rule: the
continuation $K$ is duplicated for transforming the branches and may
result in exponential explosion in the code size when nested
conditional expressions are present. The $\cpsif$ rule solves this
problem by use a variable $k$ as a placeholder of the continuation for
the branches. When transforming the condition $M_1$, it uses the
continuation
\[
  {\letexp k {\abs a {(\capp {K} a)}} {(\ifz {x_1} {M_2'} {M_3'})}}
\]
that links this placeholder to the actual continuation through a let
expression. Since there is only one copy of $K$ used in the recursive
transformation, the program grows linearly in space.

The $\cpsfix$ rule for transforming functions mostly follows the
corresponding informal description given in
Section\cspc\ref{sec:cps_overview}. To transform a function $\fix f x
M$ with the continuation $K$, we first recursively transform the
function body $M$ with the continuation $\cabs y {k\app y}$ where $k$
is the continuation argument of the transformed function to get its
CPS form $M'$. The CPS form of the original function is then the
following
\[
{(\fix f p
          {\letexp k {\fst p}
            {\letexp x {\snd p} {M'}}})}.
\]
whose second argument is a pair consisting of the continuation
argument and the original argument and whose body is essentially $M'$
except for the let expressions for selecting arguments from the pair.
Denoting it by $F$, we get the output of the transformation $(\letexp
v F {(\capp K v)})$.
The $\cpsapp$ rule for transforming function applications almost
follows its informal description given in
Section\cspc\ref{sec:cps_overview}. The only difference is that we
need to pack the continuation argument and the original argument into
a pair argument.

An important aspect of the rules in Figure\cspc\ref{fig:cps_rules} is
the freshness side conditions for variables for guaranteeing the
correctness of the CPS transformation. Such side conditions can be
classified into two categories. One is for the free variables
introduced by recursion over binding operators, including $x$ in
$\cpslet$ and $f,x$ in $\cpsfix$. These free variables must be fresh
to avoid accidental capturing of them in the recursive
transformations. Another category---and the more interesting one---is
for ``placeholder'' variables, including $k$ in $\cpsfix$ and $\cpsif$
and $x_1$ in $\cpsplus$, $\cpspair$ and $\cpsapp$. In all those rules,
we perform recursive CPS transformations with continuations containing
free variables that are placeholders for arguments that will be bound
later in construction of other continuations. To avoid accidental
capturing of these free variables in the recursive transformations, we
must place freshness constraints on them too.

\section{Implementing the Transformation in \LProlog}
\label{sec:cps_impl}

Our presentation of the implementation of the CPS transformation
consists of two parts: we first show how to encode the source (and also target)
language in \LProlog and we then present a \LProlog program for the
transformation. In describing the first part, we discuss also the
formalization of the typing rules. In describing the second part, we
shall discuss how the $\lambda$-tree syntax approach is used to
capture the reading of administrative and dynamic $\beta$-redexes and
the side conditions and show the transparent correspondence between
our \LProlog encoding and the original transformation rules.

\subsection{Encoding the language}
\label{subsec:cps_lang_encoding}

\pagelabel{ecd:cps_src_lang}
We first consider the encoding of types. We use the \LProlog type \kty
to represent the types of the language. The constructors \ktnat,
\ktunit and \kprod encode, respectively, the natural number, unit and
pair types. We represent the arrow type constructor $\to$ by
\karr. These decisions are summarized in the following \LProlog
signature.

\begin{tabbing}
\qquad\=$\karr,\kprod$ \qquad\=$:$ \quad\=\kill
\>$\ktnat,\ktunit$ \>$:$ \>$\kty$\\
\>$\karr,\kprod$ \>$:$ \>$\kty \to \kty \to \kty$
\end{tabbing}

We use the \LProlog type \ktm for encodings the terms in the
language. The particular constructors that we will use for
representing the terms themselves are the following, assuming that
\knat is a type encoding the type of natural numbers:
\begin{tabbing}
\qquad\=$\kplus,\kpair,\kapp$ \qquad\=$:$ \quad\=\kill
\>\knat \>$:$ \>$\knat \to \ktm$\\
\>$\kpred,\kfst,\ksnd$ \>$:$ \>$\ktm \to \ktm$\\
\>$\kunit$ \>$:$ \>$\ktm$\\
\>$\kplus,\kpair,\kapp$  \>$:$ \>$\ktm \to \ktm \to \ktm$\\
\>\kifz \>$:$ \>$\ktm \to \ktm \to \ktm \to \ktm$\\
\>\klet \>$:$ \>$\ktm \to (\ktm \to \ktm) \to \ktm$\\
\>\kfix \>$:$ \>$(\ktm \to \ktm \to \ktm) \to \ktm$
\end{tabbing}
The only constructors that need further explanation here are \klet and
\kfix. These encode binding constructs in the language and, as
expected, we use \LProlog abstraction to capture their binding
structure. Thus, $(\letexp x n x)$ is encoded as $(\klet\app
(\knat\app n)\app (x \mlam x))$.
Similarly, the \LProlog term $(\kfix\app (f \mlam x \mlam \kapp\app
f\app x))$ represents the source language expression $(\fix f x {f\app
  x})$.

\pagelabel{ecd:cps_src_typing}
Following Section\cspc\ref{subsec:ecd_bindings}, we represent typing
judgments as relations between terms and types, treating contexts
implicitly via dynamically added clauses that assign types to free
variables. We use the predicate symbol $\kof : \ktm \to \kty \to
\omic$ to encode typing in the language. Every typing rule in
Figure\cspc\ref{fig:cps_src_typing} is translated into a clause for
\kof. These clauses are listed as follows:
\begin{tabbing}
\qquad\=\qquad\=\kill
\>$\kof\app (\knat\app N)\app \ktnat.$\\
\>$\kof\app (\kpred\app M)\app \ktnat \limply \kof\app M\app \ktnat.$\\
\>$\kof\app (\kplus\app M_1\app M_2)\app \ktnat
    \limply \kof\app M_1\app \ktnat \scomma \kof\app M_2\app \ktnat.$\\
\>$\kof\app (\kifz\app M\app M_1\app M_2)\app T \limply
     \kof\app M\app \ktnat \scomma \kof\app M_1\app T \scomma \kof\app M_2\app T.$\\
\>$\kof\app \kunit\app \ktunit.$\\
\>$\kof\app (\kpair\app M_1\app M_2)\app (\kprod\app T_1 T_2) \limply
   \kof\app M_1\app T_1\scomma \kof\app M_2\app T_2.$\\
\>$\kof\app (\kfst\app M)\app T_1 \limply \kof\app M\app (\kprod\app T_1\app T_2).$\\
\>$\kof\app (\ksnd\app M)\app T_2 \limply \kof\app M\app (\kprod\app T_1\app T_2).$\\
\>$\kof\app (\klet\app M\app R)\app T \limply
    \kof\app M\app T_1 \scomma
    (\forallx x {\kof\app x\app T_1 \simply \kof\app (R\app x)\app T}).$\\
\>$\kof\app (\kfix\app R)\app (\karr\app T_1\app T_2) \limply$\\
\>\>$\forallx {f, x}
       {\kof\app f\app (\karr\app T_1\app T_2) \simply
        \kof\app x\app T_1 \simply
        \kof\app (R\app f\app x)\app T_2}.$\\
\>$\kof\app (\kapp\app M_1\app M_2)\app T \limply
     \kof\app M_1\app (\karr\app T_1\app T) \scomma \kof\app M_2\app T_1.$
\end{tabbing}
The only interesting clauses are the second and third ones to the last
pertaining to the binding constructs $\kfix$ and $\klet$. Note how the
required freshness constraint and the extension of the typing contexts
are realized in these clauses: take the clause for \kfix as an example,
the universal goals over $f$ and $x$ introduce new names and the
application $(R\app f\app x)$ replaces the bound variables with these
names, and the hypothetical goals dynamically introduce the typing
assignments for these names and generate the new typing judgment that
must be derived.

\subsection{Specifying the CPS transformation}
\label{subsec:specify_cps}

An important task in specifying the CPS transformation is to capture
the reading of administrative $\beta$-redexes. We shall use meta-level
$\lambda$-abstraction in \LProlog to represent the administrative
$\lambda$-abstractions. As a result, administrative $\beta$-redexes
become meta-level $\beta$-redexes and substitution denoted by them is
realized by meta-level $\beta$-reduction.

\pagelabel{ecd:cps_rules}
With this in mind, we encode the rule-based description of the CPS
transformation in Section\cspc\ref{sec:rule_based_cps} as follows. We
first identify the predicate symbol
\[
\kcps : \ktm \to (\ktm \to \ktm) \to \ktm \to \omic
\]
for representing the transformation relation such that $\kcps\app
M\app K\app M'$ holds if and only if given the input continuation $K$
the source term $M$ is transformed into the CPS term $M'$. Note here
we use a meta-level abstraction $K$ to encode the input continuation
which is an administrative $\lambda$-abstraction. Since $M$ can be an
open term, we also need to keep track of the context of the CPS
transformation that contains all the free variables in $M$. Similar to
the encoding of the typing rules, this context is implicitly
represented by the dynamic context and thus is not explicitly given as
an argument of \kcps. The transformation rules in
Figure\cspc\ref{fig:cps_rules} then translate into the following
clauses defining $\kcps$:

\begin{tabbing}
\qquad\=\qquad\=$\pi k,f,x.($\=$($\=\kill
\>$\kcps\app (\knat\app N)\app K\app (K\app (\knat\app N))$.\\
\>$\kcps\app \kunit\app K\app (K\app \kunit).$\\
\>$\kcps\app (\kpred\app M)\app K\app M' \limply
  \kcps\app M\app (x \mlam \klet\app (\kpred\app x)\app (v \mlam K\app v))\app M'$.\\
\>$\kcps\app (\kplus\app M_1\app M_2)\app K\app M' \limply$\\
\>\>$(\forallx
          {x_1}
          {\kcps\app M_2\app
            (x_2 \mlam \klet\app (\kplus\app x_1\app x_2)\app (v \mlam K\app v))\app (M_2'\app x_1)}),
     \kcps\app M_1\app M_2'\app M'.$\\
\>$\kcps\app (\kpair\app M_1\app M_2)\app K\app M' \limply$\\
\>\>$(\forallx
         {x_1}
         {\kcps\app M_2\app
            (x_2 \mlam \klet\app (\kpair\app x_1\app x_2)\app (v \mlam K\app v))\app (M_2'\app x_1)}),
     \kcps\app M_1\app M_2'\app M'.$\\
\>$\kcps\app (\kifz\app M_1\app M_2\app M_3)\app K\app M' \limply$\\
\>\>$(\forallx k {\kcps\app M_2\app (x \mlam \kapp\app k\app x)\app (M_2'\app k)}),
     (\forallx k {\kcps\app M_3\app (x \mlam \kapp\app k\app x)\app (M_3'\app k)}),$\\
\>\>$\kcps\app
         M_1\app
         (x_1 \mlam \klet\app (\kfix\app f \mlam K)\app
                              (k \mlam \kifz\app x_1\app (M_2'\app k)\app (M_3'\app k)))\app
         M'.$\\
\>$\kcps\app (\kfst\app M)\app K\app M' \limply
     \kcps\app M\app (x \mlam \klet\app (\kfst\app x)\app (v \mlam K\app v))\app M'.$\\
\>$\kcps\app (\ksnd\app M)\app K\app M' \limply
     \kcps\app M\app (x \mlam \klet\app (\ksnd\app x)\app (v \mlam K\app v))\app M'.$\\
\>$\kcps\app (\klet\app M\app R)\app K\app M' \limply$\\
\>\>$(\forallx {x,k}
            {(\kcps\app x\app k\app (k\app x)) \simply
              \kcps\app (R\app x)\app K\app (R'\app x)}),
     \kcps\app M\app R'\app M'.$\\
\>$\kcps\app
     (\kfix\app R)\app K$\\
\>\>$(\klet\app
         (\kfix\app
            (f \mlam p \mlam
               \klet\app (\kfst\app p)\app
                         (k \mlam \klet\app (\ksnd\app p)\app (x \mlam R'\app f\app k\app x))))\app
         K) \limply$\\
\>\>$\forallx
       {k,f,x}
       {(\forallx k {\kcps\app f\app k\app (k\app f)}) \simply
        (\forallx k {\kcps\app x\app k\app (k\app x)}) \simply}$\\
\>\>\>$\kcps\app (R\app f\app x)\app (y \mlam \kapp\app k\app y)\app (R'\app f\app k\app x).$\\
\>$\kcps\app (\kapp\app M_1\app M_2)\app K\app M' \limply$\\
\>\>$(\forallx {x_1}
        {\kcps\app
            M_2\app
            (x_2 \mlam \klet\app
                         (\kfix\app f \mlam K)
                         (k \mlam \klet\app (\kpair\app k\app x_2)\app (p \mlam \kapp\app x_1\app p)))}$\\
    \>\>\>\>$(M_2'\app x_1)),$\\
\>\>$\kcps\app M_1\app M_2'\app M'.$
\end{tabbing}
By using the meta-level $\lambda$-abstractions and applications to
represent their administrative counterparts, every rule in
Figure\cspc\ref{fig:cps_rules} except $\cpsvar$ transparently
translates into one clause for \kcps. Note that in the clauses
corresponding to the rules $\cpsif$ and $\cpsapp$ the term $(\kfix\app
f \mlam K)$ (\ie, $(\kfix\app f \mlam a \mlam K\app a)$) where $f$
does not occur in $K$ is used to represent the function $({\abs a
  {(\capp {K} a)}})$ in the original rules.
The freshness side conditions of the original rules are captured in a
concise and logically precise way by using the $\lambda$-tree
representation, meta-level applications and universal goals. For
example, given the function $(\kfix\app R)$ and the continuation $K$,
to derive the $\kcps\app (\kfix\app R)\app K\app M'$ for some $M'$, we
must backchain this goal on the penultimate clause for \kcps and prove
the following formula for some $R'$:
\begin{tabbing}
\qquad\=\qquad\=$\pi k,f,x.($\=$($\=\kill
\>\>$\forallx
       {k,f,x}
       {(\forallx k {\kcps\app f\app k\app (k\app f)}) \simply
        (\forallx k {\kcps\app x\app k\app (k\app x)}) \simply}$\\
\>\>\>$\kcps\app (R\app f\app x)\app
                  (y \mlam \kapp\app k\app y)\app
                  (R'\app f\app k\app x).$
\end{tabbing}
Derivation of this goal must introduce constants for $k$, $f$ and $x$
that are fresh with respect to $(\kfix\app R)$, $K$ and $R'$. Those
constants will replace $k$, $f$ and $x$ in the subsequent derivation
of $\kcps\app (R\app f\app x)\app (y \mlam \kapp\app k\app y)\app
(R'\app f\app k\app x)$ where the meta-level application $(R\app f\app
x)$ represents the body of the function and $(R'\app f\app k\app x)$
represents its CPS form. This example also illustrates how the
$\cpsvar$ rule is represented in our encoding. Similar to the encoding
of the transformation between $\lambda$-terms and de Bruijn terms in
Section\cspc\ref{subsec:hhw_mtv_exm}, we use hypothetical goals to
dynamically introduce program clauses that represent the
transformation rule for the binding variables at the point they are
introduced via universal goals. In the case of the above example, the
binding variables are $f$ and $x$ and the clauses representing
$\cpsvar$ for them are $(\forallx k {\kcps\app f\app k\app (k\app
  f)})$ and $(\forallx k {\kcps\app x\app k\app (k\app x)})$. When the
source term becomes a variable, the transformation makes progress by
backchaining on the appropriate clause for that variable in the
dynamic context.

\section{Informal Verification of the Transformation}
\label{sec:informal_verify_cps}

We give an informal description of the verification of the CPS
transformation in this section based on the ideas presented in
Section\cspc\ref{sec:vfc_approach}. It serves as the basis for the
formal verification of the CPS transformation in \Abella which we
shall discuss in Section\cspc\ref{sec:formal_verify_cps}.

\subsection{Type preservation of the transformation}
\label{subsec:cps_informal_typ_pres}
Before we start talking about the semantics preservation property of
the CPS transformation, let us first prove that the CPS transformation
preserves types. This type preservation property is essential for
composing the semantics preservation proofs and have a special use for
deriving the property of being closed terms which is useful in the
formal verification. The description of type preservation in this
section is partially based on \cite{guillemette07plpv}.

Strictly speaking, the CPS transformation changes the types of terms,
albeit in a systematic way. Because the CPS transformation embeds the
(transformed) source term into a context (which is the initial input
continuation), the type of the output of the CPS transformation is
determined by the target type of the context. Moreover, since
continuations in the transformed source term represent future
computations which always end up in evaluating the embedding of the
value of the source term in the initial context, their target type
must be the same as the target type of the context. The only
(sub)terms from the source term whose types are changed by the
transformation are functions because every function after the
transformation takes an extra argument which is a continuation that
will be called when the function returns. By this observation, we
represent a mapping between types of terms before the CPS
transformation and that after the transformation as a ternary relation
$\simeq$ such that $\cpsty S {T_1} {T_2}$ holds if the type $T_1$ of a
source term is mapped to $T_2$ given the target type $S$ of the
initial context. This relation is defined by the rules depicted in
Figure\cspc\ref{fig:cps_typ_mapping}. The last rule formalizes the
mapping between function types based the previous observation: a
function of type ${T_1 \to T_2}$ after the CPS transformation has the
type ${((T_2' \to S) \tprod T_1') \to S}$ where $T_1'$ is the type of
the original argument after the transformation and $T_2'$ is the type
of the function body after the transformation; note that the function
takes a continuation argument of type $(T_2' \to S)$ and its target
type indicates an application of the function should have the same
type as that of the continuation argument, which conforms to the way
the continuation argument is used in the function.

\begin{figure}[!ht]
  \begin{gather*}
    \infer{
      \cpsty S \tnat \tnat
    }{}
    \qquad
    \infer{
      \cpsty S \tunit \tunit
    }{}
    \qquad
    \infer{
      \cpsty S {T_1 \tprod T_2} {T_1' \tprod T_2'}
    }{
      \cpsty S {T_1} {T_1'}
      &
      \cpsty S {T_2} {T_2'}
    }
    \\
    \infer{
      \cpsty S {T_1 \to T_2}
               {((T_2' \to S) \tprod T_1') \to S}
    }{
      \cpsty S {T_1} {T_1'}
      &
      \cpsty S {T_2} {T_2'}
    }
  \end{gather*}
  \caption{The Rules for Mapping Types of Terms in the CPS Transformation}
  \label{fig:cps_typ_mapping}
\end{figure}

Now we state the type preservation property of the CPS transformation,
as follows:
\begin{mythm}\label{thm:cps_typ_pres_open}
  Let $M$ be a term whose free variables are contained in the set
  $\rho = \{x_1,\ldots,x_n\}$, $\G$ be a typing context
  $(x_1:T_1,\ldots,x_n:T_n)$ such that $\G \stseq M:T$ holds for some
  $T$, $K$ be a continuation of type $T' \to S$ such that $\cpsty S T
  {T'}$ holds, $\G'$ be the type context $(x_1:T_1',\ldots,x_n:T_n')$
  such that $\cpsty S {T_i} {T_i'}$ holds for $1 \leq i \leq n$. If
  $\cps \rho M K {M'}$ holds for some $M'$, then $\G' \stseq M':S$
  holds.
\end{mythm}
\noindent In essence, this theorem states that if a source term $M$ is
CPS transformed into $M'$ in the context $K$, then $M'$ has the target
type of $K$. We prove it by induction on the derivation of $\cps \rho
M K {M'}$ and analyzing the last rule of the derivation. When the last
rule is $\cpsnat$, $\cpsvar$ or $\cpsunit$, the proof is obvious. The
rest of the cases are proved by following a set pattern, as follows. We
examine the premises of the last rule from left and right. For each
premise, we apply the induction hypothesis to get the type of its
output. Once this type is known, it is easy to derive the type of the
input continuation in the next premise (if any). We can then repeat
the previous process on the next premise. This continues until we have
derived the type of the output of the conclusion. At that point this
case is finished.

Given Theorem\cspc\ref{thm:cps_typ_pres_open}, it is easy to show the
following type preservation property for closed terms:
\begin{mycoro}\label{coro:cps_typ_pres_closed}
  If $\stseq M:\tnat$, $K$ is a continuation of type $\tnat \to S$ and
  $\cps \emptyset M K {M'}$, then $\stseq M':S$.
\end{mycoro}
\noindent By choosing $K$ to be $\cabs x x$, we get the
following corollary:
\begin{mycoro}\label{coro:cps_typ_pres_closed_id}
  If $\stseq M:\tnat$ and
  $\cps \emptyset M {(\cabs x x)} {M'}$, then $\stseq M':\tnat$.
\end{mycoro}

\subsection{Semantics preservation of the transformation}
\label{subsec:cps_informal_sem_pres}

We informally describe the proof of semantics preservation for the CPS
transformation in this section. We first describe the operational
(evaluation) semantics of the source/target language of the
transformation, then the logical relations for denoting equivalence
between the source and target programs and their properties, and
finally the semantics preservation theorem and its proof.

\subsubsection{Operational semantics of the source/target language}

\pagelabel{txt:cps_eval}
To describe semantics preservation, we must first define the
operational semantics of the languages involved the CPS
transformation. Since the source language and the target languages of
the CPS transformation are the same, they share the same operational
semantics. The operational semantics is based on a left to right,
call-by-value evaluation strategy. We assume that this is given in
a small-step form and, we write $M \step{1} M'$ to denote that $M$
evaluates to $M'$ in one step. The evaluation rules are defined in a
standard way, as shown in Figure\cspc\ref{fig:cps_step_rules}. The
only interesting rule is the last one which shows how recursion works
in the language: evaluating an application of a recursive function to
a value causes the occurrences of the function parameter in the
function body to be replaced by the function itself.
One-step evaluation generalizes to $n$-step evaluation that we denote by $M
\step{n} M'$ and that holds if $M$ reduces to $M'$ by $n$ one-step
evaluations. We shall write $M \step{*} M'$ to denote that there
exists some $n$ such that $M \step{n} M'$ holds. Finally, we write $M
\eval V$ to denote the evaluation of $M$ to the value $V$ through $0$
or more steps.

\begin{figure}[!ht]
  \begin{gather*}
    \infer[\stpred]{
      \pred M \step{1} \pred {M'}
    }{
      M \step{1} M'
    }
    \quad
    \infer[\stpredbase]{
      \pred n \step{1} n'
    }{
      n' \text{ is the predecessor of } n
    }
    \\
    \infer[\stplusleft]{
      M_1 + M_2 \step{1} M_1' + M_2
    }{
      M_1 \step{1} M_1'
    }
    \quad
    \infer[\stplusright]{
      V_1 + M_2 \step{1} V_1 + M_2'
    }{
      M_2 \step{1} M_2'
    }
    \\
    \infer[\stplusbase]{
      n_1 + n_2 \step{1} n_3
    }{
      n_3 \text{ is the sum of } n_1 \text{ and } n_2
    }
    \\
    \infer[\stifcond]{
      \ifz {M_1} {M_2} {M_3} \step{1} \ifz {M_1'} {M_2} {M_3}
    }{
      M_1 \step{1} M_1'
    }
    \\
    \infer[\stifleft]{
      \ifz 0 {M_1} {M_2} \step{1} M_1
    }{}
    \;\;
    \infer[\stifright]{
      \ifz n {M_1} {M_2} \step{1} M_2
    }{n > 0}
    \\
    \infer[\stpairleft]{
      (M_1, M_2) \step{1} (M_1', M_2)
    }{
      M_1 \step{1} M_1'
    }
    \quad
    \infer[\stpairright]{
      (V_1, M_2) \step{1} (V_1, M_2')
    }{
      M_2 \step{1} M_2'
    }
    \\
    \infer[\stfst]{
      \fst M \step{1} \fst {M'}
    }{
      M \step{1} M'
    }
    \quad
    \infer[\stfstbase]{
      \fst {(V_1, V_2)} \step{1} V_1
    }{}
    \\
    \infer[\stsnd]{
      \snd M \step{1} \snd {M'}
    }{
      M \step{1} M'
    }
    \quad
    \infer[\stsndbase]{
      \snd {(V_1, V_2)} \step{1} V_2
    }{}
    \\
    \infer[\stletarg]{
      \letexp x {M_1} {M_2} \step{1} \letexp x {M_1'} {M_2}
    }{
      M_1 \step{1} M_1'
    }
    \\
    \infer[\stletbody]{
      \letexp x {V} {M} \step{1} M[V/x]
    }{}
    \\
    \infer[\stappfun]{
      {M_1}\app {M_2} \step{1} {M_1'}\app {M_2}
    }{
      M_1 \step{1} M_1'
    }
    \quad
    \infer[\stapparg]{
      {V_1}\app {M_2} \step{1} {V_1}\app {M_2'}
    }{
      M_2 \step{1} M_2'
    }
    \\
    \infer[\stappbase]{
      {(\fix f x M)}\app V \step{1} M[\fix f x M/f][V/x]
    }{}
  \end{gather*}
  \caption{Evaluation Rules for the Source Language of the CPS Transformation}
  \label{fig:cps_step_rules}
\end{figure}

\subsubsection{Logical relations and their properties}

Following the ideas in Section\cspc\ref{sec:vfc_approach}, we use
step-indexing logical relations to characterize the semantics
preservation property between source and target programs of the CPS
transformation. We define the mutually recursive simulation relation
$\sim$ between pairs of closed source terms and input continuations
and target terms and the equivalence relation $\approx$ between closed
source and target values, each indexed by a type and a step measure,
in Figure\cspc\ref{fig:cps_logical_relations}.
We give an intuitive explanation of $\sim$ and $\approx$ as follows.

The simulation relation $\simindex T i {M;K} M'$ holds between the
term $M$, the continuation $K$ and the term $M'$ where $K$ is the
context in which the CPS transformed $M$ will be used and $M'$ can be
thought of as the program obtained by combining the CPS form of $M$
and the application of $K$ to the value of $M$. When the relation holds,
we say $M$ simulates $M'$ in the context $K$ at the type $T$ within
$i$ steps. Its definition is interpreted as follows: if $M$ evaluates
to some value $V$, then its CPS form must evaluate to an equivalent
value $V'$ and hence $M'$ must evaluates to the result of applying $K$
to $V'$, which is represented by the administrative $\beta$-redex
$\capp K {V'}$. The uses of indexing types and steps in the definition
of $\sim$ indicate that $M$ simulates $M'$ within at least $i$ steps
of evaluation at the type $T$.

The equivalence relation $\equalindex T i V {V'}$ holds if and only if
the values $V$ and $V'$ cannot be distinguished from each other (\ie,
considered as equivalent) in any context within at least $i$ steps of
evaluation at the type $T$. Its definition is explained as follows. At
the types of natural numbers and unit, two values are equivalent
exactly when they are identical. Two pairs of values are considered
equivalent within at least $i$ steps exactly when their constituting
elements are respectively equivalent within at least $i$ steps. The
equivalence relation at the functions types holds between two
functions $\fix f x M$ and $\fix f p {M'}$ where the latter can be
thought of as the CPS form of the former whose parameter $p$ denotes a
pair consisting of the input continuation and the actual argument of
the function. The two functions are equivalent within at least $i$
steps exactly when the following holds: for any $j$ less than $i$,
given any continuation $K$ and any arguments that are equivalent
within at least $j$ steps, the result of applying $\fix f x M$ to the
source arguments simulates the result of applying $\fix f p {M'}$ to
the target arguments in the context $K$ within at least $j$ steps.
Note that the term $\cabs x {\capp K x}$ (which is a term equivalent
to $K$ in the object language) is fed to $\fix f p {M'}$ as its input
continuation, which conforms to how functions in CPS form should work:
when such a function finishes its evaluation, computation proceeds by
applying the continuation representing the remaining computation---in our
case, $K$---to the evaluation result.
Note that the definition of $\approx$ in the function case uses
$\approx$ negatively at the same type. However, it is still a
well-defined notion because the index decreases.

\begin{figure}[!ht]
\center
\begin{tabbing}
\qquad\qquad\qquad\=\quad\=\quad\=\kill
\>$\simindex T i {M;K} {M'} \iff$\\
\>\>$\rforallx
          {j \leq i}
          {\rforallx V {M \step{j} V} \rimp
            \rexistsx {V'}
                      {M' \step{*} {\capp K {V'}} \rand \equalindex T {i-j} V {V'}}};$\\
\>$\equalindex \tnat i n n;$\\
\>$\equalindex \tunit i \unit \unit;$\\
\>$\equalindex {T_1 \tprod T_2} i {(V_1,V_2)} {(V_1',V_2')} \iff
          \equalindex {T_1} i {V_1} {V_1'} \rand \equalindex {T_2} i {V_2} {V_2'};$\\
\>$\equalindex {T_1 \to T_2} i
       {\fix f x M}
       {\fix f p {M'}} \iff$\\
\>\>$\rfall j < i. \rfall V_1, V_1', V_2, V_2', K.
       \equalindex {T_1} j {V_1} {V_1'} \rimp
       \equalindex {T_1 \to T_2} j {V_2} {V_2'} \rimp$\\
\>\>\>$\simindex {T_2} j {M[V_2/f, V_1/x];K} {M'[V_2'/f, (\abs x {\capp K x}, V_1')/p]}.$
\end{tabbing}

  \caption{The Logical Relations for Verifying the CPS Transformation}
  \label{fig:cps_logical_relations}
\end{figure}

A property we will need about the above step-indexing logical
relations is that $\approx$ is closed under decreasing indexes. It is
stated as the following lemma:
\begin{mylemma}\label{lem:cps_equiv_closed}
  If $\equalindex T i V {V'}$ holds, then for any $j$ such that $j
  \leq i$, $\equalindex T j V {V'}$ holds.
\end{mylemma}
\noindent Note that we cannot prove this lemma by induction on the
relation $\equalindex T i V {V'}$ because, as we have explained in
Section\cspc\ref{sec:vfc_approach}, $\approx$ is not an inductive
definition. Instead, $\approx$ is a recursive definition based on
inductive defined indexing types and steps. We therefore prove this
lemma by a (nested) induction first on the types and then on the step
indexes of $\approx$. The proof itself is obvious. However, it is
important to keep in mind of the following recurring theme about
proving properties of logical relations: These properties cannot be
proved by induction on logical relations but instead are usually
proved by induction on their indexes or other inductively defined
relations.

We show that simulation relations can be composed similar to how the
transformation relations are built by using the transformation
rules. Those properties, commonly known as ``compatibility lemmas'',
are shown as follows:
\begin{mylemma}\label{lem:cps_sim_compat}
\begin{enumerate}
  \item If $\simindex \tnat i {M;(\cabs x {\letexp v {\pred x} {\capp K v}})} {M'}$
    then\\ $\simindex \tnat i {(\pred M); K} {M'};$
  \item If $\simindex \tnat i {M_2; (\cabs {x_2} {\letexp v {x_1 + x_2} {\capp K v}})} {M_2'}$
    and $\simindex \tnat i {M_1; (\cabs {x_1} {M_2'})} {M'}$ then
    $\simindex \tnat i {(M_1 + M_2); K} {M'}$.
  \item If $\simindex {T_1 \tprod T_2} i {M; (\cabs x {\letexp v {\fst
        x} {\capp K v}})} {M'}$ then $\simindex {T_1} i {(\fst{M}); K}
    {\fst{M'}}$.
  \item If $\simindex {T_1 \tprod T_2} i {M; (\cabs x {\letexp v {\snd
        x} {\capp K v}})} {M'}$ then $\simindex {T_2} i {(\snd{M}); K}
    {\fst{M'}}$.
  \item If $\simindex {T_2} {i} {M_2;(\cabs {x_2} {\letexp v {\pair {x_1} {x_2}} {\capp K v}})} {M_2'}$
    and $\simindex {T_1} {i} {M_1;\cabs {x_1} {M_2'}} {M'}$
    then $\simindex {T_1 \tprod T_2} {i} {\pair {M_1} {M_2}; K} {M'}.$
  \item If $\simindex T i {M_2;K} {M_2'}$, $\simindex T i
    {M_3;K} {M_3'}$ and $\simindex \tnat i {M_1;(\cabs {x_1} {\ifz {x_1} {M_2'} {M_3'}})} {M'}$,
    then $\simindex T k {(\ifz {M_1} {M_2} {M_3});K} {M'}$.
  \item If $\simindex{T_1}{i}
               {M_2;(\cabs {x_2} {\letexp k {\abs a {(\capp {K} a)}}
                                           {\letexp p {(k,x_2)} {(x_1\app p)}}})}
               {M_2'}$
   and $\simindex{T_1 \to T}{i}{M_1;\cabs{x_1}{M_2'}}{M'}$
   then $\simindex{T_2}{i}{(M_1 \app M_2); K} {M'}.$
\end{enumerate}
\end{mylemma}
\noindent These lemmas are proved by analyzing the simulation relation
and using the evaluation rules. The main complication involved in
those proofs is caused by calculating and comparing the step measures,
which is an inherent difficulty in any proof techniques that make use
of step-indexing. Some proofs of these properties need the property
that the equivalence relation is closed under decreasing indices,
which have already proved as Lemma\cspc\ref{lem:cps_equiv_closed}. For
instance, the proof of the last of these properties requires us to
consider the evaluation of the application of fixed point expressions
which involves ``feeding'' these expressions that are equivalent at a
lower step measure as arguments to their own body. These arguments are
obtained by applying Lemma\cspc\ref{lem:cps_equiv_closed}.

\subsubsection{Informal proof of semantics preservation}

Our notion of equivalence only relates closed terms. However, the CPS
transformation typically operates on open terms. To handle this
situation, we consider semantics preservation for possibly open terms
under closed substitutions. We will take substitutions in both the
source and target settings to be simultaneous mappings of closed
values for a finite collection of variables, written as
$(V_1/x_1,\ldots,V_n/x_n)$. The equivalence between substitutions
$\theta$ and $\theta'$ is indexed by a typing context $\G$ and a step
measure $i$ and written as $\equalindex \G i \theta {\theta'}$. Its
definition is given as follows:
\begin{gather*}
  \equalindex {(x_1:T_1,\ldots,x_n:T_n)} k {(V_1/x_1,\ldots,V_n/x_n)} {(V_1'/x_1,\ldots,V_n'/x_n)}
  \iff (\forall 1 \leq i \leq n. \equalindex {T_i} k {V_i} {V_i'})
\end{gather*}

The semantics preservation theorem for the CPS transformation can now
be stated as follows, which is a realization of
Property\cspc\ref{thm:sem_pres_index_open} for the CPS transformation:
\begin{mythm}\label{thm:cps_sem_pres_open}
  Let $\Gamma = (x_1:T_1,\ldots,x_n:T_n)$, $\rho = (x_1,\ldots,x_n)$,
  $\equalindex \Gamma i \theta {\theta'}$ and $\Gamma \stseq M : T$.
  If $\cps \rho M K {M'}$, then $\simindex T i {M[\theta];K[\theta']}
  {M'[\theta']}$.
\end{mythm}
\begin{proof}
We outline the main steps in the argument for this theorem:
these will guide the development of a formal proof in
Section\cspc\ref{sec:formal_verify_cps}. We proceed by induction on
the derivation of $\cps \rho M K {M'}$, analyzing the last step in it.
This obviously depends on the structure of $M$. The cases for a natural
number, the unit constructor or a variable are obvious.
In the remaining cases, other than when $M$ is of the form $\letexp x
{M_1} {M_2}$ or $\fix f x {M_1}$, the argument follows a set pattern: we
observe that substitutions distribute to the sub-components of
expressions, we invoke the induction hypothesis over the
sub-components and then we use Lemma\cspc\ref{lem:cps_sim_compat} to
conclude.
If $M$ is of the form $\letexp x {M_1} {M_2}$, then we have $\cps
{\rho,x} {M_2} K {M_2'}$ and $\cps \rho {M_1} {\cabs x {M_2'}} {M'}$
for some $M_2'$. Here again the substitutions distribute over $M_1$,
$M_2$, $M_2'$ and $M'$. We can therefore apply the inductive
hypothesis (I.H.) to $\cps \rho {M_1} {\cabs x {M_2'}} {M'}$ to get a
simulation relation $S$. However, we cannot directly apply I.H. to
$\cps {\rho,x} {M_2} K {M_2'}$ since the transformation on $M_2$
introduces $x$ as a new variable into $\rho$. For this we need to
extend the substitutions with equivalent values for $x$. These values
are derived from the previous simulation relation $S$. This case is
concluded by following the pattern for proving
Lemma\cspc\ref{lem:cps_sim_compat} after we got the simulation
relations from applying I.H. to $M_1$ and $M_2$.
Finally, if $M$ is of the form $\fix f x {M_1}$, then we have $\cps
{\rho,f,x} {M_1} {\cabs y {k\app y}} {M_1'}$. By the definition of
$\sim$, the simulation relation we need to prove reduces into an
equivalence relation $R$ on function values. Similar to the last case,
to apply I.H. to $M_1$ we need to extend the substitutions with
equivalent values for $f$ and $x$ which are obtained from the
assumptions in $R$. The process of proving $R$ after applying I.H. is
like in other cases.
Another important yet subtle point we need to show is that the
functions related by $R$ must be closed, which is equivalent to
showing that $M[\theta]$ and $M'[\theta']$ are closed. It is easy to
show that $M[\theta]$ is closed by observing that the values of
$\theta$ are closed and the domain of $\theta$ is the same as that of
$\G$ which contains all the free variables of $M$. To show
$M'[\theta']$ is closed, we need to apply the type preservation
theorem---\ie, Theorem\cspc\ref{thm:cps_typ_pres_open}---to $M$ to
show that the free variables of $M'$ are also contained in $\G$.
\end{proof}

An immediate corollary of Theorem\cspc\ref{thm:cps_sem_pres_open} is
the following which correspond to
Property\cspc\ref{thm:sem_pres_index_closed} in the setting of CPS
transformation:
\begin{mycoro}\label{coro:cps_sem_pres_closed}
  If $\stseq M : T$ and $\cps \emptyset M K {M'}$, then $\simindex T i
  {M;K} {M'}$ for any $i$.
\end{mycoro}
\noindent From this corollary, it is easy to derive the following correctness
property of the CPS transformation for closed programs at atomic types:
\begin{mycoro}\label{coro:cps_sem_pres_atom}
  If $\stseq M : \tnat$, $\cps \emptyset M K {M'}$ and $M \eval V$, then
  $M' \step{*} \capp K V$.
\end{mycoro}
\noindent By choosing $K$ to be the identity function $\cabs x x$, we
get the following corollary that corresponds to
Property\cspc\ref{thm:sem_pres_atom} in the setting of the CPS
transformation:
\begin{mycoro}\label{coro:cps_sem_pres_atom_id}
  If $\stseq M : \tnat$, $\cps \emptyset M {\cabs x x} {M'}$ and $M
  \eval V$, then $M' \eval V$.
\end{mycoro}

\section{Verifying the \LProlog Program in \Abella}
\label{sec:formal_verify_cps}

In this section, we formally verify the correctness of the CPS
transformation in \Abella based on the informal proof given in
Section\cspc\ref{sec:informal_verify_cps}. The most significant
difference between the formal and informal proof is that the binding
related properties in the former must be made explicit, often stated
and proved formally as theorems. We show that the $\lambda$-tree
syntax approach can be used to significantly alleviate this
difficulty, producing a formal correctness proof that closely follows
the informal one.

\subsection{Type preservation of the transformation}
\label{subsec:cps_formal_typ_pres}

We prove that the \LProlog implementation of the CPS transformation
given in Section\cspc\ref{subsec:specify_cps} preserves typing by
following the informal argument given in
Section\cspc\ref{subsec:cps_informal_typ_pres}.
We start by giving an inductive definition of the types of the source
(target) language, which will be useful later when induction on types
are needed. We identify a constant $\kissty : \kty \to \omic$ such
that $\kissty\app T$ holds if and only if $T$ is a well-formed
type. It is defined through the following \LProlog clauses:
\begin{tabbing}
\qquad\=\kill
\>$\kissty\app \ktnat.$\\
\>$\kissty\app \ktunit.$\\
\>$\kissty\app (\kprod\app T_1\app T_2) \limply
    \kissty\app T_1 \scomma \kissty\app T_2.$\\
\>$\kissty\app (\karr\app T_1\app T_2) \limply
    \kissty\app T_1 \scomma \kissty\app T_2.$
\end{tabbing}

The typing context of the source (target) language is characterized as
the following fixed-point definition for $\ksctx : \olist \to \prop$:
\begin{tabbing}
\qquad\=$\nablax x {\ksctx\app (\kof\app x\app T \cons L)}$
  \quad\=$\rdef$ \quad\=\kill
\>$\ksctx\app \knil$ \>$\rdef$ \>$\rtrue$\\
\>$\nablax x {\ksctx\app (\kof\app x\app T \cons L)}$
   \>$\rdef$ \>$\ksctx\app L \rand \oseq{\kissty\app T}.$
\end{tabbing}
By this definition, $\ksctx\app L$ holds if and only if $L$ consists
of formulas of the form $\kof\app x\app T$ where $T$ is well-formed
and assigns types to unique nominal constants. If $\ksctx\app L$
holds, then $\oseq{L \stseq \kof\app M\app T}$ holds if and only if $M$
is a well-formed term of type $T$ whose free variables are represented
by the nominal constants in $L$.

The mapping between types before and after the CPS transformation is
defined through the following clauses for $\kcpsty : \kty \to \kty
\to \kty \to \prop$ such that $\cpsty S {T_1} {T_2}$ holds if and
only if $\kcpsty\app S\app T_1\app T_2$ holds.
\begin{tabbing}
\qquad\=\qquad\=\kill
\>$\kcpsty\app S\app \knat\app \knat$ \quad$\rdef$ \quad$\rtrue$\\
\>$\kcpsty\app S\app \kunit\app \kunit$ \quad$\rdef$ \quad$\rtrue$\\
\>$\kcpsty\app S\app (\kprod\app T_1\app T_2)\app (\kprod\app T_1'\app T_2')$
   \quad$\rdef$ \quad
   $\kcpsty\app S\app T_1\app T_1' \rand \kcpsty\app S\app T_2\app T_2'$\\
\>$\kcpsty\app S\app (\karr\app T_1\app T_2)\app
                     (\karr\app (\kprod\app (\karr\app T_2'\app S)\app T_1')\app S)$ \quad$\rdef$\\
\>\>$\kcpsty\app S\app T_1\app T_1' \rand \kcpsty\app S\app T_2\app T_2'.$
\end{tabbing}
These clauses are transparently translated from the rules in
Figure\cspc\ref{fig:cps_typ_mapping}. The mapping between types can be
generalized to typing contexts, as follows:
\begin{tabbing}
\qquad\=\qquad\=$\rdef$\kill
\>$\kcpssctx\app S\app \knil\app \knil$ \quad$\rdef$ \quad$\rtrue$\\
\>$\nablax x {\kcpssctx\app S\app (\kof\app x\app T \cons L)\app
                 (\kof\app x\app T' \cons L')}$ \quad$\rdef$\\
\>\>$\kcpsty\app S\app T\app T' \rand \kcpssctx\app S\app L\app L'.$
\end{tabbing}

The \LProlog implementation of the transformation makes use of the
dynamic context to bind the free variables of the source term and
to represent the rules for transforming these variables. We
characterized such a dynamic context by the following definition for
$\kcctx : \olist \to \prop$:
\begin{tabbing}
\qquad\=$\nablax x {\kcctx\app ((\forallx k {\kcps\app x\app k\app (k\app x))} \cons L)}$
  \quad\=$\rdef$ \quad\=\kill
\>$\kcctx\app \knil$ \>$\rdef$ \>$\rtrue$\\
\>$\nablax x {\kcctx\app ((\forallx k {\kcps\app x\app k\app (k\app x))} \cons L)}$
   \>$\rdef$ \>$\kcctx\app L.$
\end{tabbing}

\pagelabel{ecd:cps_typ_pres_open}
We can now state the type preservation theorem of the transformation
in \Abella as follows, called \kcpstyppres:
\begin{tabbing}
\qquad\=\quad\=\quad\=\kill
\>$\rfall \TL, \CL, M, T, K, M', T', \TL', S.$\\
\>\>$\oseq{\kissty\app T} \rimp \oseq{\kissty\app S} \rimp \ksctx\app \TL \rimp \kcctx\app \CL \rimp$\\
\>\>$\oseq{\TL \stseq \kof\app M\app T} \rimp \oseq{\CL \stseq \kcps\app M\app K\app M'} \rimp$\\
\>\>$\kcpsty\app S\app T\app T' \rimp \kcpssctx\app S\app \TL\app \TL'\app \rimp$\\
\>\>\>$\oseq{\TL', \forallx x {\kof\app x\app T' \simply \kof\app (K\app x)\app S} \stseq \kof\app M'\app S}.$
\end{tabbing}
This is a direct translation from
Theorem\cspc\ref{thm:cps_typ_pres_open}. Note that $\ksctx\app \TL$
and $\oseq{\TL \stseq \kof\app M\app T}$ together assert that $M$ has type
$T$ in the context $\TL$, and $\kcctx\app \CL$ and $\oseq{\CL \stseq
  \kcps\app M\app K\app M'}$ together assert that $M$ whose free
variables are bound in $\CL$ with the input continuation $K$ is
transformed into $M'$. The (higher-order) formula $\forallx x
{\kof\app x\app T' \simply \kof\app (K\app x)\app S}$ in the dynamic
context of the conclusion asserts that $K$ has the type $T' \to S$.
The theorem is proved by induction on $\oseq{\CL \stseq \kcps\app
  M\app K\app M'}$. By the transparent encoding of \HHw in \Abella and
the transparent relation between the \LProlog implementation and the
rules its encodes, the proof corresponds to induction on the
derivation of the \HHw judgment $\CL \stseq \kcps\app M\app K\app M'$
and furthermore on the derivation using the transformation rules. It
has the same structure as its informal rendition.

From the above theorem, it is easy to prove the following rendition of
Corollary\cspc\ref{coro:cps_typ_pres_closed} in \Abella, called
\kcpstyppresclosed:
\begin{tabbing}
\qquad\=\quad\=\kill
\>$\rfall S,M,K,M'.
     \oseq{\kissty\app S} \rimp \oseq{\kof\app M\app \ktnat} \rimp
     \oseq{\kcps\app M\app K\app M'} \rimp$\\
\>\>$\oseq{\forallx x {\kof\app x\app \ktnat \simply \kof\app (K\app x)\app S}
            \stseq \kof\app M'\app S}$.
\end{tabbing}
Letting $K = x \mlam x$, it is easy to prove the following rendition
of Corollary\cspc\ref{coro:cps_typ_pres_closed_id}:
\begin{tabbing}
\qquad\=\quad\=\kill
\>$\rfall M,M'.
     \oseq{\kof\app M\app \ktnat} \rimp
     \oseq{\kcps\app M\app (x\mlam x)\app M'} \rimp
     \oseq{\kof\app M'\app \ktnat}$.
\end{tabbing}
Now we have formally proved that the implementation of the CPS
transformation preserves types.

\subsection{Semantics preservation of the transformation}
\label{subsec:cps_formal_sem_pres}

In this section, we formally develop the semantics preservation proof
of the CPS transformation in \Abella. We place a focus on showing how
the $\lambda$-tree syntax approach can alleviate the difficulties in
reasoning about binding structure. Specifically, we show how the
property of being closed can be elegantly encoded as a \LProlog
program, how the definition of substitution as a relation can be used
to derive ``boilerplate'' properties with very little effort, how the
definition of substitution is used to prove the closedness property,
and how the logical structure of binding related treatments in the
\LProlog can be exploit to greatly simplify the correctness proof of
the transformation.

\subsubsection{Formalizing the operational semantics}

\pagelabel{ecd:cps_eval}
We formalize the small-step operational semantics as a \LProlog
program. For this we need to formally define what are values. We
identify the constant $\kval : \ktm \to \omic$ such that $\kval\app V$
holds if and only if $V$ is a value. It is defined by the following
program clauses:
\begin{tabbing}
\qquad\=\kill
\>$\kval\app (\knat\app N).$\\
\>$\kval\app \kunit.$\\
\>$\kval\app (\kpair\app V_1\app V_2) \limply \kval\app V_1 \scomma \kval\app V_2.$\\
\>$\kval\app (\kfix\app R).$
\end{tabbing}
We also need to define addition and predecessor operations on natural
numbers. They are given through the following clauses with obvious
meanings:
\begin{tabbing}
\qquad\=\kill
\>$\knpred\app \kz\app \kz.$\\
\>$\knpred\app (\ks\app N)\app N.$\\
\>$\kadd\app \kz\app N\app N.$\\
\>$\kadd\app (\ks\app N_1)\app N_2\app (\ks\app N_3) \limply \kadd\app N_1\app N_2\app N_3.$
\end{tabbing}
We designate the constant $\kstep : \ktm \to \ktm \to \omic$ to
represent one-step evaluation such that $\kstep\app M_1\app M_2$ holds
if $M_1$ evaluates to $M_2$ in one step. It is defined by the
following program clauses:
\begin{tabbing}
\qquad\=\kill
\>$\kstep\app (\kpred\app M)\app (\kpred\app M') \limply \kstep\app M\app M'.$\\
\>$\kstep\app (\kpred\app (\knat\app N))\app (\knat\app N') \limply \knpred\app N\app N'.$\\
\>$\kstep\app (\kplus\app M_1\app M_2)\app (\kplus\app M_1'\app M_2) \limply
     \kstep\app M_1\app  M_1'.$\\
\>$\kstep\app (\kplus\app V_1\app M_2)\app (\kplus\app V_1\app M_2') \limply
     \kval\app V_1\scomma \kstep\app M_2\app M_2'.$\\
\>$\kstep\app (\kplus\app (\knat\app N_1)\app (\knat\app N_2))\app (\knat\app N)
     \limply \kadd\app N_1\app N_2\app N.$\\
\>$\kstep\app (\kifz\app M\app M_1\app M_2)\app (\kifz\app M'\app M_1\app M_2)
     \limply \kstep\app M\app M'.$\\
\>$\kstep\app (\kifz\app (\knat\app \kz)\app M_1\app M_2)\app M_1.$\\
\>$\kstep\app (\kifz\app (\knat\app (\ks\app N))\app M_1\app M_2)\app M_2.$\\
\>$\kstep\app (\kpair\app M_1\app M_2)\app (\kpair\app M_1'\app M_2)
     \limply \kstep\app M_1\app M_1'.$\\
\>$\kstep\app (\kpair\app V_1\app M_2)\app (\kpair\app V_1\app M_2')
     \limply \kval\app V_1\scomma \kstep\app M_2\app M_2'.$\\
\>$\kstep\app (\kfst\app M)\app (\kfst\app M') \limply \kstep\app M\app M'.$\\
\>$\kstep\app (\kfst\app (\kpair\app V_1\app V_2))\app V_1
     \limply \kval\app (\kpair\app V_1\app V_2).$\\
\>$\kstep\app (\ksnd\app M)\app (\ksnd\app M') \limply \kstep\app M\app M'.$\\
\>$\kstep\app (\ksnd\app (\kpair\app V_1\app V_2))\app V_2
     \limply \kval\app (\kpair\app V_1\app V_2).$\\
\>$\kstep\app (\klet\app M\app R)\app (\klet\app M'\app R)\app
     \limply \kstep\app M\app M'.$\\
\>$\kstep\app (\klet\app V\app R)\app (R\app V) \limply \kval\app V.$\\
\>$\kstep\app (\kapp\app M_1\app M_2)\app (\kapp\app M_1'\app M_2)
     \limply \kstep\app M_1\app M_1'.$\\
\>$\kstep\app (\kapp\app V_1\app M_2)\app (\kapp\app V_1\app M_2')
     \limply \kval\app V_1\scomma \kstep\app M_2\app M_2'.$\\
\>$\kstep\app (\kapp\app (\kfix\app R)\app V)\app (R\app (\kfix\app R)\app V)
     \limply \kval\app V.$
\end{tabbing}
These program clauses are transparently translated from the evaluation
rules in Figure\cspc\ref{fig:cps_step_rules}. Note here how
application in the meta-language realizes substitution.
We designate the constant $\knstep : \knat \to \ktm \to \ktm \to
\omic$ to represent n-step evaluation, defined as follows:
\begin{tabbing}
\qquad\=\kill
\>$\knstep\app \kz\app M\app M.$\\
\>$\knstep\app (\ks\app N)\app M\app M'' \limply
   \kstep\app M\app M' \scomma \knstep\app N\app M'\app M''.$
\end{tabbing}
Finally, we use the constant $\keval: \ktm \to \ktm \to \omic$ to
represent the evaluation relation such that $\keval\app M\app V$ if
and only if $M$ evaluates to the value $V$ in a finite number of
steps, defined by the following clause:
\begin{tabbing}
\qquad\=\kill
\>$\keval\app M\app V \limply \knstep\app N\app M\app V \scomma \kval\app V.$
\end{tabbing}

\subsubsection{Formalizing the closedness property}
A notion related to bindings that is important for our verification
task is the property of being closed terms, \ie, terms that do not
have any free variable. We have seen in
Section\cspc\ref{subsec:lts_reason} how the closed \STLC terms are
characterized in \Abella. The same idea carries over to our
setting. We first defined the predicate $\ktm : \ktm \to \omic$ such
that $\oseq{\ktm\app M}$ holds if $M$ is a well-formed term in our
source (target) language through the following program clauses in
\LProlog:
\begin{tabbing}
\qquad\=\kill
\>$\ktm\app (\knat\app N).$\\
\>$\ktm\app (\kpred\app M) \limply \ktm\app M.$\\
\>$\ktm\app (\kplus\app M_1\app M_2) \limply \ktm\app M_1\scomma \ktm\app M_2.$\\
\>$\ktm\app (\kifz\app M\app M_1\app M_2) \limply
     \ktm\app M\scomma \ktm\app M_1\scomma \ktm\app M_2.$\\
\>$\ktm\app \kunit.$\\
\>$\ktm\app (\kpair\app M_1\app M_2) \limply \ktm\app M_1\scomma \ktm\app M_2.$\\
\>$\ktm\app (\kfst\app M) \limply \ktm\app M.$\\
\>$\ktm\app (\ksnd\app M) \limply \ktm\app M.$\\
\>$\ktm\app (\klet\app M\app R) \limply
    \ktm\app M\scomma \forallx x {\ktm\app x \simply \ktm\app (R\app x)}.$\\
\>$\ktm\app (\kfix\app R) \limply
     \forallx {f,x} {\ktm\app f \simply \ktm\app x \simply \ktm\app (R\app f\app x)}.$\\
\>$\ktm\app (\kapp\app M_1\app M_2) \limply \ktm\app M_1\scomma \ktm\app M_2.$
\end{tabbing}
These clauses resembles the clauses encoding the typing rules, except
that they do not have any type information. Similar to the typing
rules, we can define the context of well-formedness as follows:
\begin{tabbing}
\qquad\=$\ktmsctx\app (\ktm\app X \cons L)$
  \quad \=$\rdef$ \quad\=\kill
\>$\ktmsctx\app \knil$ \>$\rdef$ \>$\rtrue$\\
\>$\ktmsctx\app (\ktm\app X \cons L)$ \>$\rdef$ \>$\ktmsctx\app L \rand \kname\app X.$
\end{tabbing}
Now, $\ktmsctx\app L$ and $\oseq{L \stseq \ktm\app M}$ hold if and only
if $M$ is a well formed term whose free variables are bound in $L$. We
can therefore use the judgment $\oseq{\ktm\app M}$ to denote that $M$
is closed.

We usually do not explicitly prove that a term is well-formed or
closed. Instead, because we only deal with typed terms, we can derive
well-formedness by making use of the property that well-typed terms
are also well-formed. This property is stated as the following
theorem, named \ksoftotm, and proved by induction on $\oseq{L \stseq
  \kof\app M\app T}$:
\begin{tabbing}
\qquad\=\quad\=\kill
\>$\rfall L, \Vs, \SL, M, T, \oseq{\kissty\app T} \rimp
    \ksctx\app L \rimp \kvarsofsctx\app L\app \Vs \rimp \ktmsctx\app \SL \rimp$\\
\>\>$\kvarsoftmsctx\app \SL\app \Vs \rimp
       \oseq{L \stseq \kof\app M\app T} \rimp \oseq{\SL \stseq \ktm\app M}.$
\end{tabbing}
Here the predicate constant $\kvarsoftmsctx : \olist \to \klist\app
\ktm \to \prop$ is used to collect the variables in the contexts for
$\ktm$. It is defined through the following clauses:
\begin{tabbing}
\qquad\=$\nablax x {\kvarsoftmsctx\app (\ktm\app x \cons L)\app (x \cons L')}$
  \quad\=$\rdef$ \quad\=\kill
\>$\kvarsoftmsctx\app \knil\app \knil$ \>$\rdef$ \>$\rtrue$\\
\>$\nablax x {\kvarsoftmsctx\app (\ktm\app x \cons L)\app (x \cons L')}$
  \>$\rdef$ \>$\kvarsoftmsctx\app L\app L'.$
\end{tabbing}
Similarly, the predicate constant $\kvarsofsctx : \olist \to
\klist\app \ktm \to \prop$ is used to collect variables from the
typing contexts represented by \ksctx. It is defined through the
following clauses:
\begin{tabbing}
\qquad\=$\nablax x {\kvarsofsctx\app (\kof\app x\app T \cons L)\app (x \cons L')}$
   \quad\=$\rdef$ \quad\=\kill
\>$\kvarsofsctx\app \knil\app \knil$ \>$\rdef$ \>$\rtrue$\\
\>$\nablax x {\kvarsofsctx\app (\kof\app x\app T \cons L)\app (x \cons L')}$
   \>$\rdef$ \>$\kvarsofsctx\app L\app L'.$
\end{tabbing}

An important property of closed terms that will be used very often
later is that they do not depend on any nominal constant, which is
stated as the following theorem named \ksclosedtmprune:
\begin{gather*}
\rforallx {M} {\nablax {x:\ktm}
  \oseq{\ktm\app (M\app x)} \rimp \rexistsx {M'} {M = y \mlam M'}}.
\end{gather*}
Again, this is an example of the pruning properties we have discussed
before. It is a special case of the following theorem
\begin{gather*}
\rforallx {M, L} {\nablax {x:\ktm}
  {\ktmsctx\app L \rimp \oseq{L \sseq \ktm\app (M\app x)} \rimp
    \rexistsx {M'} {M = y \mlam M'}}}.
\end{gather*}
This theorem is proved easily by induction on $\oseq{L \sseq \ktm\app
  (M\app x)}$.

\subsubsection{Formalizing the logical relations}

We describe some auxiliary predicates that are necessary for
formalizing the logical relations. We define the predicate $\kle :
\knat \to \knat \to \prop$ such that $\kle\app N_1\app N_2$ holds if
and only if $N_1 \leq N_2$ through the following clause:
\begin{tabbing}
\qquad\=\kill
\>$\kle\app N_1\app N_2 \quad\rdef\quad \rexistsx N \oseq{\kadd\app N_1\app N\app N_2}.$
\end{tabbing}
We also define a predicate $\kstep* : \ktm \to \ktm \to \prop$ as a
short cut of n-step evaluation as follows:
\begin{tabbing}
\qquad\=\kill
\>$\kstep*\app M\app M'$ \quad$\rdef$ \quad$\rexistsx N {\oseq{\knstep\app N\app M\app M'}}.$
\end{tabbing}

\pagelabel{ecd:cps_logical_relations}
We designate the predicate constants $\ksimcps : \kty \to \knat \to \ktm
\to (\ktm \to \ktm) \to \ktm \to \prop$ and $\kequivcps : \kty \to \knat
\to \ktm \to \ktm \to \prop$ to respectively represent the simulation
and equivalence relations such that $\ksimcps\app T\app I\app M\app K\app
M'$ holds if and only if $\simindex T I {M;K} {M'}$ holds and
$\kequivcps\app T\app I\app M\app M'$ holds if and only if $\equalindex T
I M M'$ holds. Note that how the meta-level abstraction $K$ is used to
represent the input continuation for $\ksimcps$. The two predicates are
defined as follows:
\begin{tabbing}
\qquad\=quad\=\quad\=\kill
\>$\ksimcps\app T\app I\app M\app K\app M'$ \quad$\rdef$\\
\>\>$\rfall J, V. \kle\app J\app I \rimp \oseq{\knstep\app J\app M\app V}
     \rimp \oseq{\kval\app V} \rimp$\\
\>\>\>$\rexistsx {N, V'} {\kstep*\app M'\app (K\app V') \rand \oseq{\kadd\app J\app N\app I}
        \rand \kequivcps\app T\app N\app V\app V'}$\\
\>$\kequivcps\app \ktnat\app I\app (\knat\app N)\app (\knat\app N)$ \quad$\rdef$\quad $\rtrue$\\
\>$\kequivcps\app \ktunit\app I\app \kunit\app \kunit$ \quad$\rdef$\quad $\rtrue$\\
\>$\kequivcps\app (\kprod\app T_1\app T_2)\app I\app
     (\kpair\app V_1\app V_2)\app (\kpair\app V_1'\app V_2')\app$ \quad$\rdef$\\
\>\>$\kequivcps\app T_1\app I\app V_1\app V_1' \rand \kequivcps\app T_2\app I\app V_2\app V_2' \rand$\\
\>\>$\oseq{\ktm\app V_1} \rand \oseq{\ktm\app V_2}
     \rand \oseq{\ktm\app V_1'} \rand \oseq{\ktm\app V_2'}$\\
\>$\kequivcps\app (\karr\app T_1\app T_2)\app \kz\app
        (\kfix f\mlam x\mlam R\app f\app x)\app (\kfix f\mlam p\mlam R'\app f\app p)$ \quad$\rdef$\\
\>\>$\oseq{\ktm\app (\kfix\app R)} \rand \oseq{\ktm\app (\kfix\app R')}$\\
\>$\kequivcps\app (\karr\app T_1\app T_2)\app (\ks\app I)\app
        (\kfix f\mlam x\mlam R\app f\app x)\app (\kfix f\mlam p\mlam R'\app f\app p)$ \quad$\rdef$\\
\>\>$\rfall V_1,V_1',V_2,V_2',K.$\\
\>\>\>$\kequivcps\app T_1\app I\app V_1\app V_1' \rimp
      \kequivcps\app (\karr\app T_1\app T_2)\app I\app V_2\app V_2' \rimp$\\
\>\>\>$\ksimcps\app T_2\app I\app (R\app V_2\app V_1)\app K\app
                 (R'\app V_2'\app (\kpair\app (\kfix\app f\mlam K)\app V_1'))$
\end{tabbing}
These clauses are translated from the definition in
Figure\cspc\ref{fig:cps_logical_relations}. Notice how the \ktm
predicate is used to enforce the closedness constraint and how
meta-level $\beta$-redexes are used to model administrative
$\beta$-redexes and substitutions. For example, the administrative
$\beta$-redex $\capp K {V'}$ in
Figure\cspc\ref{fig:cps_logical_relations} is represented by the
meta-level $\beta$-redex $(K\app V')$ in the definition of \ksimcps. Note
also that $(\kfix\app f\mlam K)$ encodes the non-recursive function
$\abs x (K\app x)$ which is a shorthand for $\fix f x (K\app x)$.

The definition of the \kequivcps relation uses itself negatively in the
last clause. As such, we cannot view this as a fixed-point definition
in \Gee. However, we use it only as a recursive definition, \ie, as a
definition based on which we can do unfolding or rewriting but not
case analysis. In fact, this is the reason why we ``build'' the
relation up over the natural numbers rather than mirroring directly
the structure of the informal definition.

The property that \kequivcps is closed under decreasing indices is stated
as follows
\begin{tabbing}
\qquad\=\quad\=\kill
\>$\rfall T, I, J, V, V'.
  \oseq{\kissty\app T} \rimp \oseq{\kisnat\app I} \rimp$\\
\>\>$\kequivcps\app T\app I\app V\app V' \rimp \kle\app J\app I \rimp
       \kequivcps\app T\app J\app V\app V'.$
\end{tabbing}
where $\kisnat: \knat \to \prop$ is defined by the following \LProlog program:
\begin{tabbing}
\qquad\=\kill
\>$\kisnat\app \kz.$\\
\>$\kisnat\app (\ks\app N) \limply \kisnat\app N$
\end{tabbing}
It is proved by a nested induction on $\oseq{\kissty\app T}$ and
$\oseq{\kisnat\app I}$ and unfolding of $\kequivcps\app T\app I\app V\app
V'$ in each case.

The compatibility lemmas for the simulation relation are formalized as
the following theorems in \Abella:
\begin{tabbing}
\qquad\=\quad\=\quad\=\kill
\>$\rfall I, M, M', K.
   \ksimcps\app\app \ktnat\app I\app M\app (x\mlam \klet\app (\kpred\app x)\app (v\mlam K\app v))\app M' \rimp$\\
\>\>$\ksimcps\app\app \ktnat\app I\app (\kpred\app M)\app K\app M'.$\\
\>$\rfall I, M_1, M_2, M_2', M', K. \nabla x_1.
  \oseq{\kisnat\app I} \rimp \ksimcps\app \ktnat\app I\app M_1\app M_2'\app M' \rimp$\\
\>\>$\ksimcps\app \ktnat\app I\app M_2\app
     (x_2\mlam \klet\app (\kplus\app x_1\app x_2)\app (v\mlam K\app v))\app (M_2'\app x_1) \rimp$\\
\>\>\>$\ksimcps\app \ktnat\app I\app (\kplus\app M_1\app M_2)\app K\app M'.$\\
\>$\rfall T_1\app T_2\app K\app I\app M\app M'.
  \oseq{\kisnat\app I} \rimp \oseq{\kissty\app (\kprod\app T_1\app T_2)} \rimp$\\
\>\>$\ksimcps\app (\kprod\app T_1\app T_2)\app I\app M\app (x\mlam \klet\app (\kfst\app x)\app (v\mlam K\app v))\app M' \rimp$\\
\>\>\>$\ksimcps\app T_1\app I\app (\kfst\app M)\app K\app M'.$\\
\>$\rfall T_1, T_2, K, I, M, M'.
  \oseq{\kisnat\app I} \rimp \oseq{\kissty\app (\kprod\app T_1\app T_2)} \rimp $\\
\>\>$\ksimcps\app (\kprod\app T_1\app T_2)\app I\app M\app (x\mlam \klet\app (\ksnd\app x)\app (v\mlam K\app v))\app M' \rimp$\\
\>\>\>$\ksimcps\app T_2\app I\app (\ksnd\app M)\app K\app M'.$\\
\>$\rfall I, T_1, T_2, M_1, M_2, M_2', M', K. \nabla x_1.
  \oseq{\kisnat\app I} \rimp \oseq{\kissty\app T_1} \rimp \oseq{\kissty\app T_2} \rimp$\\
\>\>$\ksimcps\app T_1\app I\app M_1\app M_2'\app M' \rimp$\\
\>\>$\ksimcps\app T_2\app I\app M_2\app (x_2\mlam \klet\app (\kpair\app x_1\app x_2)\app (v\mlam K\app v))\app (M_2'\app x_1) \rimp$\\
\>\>\>$\ksimcps\app (\kprod\app T_1\app T_2)\app I\app (\kpair\app M_1\app M_2)\app K\app M'.$\\
\>$\rfall I, T, K', M_1, M_2, M_3, M_2', M_3', M'.
  \oseq{\kisnat\app I} \rimp \oseq{\kissty\app T} \rimp$\\
\>\>$\ksimcps\app T\app I\app M_2\app (x\mlam \kapp\app (\kfix\app f\mlam K')\app x)\app (M_2'\app (\kfix\app f\mlam K')) \rimp$\\
\>\>$\ksimcps\app T\app I\app M_3\app (x\mlam \kapp\app (\kfix\app f\mlam K')\app x)\app (M_3'\app (\kfix\app f\mlam K')) \rimp$\\
\>\>$\ksimcps\app \ktnat\app I\app M_1\app (x_1\mlam \klet\app (\kfix\app f\mlam K')\app (k\mlam \kifz\app x_1\app (M_2'\app k)\app (M_3'\app k)))\app M' \rimp$\\
\>\>\>$\ksimcps\app T\app I\app (\kifz\app M_1\app M_2\app M_3)\app K'\app M'.$\\
\>$\rfall T_1, T, I, K', M_1, M', M_2, M_2'. \nabla x_1.
  \oseq{\kisnat\app I} \rimp \oseq{\kissty\app (\karr\app T_1\app T)} \rimp$\\
\>\>$\ksimcps\app (\karr\app T_1\app T)\app I\app M_1\app M_2'\app M' \rimp$\\
\>\>$\ksimcps\app T_1\app I\app M_2\app (x_2\mlam \klet\app (\kfix\app (f\mlam K'))\app (k\mlam \klet\app (\kpair\app k\app x_2)\app (p\mlam \kapp\app x_1\app p)))$\\
\>\>\>$\qquad\quad(M_2'\app x_1) \rimp$\\
\>\>\>$\ksimcps\app T\app I\app (\kapp\app M_1\app M_2)\app K'\app M'.$
\end{tabbing}
These theorems are directly translated from
Lemma\cspc\ref{lem:cps_sim_compat}.
%% Note how the administrative abstractions and applications are
%% represented as their meta-level counterparts in them.
They are proved by following
arguments similar to the informal ones. Therefore the closedness
property of $\kequivcps$ we proved previously is used here. 
%% In proving
%% the compatibility lemmas that contain $\nabla$-quantified variables,
%% we use \ksclosedtmprune to discharge the dependence of newly
%% introduced variables on these variables.

\subsubsection{Representing substitutions}

We treat substitutions as discussed in
Section\cspc\ref{subsec:schm_poly_exms}. We identify the type $\kmap :
\ktype \to \ktype \to \ktype$ as the type of mappings and its
constructor $\kmap : A \to B \to \kmap\app A\app B$. A substitution is
then represented as a list of objects of the form $\kmap\app x\app V$
where $x$ is a variable and $V$ is a closed value in which the
variables are distinct. The substitutions for the source language is
characterized by the predicate $\ksubst: \ktm \to \prop$ with the
following definition such that $\ksubst\app M$ holds if and only if
$M$ is a substitution:
\begin{tabbing}
\qquad\=$\nablax x {\ksubst\app (\kmap\app x\app V \cons \ML)}$ \quad\=$\rdef$ \quad\=\kill
\>$\ksubst\app \knil$ \>$\rdef$ \>$\rtrue$\\
\>$\nablax x {\ksubst\app (\kmap\app x\app V \cons \ML)}$
   \>$\rdef$ \>$\ksubst\app \ML \rand \oseq{\kval\app V} \rand \oseq{\ktm\app V}.$
\end{tabbing}

As described in Section\cspc\ref{subsec:schm_poly_exms}, application
of substitution is identified by the predicate symbol $\kappsubst:
\klist\app (\kmap\app A\app A) \to B \to B \to \prop$ defined through
the following clauses such that $\kappsubst\app S\app M\app M'$ holds
exactly when $M'$ is the result of applying the substitution $S$ to
$M$:
\begin{tabbing}
\qquad\=$\nablax x {\kappsubst\app ((\kmap\app x\app V) \cons S)\app (R\app x)\app M}$
\quad\=$\rdef$ \quad\=\kill
\>$\kappsubst\app \knil\app M\app M$ \>$\rdef$ \>$\rtrue$\\
\>$\nablax x {\kappsubst\app ((\kmap\app x\app V) \cons S)\app (R\app x)\app M}$
  \>$\rdef$ \>$\kappsubst\app S\app (R\app V)\app M$
\end{tabbing}

Given the above definitions, we can easily prove properties about
substitution. The first important property that is useful in the
semantics preservation proof is that substitution application
distributes over term structure. The following are some examples:
\begin{tabbing}
\qquad\=\quad\=\kill
\>$\rfall \ML, M, M'. \kappsubst\app \ML\app (\kpred\app M)\app M' \rimp$\\
\>\>$\rexistsx {M''} {M' = \kpred\app M'' \rand \kappsubst\app \ML\app M\app M''}.$\\
\>$\rfall \ML, M_1, M_2, M'.
    \kappsubst\app \ML\app (\kapp\app M_1\app M_2)\app M' \rimp$\\
\>\>$\rexistsx {M_1', M_2'}
       {M' = \kapp\app M_1'\app M_2' \rand
        \kappsubst\app \ML\app M_1\app M_1' \rand \kappsubst\app \ML\app M_2\app M_2'}.$\\
\>$\rfall R, M'. \kappsubst\app \ML\app (\kfix\app R)\app M' \rimp$\\
\>\>$\rexistsx {R'} {M' = \kfix\app R' \rand
     \nablax {f,x} {\kappsubst\app \ML\app (R\app f\app x)\app (R'\app f\app x)}}.$
\end{tabbing}
All such distribution lemmas have exactly the same proof. We do
induction on the only assumption and analyze its cases. The base case
is immediately concluded. In the inductive case we apply the inductive
hypothesis on the smaller derivations for $\kappsubst$ and use the
results to conclude the case. Because of the $\lambda$-tree syntax
approach, the above lemmas are just explicit statements of
corresponding distribution properties for $\beta$-reductions. So it is
no surprise they have such simple proofs.

Another important property that will be used very often is that
substitution has no effect on closed terms. It is stated as follows:
\begin{tabbing}
\qquad\=\kill
\>$\rforallx {M,ML,M'} {\oseq{\ktm\app M} \rimp
    \kappsubst\app \ML\app M\app M' \rimp M = M'}.$
\end{tabbing}
This lemma shows how the closedness property defined at the
implementation level coordinates with the reasoning. It is easily
proved by induction on $\kappsubst\app \ML\app M\app M'$ and using
\ksclosedtmprune to discharge the dependence of $M$ on the variables
in the substitution $\ML$.

The last important and useful property of substitution is that if all
free variables of a term $M$ are bound in a substitution $\ML$, then
the result of applying $\ML$ to $M$ is a closed term. This is stated
as the following theorem, called \ksubstresultclosedtm:
\begin{tabbing}
\qquad\=\quad\=\kill
\>$\rfall \ML,L,M,M',\Vs.
  \ktmsctx\app L \rimp \oseq{L \stseq \ktm\app M} \rimp \kvarsoftmsctx\app L\app \Vs \rimp$\\
\>\>$\ksubst\app \ML \rimp \kvarsofsubst\app \ML\app \Vs \rimp
      \kappsubst\app \ML\app M\app M'\app \rimp \oseq{\ktm\app M'}.$
\end{tabbing}
Here we use the predicate constant $\kvarsofsubst : \klist\app
(\kmap\app \ktm\app \ktm) \to \klist\app \ktm \to \prop$ such that
$\kvarsofsubst\app \ML\app \Vs$ holds if and only $\Vs$ is the domain of
the substitution $\ML$. Its definition is similar to $\kvarsofsctx$
and is elided here.
Again, the above theorem is easily proved by induction on
$\kappsubst\app \ML\app M\app M'$ thanks to the $\lambda$-tree syntax
representation of substitutions.

\subsubsection{The equivalence relation on substitutions}

The general form of the semantics preservation theorem requires a
notion of equivalence between substitutions. We designate the constant
$\ksubstequivcps : \olist \to \knat \to \klist\app (\kmap\app \ktm\app
\ktm) \to \klist\app (\kmap\app \ktm\app \ktm) \to \prop$ to represent
this equivalence relation such that $\ksubstequivcps\app L\app I\app
\ML\app \ML'$ holds if and only if $\equalindex \G i \theta {\theta'}$
holds where $L$, $I$, $\ML$ and $\ML'$ are respectively encodings of
$\G$, $i$, $\theta$ and $\theta'$. It is defined as follows:
\begin{tabbing}
p\qquad\=\quad\=\kill
\>$\ksubstequivcps\app \knil\app I\app \knil\app \knil \quad\rdef \quad\rtrue$\\
\>$\nablax x {\ksubstequivcps\app (\kof\app x\app T \cons L)\app I\app
               (\kmap\app x\app V \cons \ML)\app (\kmap\app x\app V' \cons \ML')}$
   \quad$\rdef$\\
\>\>$\kequivcps\app T\app I\app V\app V' \rand \ksubstequivcps\app L\app I\app \ML\app \ML'.$
\end{tabbing}

\subsubsection{Formalizing the semantics preservation theorems}

To formalize the semantics preservation theorems, we will need another
auxiliary predicate constant $\kvarsofcctx : \olist \to \klist\app
\ktm \to \prop$ for collecting variables in the transformation context
defined as follows:
\begin{tabbing}
\qquad\=$\nablax x {\kvarsofcctx\app ((\forallx k {\kcps\app x\app k\app (k\app x)}) \cons L)\app (x \cons L')}$
  \quad\=$\rdef$ \quad\=\kill
\>$\kvarsofcctx\app \knil\app \knil$ \>$\rdef$ \>$\rtrue$\\
\>$\nablax x {\kvarsofcctx\app ((\forallx k {\kcps\app x\app k\app (k\app x)}) \cons L)\app (x \cons L')}$
  \>$\rdef$ \>$\kvarsofcctx\app L\app L'.$
\end{tabbing}

\pagelabel{ecd:cps_sem_pres_open}
Theorem\cspc\ref{thm:cps_sem_pres_open} is now formalized as follows:
\begin{tabbing}
\qquad\=\quad\=\quad\=\kill
\>$\rfall \ML, \ML', \TL, \CL, \VL, M, T, K, K', M', P, P', I.$\\
\>\>$\oseq{\kisnat\app I} \rimp \oseq{\kissty\app T} \rimp$\\
\>\>$\ksctx\app \TL \rimp \kvarsofsctx\app \TL\app \VL \rimp$\\
\>\>$\kcctx\app \CL \rimp \kvarsofcctx\app \CL\app \VL \rimp$\\
\>\>$\ksubst\app \ML \rimp \ksubst\app \ML' \rimp \ksubstequivcps\app \TL\app I\app \ML\app \ML' \rimp$\\
\>\>$\oseq{\TL \stseq \kof\app M\app T} \rimp \oseq{\CL \stseq \kcps\app M\app K\app M'} \rimp$\\
\>\>$\kappsubst\app \ML\app M\app P \rimp \kappsubst\app \ML'\app M'\app P' \rimp$\\
\>\>$(\nablax x {\kappsubst\app \ML'\app (K\app x)\app (K'\app x)}) \rimp$\\
\>\>\>$\ksimcps\app T\app I\app P\app K'\app P'.$
\end{tabbing}
We prove this theorem by induction on $\oseq{\CL \stseq \kcps\app
  M\app K\app M'}$, the derivation of the CPS transformation. The proof
closely follows the informal proof for
Theorem\cspc\ref{thm:cps_sem_pres_open}. The proof for base cases is
obvious. In the remaining cases, other than when $M$ is a let
expression or a fixed-point, the proof follows a set pattern: we first
apply the distribution lemmas of \kappsubst to $M$, then apply the
inductive hypotheses to sub-expressions of $M$ to get the simulation
relation between sub-expressions, and finally conclude by applying the
compatibility lemmas of $\ksimcps$. When $M$ is a let expression or a
fixed-point, we extend the substitutions with equivalent values for
new variables introduced by recursion for applying the inductive
hypothesis on expressions under the binding operators of $M$ and use
the results to conclude the case.

As we have said, many aspects of bindings that are implicit in
implementation must be made explicit in the correctness proof, which
may blur the essential contents of the proof. Thanks to the uses of
the $\lambda$-tree syntax approach, reasoning about bindings is
greatly simplified, resulting in the closed correspondence between our
formal proof and the informal one. The uses and benefits of the
$\lambda$-tree syntax approach in our formal proof are summarized as
follows. First, because the treatments of bindings at the
specification level are transparently reflected into the reasoning
level via the two-level logic approach, properties of administrative
$\beta$-redexes are reflected into properties of meta-level
$\beta$-redexes in \Abella and are immediately available in
reasoning. As a result, minimum effort is needed to reason about
administrative $\beta$-redexes. Second, the ``boilerplate'' properties
of substitution and closed terms are captured as lemmas of \kappsubst
and \ktm and used to simplify the reasoning about them. For example,
when $M$ is a natural number, we apply the lemma that \kappsubst has
no effect on closed terms to show that it does not affect $M$. As
another example, when $M$ is a fixed-point, to prove the simulation
relation $\ksimcps\app T\app I\app P\app K'\app P'$ we need to show $P$
and $P'$ are closed terms. To show the former, we first apply
\ksoftotm to $\oseq{\TL \stseq \kof\app M\app T}$ to show that $M$ is
a well-formed term whose variables are contained in the domain of
$\ML$ and then apply \ksubstresultclosedtm to $\kappsubst\app \ML\app
M\app P$ to show that $\oseq{\ktm\app P}$ holds, \ie, $P$ is
closed. To show the latter, we first apply the type preservation lemma
to $\oseq{\TL \stseq \kof\app M\app T}$ and $\oseq{\CL \stseq \kcps\app
  M\app K\app M'}$ to show that $M'$ is well-typed, at this point we
follow the same process for proving $P$ is closed to show that
$\oseq{\ktm\app P'}$ holds.

Finally, we formalize Corollary\cspc\ref{coro:cps_sem_pres_atom} as
the following theorem in \Abella which is easily proved by applying
the above theorem:
\begin{tabbing}
\qquad\=\quad\=\kill
\>$\rfall M, K, M', V.
     \oseq{\kof\app M\app \ktnat} \rimp \oseq{\kcps\app M\app K\app M'} \rimp$\\
\>\>$\oseq{\keval\app M\app V} \rimp \kstep*\app M'\app (K\app V).$
\end{tabbing}
Letting $K = x\mlam x$, it is easy to prove the following formalized
version of Corollary\cspc\ref{coro:cps_sem_pres_atom_id}:
\begin{tabbing}
\qquad\=\quad\=\kill
\>$\rfall M, M', V.
     \oseq{\kof\app M\app \ktnat} \rimp \oseq{\kcps\app M\app (x\mlam x)\app M'} \rimp$\\
\>\>$\oseq{\keval\app M\app V} \rimp \oseq{\keval\app M'\app V}.$
\end{tabbing}

%%%%%%%%%%%%%%%%%%%%%%%%%%%%%%%%%%%%%%%%%%%%%%%%%%%%%%%%%%%%%%%%%%%%%%%%%%%%%%%%

% The closure conversion transformation
%%%%%%%%%%%%%%%%%%%%%%%%%%%%%%%%%%%%%%%%%%%%%%%%%%%%%%%%%%%%%%%%%%%%%%%%%%%%%%%
% closure.tex: The closure conversion transformation
%%%%%%%%%%%%%%%%%%%%%%%%%%%%%%%%%%%%%%%%%%%%%%%%%%%%%%%%%%%%%%%%%%%%%%%%%%%%%%%%
\chapter{The Closure Conversion Transformation}
\label{ch:closure}

An aspect that complicates the compilation of functional programs is
the presence of nested functions: since such functions can use
non-local variables, their invocation must be parameterized by the
context in which they appear. The closure conversion transformation
makes the kind of parameterization that is needed explicit.
It does so by transforming every
function into a \emph{closure} which consists of a closed function
called its code part and an environment part. For every function,
closure conversion identifies its free variables and constructs
the environment that contains bindings for these free variables in the
context of the function. It also abstracts the original function over
an extra environment parameter and replaces the occurrences of free
variables in the function with appropriate references to the
environment parameter to form the closed function. The closed
function and the constructed environment constitute the closure for
the original function. Because functions become closed after closure
conversion, they can be freely moved around, enabling other
transformations. One such transformation is code hoisting, which
eliminates nested functions.

The closure conversion transformation we develop in this chapter is
the second phase of the compiler described in
Section\cspc\ref{sec:vfc_exercise}. Following the approach
described in Section\cspc\ref{sec:vfc_approach}, we can describe
closure conversion in a rule-based and relational style and prove its
semantics preservation property based on logical relations. To get a
verified implementation of closure conversion, we follow the approach
given in Section\cspc\ref{sec:fm_vfc} to implement the rule-based
relational descriptions of closure conversion as a \LProlog program
and prove the correctness of the implementation in \Abella. Similar to
the CPS transformation described in Chapter\cspc\ref{ch:cps}, we
exploit the $\lambda$-tree syntax approach to simplify both the
implementation and verification.

A major difficulty in formalizing closure conversion is to formally
describe the computation of free variables. This involves non-trivial
analysis of the binding structure of functional terms. Moreover, in
formally proving the correctness of the closure conversion, we need to
prove non-trivial properties about such analysis. If not handled
properly, the proof of such properties can significantly complicate the main
verification effort. We shall see that by using the $\lambda$-tree
syntax approach we can implement the identification of free variables
as a \LProlog program and give concise proofs to the desired
properties of this program in \Abella.

We describe the implementation and verification of closure conversion
in the rest of this chapter. Again, our discussion focuses on showing
how the $\lambda$-tree syntax approach can be effectively exploited in
both the implementation and verification of closure conversion. Because the
structure and contents of this chapter share a lot of similarity with
that in Chapter\cspc\ref{ch:cps}, we shall omit discussion of the
similar contents and focus on the distinct aspects of the
implementation and verification of closure conversion. We start by
giving an overview of closure
conversion in Section\cspc\ref{sec:cc_overview}. We then present the
source and target languages of the transformation and the rule-based
relational descriptions of the transformation in
Section\cspc\ref{sec:rule_based_cc}. In Section\cspc\ref{sec:cc_impl}
we present the implementation of the rule-based description as an
\LProlog program for closure conversion. In
Section\cspc\ref{sec:informal_verify_cc} we describe the informal
verification of the closure conversion transformation. In
Section\cspc\ref{sec:formal_verify_cc} we present the formalization of
the verification.

\section{An Overview of the Transformation}
\label{sec:cc_overview}

The formulation of the closure conversion transformation that we use
here is based on \cite{minamide95tr}.
%
%% It is designed to replace (possibly nested) functions in a program by
%% \emph{closures} that each consist of a function and an
%% environment. The function or code part is a transformation of the
%% original function into a form where its free variables are replaced by
%% projections into a new environment parameter. The environment
%% component, on the other hand, encodes the construction of a tuple for
%% this parameter in the enclosing context. For example, when this
To illustrate this transformation concretely, when it is
applied to the following pseudo OCaml code segment
\begin{tabbing}
\qquad\=\quad\=\kill
\>$\kwd{let}\app x = 2\app \kwd{in}\app \kwd{let}\app y = 3\app \kwd{in}$\\
\>\>$\kwd{fun}\app z \rightarrow z + x + y$
\end{tabbing}
it will yield
\begin{tabbing}
\qquad\=\quad\=\kill
\>$\kwd{let}\app x = 2\app \kwd{in}\app \kwd{let}\app y = 3\app \kwd{in}$\\
\>\>$\clos{(\kwd{fun}\app z\app e \rightarrow z + e.1 + e.2)}{(x,y)}$
\end{tabbing}
We write $\clos F E$ here to represent a closure whose code part is
$F$ and environment part is $E$, and $e.i$ to represent the $i$-th
projection applied to an ``environment parameter'' $e$. A closure can
be thought as a partial application of the code part to its
environment part, which is suspended until the actual argument of the
function is provided at run-time. This transformation makes the
function part independent of the context in which it appears, thereby
allowing it to be extracted out to the top-level of the program.

Closure conversion is performed recursively on the structure of
terms. In the general case when such terms contain nested functions,
closure conversion needs to identify free variables and transform
occurrences of free variables under the nested binders as well.
As an illustration, consider the following pseudo OCaml code:
\begin{tabbing}
\qquad\=\quad\=\kill
\>$\kwd{let}\app x = 3\app \kwd{in}$\\
\>\>$\kwd{fun}\app y \rightarrow \kwd{fun}\app z \rightarrow x + y + z.$
\end{tabbing}
Closure conversion transforms it into the following code:
\begin{tabbing}
\qquad\=\quad\=\kill
\>$\kwd{let}\app x = 3\app \kwd{in}$\\
\>\>$\clos
       {(\kwd{fun}\app y\app e_1 \rightarrow
           \clos{(\kwd{fun}\app z\app e_2 \rightarrow e_2.1 + e_2.2 + z)}
                {(e_1.1,y)})}
       {(x)}.$
\end{tabbing}
In the first phase, it transforms the outer function.
Specifically, it determines that the only free variable at that level
is $x$. It then constructs the environment $(x)$ for the function and
parameterizes that function with the environment argument $e_1$.
It must then transform the body of the outer function, which is
exactly the inner function. Observe, however, that when it effects
this transformation, the variable $x$ that appears free in it must not
be referred to directly but as $e_1.1$. The consequence of this
observation can be seen in the code that is shown for the transformed
version of the inner function.

From the previous example, it is obvious that the essential work of
closure conversion is to compute free variables and to replace
occurrences of free variables with references to environment
parameters under nested binders. We shall discuss how to precisely describe
such manipulation and analysis of bindings and how to prove properties
about them in the rest of the chapter.

\section{A Rule-Based Description of the Transformation}
\label{sec:rule_based_cc}

We give a rule-based relational description of the closure conversion
transformation in this section. We first describe the source and
target languages of the transformation, including their typing rules,
and then present the transformation rules.

\subsection{The source and target languages}
\label{sec:cc_langs}

The source language of closure conversion is the same as the target
language of the CPS transformation. Its syntax and typing rules are
already given in Figures\cspc\ref{fig:cps_src_lang} and
\ref{fig:cps_src_typing}. The syntax of the target language is
depicted in Figure\cspc\ref{fig:cc_targ_lang}. One important difference between
the target language and the source language is that the former
includes constructs for dealing with
closures. Compared to the source language, the target language
includes the expressions $\clos {M_1} {M_2}$ and $(\open {x_f} {x_e}
{M_1} {M_2})$ representing the formation and application of
closures. Another important difference is that
the target language does not have an explicit fixed point
constructor. Instead, recursion is realized by parameterizing the
function part of a closure with a function component; this treatment
should become clear from the rules for typing closures and the
transformation that we present below. The usual forms of abstraction
and application are included in the target language to simplify the
presentation of the transformation. Note that the usual function type
is reserved for closures. The abstractions in the target language are
given the type ${T_1} \carr {T_2}$ in the target language. We
abbreviate $\pair {M_1} {\ldots\pair {M_n} \unit}$ by $(M_1,\ldots,M_n)$ and
$\fst {(\snd {(\ldots(\snd M))})}$ where $\mathbf{snd}$ is applied $i-1$
times for $i \geq 1$ by $\pi_i(M)$.

\begin{figure}[ht!]
  \begin{tabbing}
    \qquad\qquad\qquad\qquad\=$T$ \quad\=$::=$ \quad\=\kill
    \>$T$ \>$::=$ \>$\tnat \sep {T_1} \to {T_2} \sep {T_1} \carr {T_2} \sep \tunit \sep  {T_1} \tprod {T_2}$
    \\[1ex]
    \>$M$ \>$::=$ \>$n \sep x \sep \pred M \sep M_1 + M_2 \sep$\\
    \>\>\>$\ifz {M_1} {M_2} {M_3} \sep$ \\
    \>\>\>$\unit \sep \pair {M_1} {M_2} \sep \fst M \sep \snd M \sep$\\
    \>\>\>$\letexp x {M_1} {M_2} \sep \ \abs x M \sep (M_1 \app M_2) \sep$\\
    \>\>\>$\clos{M_1}{M_2} \sep \open {x_f} {x_e} {M_1} {M_2}$
    \\[1ex]
    \>$V$ \>$::=$ \>$n \sep \abs x M \sep () \sep \pair {V_1} {V_2} \sep \clos {V_1} {V_2}$
  \end{tabbing}
  \caption{The Syntax of The Target Language of Closure Conversion}
  \label{fig:cc_targ_lang}
\end{figure}

\pagelabel{txt:cc_targ_typing}
Typing judgments for the target language are written as $\Gamma \stseq
M : T$, where $\Gamma$ is a list of type assignments for
variables. Note that here we overload the syntax of the typing
judgments in Section\cspc\ref{sec:cps_langs}. The exact meaning of such a
judgment is inferred from the context if it is not explained
explicitly. Many of the program constructs in the source language are
present also in the target language and the rules in
Figure\cspc\ref{fig:cps_src_typing} for typing them carry over also to
the target language. The only exceptions are those for typing
abstractions and applications. In addition, we need rules for introducing
and eliminating closures that are present only in the target
language. Closures and abstractions both have a ``function'' type but
the specific way in which they figure in the target language is
different. We use types of two different forms, ${T_1} \to {T_2}$ and
${T_1} \carr {T_2}$, to distinguish between the roles of these
expressions. The typing rules, whose spirit is borrowed from the
presentation in \cite{belanger13cpp}, should help in explaining the
intended roles of the different function-values expressions.

The type used for abstractions has the form ${T_1} \carr {T_2}$. The
rules for typing them and the associated applications are shown below:
\begin{gather*}
  \infer[x\not\in\dom{\G}\; \cofabs]{
    \Gamma \stseq {\abs x M} : {T_1 \carr T_2}
  }{
    {\Gamma, x: T_1} \stseq M : {T_1}
  }
  \quad
  \infer[\cofapp]{
    \Gamma \stseq {{M_1}\app {M_2}} : T
  }{
    \Gamma \stseq {M_1} : {T_1 \carr T}
    &
    \Gamma \stseq {M_2} : {T_1}
  }
\end{gather*}
A closure differs from an abstraction in the sense that it represents
a partially applied function: it is packaged with an environment that
must be coupled with the ``real'' argument to complete the
application. We use a type of the form ${T_1} \to {T_2}$ to encode
this kind of partial application.
The specific rules for introducing and eliminating closures are shown
below:
\begin{gather*}
  \infer[\cofclos]{
    \Gamma \stseq {\clos {M_1} {M_2}} : {T_1 \to T_2}
  }{
    \stseq {M_1} : {((T_1 \to T_2) \tprod T_1 \tprod T_e) \carr T_2}
    &
    \Gamma \stseq {M_2} : {T_e}
  }
  \\[5pt]
  \infer[\cofopen]{
    \Gamma \stseq  {\open {x_f} {x_e} {M_1} {M_2}} : T
  }{
    \Gamma \stseq {M_1} : {T_1 \to T_2}
    &
    {\Gamma, x_f:((T_1 \to T_2) \tprod T_1 \tprod l) \carr T_2, x_e:l} \stseq {M_2} : T
  }
\end{gather*}
In $\cofopen$,
$x_f$, $x_e$ must be names that are new to $\Gamma$.  This rule also
uses a ``type'' $l$ that is to be interpreted as a new type constant,
different from $\tnat$ and $\unit$ and any other type constant used in
the typing derivation.
Intuitively, the rule $\cofclos$ states that for a closure
$\clos{M_1}{M_2}$ to have the type $T_1 \to T_2$ there must exists
some type $T_e$ for its environment part and $M_1$ must be a closed
function of type ${((T_1 \to T_2) \tprod T_1 \tprod T_e) \carr T_2}$;
the requirement that $M_1$ is closed follows from the fact that it
must be typable in an empty context. The type of $M_1$ indicates
how recursion is realized in the target language: the function part of
a closure takes the closure itself as an input besides the actual and
environment arguments. The rule $\cofopen$ conforms to how closure
applications work. In an expression of the form ${\open {x_f} {x_e}
  {M_1} {M_2}}$, $M_1$ is must evaluate to a closure whose
function and environment parts are extracted into $x_f$ and $x_e$ and
then used in $M_2$ through the occurrences of these variables in that
expression. Another interesting aspect to note about this rule is that
it enforces an opaqueness criterion on the environment component that
is extracted. This follows from the use of the fresh type constant
$l$ to represent the type of the environment.
%The use of $l$ in
%$\cofopen$ is also an example of existential
%types\cspc\cite{mitchell88toplas}.

\subsection{The transformation rules}
\label{subsec:cc_rules}

\begin{figure}[ht!]
\begin{gather*}
  \infer[\ccnat]{
    \cc {\rho} n n
  }{}
  \quad
  \infer[\ccvar]{
    \cc {\rho} x M
  }{
    (x \mapsto M) \in \rho
  }
  \\
  \infer[\ccfvs]{
    \ccenv {\rho} {(x_1,\ldots,x_n)} {(M_1,\ldots,M_n)}
  }{
    \cc {\rho} {x_1} {M_1}
    &
    \ldots
    &
    \cc {\rho} {x_n} {M_n}
  }
  \\
  \infer[\ccpred]{
    \cc {\rho} {\pred M} {\pred M'}
  }{
    \cc{\rho} M {M'}
  }
  \quad
  \infer[\ccplus]{
    \cc {\rho} {M_1 + M_2} {M_1' + M_2'}
  }{
    \cc \rho {M_1} {M_1'}
    &
    \cc \rho {M_2} {M_2'}
  }
  \\
  \infer[\ccifz]{
    \cc {\rho} {\ifz {M} {M_1} {M_2}} {\ifz {M'} {M_1'} {M_2'}}
  }{
    \cc \rho {M} {M'}
    &
    \cc \rho {M_1} {M_1'}
    &
    \cc \rho {M_2} {M_2'}
  }
  \\
  \infer[\ccunit]{
    \cc \rho \unit \unit
  }{}
  \qquad
  \infer[\ccpair]{
    \cc {\rho} {\pair {M_1} {M_2}} {\pair {M_1'} {M_2'}}
  }{
    \cc \rho {M_1} {M_1'}
    &
    \cc \rho {M_2} {M_2'}
  }
  \\
  \infer[\ccfst]{
    \cc {\rho} {\fst M} {\fst M'}
  }{
    \cc{\rho} M {M'}
  }
  \qquad
  \infer[\ccsnd]{
    \cc {\rho} {\snd M} {\snd M'}
  }{
    \cc{\rho} M {M'}
  }
  \\
  \infer[{\small(y\ \mbox{\rm is fresh})} \cclet]{
    \cc \rho {\letexp x {M_1} {M_2}} {\letexp y {M_1'} {M_2'}}
  }{
    \cc \rho {M_1} {M_1'}
    &
    \cc {\rho, x \mapsto y} {M_2} {M_2'}
  }
  \\
  \infer[{\small(g\ \mbox{\rm is fresh})} \ccapp]{
    \cc {\rho} {M_1 \app M_2}
        {\letexp g {M_1'}
          {\open {x_f} {x_e} {g} {x_f \app (g,M_2',x_e)}}}
  }{
    \cc {\rho} {M_1} {M_1'}
    &
    \cc {\rho} {M_2} {M_2'}
  }
  \\
  \infer[\ccfix]{
    \begin{align*}
    & \rho \triangleright {\fix f x M} \leadsto_{cc}\\
    & \qquad
      {\clos
          {\abs p {\letexp g {\pi_1(p)}
                   {\letexp y {\pi_2(p)}
                    {\letexp {x_e} {\pi_3(p)} {M'}}}}}
          {M_e}}
    %% \cc {\rho} {\fix f x M}
    %%     {\clos
    %%       {\abs p {\letexp g {\pi_1(p)}
    %%                {\letexp y {\pi_2(p)}
    %%                 {\letexp {x_e} {\pi_3(p)} {M'}}}}}
    %%       {M_e}}
    \end{align*}
  }{
    (x_1,\ldots,x_n) \supseteq \fvars {\fix f x M}
    &
    \ccenv {\rho} {(x_1,\ldots,x_n)} {M_e}
    &
    \cc {\rho'} M {M'}
  }
  \\
  \mbox{\small (where $\rho' = (x \mapsto y, f \mapsto g, x_1 \mapsto
    \pi_1(x_e), \ldots, x_n \mapsto \pi_n(x_e))$}
  \\
  \mbox{\small and $p, g, y,$ and $x_e$ are fresh variables)}
\end{gather*}
\caption{The Rules for Closure Conversion}
\label{fig:cc_rules}
\end{figure}

\noindent In the general case, we must transform terms under mappings for their
free variables: for a function term, this mapping represents the
replacement of the free variables by projections from the environment
variable for which a new abstraction will be introduced into the term.
Accordingly, we specify the transformation as a 3-place relation
written as $\cc \rho M {M'}$, where $M$ and $M'$ are source and target
language terms and $\rho$ is a mapping from source language variables
to target language terms. We write $(\rho, x \mapsto M)$ to denote the
extension of $\rho$ with a mapping for $x$ and $(x \mapsto M) \in
\rho$ to mean that $\rho$ contains a mapping of $x$ to $M$.
Figure\cspc\ref{fig:cc_rules} defines the $\cc \rho M {M'}$ relation
in a rule-based fashion; these rules use the auxiliary relation
$\ccenv \rho {(x_1,\ldots,x_n)} {M_e}$ that determines an environment
corresponding to a tuple of variables. The $\cclet$ and $\ccfix$ rules
have a proviso: the bound variables, $x$ and $f, x$ respectively,
should have been renamed to avoid clashes with the domain of
$\rho$.
Most of the rules have an obvious structure. We comment only on the
ones for transforming fixed point expressions and applications. The
former translates into a closure. The function part of the closure is
obtained by transforming the body of the abstraction, but under a new
mapping $\rho'$ that maps $(x_1,\ldots,x_n)$ to appropriate
projections from the environment parameter; the expression
$(x_1,\ldots,x_n) \supseteq \fvars{\fix f x M}$ means that all the
free variables of $(\fix f x M)$ appear in the tuple. The environment
part of the closure correspondingly contain mappings for the variables
in the tuple that are determined by the enclosing context.
Note also that the parameter for the function part of the closure is
expected to be a triple, the first item of which corresponds to the
function being defined recursively in the source language expression.
The transformation of a source language application makes clear how
this structure is used to realize recursion: the
constructed closure application has the effect of feeding the closure
to its function part as the first component of its argument.

\section{Implementing the Transformation in \LProlog}
\label{sec:cc_impl}

Our presentation of the implementation of closure conversion has two
parts: we first show how to encode the source and target languages and
we then present a \LProlog specification of the transformation. In the
first part, we discuss also the formalization of typing relations;
these will be used in the correctness proofs that we develop later.

\subsection{Encoding the language}
\label{subsec:cc_lang_encoding}

The source language of closure conversion is the target language of
the CPS transformation. The encoding of its syntax and typing
relations are already given in
Section\cspc\ref{subsec:cps_lang_encoding}. We are left with the
encoding of the target language.

\pagelabel{ecd:cc_targ_lang}
We first consider the encoding of types in the target language. We use
the \LProlog type \kty, the same type we used for encoding the source
language, for representing the types of the target language. The
constants for encoding the natural number, unit and pair types are
also the same as those for encoding the source language. We represent
the new arrow type constructor $\carr$ by \kcarr.

We use the \LProlog type \kctm for encodings of target language
terms. To represent the constructs the target language shares with the
source language, we will use ``primed'' versions of the \LProlog
constants seen in Section\cspc\ref{subsec:cps_lang_encoding}; \eg,
\kcunit of type \kctm will represent the unit constructor. Of course,
there will be no constructor corresponding to \kfix. We will also
need the following additional constructors:
\begin{tabbing}
\qquad\=\kill
\>$\kcabs : (\kctm \to \kctm) \to \kctm$\\
\>$\kclos : \kctm \to \kctm \to \kctm$\\
\>$\kopen : \kctm \to (\kctm \to \kctm \to \kctm) \to \kctm$
\end{tabbing}
Here, \kcabs encodes $\lambda$-abstraction and \kclos and \kopen
encode closures and their applications. Note again the $\lambda$-tree
syntax representation for binding constructs.

\pagelabel{ecd:cc_targ_typing}
We use the predicates $\kcof : \kctm \to \kty \to \prop$ to encode
typing in the target language.  The clauses defining this predicate
are routine and we show only the encoding of typing rules for closures
and closure applications. The following clauses encode these rules.
\begin{tabbing}
\qquad\=\quad\=$\sfall f,e,l.$\=\kill
\>$\kcof\app (\kclos\app F\app E)\app (\karr\app T_1\app T_2) \quad \limply$\\
\>\>$\kcof\app F\app
       (\kcarr\app (\kprod\app (\karr\app T_1\app T_2)\app (\kprod\app T_1\app \TL))\app T_2)
       \scomma \kcof\app E\app \TL.$\\
\>$\kcof (\kopen\app M\app R)\app T \quad\limply$\\
\>\>$\kcof\app M\app (\karr\app T_1\app T_2) \scomma$\\
\>\>$\sfall f,e,l.
       \kcof\app f\app (\kcarr\app (\kprod\app (\karr\app T_1\app T_2)\app (\kprod\app T_1\app l))\app T_2) \simply$\\
\>\>\>$\kcof\app e\app l \simply \kcof\app (R\app f\app e)\app T.$
\end{tabbing}
Here again we use universal quantifiers in goals to encode the
freshness constraint.
Note especially how the universal quantifier over the variable $l$
captures the opaqueness quality of the type of the environment of the
closure involved in the construct.

We should note that the clause that encodes the typing  of closures is
not an entirely accurate rendition of the typing rule presented in
Section\cspc\ref{sec:cc_langs}. In particular, it does not capture the
fact that the function part must be typed in an empty typing
environment. However, the fact that it {\it is} typed in an empty
typing environment can be encoded in a suitable theorem about the
implementation of the transformation and this theorem can be proved
using Abella.

%%  Strictly speaking, the first clause does not
%% accurately represent the rule for typing closures because that the
%% closedness of the function parts of closures is not enforced by it.
%% %
%% This is not a problem for our implementation and verification tasks
%% because this constraint will be enforced on an as-needed basis by
%% using $\lambda$-tree syntax (See the encoding of code hoisting in
%% Chapter\cspc\ref{ch:codehoist} for an example).

\subsection{Specifying the closure conversion transformation}
\label{subsec:specify_cc}

We first introduce an auxiliary predicate $\kcombine : \klist\app A
\to \klist\app A \to \klist\app A \to \omic$ that holds between three
lists when the last is composed of the elements of the first
two. It is defined as follows where $\kmemb$ is the membership
relation defined in Section\cspc\ref{subsec:schm_poly}:
\begin{tabbing}
\qquad\=\kill
\>$\kcombine\app \knil\app L\app L.$\\
\>$\kcombine\app (X \cons L_1)\app L_2\app L
  \app\limply\app \kmemb\app X\app L_2 \scomma \kcombine\app L_1\app L_2\app L.$\\
\>$\kcombine\app (X \cons L_1)\app L_2\app (X \cons L) \app\limply\app
     \kcombine\app L_1\app L_2\app L.$
\end{tabbing}
Observe that the definition of \kcombine does not enforce a possible
additional requirement that the third argument does not contain
repeated elements; indeed, such a requirement, which involves a
negative constraint, is impossible to encode using the logical
apparatus of \HHw. However, if the first two arguments to a \kcombine
goal are specific lists of distinct elements and the third is an
unspecified one, then at least one of the values for the last argument
that satisfies the goal is a list in which each of the elements is
distinct. Indeed, this would be the first ``solution'' that will be
found for the goal by the logic programming style interpretation
embedded in \LProlog.

The crux in formalizing the definition of closure conversion is
capturing the content of the $\ccfix$ rule. A key part of this rule is
identifying the free variables in a given source language term. We
realize the requirement by defining a predicate constant $\kfvars :
\ktm \to \klist\app \ktm \to \klist\app \ktm \to \omic$ such that if
$L_1$ is a list that includes all the free variables of $M$ and
$(\kfvars\app M\app L_1\app L_2)$ holds, then $L_2$ is another list
that contains exactly the free variables of $M$. We show the
definition of this predicate as follows:
\begin{tabbing}
\qquad\=\quad\=\kill
\>$\kfvars\app X\app \Vs\app \knil \app\limply\app \knotfree\app X.$\\
\>$\kfvars\app Y\app \Vs\app (Y \cons \knil) \app\limply\app \kmember\app Y\app \Vs.$\\
\>$\kfvars\app (\knat\app N)\app \Vs\app \knil.$\\
\>$\kfvars\app (\kplus\app M_1\app M_2)\app \Vs\app \FVs \app\limply\app$\\
\>\>$\kfvars\app M_1\app \Vs\app \FVs_1 \scomma
     \kfvars\app M_2\app \Vs\app \FVs_2 \scomma
     \kcombine\app \FVs_1\app \FVs_2\app \FVs.$\\
\>$\kfvars\app (\kifz\app M\app M_1\app M_2)\app \Vs\app \FVs' \app\limply\app$\\
\>\>$\kfvars\app M\app \Vs\app \FVs \scomma
     \kfvars\app M_1\app \Vs\app \FVs_1 \scomma \kfvars\app M_2\app \Vs\app \FVs_2,$\\
\>\>$\kcombine\app \FVs\app \FVs_1\app \FVs_1' \scomma
      \kcombine\app \FVs_1'\app \FVs_2\app \FVs'.$\\
\>$\kfvars\app \kunit\app \_\app \knil.$\\
\>$\kfvars\app (\kpair\app M_1\app M_2)\app \Vs\app \FVs \app\limply\app$\\
\>\>$\kfvars\app M_1\app \Vs\app \FVs_1 \scomma
     \kfvars\app M_2\app \Vs\app \FVs_2 \scomma
     \kcombine\app \FVs_1\app \FVs_2\app \FVs.$\\
\>$\kfvars\app (\kfst\app M)\app \Vs\app \FVs \app\limply\app \kfvars\app M\app \Vs\app \FVs.$\\
\>$\kfvars\app (\ksnd\app M)\app \Vs\app \FVs \app\limply\app \kfvars\app M\app \Vs\app \FVs.$\\
\>$\kfvars\app (\klet\app M\app R)\app \Vs\app \FVs \app\limply\app \kfvars\app M\app \Vs\app \FVs_1 \scomma$\\
\>\>$(\forallx x {\knotfree\app x \simply \kfvars\app (R\app x)\app \Vs\app \FVs_2}) \scomma
      \kcombine\app \FVs_1\app \FVs_2\app \FVs.$\\
\>$\kfvars\app (\kfix\app R)\app \Vs\app \FVs \app\limply\app$\\
\>\>$\forallx {f,x} \knotfree\app f \simply \knotfree\app x \simply
       \kfvars\app (R\app f\app x)\app \Vs\app \FVs.$
\end{tabbing}
The essence of the definition of \kfvars is in the treatment of
binding constructs. Viewed operationally, the body of such a construct
is descended into after instantiating the binder with a new variable
marked $\knotfree : \ktm \to \omic$. Thus, the variables that are
marked in this way correspond to exactly those that are explicitly
bound in the term.  When the first argument of \kfvars is such a
variable it will not be collected as a free variable as indicated by
the first clause. Only those that are not marked \knotfree are
collected through the second clause and combination of the results of
recursive calls to \kfvars on subterms. It is important also to note
that the specification of \kfvars has a completely logical structure;
this fact can be exploited during verification.

The $\ccfix$ rule requires us to construct an environment representing
the mappings for the variables found by \kfvars. The predicate
$\kmapenv : \klist\app \ktm \to \klist\app (\kmap\app \ktm\app \kctm)
\to \kctm \to \omic$. specified by the following clauses provides this
functionality.
\begin{tabbing}
\qquad\=\quad\=\kill
\>$\kmapenv\app \knil\app \_\app \kunit.$\\
\>$\kmapenv\app (X \cons L)\app \Map\app (\kcpair\app M\app \ML) \app\limply\app$\\
\>\>$\kmember\app (\kmap\app X\app M)\app \Map \scomma \kmapenv\app L\app \Map\app \ML.$
\end{tabbing}
The $\ccfix$ rule also requires us to create a new mapping from the
variable list to projections from an environment variable.
Representing the list of projection mappings as a function from the
environment variable, this relation is given by the predicate
$\kmapvar : \klist\app \ktm \to (\kctm \to \klist\app (\kmap\app
\ktm\app \kctm)) \to \omic$ that is defined by the following clauses.
\begin{tabbing}
\qquad\=\kill
\>$\kmapvar\app \knil\app (e \mlam \knil).$\\
\>$\kmapvar\app (X \cons L)\app
      (e \mlam (\kmap\app X\app (\kcfst\app e)) \cons (\Map\app (\kcsnd\app e))) \app\limply\app
   \kmapvar\app L\app \Map.$
\end{tabbing}

\pagelabel{ecd:cc_rules}
We can now specify the closure conversion transformation. We provide
clauses below that define the predicate $\kcc : \klist\app (\kmap\app
\ktm\app \kctm) \to \klist\app \ktm \to \ktm \to \kctm \to
\omic$ such that $(\kcc\app \Map\app \Vs\app M\app M')$ holds if $M'$
is a transformed version of $M$ under the mapping $\Map$ for the
variables in $\Vs$; we assume that $\Vs$ contains all the free
variables of $M$.
\begin{tabbing}
\qquad\=\quad\=\quad\=\kill
\>$\kcc\app \Map\app \Vs\app (\knat\app N)\app (\kcnat\app N).$\\
\>$\kcc\app \Map\app \Vs\app X\app M \app\limply\app \kmember\app (\kmap\app X\app M)\app \Map.$\\
\>$\kcc\app \Map\app \Vs\app (\kpred\app M)\app (\kcpred\app M') \app\limply\app
     \kcc\app \Map\app \Vs\app M\app M'.$\\
\>$\kcc\app \Map\app \Vs\app (\kplus\app M_1\app M_2)\app (\kcplus\app M_1'\app M_2') \app\limply\app$\\
\>\>$\kcc\app \Map\app \Vs\app M_1\app M_1' \scomma \kcc\app \Map\app \Vs\app M_2\app M_2'.$\\
\>$\kcc\app \Map\app \Vs\app (\kifz\app M\app M_1\app M_2)\app (\kcifz\app M'\app M_1'\app M_2') \app\limply\app$\\
\>\>$\kcc\app \Map\app \Vs\app M\app M' \scomma
     \kcc\app \Map\app \Vs\app M_1\app M_1' \scomma
     \kcc\app \Map\app \Vs\app M_2\app M_2'.$\\
\>$\kcc\app \Map\app \Vs\app \kunit\app \kcunit.$\\
\>$\kcc\app \Map\app \Vs\app (\kpair\app M_1\app M_2)\app (\kcpair\app M_1'\app M_2') \app\limply\app$\\
\>\>$\kcc\app \Map\app \Vs\app M_1\app M_1' \scomma
     \kcc\app \Map\app \Vs\app M_2\app M_2'.$\\
\>$\kcc\app \Map\app \Vs\app (\kfst\app M)\app (\kcfst\app M') \app\limply\app
      \kcc\app \Map\app \Vs\app M\app M'.$\\
\>$\kcc\app \Map\app \Vs\app (\ksnd\app M)\app (\kcsnd\app M') \app\limply\app
      \kcc\app \Map\app \Vs\app M\app M'.$\\
\>$\kcc\app \Map\app \Vs\app (\klet\app M\app R)\app (\kclet\app M'\app R') \app\limply\app
     \kcc\app \Map\app \Vs\app M\app M' \scomma$\\
\>\>$\forallx {x,y}
       {\kcc\app ((\kmap\app x\app y) \cons \Map)\app (x \cons \Vs)\app (R\app x)\app (R'\app y)}.$\\
\>$\kcc\app \Map\app \Vs\app (\kfix\app R)$\\
\>\>$(\kclos\app (\kcabs\app (p \mlam \kclet\app (\kcfst\app p)\app (g \mlam$\\
\>\>\>$\kclet\app (\kcfst\app (\kcsnd\app p))\app (y \mlam $\\
\>\>\>$\kclet\app (\kcsnd\app (\kcsnd\app p))\app (e \mlam R'\app g\app y\app e)))))\app E) \app\limply\app$\\
\>\>$\kfvars\app (\kfix\app R)\app \Vs\app \FVs \scomma
     \kmapenv\app \FVs\app \Map\app E \scomma
     \kmapvar\app \FVs\app \NMap \scomma$\\
\>\>$\sfall {f,x,g,y,e}.$\\
\>\>\>$\kcc\app ((\kmap\app x\app y) \cons (\kmap\app f\app g) \cons (\NMap\app e))\app 
         (x \cons f \cons \FVs)\app (R\app f\app x)\app (R'\app g\app y\app e).$\\
\>$\kcc\app \Map\app \Vs\app (\kcapp\app M_1\app M_2)$\\
\>\>$(\klet\app M_1'\app (g \mlam \kopen\app g\app
        (f \mlam e \mlam \kcapp\app f\app (\kcpair\app g\app (\kcpair\app M_2'\app e))))) \app\limply\app$\\
\>\>\>$\kcc\app \Map\app \Vs\app M_1\app M_1' \scomma \kcc\app \Map\app \Vs\app M_2\app M_2'$.
\end{tabbing}
These clauses correspond very closely to the rules in
Figure\cspc\ref{fig:cc_rules}.
Note especially the clause for transforming an expression of the form
$(\kfix\app R)$ that encodes the content of the $\ccfix$ rule. In the
body of this clause, \kfvars is used to identify the free variables of
the expression, and \kmapenv and \kmapvar are used to create the
reified environment and the new mapping. In both this clause and in
the one for transforming a \klet expression, the $\lambda$-tree
representation, universal goals and (meta-language) applications are
used to encode freshness and renaming requirements related to bound
variables in a concise and logically precise way.

\section{Informal Verification of the Transformation}
\label{sec:informal_verify_cc}

We informally describe the verification of closure conversion in this
section based on the ideas presented in
Section\cspc\ref{sec:vfc_approach}. Similar to the informal
verification of the CPS transformation discussed in
Section\cspc\ref{sec:informal_verify_cps}, we first discuss the type
preservation and then the semantics preservation of closure conversion. 
The proofs we present in this section is based on the
proofs in \cite{minamide95tr}. In particular, we have adapted the
proof of semantics preservation in \cite{minamide95tr} which is based
on regular logical relations to be based on step-indexing logical
relations.

\subsection{Type preservation of the transformation}
\label{subsec:cc_informal_typ_pres}

In proving that closure conversion preserves typing, the most
difficult step is to ensure that this property holds at the points of
the transformation where we cross a function boundary. For this we
need the following strengthening lemma:
\begin{mylemma}\label{lem:cc_typ_str}
  If $\Gamma \stseq M:T$, $\{x_1,\ldots,x_n\} \supseteq \fvars M$ and
  $x_i:T_i \in \Gamma$ for $1 \leq i \leq n$, then $x_1:T_1
  ,\ldots,x_n:T_n \stseq M :T$.
\end{mylemma}
\noindent The type preservation property is then stated as follows:
\begin{mythm}\label{thm:cc_typ_pres_open}
Let $\rho$ be the mapping $(x_1 \mapsto M_1,\ldots,x_n \mapsto M_n)$ from
source variables to target terms, $\G$ be the typing context $(x_1 :
T_1,\ldots,x_n:T_n)$ in the source language, $\G'$ be a typing context in
the target language such that $\G' \stseq M_i : T_i$ for $1 \leq i
\leq n$. If $\G \stseq M : T$ and $\cc \rho M {M'}$, then $\G' \stseq
M' : T$.
\end{mythm}
\noindent This theorem states that if $\rho$ is a type preserving mapping and
$M$ is transformed into $M'$ under $\rho$, then the type of $M'$ is
preserved. It is proved by induction on the derivation of $\cc \rho M
{M'}$ and analyzing its last rule. When the last rule is $\ccnat$,
$\ccvar$ or $\ccunit$, the proof is obvious. The rest of the cases, except
when the last rule is $\ccfix$, are proved by following a set pattern:
We analyze the derivation of $\G \stseq M : T$, apply the inductive
hypothesis on the subterms of $M$ and conclude from the results. When
the last rule is $\ccfix$, we apply Lemma\cspc\ref{lem:cc_typ_str} to
get a typing judgment that can be used for applying the inductive
hypothesis. The rest of the proof follows the set pattern.

Given Theorem\cspc\ref{thm:cc_typ_pres_open}, it is easy to show the
following type preservation property for closed terms:
\begin{mycoro}\label{coro:cc_typ_pres_closed}
If $\emptyset \stseq M : T$ and $\cc \emptyset M {M'}$, then
$\emptyset \stseq M' : T$.
\end{mycoro}

\subsection{Semantics preservation of the transformation}
\label{subsec:cc_informal_sem_pres}

We give an informal description of semantics preservation for closure
conversion in this section. We first describe the operational
semantics of the source and target languages of the transformation,
then the logical relations for denoting equivalence between the source
and target programs and their properties, and finally the semantics
preservation theorem and its proof.

\subsubsection{Operational semantics of the source and target languages}

\pagelabel{txt:cc_eval}
The operational semantics of the source language is already given in
Section\cspc\ref{subsec:cps_informal_sem_pres}.
The operational semantics of the target language is based on a left to right,
call-by-value evaluation strategy and is presented in small-step
form. We overload the syntax of evaluation judgments described in
Section\cspc\ref{subsec:cps_informal_sem_pres} for representing the
evaluation judgments for the target language. That is, $M \step{1} M'$ if
and only if $M$ evaluates to $M'$ in one-step, $M \step{n} M'$ if $M$
evaluates to $M'$ in $n$ steps, $M \step{*} M'$ if there exists some
$n$ such that $M \step{n} M'$, and $M \eval V$ if $M \step{*} V$ and
$V$ is a value. The designations of such judgments are inferred from
the context if they are not given explicitly.
The one-step evaluation rules for the target language are mostly the same as that for the
source language. The only evaluation rules that may be non-obvious are
the ones for closures and closure applications. They are the following:
\begin{gather*}
  \infer[]{
    \clos{M_1}{M_2} \step{1} \clos{M_1'}{M_2}
  }{
    M_1 \step{1} M_1'
  }
  \qquad
  \infer[]{
    \clos{V_1}{M_2} \step{1} \clos{V_1}{M_2'}
  }{
    M_2 \step{1} M_2'
  }
  \\
  \infer[]{
    {\open {x_f} {x_e} {M_1} {M_2}} \step{1} {\open {x_f} {x_e} {M_1'} {M_2}}
  }{
    {M_1} \step{1} {M_1'}
  }
  \\[7pt]
  \infer[]{
    {\open {x_f} {x_e} {\clos {V_f} {V_e}} {M_2}} \step{1}
    {M_2[V_f/x_f, V_e/x_e]}
  }{}
\end{gather*}

\subsubsection{Logical relations and their properties}

Following the ideas in Section\cspc\ref{sec:vfc_approach}, we use
step-indexing logical relations to characterize the semantics
preservation property of closure conversion. Specifically, we define
the mutually recursive simulation relation $\sim$ between closed
source and target terms and equivalence relation $\approx$ between
closed source and target values, each indexed by a type and a step
measure, in Figure\cspc\ref{fig:cc_logical_relations}.
By definition, the simulation relation $\simindex T i M M'$ holds if
and only if $M$ simulates $M'$ within at least $i$ steps of evaluation
at the type $T$. The equivalence relation $\equalindex T i V {V'}$
holds if and only if the values $V$ and $V'$ cannot be distinguished
from each other (\ie, considered as equivalent) in any context within
at least $i$ steps of evaluation at the type $T$.
The rules in Figure\cspc\ref{fig:cc_logical_relations} except the last
one have obvious meaning. The last rule states that a function ${(\fix
  f x M)}$ is equivalent to a closure ${\clos {V'} {V_e}}$ at type
$T_1 \to T_2$ for $i$ steps if and only if given any step $j$ smaller
than $i$ and any argument $V_1,V_1',V_2,V_2'$ such that $V_1$ and
$V_1'$ are equivalent at $T_1$ and $V_2$ and $V_2$ are equivalent at
$T_1 \to T_2$ for $j$ steps, the application of ${(\fix f x M)}$ to
$V_1$ and $V_2$ simulates the application of ${\clos {V'} {V_e}}$ to
$V_1'$ and $V_2'$.
Note that the definition of $\approx$ in the last rule uses $\approx$
negatively at the same type. However, it is still a well-defined
notion because the index decreases.
\begin{figure}[!ht]
\center
\begin{tabbing}
\qquad\qquad\qquad\=\quad\=\quad\=\kill
\>$\simindex T k M M' \iff$\\
\>\>$\forall j \leq k. \forall V. M \step{j} V \rimp
      \exists V'. {M'} \eval {V'} \rand \equalindex T {k-j} V {V'};$\\
\>$\equalindex \tnat k n n;$\\
\>$\equalindex \tunit k \unit \unit;$\\
\>$\equalindex {(T_1 \tprod T_2)} k {\pair {V_1} {V_2}} {\pair {V_1'} {V_2'}} \iff
       \equalindex {T_1} k {V_1} {V_1'} \rand \equalindex {T_2} k {V_2} {V_2'};$\\
\>$\equalindex {T_1 \to T_2} k
           {(\fix f x M)}
           {\clos {V'} {V_e}} \iff$\\
\>\>$\forall j < k. \forall V_1, V_1', V_2, V_2'.
        \equalindex {T_1} j {V_1} {V_1'} \rimp
        \equalindex {T_1 \to T_2} j {V_2} {V_2'} \rimp$\\
\>\>\>$\simindex {T_2} j {M[V_2/f, V_1/x]} {V' \app (V_2', V_1', V_e)}.$
\end{tabbing}

  \caption{The Logical Relations for Verifying Closure Conversion}
  \label{fig:cc_logical_relations}
\end{figure}

A property we will need later is that $\approx$ is closed under
decreasing indexes. It is stated as the following lemma:
\begin{mylemma}\label{lem:cc_equiv_closed}
  If $\equalindex T i V {V'}$ holds, then for any $j$ such that $j
  \leq i$, $\equalindex T j V {V'}$ holds.
\end{mylemma}
\noindent We prove this lemma by a (nested) induction first on the types and
then on the step indexes of $\approx$. The proof itself is obvious.

Analyzing the simulation relation and using the evaluation rules,
we can show the following ``compatibility'' lemmas for various
constructs in the source language.
\begin{mylemma}\label{lem:cc_sim_compat}
\begin{enumerate}
  \item If $\simindex \tnat k {M} {M'}$ then $\simindex \tnat k {\pred M}
  {\pred M'}$. If also $\simindex \tnat k {N} {N'}$ then $\simindex
  \tnat k {M + N} {M' + N'}$.

  \item  If $\simindex {T_1 \tprod T_2} k {M} {M'}$ then $\simindex {T_1} k
        {\fst{M}} {\fst{M'}}$ and $\simindex {T_2} k {\snd{M}}
        {\snd{M'}}$.

  \item If $\simindex {T_1} {k} {M} {M'}$ and $\simindex {T_2} {k} {N}
        {N'}$ then $\simindex {T_1 \tprod T_2} {k} {(M,N)} {(M',N')}.$

  \item If $\simindex \tnat k M {M'}$, $\simindex T k {M_1} {M_1'}$ and
      $\simindex T k {M_2} {M_2'}$, then\\
      $\simindex T k {\ifz M {M_1} {M_2}} {\ifz {M'} {M_1'} {M_2'}}$.

  \item If $\simindex{T_1 \to T_2}{k}{M_1}{M_1'}$ and
  $\simindex{T_1}{k}{M_2}{M_2'}$ then\\
  $\simindex{T_2}{k} {M_1 \app M_2} {\letexp g {M_1'}
     {\open {x_f} {x_e} {g}{x_f \app (g,M_2',x_e)}}}.$
\end{enumerate}
\end{mylemma}
\noindent These lemmas are proved by analyzing the simulation
relations. Some proofs of these properties need the property that the
equivalence relation is closed under decreasing indices, which we have
already proved as Lemma\cspc\ref{lem:cps_equiv_closed}.
The proof of the last of these properties requires us to consider the
evaluation of the application of a fixed point expression which
involves ``feeding'' the expression to its own body, albeit at a
smaller step measure. We apply Lemma\cspc\ref{lem:cps_equiv_closed} to
obtain such expressions.

\subsubsection{Informal proof of semantics preservation}

Similar to semantics preservation of the CPS transformation described
in Section\cspc\ref{subsec:cps_informal_sem_pres}, we consider
semantics preservation for possibly open terms under closed
substitutions. We overload the notation $(V_1/x_1,\ldots,V_n/x_n)$ for
representing substitution of closed values for variables for the
target language.
In defining a correspondence between source and target
language substitutions, we need to consider the possibility that a
collection of free variables in the first may be reified into an
environment variable in the second. This motivates the following
definition in which $\gamma$ represents a source language
substitution:
\begin{tabbing}
\qquad\=\kill
\>$\equalindex
           {x_m:T_m, \ldots, x_1:T_1}
           k
           {\gamma}
           {(V_1,\ldots,V_m)}
    \iff
    \forall 1 \leq i \leq m. \equalindex {T_i} k {\gamma(x_i)} {V_i}.$
\end{tabbing}
Writing $\substconcat{\gamma_1}{\gamma_2}$ for the concatenation of
two substitutions viewed as lists, equivalence between substitutions
is then defined as follows:
\begin{tabbing}
\qquad\=\quad\=\kill
\>$\equalindex
           {\Gamma, x_n:T_n, \ldots,x_1:T_1}
           k
           {\substconcat{(V_1/x_1,\ldots,V_n/x_n) }{\gamma}}
           {(V_1'/y_1, \ldots, V_n'/y_n, V_e/x_e)}$\\
\>\>$\iff
    (\forall 1 \leq i \leq n. \equalindex {T_i} k {V_i} {V_i'}) \rand
    \equalindex \Gamma k \gamma {V_e}.$
\end{tabbing}
Note that both relations are indexed by a source language typing
context and a step measure. The second relation allows the
substitutions to be for different variables in the source and target
languages. A relevant mapping will determine a correspondence between
these variables when we use the relation.

The first part of the following lemma, proved by an easy use of the
definitions of $\approx$ and evaluation, provides the basis for
justifying the treatment of free variables via their transformation
into projections over environment variables introduced at function
boundaries in the closure conversion transformation. The second part
of the lemma is a corollary of the first part that relates a source
substitution and an environment computed during the closure conversion
of fixed points.
\begin{mylemma}\label{lem:cc_var_sem_pres}
  Let $\delta = \substconcat{(V_1/x_1,\ldots,V_n/x_n)}{\gamma}$ and
  $\delta' = (V_1'/y_1,\ldots,$ $V_n'/y_n,V_e/x_e)$
  be source and target language substitutions and let $\Gamma =
  (x_m':T_m',\ldots,x_1':T_1',x_n:T_n,\ldots,x_1:T_1)$ be a source language
  typing context such that $\equalindex \Gamma k \delta
  {\delta'}$. Further, let $\rho = (x_1  \mapsto y_1,\ldots,x_n \mapsto
  y_n, x_1' \mapsto \pi_1(x_e), \ldots, x_m' \mapsto \pi_m(x_e))$.
\begin{enumerate}
\item If $x : T \in \Gamma$ then there exists a value $V'$ such that
  $(\rho(x))[\delta'] \eval V'$ and $\equalindex T k {\delta(x)} {V'}$.

\item   If $\Gamma' = (z_1:T_{z_1},\ldots,z_j:T_{z_j})$ for $\Gamma'
  \subseteq \Gamma$ and $\ccenv \rho {(z_1,\ldots,z_j)} M$, then
  there exists $V_e'$ such that $M[\delta'] \eval V_e'$ and $\equalindex {\Gamma'} k
  \delta {V_e'}$.
\end{enumerate}
\end{mylemma}

The semantic preservation theorem in its most general form can now be
stated as follows, which is a realization of
Property\cspc\ref{thm:sem_pres_index_open} in the setting of closure
conversion:
\begin{mythm}\label{thm:cc_sem_pres_open}
  Let $\delta = \substconcat{(V_1/x_1,\ldots,V_n/x_n)}{\gamma}$ and
  $\delta' = (V_1'/y_1,\ldots,$ $V_n'/y_n,V_e/x_e)$ be source and
  target language substitutions and let $\Gamma =
  (x_m':T_m',\ldots,x_1':T_1',x_n:T_n,\ldots,x_1:T_1)$ be a source
  language typing context such that $\equalindex \Gamma k \delta
  {\delta'}$. Further, let $\rho = (x_1 \mapsto y_1,\ldots,x_n \mapsto
  y_n, x_1' \mapsto \pi_1(x_e), \ldots, x_m' \mapsto \pi_m(x_e))$.  If
  $\Gamma \stseq M:T$ and $\cc \rho M M'$, then $\simindex T k
  {M[\delta]} {M'[\delta']}$.
\end{mythm}
\noindent We outline the main steps in the argument for this theorem:
these will guide the development of a formal proof in
Section\cspc\ref{sec:formal_verify_cc}. We proceed by induction on
the derivation of $\cc \rho M M'$, analyzing the last step in it.
This obviously depends on the structure of $M$. The cases for a number
or the unit constructor are obvious and for a variable we use
Lemma\cspc\ref{lem:cc_var_sem_pres}.1.
In the remaining cases, other than when $M$ is of the form $(\letexp x
{M_1} {M_2})$ or $(\fix f x {M_1})$, the argument follows a set
pattern: we observe that substitutions distribute to the
sub-components of expressions, we invoke the induction hypothesis over
the sub-components and then we use Lemma\cspc\ref{lem:cc_sim_compat}
to conclude.
If $M$ is of the form $(\letexp x {M_1} {M_2})$, then $M'$ must be of
the form $(\letexp y {M_1'} {M_2'})$. Here again the substitutions
distribute to $M_1$ and $M_2$ and to $M_1'$ and $M_2'$,
respectively. We then apply the induction hypothesis first to $M_1$
and $M_1'$ and then to $M_2$ and $M_2'$; in the latter case, we need
to consider extended substitutions with equivalent values from the
former case.
Finally, if $M$ is of the form $(\fix f x {M_1})$, then $M'$ must have
the form $\clos {M_1'} {M_2'}$.
We can prove that the abstraction $M_1'$ is closed by using the type
preservation theorem (Theorem\cspc\ref{thm:cc_typ_pres_open}) and
therefore that $M'[\sigma'] = \clos {M_1'} {M_2'[\sigma']}$.
We then apply the induction hypothesis on $M_1$. In order to do so, we
generate the appropriate typing judgment using
Lemma\cspc\ref{lem:cc_typ_str} and a new pair of equivalent
substitutions (under a suitable step index) using
Lemma\cspc\ref{lem:cc_var_sem_pres}.2.
The case is easily concluded from the result of the previous
application of the inductive hypothesis.

An immediate corollary of Theorem\cspc\ref{thm:cc_sem_pres_open} is
the following which is a realization of
Property\cspc\ref{thm:sem_pres_index_closed} in the setting of closure
conversion:
\begin{mycoro}\label{coro:cc_sem_pres_closed}
  If $\emptyset \stseq M : T$ and $\cc \emptyset M {M'}$, then
  $\simindex T i {M} {M'}$ for any $i$.
\end{mycoro}
\noindent From this corollary, it is easy to derive the following correctness
property of the CPS transformation for closed programs at atomic types
that correspond to Property\cspc\ref{thm:sem_pres_atom}:
\begin{mycoro}\label{coro:cc_sem_pres_atom}
  If $\emptyset \stseq M : \tnat$, $\cc \emptyset M {M'}$ and $M \eval V$, then
  $M' \eval V$.
\end{mycoro}

\section{Verifying the \LProlog Program in \Abella}
\label{sec:formal_verify_cc}

In this section, we formalize the verification of closure conversion
described in Section\cspc\ref{sec:informal_verify_cc} in \Abella. For
this we need to make explicit the proofs of all the binding related
properties. We show that the $\lambda$-tree syntax approach can be
used to significantly alleviate this effort, leading to formal proofs
that closely follows the informal ones.

\subsection{Type preservation of the transformation}
\label{subsec:cc_formal_typ_pres}

We prove that the \LProlog implementation of closure conversion
preserves typing by following the informal argument given in
Section\cspc\ref{subsec:cc_informal_typ_pres}. First we characterize
the typing contexts for the source and target languages. The typing
context of the source language is defined through the following
clauses for $\kctx : \olist \to \prop$:
\begin{tabbing}
\qquad\=$\kctx\app (\kof\app X\app T \cons L)$ \quad\=$\rdef$ \quad\=\quad\=\kill
\>$\kctx\app \knil$ \>$\rdef$ \>$\rtrue$\\
\>$\kctx\app (\kof\app X\app T \cons L)$ \>$\rdef$
     \>$\kctx\app L \rand \kname\app X \rand \oseq{\kissty\app T} \rand$\\
\>\>\>\>$\rforallx {T'} {(\kmember\app (\kof\app X\app T')\app L \rimp T = T')}.$
\end{tabbing}
Typing contexts often encode a constraint that every element in the
context pertains to a distinct variable.
The definition here does not enforce such a requirement.
The reason for not doing so is that we may in fact have to consider
contexts in which there are multiple entries for the same variable;
the reason for this is that the \kcombine predicate that we used in
the specification of closure conversion does not force uniqueness in
the listing of the free variables.
What our definition of \kctx does ensure, though, is that every
assignment for a given variable assigns it the same type.

%% To accord with the way closure conversion is formalized, we allow
%% multiple assignments of types for a given variable in a typing context
%% for the source language, but we require all of them to be of the same
%% type.
%% %
The typing context of the target language is defined through the
following clauses for $\kdctx : \olist \to \prop$:
\begin{tabbing}
\qquad\=$\nablax x {\kdctx (\kcof\app x\app T \cons L)}$ \quad\=$\rdef$ \quad\=\kill
\>$\kdctx\app \knil$ \>$\rdef$ \>$\rtrue$\\
\>$\nablax x {\kdctx (\kcof\app x\app T \cons L)}$
  \>$\rdef$ \>$\kdctx\app L \rand \oseq{\kiscty\app T}.$
\end{tabbing}
Here the predicate constant $\kiscty : \kty \to \omic$ defines
well-formed types in the target language and is given by the following
\LProlog clauses:
\begin{tabbing}
\qquad\=$\kiscty\app (\kprod\app T_1\app T_2)$ \quad\=$\limply$ \quad\=\kill
\>$\kiscty\app \ktnat.$\\
\>$\kiscty\app \ktunit.$\\
\>$\kiscty\app (\kprod\app T_1\app T_2)$ \>$\limply$
  \>$\kiscty\app T_1 \scomma \kiscty\app T_2.$\\
\>$\kiscty\app (\karr\app T_1\app T_2)$ \>$\limply$
   \>$\kiscty\app T_1 \scomma \kiscty\app T_2.$\\
\>$\kiscty\app (\kcarr\app T_1\app T_2)$ \>$\limply$
  \>$\kiscty\app T_1 \scomma \kiscty\app T_2.$
\end{tabbing}

We need to prove the strengthening lemma of typing for the source
language (\ie, Lemma\cspc\ref{lem:cc_typ_str}). We define the
strengthening of typing context through the following clauses for the
predicate constant $\kprunectx : \klist\app \ktm \to \olist \to \olist
\to \prop$:
\begin{tabbing}
\qquad\=\quad\=\kill
\>$\kprunectx\app \knil\app L\app \knil$ \quad$\rdef$ \quad$\rtrue$\\
\>$\kprunectx\app (X \cons Vs)\app L\app (\kof\app X\app T \cons L')$ \quad$\rdef$\\
\>\>$\kmember\app (\kof\app X\app T)\app L \rand \kprunectx\app \Vs\app L\app L'.$
\end{tabbing}
By definition, $\kprunectx\app \Vs\app L\app L'$ if and only if $\Vs$
is a sequence of variables in the domain of the typing context $L$ and
$L'$ is the typing context obtained by strengthening $L$ to the domain
$\Vs$. Then Lemma\cspc\ref{lem:cc_typ_str} is formalized as follows:
\begin{tabbing}
\qquad\=\quad\=\kill
\>$\rfall L, \Vs, M, T, \FVs.
  \kctx\app L \rimp \kvarsofctx\app L\app \Vs \rimp \oseq{L \stseq \kof\app M\app T} \rimp$\\
\>\>$\oseq{\kfvars\app M\app \Vs\app \FVs} \rimp
      \rexistsx {L'} {\kprunectx\app \FVs\app L\app L' \rand \oseq{L' \stseq \kof\app M\app T}}.$
\end{tabbing}
We may think of proving it by induction on $\oseq{\kfvars\app M\app
  \Vs\app \FVs}$. However, this will not work because \kfvars may go
under binders in $M$ and introduce new assumptions into the dynamic
context for marking bound variables. Thus, we need to generalize the
above theorem to accommodate such possibility. We define the predicate
$\kbvars : \olist \to \prop$ through the following clauses for
characterizing the assumptions dynamically introduced by \kfvars:
\begin{tabbing}
\qquad\=$\nablax x {\kbvars\app (\knotfree\app x \cons L)}$ \quad\=$\rdef$ \quad\=\kill
\>$\kbvars\app \knil$ \>$\rdef$ \>$\rtrue$\\
\>$\nablax x {\kbvars\app (\knotfree\app x \cons L)}$ \>$\rdef$ \>$\kbvars\app L.$
\end{tabbing}
We designate the constant $\kctxbvars : \olist \to \olist \to \prop$
for relating the type assignments with the \knonfree assumptions:
\begin{tabbing}
\qquad\=$\kctxbvars\app (\kof\app X\app T \cons L)\app (\knotfree\app X \cons L')$
  \quad\=$\rdef$ \quad\=\kill
\>$\kctxbvars\app \knil\app \knil$ \>$\rdef$ \>$\rtrue$\\
\>$\kctxbvars\app (\kof\app X\app T \cons L)\app (\knotfree\app X \cons L')$
  \>$\rdef$ \>$\kctxbvars\app L\app L'.$
\end{tabbing}
We now generalize the theorem to the following:
\begin{tabbing}
\qquad\=\quad\=\kill
\>$\rfall L, L_1, L_2, \Ps, \Vs, \FVs, M, T.$\\
\>\>$\kctx\app L \rimp \kappend\app L_1\app L_2\app L \rimp \kctxbvars\app L_1\app \Ps \rimp
       \kvarsofctx\app L_2\app \Vs \rimp$\\
\>\>$\kbvars\app \Ps \rimp \oseq{\Ps \stseq \kfvars\app M\app \Vs\app \FVs} \rimp
     \oseq{L \stseq \kof\app M\app T} \rimp$\\
\>\>$\rexistsx {L', L_2'}
        {\kprunectx\app \FVs\app L_2\app L_2' \rand
         \kappend\app L_1\app L_2'\app L' \rand \kctx\app L' \rand
         \oseq{L' \stseq \kof\app M\app T}}.$
\end{tabbing}
Note here that the typing context $L$ consists of two parts: $L_1$
contains variables that are marked as non-free by \kfvars and $L_2$
contains the initially free variables. The theorem is proved by
induction on $\oseq{\Ps \stseq \kfvars\app M\app \Vs\app \FVs}$. The
proof itself is easy to construct by inspecting the logical structure
of \kfvars derivations. It is easy to see that the formalized version
of Lemma\cspc\ref{lem:cc_typ_str} is just a special case of the above
theorem when $L_1$ is empty.

\pagelabel{ecd:cc_typ_pres_open}
To formally state the type preservation theorem, we need to
characterize the type preserving mappings from source variables to
target terms. This is done through defining the predicate $\kgoodmap :
\olist \to \klist\app (\kmap\app \ktm\app \kctm) \to \olist \to \prop$,
as follows:
\begin{tabbing}
\qquad\=\quad\=\kill
\>$\kgoodmap\app \CL\app \knil\app \knil \quad\rdef\quad \rtrue$\\
\>$\kgoodmap\app \CL\app (\kmap\app X\app M \cons \ML)\app (\kof\app X\app T \cons \SL) \quad\rdef$\\
\>\>$\kname\app X \rand \oseq{CL \stseq \kcof\app M\app T} \rand \kgoodmap\app \CL\app \ML\app \SL.$
\end{tabbing}
The type preserving theorem is now formally stated as follows which
corresponds to Theorem\cspc\ref{thm:cc_typ_pres_open}:
\begin{tabbing}
\qquad\=\quad\=\kill
\>$\rfall \SL, \CL, \Map, M, \Vs, M', T.$\\
\>\>$\kctx\app \SL \rimp \kdctx \CL \rimp \kgoodmap\app \CL\app \Map\app \SL
    \rimp \oseq{\SL \stseq \kof\app M\app T} \rimp$\\
\>\>$\kvarsofctx\app \SL\app \Vs \rimp \oseq{\kcc\app \Map\app \Vs\app M\app M'}
      \rimp \oseq{\CL \stseq \kcof\app M'\app T}.$
\end{tabbing}
This theorem is proved by induction on $\oseq{\kcc\app \Map\app
  \Vs\app M\app M'}$. The argument closely follows the informal proof
for Theorem\cspc\ref{thm:cc_typ_pres_open}. Given this theorem, it is
easy to prove the following formalized version of
Corollary\cspc\ref{coro:cc_typ_pres_closed}:
\begin{tabbing}
\qquad\=\quad\=\kill
\>$\rfall M, M', T.
      \oseq{\kof\app M\app T} \rimp \oseq{\kcc\app \knil\app \knil\app M\app M'}
      \rimp \oseq{\kcof\app M'\app T}.$
\end{tabbing}

\subsection{Semantics preservation of the transformation}
\label{subsec:cc_formal_sem_pres}

In this section, we formalize the semantics preservation proof of
closure conversion described in
Section\cspc\ref{subsec:cc_informal_sem_pres} in \Abella. Again, we
focus on showing the benefits of the $\lambda$-tree syntax approach in
this formalization. The formal treatments of binding related
properties such as closedness and substitution described in
Section\cspc\ref{subsec:cps_formal_sem_pres} carries over to this
setting. We avoid a detailed discussion of them and instead focus on
showing how reasoning about the unique features of closure conversion
such as computation of free variables can benefit from the
$\lambda$-tree syntax approach.

\subsubsection{Formalizing the operational semantics}

\pagelabel{ecd:cc_eval}
The operational semantics of the source language has already been
formalized in Section\cspc\ref{subsec:cps_formal_sem_pres}. We are
left with the operational semantics of the target language.
We encode
the one step evaluation rules for the target language using the
predicate constant $\kcstep : \kctm \to \kctm \to \omic$. We again consider
only a few interesting cases in their definition. Assuming that
$\kcval : \kctm \to \omic$ recognize values in the target language
which has an obvious definition in \LProlog, the clauses for
evaluating the application of closures and closure applications are
the following.
\begin{tabbing}
\qquad\=$\kcstep\app (\kopen\app (\kclos\app F\app E)\app R)\app (R\app F\app E)$
  \quad\=$\limply$ \quad\=\kill
\>$\kcstep\app (\kclos\app F\app E)\app (\kclos\app F'\app E)$
  \>$\limply$ \>$\kcstep\app F\app F'.$\\
\>$\kcstep\app (\kclos\app F\app E)\app (\kclos\app F\app E')$
  \>$\limply$ \>$\kcval\app F \scomma \kcstep\app E\app E'.$\\
\>$\kcstep\app (\kopen\app M\app R)\app (\kopen\app M'\app R)$
  \>$\limply$ \>$\kcstep\app M\app M'.$\\
\>$\kcstep\app (\kopen\app (\kclos\app F\app E)\app R)\app (R\app F\app E)$
  \>$\limply$ \>$\kcval\app (\kclos\app F\app E).$
\end{tabbing}
Note here how application in the meta-language realizes substitution.

We use the predicates $\kcnstep : \knat \to \kctm \to \kctm \to \omic$
and $\keval : \kctm \to \kctm \to \omic$ to represent the $n$-step and
full evaluation relations for the target language, respectively.
These predicates have definitions similar to their counterparts in
Section\cspc\ref{subsec:cps_formal_sem_pres}.

\subsubsection{Formalizing the closedness property}

The closedness property for terms in the source language has already
been defined through $\ktm : \ktm \to \omic$ in
Section\cspc\ref{subsec:cps_formal_sem_pres}. We further identify the
predicate $\kctm : \kctm \to \omic$ such that $\oseq{\kctm\app M}$
holds if $M$ is a well-formed term in the target language. It is
defined through a \LProlog program similar to that defining \ktm. The
only interesting clauses in the definition are the following ones for
closures and closure applications:
\begin{tabbing}
\qquad\=$\kctm\app (\kclos\app M_1\app M_2)$ \quad\=$\limply$ \quad\=\kill
\>$\kctm\app (\kclos\app M_1\app M_2)$ \>$\limply$ \>$\kctm\app M_1 \scomma \kctm\app M_2.$\\
\>$\kctm\app (\kopen\app M\app R)$ \>$\limply$ \>$\kctm\app M \scomma
     \forallx {f,e} {\kctm\app f \simply \kctm\app e \simply \kctm\app (R\app f\app e)}.$
\end{tabbing}
Similar to \ktm, we can easily prove the properties about well-formed
or closed terms in the target language through the definition of
\kctm, including that a well-typed term is also well-formed and that a
closed term cannot contain any nominal constant.

\subsubsection{Formalizing the logical relations}

\pagelabel{ecd:cc_logical_relations}
The following clauses define the simulation relation represented by
$\ksimcc : \kty \to \knat \to \ktm \to \kctm \to \prop$ and the
equivalence relations represented by $\kequivcc : \kty \to \knat \to
\ktm \to \kctm \to \prop$.
\begin{tabbing}
\qquad\=\quad\=\quad\=\kill
\>$\ksimcc\app T\app K\app M\app M'$ \quad$\rdef$\quad
   $\rfall J, V. \kle\app J\app K \rimp \oseq{\knstep\app J\app M\app V} \rimp \oseq{\kval\app V} \rimp$\\
\>\>$\rexistsx {V',N} {\oseq{\kceval\app M'\app V'} \rand
                        \oseq{\kadd\app J\app N\app K} \rand \kequivcc\app T\app N\app V\app V'}$\\
\>$\kequivcc\app \ktnat\app K\app (\knat\app N)\app (\kcnat N) \quad\rdef\quad \rtrue$\\
\>$\kequivcc\app \ktunit\app K\app \kunit\app \kcunit \quad\rdef\quad \rtrue$\\
\>$\kequivcc\app (\kprod\app T_1\app T_2)\app K\app (\kpair\app V_1\app V_2)\app (\kcpair\app V_1'\app V_2') \quad\rdef\quad$\\
\>\>$\kequivcc\app T_1\app K\app V_1\app V_1' \rand \kequivcc\app T_2\app K\app V_2\app V_2' \rand$\\
\>\>$\oseq{\ktm\app V_1} \rand \oseq{\ktm\app V_2} \rand \oseq{\kctm\app V_1'} \rand \oseq{\kctm\app V_2'}$\\
\>$\kequivcc\app (\karr\app T_1\app T_2)\app \kz\app (\kfix\app R)\app (\kclos\app (\kcabs\app R')\app \VE) \quad\rdef$\\
\>\>$\oseq{\kcval\app \VE} \rand \oseq{\ktm\app (\kfix\app R)} \rand
     \oseq{\kctm\app (\kclos\app (\kcabs\app R')\app \VE)}$\\
\>$\kequivcc\app (\karr\app T_1\app T_2)\app (\ks\app K)\app (\kfix\app R)\app (\kclos\app (\kcabs\app R')\app \VE) \quad\rdef\quad$\\
\>\>$\kequivcc\app (\karr\app T_1\app T_2)\app K\app (\kfix\app R)\app (\kclos\app (\kcabs\app R')\app \VE) \rand$\\
\>\>$\rfall V_1, V_1', V_2, V_2'. \kequivcc\app T_1\app K\app V_1\app V_1' \rimp \kequivcc\app (\karr\app T_1\app T_2)\app K\app V_2\app V_2' \rimp$\\
\>\>\>$\ksimcc\app T_2\app K\app (R\app V_2\app V_1)\app (R'\app (\kcpair\app V_2'\app (\kcpair\app V_1'\app \VE))).$
\end{tabbing}
The formula $\ksimcc\app T\app K\app M\app M'$ is intended to mean that
$M$ simulates $M'$ at type $T$ in at least $K$ steps; $\kequivcc\app
T\app K\app V\app V'$ has a similar interpretation.
Note the exploitation of $\lambda$-tree syntax, specifically the use
of application, to realize substitution in the definition of
\kequivcc.
It is easily shown that \ksimcc holds only between closed source and
target terms and similarly \kequivcc holds only between closed source
and target values. Similar to the formalization of equivalence in the
CPS transformation,
the definition of \kequivcc uses itself negatively in the last clause
and thereby cannot be treated as a fixed-point definition. However,
since the use is at a decreased step-index, we can treat it as a
recursive definition inductively defined on the types and the step
measures. In fact, this is the reason why we ``build'' the relation up
over the natural numbers rather than mirroring directly the structure
of the informal definition.

The property that the equivalence relation is closed under decreasing
step-indexes is formalized as follows:
\begin{tabbing}
\qquad\=\quad\=\kill
\>$\rfall T, K, J, V, V'.
  \oseq{\kissty\app T} \rimp \oseq{\kisnat\app K} \rimp$\\
\>\>$\kequivcc\app T\app K\app V\app V' \rimp \kle\app J\app K \rimp
       \kequivcc\app T\app J\app V\app V'.$
\end{tabbing}
It is proved by induction on $\oseq{\kissty\app T}$ and
$\oseq{\kisnat\app K}$.

Compatibility lemmas in the style of Lemma\cspc\ref{lem:cc_sim_compat}
are easily stated for \ksimcc. For example, the one for pairs is the
following.
\begin{tabbing}
\qquad\=\quad\=\kill
\>$\rfall T_1, T_2, K, M_1, M_2, M_1', M_2'.
    \oseq{\kisnat\app K} \rimp \oseq{\kissty\app T_1} \rimp \oseq{\kissty\app T_2} \rimp$\\
\>\>$\ksimcc\app T_1\app K\app M_1\app M_1' \rimp \ksimcc\app T_2\app K\app M_2\app M_2' \rimp$\\
\>\>$\ksimcc\app (\kprod\app T_1\app T_2)\app K\app (\kpair\app M_1\app M_2)\app (\kcpair\app M_1'\app M_2').$
\end{tabbing}
The proofs of these lemmas follows from the informal ones in a
straightforward manner.

\subsubsection{Representing substitutions}
A substitution is represented as a list of mappings from source
variables to closed target values. The substitutions for the source
language have already been given through \ksubst in
Section\cspc\ref{subsec:cps_formal_sem_pres}. The substitutions of
target values are represented by $\kcsubst : \klist\app (\kmap\app
\kctm\app \kctm) \to \prop$ which is defined as follows:
\begin{tabbing}
\qquad\=$\nablax x {\kcsubst\app (\kmap\app x\app V \cons \ML)}$ \quad\=$\rdef$ \quad\=\kill
\>$\kcsubst\app \knil$ \>$\rdef$ \>$\rtrue$\\
\>$\nablax x {\kcsubst\app (\kmap\app x\app V \cons \ML)}$
   \>$\rdef$ \>$\kcsubst\app \ML \rand \oseq{\kcval\app V} \rand \oseq{\kctm\app V}.$
\end{tabbing}
The application of substitutions has already been given by the
polymorphic definition of \kappsubst. As before, we can easily prove
properties about substitution application based on this definition
such as that such an application distributes over term structures and
that closed terms are not affected by substitution.

\subsubsection{The equivalence relation on substitutions}
We first define the relation \ksubstenvequivcc between source
substitutions and target environments:
\begin{tabbing}
\qquad\=\quad\=\quad\=\kill
\>$\ksubstenvequivcc\app \knil\app K\app \ML\app \kcunit$ \quad$\rdef$\quad $\rtrue$\\
\>$\ksubstenvequivcc\app (\kof\app X\app T \cons L)\app K\app \ML\app (\kcpair\app V'\app \VE)
   \quad\rdef\quad$\\
\>\>$\rexst V. \ksubstenvequivcc\app L\app K\app \ML\app \VE \rand$\\
\>\>\>$\kmember\app (\kmap\app X\app V)\app \ML \rand
                  \kequivcc\app T\app K\app V\app V'.$
\end{tabbing}
Using \ksubstenvequivcc, the needed relation between source and target
substitutions is defined as follows.
\begin{tabbing}
\qquad\=\quad\=\kill
\>$\nablax e {\ksubstequivcc\app L\app K\app \ML\app (\kmap\app e\app \VE \cons \knil)} \quad\rdef$\\
\>\>$\ksubstenvequivcc\app L\app K\app \ML\app \VE$\\
\>$\nablax {x,y} {\ksubstequivcc\app (\kof\app x\app T \cons L)\app K\app
                                   (\kmap\app x\app V \cons \ML)\app
                                   (\kmap\app y\app V' \cons \ML')} \quad\rdef$\\
\>\>$\kequivcc\app T\app K\app V\app V' \rand \ksubstequivcc\app L\app K\app \ML\app \ML'.$
\end{tabbing}

\subsubsection{Lemmas about \kmapvar and \kmapenv.}

A formalization of Lemma \ref{lem:cc_var_sem_pres} is needed for the main
theorem. We start with a lemma about \kmapvar.
\begin{tabbing}
\qquad\=\quad\=\quad\=\kill
\>$\rfall L, \Vs, \Map, \ML, K, \VE, X, T, M', V. \nabla e.$\\
\>\>$\oseq{\kisnat\app K} \rimp \kctx\app L \rimp \kvarsofctx\app L\app \Vs \rimp$\\
\>\>$\ksubst\app \ML \rimp \ksubstenvequivcc\app L\app K\app \ML\app \VE \rimp$\\
\>\>$\kmember\app (\kof\app X\app T)\app L \rimp
     \oseq{\kmapvar\app \Vs\app \Map} \rimp$\\
\>\>$\kappsubst\app \ML\app X\app V \rimp
       \oseq{\kmember\app (\kmap\app X\app (M'\app e))\app (Map\app e)} \rimp$\\
\>\>\>$\rexistsx {V'} {\oseq{\kceval (M'\app \VE)\app V'} \rand \kequivcc\app T\app K\app V\app V'}.$
\end{tabbing}
In words, this lemma states the following. If $L$ is a source typing
context for the variables $(x_1,\ldots,x_n)$, $\ML$ is a source
substitution and $\VE$ is an environment equivalent to $\ML$ at $L$,
then \kmapvar determines a mapping for $(x_1,\ldots,x_n)$ that are
projections over an environment with the following character: if the
environment is taken to be $\VE$, then, for $1 \leq i \leq n$, $x_i$
is mapped to a projection that must evaluate to a value equivalent to
the substitution for $x_i$ in $\ML$. The lemma is proved by induction
on the derivation of $\oseq{\kmapvar\app \Vs\app \Map}$.

Lemma\cspc\ref{lem:cc_var_sem_pres} is now formalized as follows.
\begin{tabbing}
\qquad\=\quad\=\quad\=\kill
\>$\rfall L, \ML, \ML', K, \Vs, \Vs', \Map.
   \oseq{\kisnat\app K} \rimp$\\
\>\>$\kctx\app L \rimp \kvarsofctx\app L\app \Vs \rimp$\\
\>\>$\ksubst\app \ML \rimp \kcsubst\app \ML' \rimp \ksubstequivcc\app L\app K\app \ML\app \ML' \rimp$\\
\>\>$\kcvarsofsubst\app \ML'\app \Vs' \rimp \ktomapping\app \Vs\app \Vs'\app \Map \rimp$\\
\>\>$(\rfall X, T, V, M', M''.$\\
\>\>\>$\kmember\app (\kof\app X\app T)\app L \rimp
       \oseq{\kmember\app (\kmap\app X\app M')\app \Map} \rimp$\\
\>\>\>$\kappsubst\app \ML\app X\app V \rimp \kappsubst\app \ML'\app M'\app M'' \rimp$\\
\>\>\>$\rexistsx {V'} {\oseq{\kceval\app M''\app V'} \rand \kequivcc\app T\app K\app V\app V'})
         \quad \rand$\\
\>\>$(\rfall L', \NFVs, E, E'.$\\
\>\>\>$\kprunectx\app \NFVs\app L\app L' \rimp \oseq{\kmapenv\app \NFVs\app \Map\app E} \rimp$\\
\>\>\>$\kappsubst\app \ML'\app E\app E' \rimp$\\
\>\>\>$\rexistsx {\VE'} {\oseq{\kceval\app E'\app \VE'} \rand
         \ksubstenvequivcc\app L'\app K\app \ML\app \VE'}).$
\end{tabbing}
Two new predicates are used here. The judgment $(\kcvarsofsubst\app
\ML'\app \Vs')$ ``collects'' the variables in the target substitution
$\ML'$ into $\Vs'$. The predicate $\ktomapping$ is defined as follows:
\begin{tabbing}
\qquad\=\quad\=\kill
\>$\ktomapping\app \Vs\app (E \cons \knil)\app (\Map\app E) \quad\rdef
     \quad\oseq{\kmapvar\app \Vs\app \Map}$\\
\>$\ktomapping\app (X \cons \Vs)\app (X' \cons \Vs')\app (\kmap\app X\app X' \cons \Map)$
   \quad$\rdef$\\
\>\>$\ktomapping\app \Vs\app \Vs'\app \Map.$
\end{tabbing}
By this definition, given source variables $\Vs =
(x_1,\ldots,x_n,x_1',\ldots,x_m')$ and target variables $\Vs' =
(y_1,\ldots,y_n,x_e)$, the predicate \ktomapping creates in $\Map$ the
mapping
\begin{tabbing}
\qquad\=\kill
\>$(x_1  \mapsto y_1,\ldots,x_n \mapsto y_n,
x_1' \mapsto \pi_1(x_e), \ldots, x_m' \mapsto \pi_m(x_e))$.
\end{tabbing}
The conclusion of the lemma is a conjunction representing the two
parts of Lemma\cspc\ref{lem:cc_var_sem_pres}. The first part is proved
by induction on $\oseq{\kmember\app (\kmap\app X\app M')\app \Map}$,
using the lemma for \kmapvar when $X$ is some $x_i' (1 \leq i \leq
m)$. The second part is proved by induction on $\oseq{\kmapenv\app
  \NFVs\app \Map\app E}$ using the first part.

\subsubsection{Formalizing the semantics preservation theorems}
\pagelabel{ecd:cc_sem_pres_open}
Theorem\cspc\ref{thm:cc_sem_pres_open} is now formalized as follows:
\begin{tabbing}
\qquad\=\quad\=\quad\=\kill
\>$\rfall L, \ML, \ML', K, \Vs, \Vs', \Map, T, P, P', M, M'.$\\
\>\>$\oseq{\kisnat\app K} \rimp \kctx\app L \rimp \kvarsofctx\app L\app \Vs \rimp$\\
\>\>$\ksubst\app \ML \rimp \kcsubst\app \ML' \rimp \ksubstequivcc\app L\app K\app \ML\app \ML' \rimp$\\
\>\>$\kcvarsofsubst\app \ML'\app \Vs' \rimp \ktomapping\app \Vs\app \Vs'\app \Map \rimp$\\
\>\>$\oseq{L \stseq \kof\app M\app T} \rimp
     \oseq{\kcc\app \Map\app \Vs\app M\app M'} \rimp$\\
\>\>$\kappsubst\app \ML\app M\app P \rimp \kappsubst\app \ML'\app M'\app P' \rimp$\\
\>\>\>$\ksimcc\app T\app K\app P\app P'.$
\end{tabbing}
We use an induction on $\oseq{\kcc\app \Map\app \Vs\app M \app M'}$,
the closure conversion derivation, to prove this theorem. As should be
evident from the preceding development, the proof in fact closely
follows the structure we outlined in
Section\cspc\ref{subsec:cc_informal_sem_pres}. A particular point to
note is that when $M$ is a function and $M'$ is a closure
$\clos{F}{E}$, we need to show that $F$ is closed and substitution has
no effect on it. This is achieved by applying the type preservation
lemma to $M$ to show that $F$ is typable in an empty context and hence
$\oseq{\kctm\app F}$ holds. By $\oseq{\kctm\app F}$ and the definition
of \kappsubst, we can easily show that substitution has no effect on
$F$.

Finally, from the above theorem it is easy to prove the following
properties which correspond to
Corollaries\cspc\ref{coro:cc_sem_pres_closed} and
\ref{coro:cc_sem_pres_atom}:
\begin{tabbing}
\qquad\=\quad\=\kill
\>$\rfall K, T, M, M'. \oseq{\kof\app M\app T} \rimp \oseq{\kcc\app \knil\app \knil\app M\app M'}
    \rimp \ksimcc\app T\app K\app M\app M'.$\\
\>$\rfall K, T, M, M'. \oseq{\kof\app M\app \ktnat} \rimp
    \oseq{\kcc\app \knil\app \knil\app M\app M'} \rimp$\\
\>\>$\oseq{\keval\app M\app (\knat\app N)} \rimp
     \oseq{\kceval\app M'\app (\kcnat\app N)}$
\end{tabbing}

%%%%%%%%%%%%%%%%%%%%%%%%%%%%%%%%%%%%%%%%%%%%%%%%%%%%%%%%%%%%%%%%%%%%%%%%%%%%%%%%

% The code hoisting transformation
%%%%%%%%%%%%%%%%%%%%%%%%%%%%%%%%%%%%%%%%%%%%%%%%%%%%%%%%%%%%%%%%%%%%%%%%%%%%%%%
% codehoist.tex: The code hoisting transformation
%%%%%%%%%%%%%%%%%%%%%%%%%%%%%%%%%%%%%%%%%%%%%%%%%%%%%%%%%%%%%%%%%%%%%%%%%%%%%%%%
\chapter{The Code Hoisting Transformation}
\label{ch:codehoist}

Code hoisting is a general transformation that has the effect of
moving code that appears in a nested context but that is not dependent
on that context to a place outside of it. 
In the compilation process we are considering, this transformation is
useful in moving nested functions that have been converted into a
closed form by closure conversion to the outermost level in the
program. 
Note, however, that code hoisting has a justification independently of
closure conversion and can be used even in situations where the latter
transformation has not been applied. 
Despite this fact, these two transformations have been considered only in
combination in many efforts at verified compilation of functional
programs; see, for example,\cspc\cite{belanger13cpp}
and\cspc\cite{chlipala08icfp}. 
The reason for this phenomenon is the following: the
correctness of code hoisting relies on the expression being hoisted
out being closed, an aspect that is easily verified for the
function expressions that result from closure conversion. 

The situation described above is somewhat unfortunate: it is
relatively easy to write code to determine that an expression 
is independent of the context in which it appears, but we are compelled to 
combine its movement with another process because of the difficulty in
verifying this fact. 
The $\lambda$-tree syntax approach that we are elucidating in this
thesis provides a way to overcome this problem and thereby to describe
code hoisting as a transformation in its own right.
The idea we use is impressive in its simplicity.
The independence of a subpart of a particular abstraction can be
characterized as a logical property of the binding structure of the
abstraction. 
Then, since independence is given a logical characterization, this
information can also be used in the process of proving the correctness
of the transformation.  

We elaborate on these ideas in this chapter. We start with an an
overview of the code hoisting transformation in
Section\cspc\ref{sec:ch_overview}. We then describe it in a rule-based
and relational style in Section\cspc\ref{sec:rule_based_ch} and
discuss its implementation in \LProlog in
Section\cspc\ref{sec:ch_impl}. We finally discuss the informal and
formal verification of the transformation in
Sections\cspc\ref{sec:informal_verify_ch} and
\ref{sec:formal_verify_ch}.

\section{An Overview of the Transformation}
\label{sec:ch_overview}

The code hoisting transformation that we will consider is based on
\cite{belanger13cpp}. 
In that context, the transformation is focused
on moving nested functions out to the top-level in the program.
Its particular content is most easily understood through examples. The 
following code is the output of closure conversion in the first
example in Section\cspc\ref{sec:cc_overview}:
\begin{tabbing}
\qquad\=\quad\=\kill
\>$\kwd{let}\app x = 2\app \kwd{in}\app \kwd{let}\app y = 3\app \kwd{in}$\\
\>\>$\clos{(\kwd{fun}\app z\app e \rightarrow z + e.1 + e.2)}{(x,y)}$
\end{tabbing}
As we can see, the function part of the environment is closed. Code
hoisting extracts this function to the top-level, resulting in the
following code:
\begin{tabbing}
\qquad\=\quad\=\kill
\>$\kwd{let}\app f = (\kwd{fun}\app z\app e \rightarrow z + e.1 + e.2)\app \kwd{in}$\\
\>\>$\kwd{let}\app x = 2\app \kwd{in}\app \kwd{let}\app y = 3\app \kwd{in}
      \clos{f}{(x,y)}$
\end{tabbing}

In the general case, the transformation of interest has to treat
functions that have an arbitrarily nested structure. 
Towards this end, it can be implemented as a recursive procedure:
given a function $(\abs x 
M)$, the procedure is applied to the subterms of $M$ and the extracted
functions are then moved out of $(\abs x M)$. Of course, for this
movement to be possible, it must be the case that the variable $x$
does not appear in the functions that are candidates for
extraction. This is the case for all the nested functions that are
produced by closure conversion.
For example, the following code is the output of closure conversion in
the second example in Section\cspc\ref{sec:cc_overview}:
\begin{tabbing}
\qquad\=\quad\=\kill
\>$\kwd{let}\app x = 3\app \kwd{in}$\\
\>\>$\clos
       {(\kwd{fun}\app y\app e_1 \rightarrow
           \clos{(\kwd{fun}\app z\app e_2 \rightarrow e_2.1 + e_2.2 + z)}
                {(e_1.1,y)})}
       {(x)}.$
\end{tabbing}
Here both functions are closed. Therefore they can be extracted to the
top-level. Note also that the outer function depends on the inner
one. To break the dependence, code hoisting introduces an extra
argument to the outer function which is bound by an application to the
inner function at the point where the outer function occurs. The result of
code hoisting on the above code is as follows:
\begin{tabbing}
\qquad\=\quad\=\kill
\>$\kwd{let}\app f_1 = \kwd{fun}\app z\app e_2 \rightarrow e_2.1 + e_2.2 + z\app \kwd{in}$\\
\>$\kwd{let}\app f_2 = \kwd{fun}\app f_1\app y\app e_1 \rightarrow \clos{f_1}{(e_1.1,y)}\app \kwd{in}$\\
\>\>$\kwd{let}\app x = 3\app \kwd{in}\app
      \clos{(f_2\app f_1)}{(x)}$
\end{tabbing}

From the above example, we can see that the key to specifying code
hoisting is to capture the constraint that the functions extracted from
$M$ in $(\abs x M)$ do not depend on $x$. If we combine code hoisting
with closure conversion, we have an over-arching property that 
every function is closed and therefore cannot depend on any variable
bound by an external abstraction. However, we can also think of
checking the constraint that must be satisfied directly as a
structural property of the expressions being considered. This
is the approach we take in the describing closure conversion in the
next section.

\section{A Rule-Based Description of the Transformation}
\label{sec:rule_based_ch}

We give a rule-based relational description of the code hoisting
transformation in this section. We first describe the source and
target languages of the transformation, including their typing rules,
and then present the transformation rules.

\subsection{The source and target languages}
\label{sec:ch_langs}

\pagelabel{txt:ch_targ_lang}
The source language of the transformation is the target language of
closure conversion, depicted in Figure\cspc\ref{fig:cc_targ_lang}. The
target language of the transformation is the same as the source
language. The result of code hoisting will be a term of the form
\begin{gather*}
\letexp {f_1} {M_1} {\ldots \letexp {f_n} {M_n} M}
\end{gather*}
where, for $1 \leq i \leq n$, $M_i$ corresponds to an extracted function.
We will write this term as $(\letfun {\vec{f} = \vec{M}} M)$ where
$\vec{f} = (f_1,\ldots,f_n)$ and $\vec{M} = (M_1,\ldots,M_n)$.

\subsection{The transformation rules}
\label{subsec:cc_rules}

We write the judgment of code hoisting as $(\ch \rho M {M'})$ where
$\rho$ has the form $(x_1,\ldots,x_n)$.
This judgment asserts that $M'$ is the result of extracting all
functions in $M$ to the top level, assuming that $\rho$ contains all
the bound variables in the context in which $M$ appears. The
transformation rules are summarized in Figure\cspc\ref{fig:ch_rules}
where we write $(\vec{f},\vec{g})$ and $(\vec{F},\vec{G})$ to
represent the concatenation of tuples of variables and terms. The
transformation judgment is defined by recursion on the structure of
$M$. In the base case, a (bound) variable transforms to itself. Most
of the recursive cases have the following general structure: given a
source term $M$, code hoisting is applied recursively to subterms in
$M$ and the output is formed by combining the results of
recursion. One of the main rule that deserves further discussion is
$\chabs$ for transforming functions. Intuitively, the term $(\abs x
M)$, is transformed by extracting the functions from within $M$ and
then moving them further out of the scope of $x$. For this, this rule
has a side condition that the function argument $x$ must not occur in
$\vec{F}$ which are functions extracted from $M$. The resulting body
is made independent of the extracted functions by converting it into a
form where it can be applied to these functions. Similarly, the
$\chlet$ rule also has a side condition that the extracted functions
$\vec{G}$ cannot depend on the binding variable $x$. Another point to 
note is that the rule $\chopen$ works for applications of closures with 
a particular structure, which suffices for extracting functions from
applications of closures in our setting of compilation.

\begin{figure}[!ht]
\begin{gather*}
    \infer[\chvar]{
      \ch \rho {x} {\letfun {() = ()} x}
    }{
      x \in \rho
    }
    \quad
    \infer[\chunit]{
      \ch \rho {\unit} {\letfun {() = ()} \unit}
    }{}
    \\
    \infer[\chnat]{
      \ch \rho {n} {\letfun {() = ()} n}
    }{}
    \quad
    \infer[\chpred]{
      \ch \rho {\pred M}
          {\letfun {\vec{f} = \vec{F}} {\pred {M'}}}
    }{
      \ch \rho M {\letfun {\vec{f} = \vec{F}} M'}
    }
    \\
    \infer[\chplus]{
      \ch \rho {{M_1} + {M_2}}
          {\letfun {(\vec{f},\vec{g}) = (\vec{F},\vec{G})}
                   {{M_1'} + {M_2'}}}
    }{
      \ch \rho {M_1} {\letfun {\vec{f} = \vec{F}} {M_1'}}
      &
      \ch \rho {M_2} {\letfun {\vec{g} = \vec{G}} {M_2'}}        
    }
    \\
    \infer[\chif]{
      \begin{array}{l}
        \rho \triangleright  
             {\ifz {M_1} {M_2} {M_3}} \leadsto_{ch}\\
        \qquad {{\letfun {(\vec{f},\vec{g},\vec{h}) = (\vec{F},\vec{G},\vec{H})}
                   {\ifz {M_1'} {M_2'} {M_3'}}}}
      \end{array}
    }{
      \begin{array}{c}
      \ch \rho {M_1} {\letfun {\vec{f} = \vec{F}} {M_1'}}
      \\
      \ch \rho {M_2} {\letfun {\vec{g} = \vec{G}} {M_2'}}        
      \qquad
      \ch \rho {M_3} {\letfun {\vec{h} = \vec{H}} {M_3'}}
      \end{array}
    }
    \\
    \infer[\chpair]{
      \ch \rho {\pair {M_1} {M_2}}
          {\letfun {(\vec{f},\vec{g}) = (\vec{F},\vec{G})}
                   {\pair {M_1'} {M_2'}}}
    }{
      \ch \rho {M_1} {\letfun {\vec{f} = \vec{F}} {M_1'}}
      &
      \ch \rho {M_2} {\letfun {\vec{g} = \vec{G}} {M_2'}}        
    }
    \\
    \infer[\chfst]{
      \ch \rho {\fst M}
          {\letfun {\vec{f} = \vec{F}} {\fst {M'}}}
    }{
      \ch \rho M {\letfun {\vec{f} = \vec{F}} M'}
    }
    \\
    \infer[\chsnd]{
      \ch \rho {\snd M}
          {\letfun {\vec{f} = \vec{F}} {\snd {M'}}}
    }{
      \ch \rho M {\letfun {\vec{f} = \vec{F}} M'}
    }
    \\
    \infer[\chlet]{
      \ch \rho {(\letexp x {M_1} {M_2})}
          {\letfun {(\vec{f},\vec{g}) = (\vec{F},\vec{G})}
                   {(\letexp x {M_1'} {M_2'})}}
    }{
      \ch \rho {M_1} {\letfun {\vec{f} = \vec{F}} {M_1'}}
      &
      \ch {\rho,x} {M_2} {\letfun {\vec{g} = \vec{G}} {M_2'}}        
    }
    \\
    \mbox{\small where $x$ is not already in $\rho$ 
      and is not a free variable of $\vec{G}$}
    \\
    \infer[\chabs]{
      \ch \rho {\abs x M} 
          {\letfun {(\vec{f},g) = (\vec{F},\abs {\vec{f}} {\abs x {M'}})}
            {g~\vec{f}}}
    }{
      \ch {\rho,x} M {\letfun {\vec{f} = \vec{F}} {M'}}
    }
    \\
    \mbox{\small where $x$ is not already in $\rho$
      and is not a free variable of $\vec{F}$}
    \\
    \infer[\chapp]{
      \ch \rho {M_1 ~ M_2}
          {\letfun {(\vec{f},\vec{g}) = (\vec{F},\vec{G})}
                   {{M_1'}~ {M_2'}}}
    }{
      \ch \rho {M_1} {\letfun {\vec{f} = \vec{F}} {M_1'}}
      &
      \ch \rho {M_2} {\letfun {\vec{g} = \vec{G}} {M_2'}}        
    }
    \\
    \infer[\chclos]{
      \ch \rho {\clos {M_1} {M_2}}
          {\letfun {(\vec{f},\vec{g}) = (\vec{F},\vec{G})}
                   {\clos {M_1'} {M_2'}}}
    }{
      \ch \rho {M_1} {\letfun {\vec{f} = \vec{F}} {M_1'}}
      &
      \ch \rho {M_2} {\letfun {\vec{g} = \vec{G}} {M_2'}}        
    }
    \\
    \infer[\chopen]{
      \begin{array}{l}
      \rho \triangleright {\open {x_f} {x_e} {M_1} {x_f \app (M_1,M_2,x_e)}} \leadsto_{ch}\\
       \qquad {\letfun {(\vec{f},\vec{g}) = (\vec{F},\vec{G})}
                {\open {x_f} {x_e} {M_1'} {x_f \app (M_1',M_2',x_e)}}}
      \end{array}
    }{
      \ch \rho {M_1} {\letfun {\vec{f} = \vec{F}} {M_1'}}
      &
      \ch \rho {M_2} {\letfun {\vec{g} = \vec{G}} {M_2'}}
    }
\end{gather*}
\caption{The Rules for Code Hoisting}
\label{fig:ch_rules}
\end{figure}

%% Another point to note is that the rule $\chopen$ for transforming
%% closure applications only works on terms that have the form of the
%% results of closure conversion. Ideally, this rule should look like
%% the following which subsumes the rule $\chopen$:
%% %
%% \begin{gather*}
%%     \infer[]{
%%       \mbox{\small 
%%           \stackanchor{$\rho \triangleright {\open {x_f} {x_e} {M_1} {M_2}} \leadsto_{ch}$}
%%           {$\letfun {(\vec{f},\vec{g}) = (\vec{F},\vec{G})}
%%                    {\open {x_f} {x_e} {M_1'} {M_2'}}$}}
%%     }{
%%       \ch \rho {M_1} {\letfun {\vec{f} = \vec{F}} {M_1'}}
%%       &
%%       \ch {\rho,x_e,x_f} {M_2} {\letfun {\vec{g} = \vec{G}} {M_2'}}        
%%     }
%%     \\
%%     \mbox{\small where $x_e$ and $x_f$ are not already in $\rho$
%%       and are not free variables of $\vec{G}$}
%% \end{gather*}
%% %
%% We use a more restricted version of the rule because in order to
%% verify the correctness of code hoisting using the general rule based
%% on logical relations, we need to consider a much more complicated
%% class of logical relations that takes into account existential
%% types\cspc\cite{ahmed06esop}. This is beyond the scope of this thesis.

\section{Implementing the Transformation in \LProlog}
\label{sec:ch_impl}

Our presentation of the implementation of code hoisting has two parts:
we first show the encoding of the syntax and typing rules of the
source and target languages and then present a \LProlog implementation
of the transformation.

\subsection{Encoding the language}
\label{subsec:ch_lang_encoding}

\pagelabel{ecd:ch_targ_lang}
The only issue here is to provide a convenient representation for the
output of code hoisting, or \emph{hoisted terms}. Towards this end, we
identify the constants $\khbase: \kctm \to \kctm$, $\khabs: (\kctm \to
\kctm) \to \kctm$ and $\khtm: \klist\app \kctm \to \kctm \to \kctm$ for
representing hoisted terms.
Using these constants, a hoisted term of the form
\[
\letfun {(f_1,\ldots,f_n) = (M_1,\ldots,M_n)} M
\]
will be represented by
\[
\khtm\app (M_1 \cons \ldots \cons M_n \cons \knil)\app
       (\khabs\app (f_1 \mlam \ldots (\khabs\app (f_n \mlam \khbase\app M)))).
\]
We often write $(\khabs\app (f_1,\ldots,f_n) \mlam M)$ for $\khabs\app
(f_1 \mlam \ldots (\khabs\app (f_n \mlam \khbase\app M)))$.

\pagelabel{ecd:ch_targ_typing}
We also identify the constant $\kccof : \kctm \to \kty \to \omic$ for
typing the output of code hoisting such that $\kccof\app M\app T$
holds if and only if $M$ has type $T$. They are defined by the
following clauses which simply restate the rules for typing the output
of code hoisting:
\begin{tabbing}
\qquad\=\quad\=\kill
\>$\kccof\app (\khtm\app \knil\app (\khbase\app M))\app T \limply \kcof\app M\app T.$\\
\>$\kccof\app (\khtm\app (M \cons L)\app (\khabs\app R))\app T \limply$\\
\>\>$\kcof\app M\app T_1 \scomma 
       \forallx x {\kcof\app x\app T_1 \simply \kccof\app (\khtm\app L\app (R\app x))\app T}.$
\end{tabbing}

\subsection{Specifying the code hoisting transformation}
\label{subsec:specify_ch}

\pagelabel{ecd:ch_rules}
We use the predicate $\kch : \kctm \to \kctm \to \omic$ to represent
the relation of code hoisting and translate the code hoisting rules in
Figure\cspc\ref{fig:ch_rules} into the following clauses. They define
$\kch$ such that $(\kch\app M\app M')$ holds if and only if $M$ is
transformed into $M'$ by code hoisting.
\begin{tabbing}
\qquad\=\quad\=\quad\=\kill
\>$\kch\app (\kcnat\app N)\app (\khtm\app \knil\app (\khbase\app (\kcnat\app N))).$\\
\>$\kch\app \kcunit\app (\khtm\app \knil\app (\khbase\app \kcunit)).$\\
\>$\kch\app (\kpred\app M)\app (\khtm\app \FE\app M'') \limply
  \kch\app M\app (\khtm\app \FE\app M')\scomma \khconstr\app M'\app (x \mlam \kcpred\app x)\app M''.$\\
\>$\kch\app (\kcplus\app M_1\app M_2)\app (\khtm\app \FE\app M') \limply
  \kch\app M_1\app (\khtm\app \FE_1\app M_1')\scomma \kch\app M_2\app (\khtm\app \FE_2\app M_2') \scomma$\\
\>\>$\kappend\app \FE_1\app \FE_2\app \FE\scomma
       \khcombine\app M_1'\app M_2'\app (x \mlam y \mlam \kcplus\app x\app y)\app M'.$\\
\>$\kch\app (\kcifz\app M_1\app M_2\app M_3)\app (\khtm\app \FE\app M') \limply$\\
\>\>$\kch\app M_1\app (\khtm\app \FE_1\app M_1')\scomma 
     \kch\app M_2\app (\khtm\app \FE_2\app M_2') \scomma
     \kch\app M_3\app (\khtm\app \FE_3\app M_3') \scomma$\\
\>\>$\kappend\app \FE_1\app \FE_2\app \FE_{12}\scomma
     \kappend\app \FE_{12}\app \FE_3\app \FE\scomma$\\
\>\>$\khcombinethree\app M_1'\app M_2'\app M_3'\app 
       (x \mlam y \mlam z \mlam \kcifz\app x\app y\app z)\app M'.$\\
\>$\kch\app (\kcpair\app M_1\app M_2)\app (\khtm\app \FE\app M') \limply
  \kch\app M_1\app (\khtm\app \FE_1\app M_1')\scomma \kch\app M_2\app (\khtm\app \FE_2\app M_2') \scomma$ \\
\>\>$\kappend\app \FE_1\app \FE_2\app \FE\scomma
       \khcombine\app M_1'\app M_2'\app (x \mlam y \mlam \kcpair\app x\app y)\app M'.$\\
\>$\kch\app (\kcfst\app M)\app (\khtm\app \FE\app M'') \limply
  \kch\app M\app (\khtm\app \FE\app M')\scomma \khconstr\app M'\app (x \mlam \kcfst\app x)\app M''.$\\
\>$\kch\app (\kcsnd\app M)\app (\khtm\app \FE\app M'') \limply
  \kch\app M\app (\khtm\app \FE\app M')\scomma \khconstr\app M'\app (x \mlam \kcsnd\app x)\app M''.$\\
\>$\kch\app (\kclet\app M\app R)\app (\khtm\app \FE\app M'') \limply
  \kch\app M\app (\khtm\app FE_1\app M')\scomma$\\
\>\>$(\forallx x {\kch\app x\app (\khtm\app \knil\app (\khbase\app x)) \simply \kch\app (R\app x)\app (\khtm\app \FE_2\app (R'\app x))})\scomma$ \\
\>\>$\kappend\app \FE_1\app \FE_2\app \FE\scomma \khcombineabs\app M'\app R'\app (x \mlam y \mlam \kclet\app x\app y)\app M''.$\\
\>$\kch\app (\kcabs\app R)\app (\khtm\app ((\kcabs\app F) :: \FE)\app (\khabs\app R'')) \limply$\\
\>\>$(\forallx x {\kch\app x\app (\khtm\app \knil\app (\khbase\app x)) \simply \kch\app (R\app x)\app (\khtm\app \FE\app (R'\app x))})\scomma$\\
\>\>$\kabstract\app R'\app R''\app F.$\\
\>$\kch\app (\kcapp\app M_1\app M_2)\app (\khtm\app \FE\app M') \limply
  \kch\app M_1\app (\khtm\app \FE_1\app M_1')\scomma \kch\app M_2\app (\khtm\app FE_2\app M_2')\scomma$\\
\>\>$\kappend\app \FE_1\app \FE_2\app \FE\scomma \khcombine\app M_1'\app M_2'\app (x \mlam y \mlam \kcapp\app x\app y)\app M'.$\\
\>$\kch\app (\kclos\app M_1\app M_2)\app (\khtm\app \FE\app M') \limply
  \kch\app M_1\app (\khtm\app \FE_1\app M_1')\scomma \kch\app M_2\app (\khtm\app \FE_2\app M_2')\scomma $\\
\>\>$\kappend\app \FE_1\app \FE_2\app \FE\scomma \khcombine\app M_1'\app M_2'\app (x \mlam y \mlam \kclos\app x\app y)\app M'.$\\
\>$\kch\app (\kopen\app M_1\app (f \mlam e \mlam \kcapp\app f\app (\kcpair\app M_1\app (\kcpair\app M_2\app e))))\app (\khtm\app \FE\app M') \limply$\\
\>\>$\kch\app M_1\app (\khtm\app \FE_1\app M_1')\scomma \kch\app M_2\app (\khtm\app \FE_2\app M_2') \scomma \kappend\app \FE_1\app \FE_2\app \FE\scomma$\\
\>\>$\khcombine\app M_1'\app M_2'\app (x \mlam y \mlam (\kopen\app x\app (f \mlam e \mlam \kcapp\app f\app (\kcpair\app x\app (\kcpair\app y\app e)))))\app M'.$
%% \>$\kch\app (\kopen\app M\app R)\app (\khtm\app \FE\app M') \limply
%%      \kch\app M\app (\khtm\app \FE_1\app M_1')\scomma$ \\
%% \>\>$(\kpi \app f \mlam \kpi \app e \mlam
%%      \kch\app f\app (\khtm\app \knil\app (\khbase\app f)) \simply
%%      \kch\app e\app (\khtm\app \knil\app (\khbase\app e)) \simply$\\
%% \>\>$\kch\app (R\app f\app e)\app (\khtm\app \FE_2\app (R'\app f\app e)))\scomma$\\
%% \>$\kappend\app \FE_1\app \FE_2\app \FE\scomma
%%   \khcombineabstwo\app M_1'\app R'\app (x \mlam y \mlam \kopen\app x\app y)\app M'.$
\end{tabbing}
These clauses are transparently translated from the rules in
Figure\cspc\ref{fig:ch_rules}. The predicates $\khconstr$,
$\khcombine$, $\khcombinethree$, $\khcombineabs$, $\khcombineabstwo$
are defined to combine the results of code hoisting on subterms to
form the outputs. For instance, $\khcombine$ is defined through the
following clauses:
\begin{tabbing}
\qquad\=\quad\=\kill
\>$\khcombine\app (\khbase\app M_1)\app (\khbase\app M_2)\app 
    C\app (\khbase\app (C\app M_1\app M_2)).$\\
\>$\khcombine\app (\khabs\app R)\app M\app C\app (\khabs\app R') \limply
    \forallx f {\khcombine\app (R\app f)\app M\app C\app (R'\app f)}.$\\
\>$\khcombine\app (\khbase\app M)\app (\khabs\app R)\app C\app (\khabs\app R') \limply$\\
\>\>$\forallx f {\khcombine\app (\khbase\app M)\app (R\app f)\app C\app (R'\app f)}.$
\end{tabbing}
By this definition, if $M_1'$ is $(\khabs\app (f_1,\ldots,f_n) \mlam
{M_1})$ and $M_2'$ is $(\khabs\app (g_1,\ldots,g_m) \mlam {M_2})$,
then $(\khcombine\app M_1'\app M_2'\app (x \mlam y \mlam \kpair'\app
x\app y)\app M')$ holds when $M'$ is
\[
(\khabs\app (f_1,\ldots,f_n,g_1,\ldots,g_m) \mlam {\kpair'\app M_1\app
  M_2}).
\]

Again, we use universal goals to perform recursion over binding
operators and hypothetical goals to introduce the rules for
transforming binding variables. As a result, the context of code
hoisting is represented implicitly by the dynamic context in \HHw
sequents. 
Note that the side condition that the extracted functions in the rules
$\chabs$ do not contain free occurrences of the binding variable which
they are hoisted over has a completely logical encoding: it is
statically captured by the ordering of quantifiers and dynamically via
unification. For instance, by the fourth last clause above which encodes
$\chabs$, to perform code hoisting on a function $(\kcabs\app R)$, we
need to derive the following goal:
\[
\forallx x {\kch\app x\app (\khtm\app \knil\app (\khbase\app x)) \simply
  \kch\app (R\app x)\app (\khtm\app \FE\app (R'\app x))}
\]
Here the variable $\FE$ which represents the functions extracted from
$R$ cannot depend on $x$ because $\FE$ is bound out side of $x$. Thus,
this goal is derivable only if $\FE$ matches with some term not
containing $x$ as a free variable. The same observation can be made for
the clause encoding the rules $\chlet$.

We have used the predicate \kabstract to build the final result of the
transformation on functions. Intuitively, \kabstract eliminates the
dependence of a function on the functions nested in its body by
abstracting it over a tuple that binds these functions and generates
an application expression for representing the original function.
Specifically, let $(\khtm\app \FE\app (R'\app x))$ be the result of
recursive code hoisting on the body of a function where $x$ is the
function argument. Here $R'$ is a meta-level abstraction of the form
\[
x \mlam (\khabs\app (f_1,\ldots,f_n) \mlam {(R\app f_1\app \ldots\app f_n\app x)}).
\]
where $f_1, \ldots, f_n$ are binders for the functions in $\FE$. Then
$(\kabstract\app R'\app R''\app F)$ is derivable if and only if $F$ is
\[
l \mlam {\letexp {f_1} {\pi_1(l)} {\ldots \letexp{f_n} {\pi_n(l)}
    {\kcabs\app (\abs x {R\app f_1\app \ldots\app f_n\app x})}}}
\]
and $R''$ is 
\[
f \mlam (\khabs\app (f_1,\ldots,f_n) \mlam (f\app (f_1,\ldots,f_n))).
\]
As a result, $(\kcabs\app F)$ represents a closed function obtained by
abstracting the body of $R'$ over $x$ and $(f_1,\ldots,f_n)$. As
indicated in the final output 
\[
(\khtm\app ((\kcabs\app F) \cons \FE)\app (\khabs\app R'')),
\]
the function $(\kcabs\app F)$ is hoisted to the top-level and bound by
$f$ in $R''$ and the application $(f\app (f_1,\ldots,f_n))$ in $R''$
represents the original function.
The definition of \kabstract is easy to construct and is not provided
here.

\section{Informal Verification of the Transformation}
\label{sec:informal_verify_ch}

We informally describe the verification of code hoisting in this
section based on the ideas presented in
Section\cspc\ref{sec:vfc_approach}. Similar to the informal
verification of the CPS transformation and closure conversion, we
first discuss the type preservation and then the semantics
preservation.

\subsection{Type preservation of the transformation}
\label{subsec:ch_informal_typ_pres}

The type preservation property of code hoisting is stated as follows:
\begin{mythm}\label{thm:ch_typ_pres_open}
  Let $\rho = (x_1,\ldots,x_n)$ and $\G = (x_1:T_1,\ldots,x_n:T_n)$. If $\G
  \stseq M:T$ and $\ch \rho M {M'}$, then $\G \stseq M' : T$.
\end{mythm}
\noindent This theorem is easily proved by induction on the derivation of $\ch
\rho M {M'}$, from which we can derive the following corollary:
\begin{mycoro}\label{coro:ch_typ_pres_closed}
  If $\emptyset \stseq M:T$ and $\ch \emptyset M {M'}$, then
  $\emptyset \stseq M':T'$.
\end{mycoro}

\subsection{Semantics preservation of the transformation}
\label{subsec:ch_informal_sem_pres}

We give an informal description of semantics preservation for code
hoisting in this section. The operational semantics of the source and
target languages is already known. In the rest of this section, we
present the logical relations for denoting equivalence between the
source and target programs, their properties, the semantics
preservation theorem and its proof.

\subsubsection{An alternative typing rule for closure applications}

The rule $\cofopen$ for typing the closure applications described in
Section~\ref{sec:cc_langs} introduces new type constants as the types
of the environments of the closures. Since those type constants stand
for ``unknown'' types, logical relations cannot be easily defined by
recursion on their structures. To circumvent this problem, notice that
the code hoisting transformation only deals with closure applications
of the following form
\[
\open {x_f} {x_e} {M_1} {x_f \app (M_1,M_2,x_e)}.
\]
In this case, we can use the following typing rule equivalent to
$\cofopen$ that does not introduce new constants:
\[
  \infer[]{
    \Gamma \stseq  {\open {x_f} {x_e} {M_1} {x_f \app (M_1,M_2,x_e)}} : T_2
  }{
    \Gamma \stseq {M_1} : {T_1 \to T_2}
    &
    \Gamma \stseq {M_2} : T_1
  }
\]
We shall assume that this typing rule is used instead of $\cofopen$ in
the following discussion of the semantics preservation proof for code
hoisting.

\subsubsection{Logical relations and their properties}

The step-indexing logical relations for code hoisting are depicted in
Figure\cspc\ref{fig:ch_logical_relations}. The second to last rule in
this figure defines the equivalence relation between abstraction
values and the last rule defines the equivalence relation between
closure values which function like recursive functions. 
Similar to closure conversion, we can prove the property that
$\approx$ is closed under decreasing indexes and compatibility lemmas
for program constructs in the source language.

\begin{figure}[!ht]
\center
\begin{tabbing}
\qquad\qquad\qquad\=\quad\=\quad\=\kill
\>$\simindex T k M M' \iff$\\
\>\>$\forall j \leq k. \forall V. M \step{j} V \rimp
      \exists V'. {M'} \eval {V'} \rand \equalindex T {k-j} V {V'};$\\
\>$\equalindex \tnat k n n;$\\
\>$\equalindex \tunit k \unit \unit;$\\
\>$\equalindex {(T_1 \tprod T_2)} k {\pair {V_1} {V_2}} {\pair {V_1'} {V_2'}} \iff
       \equalindex {T_1} k {V_1} {V_1'} \rand \equalindex {T_2} k {V_2} {V_2'};$\\
\>$\equalindex {(T_1 \carr T_2)} k {(\abs x R)} {(\abs x {R'})} \iff$\\
\>\>$\forall j < k. \forall V, V'.
        \equalindex {T_1} j {V} {V'} \rimp
        \simindex {T_2} j {R[V/x]} {R'[V'/x]}$.\\
\>$\equalindex {(T_1 \to T_2)} k {\clos{\abs p R}{\VE}} {\clos{\abs p R'}{\VE'}} \iff$\\
\>\>$\forall j < k. \forall V_1, V_1', V_2, V_2'.
        \equalindex {T_1} j {V_1} {V_1'} \rimp
        \equalindex {T_1 \to T_2} j {V_2} {V_2'} \rimp$\\
\>\>\>$\simindex {T_2} j {R[(V_2,V_1,\VE)/p]} {R'[(V_2',V_1',\VE')/p]}$.
\end{tabbing}

  \caption{The Logical Relations for Verifying Code Hoisting}
  \label{fig:ch_logical_relations}
\end{figure}

\subsubsection{Informal proof of semantics preservation}

Similar to the CPS transformation and closure conversion, we consider
semantics preservation for possibly open terms under closed
substitutions. A substitution has the form $(V_1/x_1,\ldots,V_n/x_n)$
where $V_i$ are closed values. The equivalence relation between source
and target substitutions is based on step-indexing logical relation
and given as follows:
\begin{gather*}
  \equalindex {(x_1:T_1,\ldots,x_n:T_n)} k {(V_1/x_1,\ldots,V_n/x_n)} {(V_1'/x_1,\ldots,V_n'/x_n)}
  \iff (\forall 1 \leq i \leq n. \equalindex {T_i} i {V_i} {V_i'})
\end{gather*}

The semantics preservation theorem for the CPS transformation can now
be stated as follows:
\begin{mythm}\label{thm:ch_sem_pres_open}
  Let $\Gamma = (x_1:T_1,\ldots,x_n:T_n)$, $\rho = (x_1,\ldots,x_n)$,
  $\equalindex \Gamma k \theta {\theta'}$ and $\Gamma \stseq M : T$.
  If $\ch \rho M {M'}$, then $\simindex T k {M[\theta]} {M'[\theta']}$.
\end{mythm}
\noindent We outline the main steps in the argument for this
theorem. We proceed by induction on the derivation of $\ch \rho M
{M'}$, analyzing the last step in it. This obviously depends on the
structure of $M$. The cases for a natural number, the unit constructor
or a variable are obvious. In the remaining case, other than when $M$
is a let expression or a function, the argument follows a set pattern:
we observe that substitutions distribute to the sub-components of
expressions, we invoke the induction hypothesis over the
sub-components and then we use the compatibility lemmas to conclude.
When $M$ is a let expression or a function, we need to extend the
substitutions with equivalent values for the binding variable for
applying the inductive hypothesis over the body of $M$. Note that to
conclude the proof, we need to make use of the side condition that the
binding variable does not occur in the functions extracted from the
body of $M$ to show that the extensions to substitutions have no
effect on these functions.

From Theorem\cspc\ref{thm:ch_sem_pres_open}, it is easy to derive the
following corollaries of semantic preservation for closed terms and
closed terms at atomic types:
\begin{mycoro}\label{coro:ch_sem_pres_closed}
  If $\emptyset \stseq M : T$ and $\ch \emptyset M {M'}$, then
  $\simindex T i {M} {M'}$ for any $i$.
\end{mycoro}
\begin{mycoro}\label{coro:ch_sem_pres_atom}
  If $\emptyset \stseq M : \tnat$, $\ch \emptyset M {M'}$ and $M \eval V$, then
  $M' \eval V$.
\end{mycoro}

\section{Verifying the \LProlog Program in \Abella}
\label{sec:formal_verify_ch}

In this section, we formalize the verification of code hoisting
described in Section\cspc\ref{sec:informal_verify_ch} in \Abella. In
this discussion, we pay particular attention to showing how the
logical treatment of the side condition that describes the
independence of extracted functions on the arguments of enclosing
functions in the \LProlog program is exploited to simplify the proofs.

\subsection{Type preservation of the transformation}
\label{subsec:cc_formal_typ_pres}

\pagelabel{ecd:ch_typ_pres_open}
In the \LProlog program, the contexts of code hoisting is represented
by the dynamic contexts of \HHw sequents that encode the rule for
transforming free variables. Such contexts are represented by the
predicate $\kchctx : \olist \to \prop$ given by the following clauses:
\begin{tabbing}
\qquad\=$\nablax x {\kchctx\app ((\kch\app x\app (\khtm\app \knil\app (\khbase\app x))) \cons L)}$
   \quad\=$\rdef$ \quad\=\kill
\>$\kchctx\app \knil$ \>$\rdef$ \>$\rtrue$\\
\>$\nablax x {\kchctx\app ((\kch\app x\app (\khtm\app \knil\app (\khbase\app x))) \cons L)}$
   \>$\rdef$ \>$\kchctx\app L.$
\end{tabbing}
Theorem\cspc\ref{thm:ch_typ_pres_open} is then formalized as follows
where \kvarsofchctx and \kcvarsofctx are respectively used to collect
variables in the contexts of code hoisting and variables in the typing
contexts:
\begin{tabbing}
\qquad\=\quad\=\kill
\>$\rfall L, \CL, \Vs, M, M', T.$\\
\>\>$\kdctx\app L \rimp \kchctx\app \CL \rimp \kcvarsofctx\app L\app \Vs \rimp
     \kvarsofchctx\app \CL\app \Vs \rimp$\\
\>\>$\oseq{L \stseq \kcof\app M\app T} \rimp 
     \oseq{\CL \stseq \kch\app M\app M'} \rimp
     \oseq{L \stseq \kccof\app M'\app T}.$
\end{tabbing}
It is proved by induction on $\oseq{\CL \stseq \kch\app M\app M'}$ in
an obvious manner. From this theorem, it is easy to prove the
following theorem corresponding to
Corollary\cspc\ref{coro:ch_typ_pres_closed}:
\begin{tabbing}
\qquad\=\quad\=\kill
\>$\rforallx {M,M',T} {
   \oseq{\kcof\app M\app T} \rimp 
   \oseq{\kch\app M\app M'} \rimp
   \oseq{\kccof\app M'\app T}}.$
\end{tabbing}

\subsection{Semantics preservation of the transformation}
\label{subsec:ch_formal_sem_pres}

In this section, we formalize the semantics preservation proof of
closure conversion described in
Section\cspc\ref{subsec:cc_informal_sem_pres} in \Abella.

\subsubsection{Formalizing the operational semantics}

\pagelabel{ecd:cc_eval}
The operational semantics of the source and target languages has
already been given in Section\cspc\ref{subsec:cc_formal_sem_pres}. The
only issue here is to encode the operational semantics for the
output of code hoisting. Towards this end, we introduce
the constant $\kcceval : \kctm \to \kctm \to \omic$ such that
$\kcceval\app (\khtm\app FE\app M)\app V$ if and only if $(\khtm\app
FE\app M)$ evaluates to the value $V$. Its definition is given as
follows which simply restates the evaluation rules for the term
represented by $(\khtm\app FE\app M)$:
\begin{tabbing}
\qquad\=\kill
\>$\kcceval\app (\khtm\app \knil\app (\khbase\app M))\app V \limply \kceval\app M\app V.$\\
\>$\kcceval\app (\khtm\app (F \cons FE)\app (\khabs\app R))\app V 
    \limply \kceval\app (\khtm\app \FE\app (R\app F))\app V.$
\end{tabbing}

\subsubsection{Formalizing the logical relations}

\pagelabel{ecd:ch_logical_relations}
The simulation and equivalence relations are represented by the
predicate constants $\ksimch : \kty \to \knat \to \kctm \to \kctm \to
\prop$ and $\kequivch : \kty \to \knat \to \kctm \to \kctm \to \prop$,
which are defined as follows:
\begin{tabbing}
\qquad\=\quad\=\quad\=$\ksimch\app T_2\app K\app $\=\kill
\>$\ksimch\app T\app K\app M\app M'$ \quad$\rdef$\quad
   $\rfall J, V. \kle\app J\app K \rimp \oseq{\kcnstep\app J\app M\app V} \rimp \oseq{\kcval\app V} \rimp$\\
\>\>$\rexistsx {V',N} {\oseq{\kcceval\app M'\app V'} \rand 
                        \oseq{\kadd\app J\app N\app K} \rand \kequivch\app T\app N\app V\app V'}$\\
\>$\kequivch\app \ktnat\app K\app (\kcnat\app N)\app (\kcnat N) \quad\rdef\quad \rtrue$\\
\>$\kequivch\app \ktunit\app K\app \kcunit\app \kcunit \quad\rdef\quad \rtrue$\\
\>$\kequivch\app (\kprod\app T_1\app T_2)\app K\app (\kcpair\app V_1\app V_2)\app (\kcpair\app V_1'\app V_2') \quad\rdef\quad$\\
\>\>$\kequivch\app T_1\app K\app V_1\app V_1' \rand \kequivch\app T_2\app K\app V_2\app V_2'$\\
\>$\kequivch\app (\kcarr\app T_1\app T_2)\app \kz\app (\kcabs\app R)\app (\kcabs\app R') \;\rdef$\\
\>\>$\oseq{\kctm\app (\kcabs\app R)} \rand \oseq{\kctm\app (\kcabs\app R')}$\\
\>$\kequivch\app (\kcarr\app T_1\app T_2)\app (\ks\app K)\app (\kcabs\app R)\app (\kcabs\app R') \quad\rdef$\\
\>\>$\kequivch\app (\kcarr\app T_1\app T_2)\app K\app (\kcabs\app R)\app (\kcabs\app R') \rand$\\
\>\>$\rfall V, V'. \kequivch\app T_1\app K\app V\app V' \rimp 
       \ksimch\app T_2\app K\app (R\app V)\app (\khtm\app \knil\app (\khbase\app (R'\app V')))$\\
\>$\kequivch\app (\karr\app T_1\app T_2)\app \kz\app (\kclos (\kcabs\app R)\app \VE)\app
                      (\kclos (\kcabs\app R')\app \VE') \quad\rdef$\\
\>\>$\oseq{\kctm\app (\kclos (\kcabs\app R)\app \VE)} \rand
     \oseq{\kctm\app (\kclos (\kcabs\app R')\app \VE')} \rand$\\
\>\>$\oseq{\kcval\app \VE} \rand \oseq{\kcval\app \VE'}$\\
\>$\kequivch\app (\karr\app T_1\app T_2)\app (\ks\app K)\app (\kclos (\kcabs\app R)\app \VE)\app
                          (\kclos (\kcabs\app R')\app \VE') \quad\rdef$\\
\>\>$\kequivch\app (\karr\app T_1\app T_2)\app K\app 
       (\kclos (\kcabs\app R)\app \VE)\app (\kclos (\kcabs\app R')\app \VE') \rand$\\
\>\>$\rfall V_1, V_1', V_2, V_2'.
      \kequivch\app T_1\app K\app V_1\app V_1' \rimp 
      \kequivch\app (\karr\app T_1\app T_2)\app K\app V_2\app V_2' \rimp$\\
\>\>\>$\ksimch\app T_2\app K\app (R\app (\kcpair\app V_2\app (\kcpair\app V_1\app \VE)))\app$\\
\>\>\>\>$(\khtm\app \knil\app (\khbase\app (R'\app (\kcpair\app V_2'\app (\kcpair\app V_1'\app \VE'))))).$
\end{tabbing}
The formula $\ksimch\app T\app K\app M\app M'$ is intended to mean that
$M$ simulates $M'$ at type $T$ in at least $K$ steps; $\kequivch\app
T\app K\app V\app V'$ has a similar interpretation. Note that the last
argument of \ksimch must be an output of code hoisting and therefore
must have the form $(\khtm\app \FE\app M'')$; this is also manifested
in the uses of \ksimch in the above definition. 
%% Note the exploitation of
%% $\lambda$-tree syntax, specifically the use of application, to realize
%% substitution in the definition of \kequivch. 

Using the above definition, it is easy to formally prove the property
that the equivalence relation is closed under decreasing step measures
and the compatibility lemmas, as we have done for the CPS
transformation and closure conversion. The structures of those theorems and
proofs are easy to predict and we omit these details here.

\subsubsection{The equivalence relation on substitutions}

We designate the constant $\ksubstequivch : \olist \to \knat \to
\klist\app (\kmap\app \kctm\app \kctm) \to \klist\app (\kmap\app
\kctm\app \kctm) \to \prop$ to represent this equivalence relation
between substitutions. It is defined as follows:
\begin{tabbing}
\qquad\=\quad\=\kill
\>$\ksubstequivch\app \knil\app K\app \knil\app \knil \quad\rdef \quad\rtrue$ \\
\>$\nablax x {\ksubstequivch\app (\kcof\app x\app T \cons L)\app K\app
               (\kmap\app x\app V \cons \ML)\app (\kmap\app x\app V' \cons \ML')}$ 
   \quad$\rdef$\\
\>\>$\kequivch\app T\app K\app V\app V' \rand \ksubstequivch\app L\app K\app \ML\app \ML'.$
\end{tabbing}

\subsubsection{Formalizing the semantics preservation theorems}

\pagelabel{ecd:ch_sem_pres_open}
Theorem\cspc\ref{thm:ch_sem_pres_open} is now formalized as follows:
\begin{tabbing}
\qquad\=\quad\=\quad\=\kill
\>$\rfall L, K, \CL, \ML, \ML', M, M', T, \FE, \FE', P, P', \Vs. \oseq{\kisnat\app K} \rimp$\\
\>\>$\kdctx\app L \rimp \kcvarsofctx\app L\app \Vs \rimp$\\
\>\>$\kchctx\app \CL \rimp \kvarsofchctx\app \CL\app \Vs \rimp$\\
\>\>$\kcsubst\app \ML \rimp \kcsubst\app \ML' \rimp \ksubstequivch\app L\app K\app \ML\app \ML' \rimp$\\
\>\>$\oseq{L \stseq \kcof\app M\app T} \rimp \oseq{\CL \stseq \kch\app M\app (\khtm\app \FE\app M')} \rimp$\\
\>\>$\kappsubst\app \ML\app M\app P \rimp \kappsubst\app \ML'\app (\khtm\app \FE\app M')\app (\khtm\app \FE'\app P') \rimp$\\
\>\>\>$\ksimch\app T\app K\app P\app (\khtm\app \FE'\app P').$
\end{tabbing}
This theorem is proved by induction on $\oseq{\CL \stseq \kch\app
  M\app (\khtm\app \FE\app M')}$. The proof closely follows the
informal argument we provided for
Theorem\cspc\ref{thm:ch_sem_pres_open}. Note that when $M$ is a
function $(\kcabs\app R)$, we need to apply the inductive hypothesis
on $(R\app x)$ where $x$ is the function argument with substitutions
$(\kmap\app x\app V \cons \ML)$ and $(\kmap\app x\app V' \cons \ML')$
where $V$ and $V'$ are equivalent values for $x$. A critical step for
proving this case is to show the extension for $x$ in the second
substitution has no effect on the functions $\FE$ extracted from
$(R\app x)$, \ie, $\kappsubst\app (\kmap\app x\app V' \cons \ML')\app
\FE\app \FE'$ if and only if $\kappsubst\app \ML'\app \FE\app \FE'$
for some $\FE'$. By observing the logical structure of the clause of
$\kch$ for this case, \Abella knows that $\FE$ cannot depend on
$x$. By the definition of $\kappsubst$, it is immediate that the above
relation holds.

From the above theorem, we can easily prove the following theorems
that formalize Corollaries\cspc\ref{coro:ch_sem_pres_closed} and
\ref{coro:ch_sem_pres_atom}:
\begin{tabbing}
\qquad\=\quad\=\kill
\>$\rfall K, T, M, M'.\oseq{\kcof\app M\app T} \rimp \oseq{\kch\app M\app M'}
    \rimp \ksimch\app T\app K\app M\app M'.$\\
\>$\rfall M, K, M', V.
     \oseq{\kcof\app M\app \ktnat} \rimp \oseq{\kch\app M\app (\khtm\app \FE\app M')} \rimp$\\
\>\>$\oseq{\kceval\app M\app (\kcnat\app N)} \rimp 
     \oseq{\kcceval\app (\khtm\app \FE\app M')\app (\kcnat\app N)}$
\end{tabbing}

%%%%%%%%%%%%%%%%%%%%%%%%%%%%%%%%%%%%%%%%%%%%%%%%%%%%%%%%%%%%%%%%%%%%%%%%%%%%%%%%

% The complete compilation process
%%%%%%%%%%%%%%%%%%%%%%%%%%%%%%%%%%%%%%%%%%%%%%%%%%%%%%%%%%%%%%%%%%%%%%%%%%%%%%%
% complete.tex: The complete compilation process
%%%%%%%%%%%%%%%%%%%%%%%%%%%%%%%%%%%%%%%%%%%%%%%%%%%%%%%%%%%%%%%%%%%%%%%%%%%%%%%%
\chapter{Completing the Compilation Process}
\label{ch:complete}

After the CPS transformation, closure conversion and code hoisting,
the higher-order functional programs have been transformed into a
first-order form in which all the functions are closed and in a flat
space at the top-level. It is now easy to transform programs in this
form into an intermediate language to which conventional techniques
for compiling imperative languages can be applied. We call this
transformation the \emph{code generation} phase. The intermediate
language we choose is closely related to Cminor\cspc\cite{leroy09jar},
the input language for the back-end of the well-known CompCert
compiler. Our investigation of verified compilation of functional
languages stops after code generation because for implementation and
verification of the transformations after code generation: 1) the
higher-order abstract syntax does not have obvious benefits since they
only deal with programs in first order forms and do not perform
complicated manipulation of bindings anymore; 2) they have been
studied extensively in the area of verifying compilers for imperative
languages.

In Section~\ref{sec:codegen}, we give a rule-based relational
description of the code generation phase and discuss its
implementation and verification using our framework. In
Section~\ref{sec:full_correctness}, we discuss how the semantics
preservation proofs for the individual phases are composed to form the
correctness proof for the full compiler, which concludes our exercise
on verified compilation of functional languages.

\section{The Code Generation Transformation}
\label{sec:codegen}

The code generation transformation takes the output of code hoisting
and generates programs in a Cminor-like language in which every
operation is applied to variables or constants and the allocation of
pairs and closures and accesses to their elements are made
explicit. In the following successive subsections, we shall introduce
the target language of code generation, the memory model of the target
language, the rule-based description of the transformation, its
implementation in \LProlog, its informal verification and its
verification in \Abella.

\subsection{The target language of the transformation}
\label{sec:cg_targ_lang}

The target language of the code generation transformation, which is
also the target language of our compiler, is a typeless language. Its
syntax is given in Figure\cspc\ref{fig:cg_targ_lang}. We use $x$ to
denote variables, $n$ to denote natural numbers, $C$ to denote
constants or variables, $E$ to denote expressions, $S$ to denote
statements which represent a sequence of operations ending with an
expression, $F$ to denote functions and $P$ to denote programs. Note that
$P$ has the form of programs that result from code hoisting. That is,
every function resides at the top-level and takes two arguments where
the first argument refers to a sequence of functions it depends on and
the second argument is the actual argument of the function. Note also
that we use nested let expressions to represent a sequence of instructions
in statements.

\begin{figure}[ht!]
  \center
  \begin{tabbing}
    \qquad\qquad\qquad\qquad\=$T$ \quad\=$::=$ \quad\=\kill
    \>$C$ \>$::=$ \>$n \sep x \sep \unit$\\
    \>$E$ \>$::=$ \>$C \sep \pred C \sep C_1 + C_2 \sep
                  \ifz {C} {S_1} {S_2} \sep$ \\
    \>\>\>$(C_1\app C_2) \sep \alloc n \sep \move {C_1} n {C_2} \sep \load {C} n$
    \\
    \>$S$ \>$::=$ \>$E \sep \letexp x E S$
    \\[2ex]
    \>$F$ \>$::=$ \>$\abs {\vec{f}} {\abs x S}$
    \\
    \>$P$ \>$::=$ \>$\letfun {\vec{f}=\vec{F}} S$
    %% \\[1ex]
    %% \>$V$ \>$::=$ \>$n \sep \abs x S \sep \unit \sep \pair {V_1} {V_2}$
  \end{tabbing}
  \caption{The Syntax of the Target Language of Code Generation}
  \label{fig:cg_targ_lang}
\end{figure}

\subsection{The memory model of the target language}
\label{subsec:memory_model}

The only interesting components of the syntax in
Figure\cspc\ref{fig:cg_targ_lang} are expressions for manipulating
memory, \ie, $(\alloc n)$ for memory allocation, $(\move {C_1} n
{C_2})$ for writing to memory and $(\load {C} n)$ for reading from
memory. To understand them, we need to understand the memory model
of the language. We assume a very simple model of memory in which
there is a single heap consisting of an infinite number of memory
cells indexed by natural numbers. The memory cells are allocated from
lower to higher indexes\footnote{We do not consider garbage collection
  in this thesis. It is left for future work}. We use $i_{max}$ to
represent the index such that the cells indexed by natural numbers up
to (but not including) $i_{max}$ are allocated and accessible. We use
the expression $(\loc i)$ to represent a reference to the memory cell
indexed by $i$. The instruction $(\alloc n)$ is used to allocate the
$n$ cells indexed from $i_{max}$ to $(i_{max}+n-1)$. Letting $C_1$ be
$(\loc\app i)$ for some $i$ and $C_2$ be a value, $(\move {C_1} n
{C_2})$ is used to assign $C_2$ to the memory cell at $i+n$ and
$(\load {C_1} n )$ is used to read the value in that cell.

\subsection{A rule-based description of the transformation}
\label{sec:rule_based_cg}

\begin{figure}[ht!]
\begin{gather*}
  \infer[\cgennat]{
    \cgen \rho n K {\capp K n}
  }{}
  \quad
  \infer[\cgenvar]{
    \cgen \rho x K {\capp K x}
  }{
    x \in \rho
  }
  \\
  \infer[\cgenunit]{
    \cgen \rho \unit K {\capp K \unit}
  }{}
  \\
  \infer[\cgenpred]{
    \cgen \rho {\pred M} K {S}
  }{
    \cgen \rho M {\cabs x {\letexp v {\pred x} {(\capp K v)}}} {S}
  }
  \\
  \infer[\cgenplus]{
    \cgen \rho {M_1 + M_2} K {S'}
  }{
    \cgen \rho {M_2} {\cabs {x_2} {\letexp v {x_1+x_2} {(\capp K v)}}} {S}
    &
    \cgen \rho {M_1} {\cabs {x_1} {S}} {S'}
  }
  \\
  \infer[\cgenpair]{
    \cgen \rho {(M_1,M_2)} K {S'}
  }{
    \cgen \rho {M_2} {K'} {S}
    &
    \cgen \rho {M_1} {\cabs {x_1} {S}} {S'}
  }
  \\
   \mbox{\small where $K' = {(\cabs {x_2} 
                       {\letexp p {\alloc 2} {
                        \letexp {v_1} {\move p 0 {x_1}} {
                        \letexp {v_2} {\move p 1 {x_2}} {
                        {(\capp K p)}}
                       }}})}$}
  \\
  \infer[\cgenfst]{
    \cgen \rho {\fst M} K {S}
  }{
    \cgen \rho M {\cabs x {\letexp v {\load x 0} {(\capp K v)}}} {S}
  }
  \\
  \infer[\cgensnd]{
    \cgen \rho {\snd M} K {S}
  }{
    \cgen \rho M {\cabs x {\letexp v {\load x 1} {(\capp K v)}}} {S}
  }
  \\
  \infer[\cgenif]{
    \cgen \rho {\ifz {M_1} {M_2} {M_3}} K {S}
  }{
    \begin{array}{c c}
      \cgen \rho {M_2} {\cabs x x} {S_2}
      &
      \cgen \rho {M_3} {\cabs x x} {S_3}
      \\
      \multicolumn{2}{c}{
        \cgen \rho {M_1} 
             {\cabs 
               {x_1} 
               {\letexp a {(\ifz {x_1} {S_2} {S_3})} {(\capp {K} a)}}}
             {S}
      }
    \end{array}
  }
  \\
  \mbox{\small (provided $k$ does not occur in $\rho$, $M_2$ and $M_3$)}
  \\
  \infer[\cgenlet]{
    \cgen \rho {\letexp x {M_1} {M_2}} K {S'}
  }{
    \cgen {\rho,x} {M_2} K {S}
    &
    \cgen \rho {M_1} {\cabs x {S}} {S'}
  }
  \\
  \mbox{\small (provided $x$ does not occur in $\rho$ and $K$)}
  \\
  \infer[\cgenapp]{
    \cgen \rho {M_1 \app M_2} K {S'}
  }{
    \cgen \rho {M_2} {\cabs {x_2} {\letexp v {x_1\app x_2} {(\capp K v)}}} {S}
    &
    \cgen \rho {M_1} {\cabs {x_1} {S}} {S'}
  }
  \\
  \infer[\cgenclos]{
    \cgen \rho {\clos {M_1} {M_2}} K {S'}
  }{
    \cgen \rho {\pair {M_1} {M_2}} K {S'}
  }
  \\
  \infer[\cgenopen]{
    \cgen \rho {\open {x_f} {x_e} {M_1} {x_f\app (M_1, M_2, x_e)}} K {S'}
  }{
    \cgen \rho {M_2} {K_1} {S}
    &
    \cgen \rho {M_1} {K_2} {S'}
  }
\end{gather*}
\vspace{-1cm}
{\small
\begin{tabbing}
\qquad\=where $K_1 = (\hat{\lambda} x_2.{\mkern 3mu}$\=\kill
\>where $K_1 = (\hat{\lambda} x_2.{\mkern 3mu}\mlet\;p_1=\alloc 2\;\mathbf{in}\;
                \mlet\;v_1={\move {p_1} 0 {x_2}}\;\mathbf{in}\;
                \mlet\;v_2={\move {p_1} 1 {x_e}}\;\mathbf{in}$\\
\>\>$\mlet\;p_2=\alloc 2\;\mathbf{in}\;
     \mlet\;v_3={\move {p_2} 0 {x_1}}\;\mathbf{in}\;
     \mlet\;v_4={\move {p_2} 1 {p_1}}\;\mathbf{in}$\\
\>\>$\mlet\;v=x_f\app p_2\;\mathbf{in}\; \capp K v)$\\
\>and $K_2 = {(\cabs {x_1} 
                     { \letexp {x_f} {\load {x_1} 0}
                      {\letexp {x_e} {\load {x_1} 1} {S}}} )}$
\end{tabbing}}

\caption{The Rules for Generating Statements}
\label{fig:cg_rules}
\end{figure}

We first describe how terms not containing functions in the source
language are translated into statements in the target language. We
identify the relation $\cgen \rho M K {S'}$ such that it holds if $M$
is a source term whose free variables are contained in the set of
variables $\rho$ and that does not contain functions, $K$ is a context
or continuation of the form $\cabs x S$, and $S'$ is a statement
resulting from applying code generation to $M$ in the context
$K$. The transformation rules of $\cgen \rho M K {S'}$ are given in
Figure\cspc\ref{fig:cg_rules}. The rules defining this relation are very
similar to those of the CPS transformation (described in
Figure\cspc\ref{fig:cps_rules}): they recursively translate the
sub-expressions of the source program and accumulates the results in
$K$ in an order that reflects the control flow of the source program;
the substitution operations in the transformation are represented by
administrative $\beta$-redexes like in the CPS transformation. 
The rules $\cgenpair$, $\cgenplus$ and
$\cgenapp$ have the side condition that $x_1$ does not occur in
$\rho$, $M_2$ and $K$. The rule $\cgenopen$ has the side condition that
$x_f$,$x_e$ and $x_1$ do not occur in $\rho$, $M_2$ and $K$. 
The $\cgenpair$ rule translates a pair expression into a
program fragment for explicitly allocating two consecutive memory
cells and assigning the elements of the pair to the cells. The
$\cgenfst$ and $\cgensnd$ rules translate the access of elements of a
pair into instructions for loading values from memory cells allocated
for the pair. The $\cgenclos$ rule transforms the
closures like pairs. The $\cgenopen$ rule makes explicit 
the selection of function and environment parts from closures and 
the application of the function part to its arguments 
by using the memory operators.

We then describe the transformation of programs and functions in the
source language. Let $M$ be a program resulting from code hoisting,
\ie, $M$ has the form $(\letfun {\vec{f} = \vec{F}} M')$ where $M'$
does not contain any function and $\vec{F}$ is a sequence of functions
of the form $\abs {\vec{f}} {\abs x M''}$ where $M''$ does not contain
any function.  We identify the relation $\cgenp \rho M P$ such that it
holds if $P$ is the result of applying code generation to $M$. We also
identify the relation $\cgenf \rho F {F'}$ such that it holds if $F$
is a function extracted by code hoisting (\ie, $F$ has the form $\abs
{\vec{f}} {\abs x M''}$ where $M''$ does not contain any function)
whose free variables are contained in $\rho$ and $F'$ is the result of
applying code generation to $F$. The rules defining $\cgenp \rho M P$
and $\cgenf \rho F {F'}$ are given in Figure\cspc\ref{fig:cgpf_rules}.
Note that the two rules make use of $\cgen \rho M K {S'}$ where $K$ is
an identity context for generating statements from the bodies of
functions and programs.

\begin{figure}[ht!]
\begin{gather*}
  \infer[\cgenabs]{
    \cgenf \rho {(\abs {\vec{f}} {\abs x M})} {(\abs {\vec{f}} {\abs x S})}
  }{
    \cgen {\rho,\vec{f},x} M {\cabs{x}{x}} {S}
  }
  \\
  \mbox{\small (provided $x$ do not occur in $\rho$ and $M$)}
  \\
  \infer[\cgenprog]{
    \cgenp \rho {(\letfun {\vec{f} = \vec{F}} M)} 
                {(\letfun {\vec{f} = \vec{F'}} S)}
  }{
    \{\cgenf \rho {F_i} {F_i'} \sep 1 \leq i \leq n\}
    &
    \cgen {\rho,\vec{f}} M {\cabs x x} S
  }
  \\
  \mbox{\small (where $f = (f_1,\ldots,f_n)$, $\vec{F} = (F_1,\ldots,F_n)$ and $\vec{F'} = (F_1',\ldots,F_n')$)}
\end{gather*}
\caption{The Rules for Generating Programs and Functions}
\label{fig:cgpf_rules}
\end{figure}

\subsection{Implementing the transformation in \LProlog}
\label{subsec:cg_impl}

\pagelabel{ecd:cg_targ_lang}
We first consider encoding terms in the target language. We use
$\kctm$ to represent the type of terms in the target language. For the
constructors of the target language that also occur in the source
language, we reuse the constants for encode them in the source
language. For example, $\kcpred$ is used to encode $\pred C$ and
$\kcpair$ is used to encode $\pair {C_1} {C_2}$.
We further introduce the constants $\kloc : \knat \to \kctm$, $\kalloc
: \knat \to \kctm$, $\kmove : \kctm \to \knat \to \kctm \to \kctm$ and
$\kload : \kctm \to \knat \to \kctm$ for encoding the expressions for
manipulating the memory. 
%
%% Obviously, terms of type $\kctm$ constructed arbitrarily may not
%% correspond to any kind of terms in Figure\cspc\ref{fig:cg_targ_lang}.
%% To identify the well-formed constants, expressions, statements,
%% functions and programs in the target language, we respectively use the
%% predicate constants $\kcst$, $\kexp$, $\kstmt$, $\kfun$ and $\kprg$ of
%% the type $\kctm \to \omic$. Their are defined through a \LProlog
%% program in an obvious way.
%

\pagelabel{ecd:cg_rules}
We then describe the encoding of the transformation rules. We identify
the following predicate constant to represent the transformation
relation $\cgen \rho M K {S'}$:
\begin{tabbing}
\qquad\=\kill
\>$\kcgen :  \kctm \to (\kctm \to \kctm) \to \kctm \to \omic.$
\end{tabbing}
Similar to the encoding of the CPS transformation, we use meta-level
$\beta$-redexes to represent administrative $\beta$-redexes.
The transformation rules in Figure\cspc\ref{fig:cg_rules}
transparently translate into clauses for \kcgen. Most of the 
clauses have similar structures as those for the
CPS transformation. We only talk about the interesting ones, including
the clauses representing $\cgenpair$, $\cgenfst$, $\cgensnd$,
$\cgenclos$ and $\cgenopen$ listed as follows:
\begin{tabbing}
\qquad\=\quad\=$(\sfall {x_1}. \kcgen\app $\=$f,e$\=\kill
\>$\kcgen\app (\kcpair\app M_1\app M_2)\app K\app M' \limply$\\
\>\>$(\sfall {x_1}. \kcgen\app M_2$\\
\>\>\>$(x_2 \mlam \kclet\app (\kalloc\app (\ks\app (\ks\app \kz)))$\\
\>\>\>$\quad(p\mlam \kclet\app (\kmove\app p\app \kz\app x_1)\app$\\
\>\>\>$\quad(u_1\mlam \kclet\app (\kmove\app p\app (\ks\app \kz)\app x_2)$\\
\>\>\>$\quad(u_2\mlam K\app p))))$\\
\>\>\>$(M_2'\app x_1)) \scomma$\\
\>\>$\kcgen\app M_1\app M_2'\app M'.$\\
\>$\kcgen\app (\kcfst\app M)\app K\app M' \limply
  \kcgen\app M\app (x\mlam \kclet\app (\kload\app x\app \kz)\app (v\mlam K\app v))\app M'.$\\
\>$\kcgen\app (\kcsnd\app M)\app K\app M' \limply
    \kcgen\app M\app (x\mlam \kclet\app (\kload\app x\app (\ks\app \kz))\app (v\mlam K\app v))\app M'.$\\
\>$\kcgen\app (\kclos\app M_1\app M_2)\app K\app M' \limply 
     \kcgen\app (\kcpair\app M_1\app M_2)\app K\app M'.$\\
\>$\kcgen\app (\kopen\app M_1\app 
                (f\mlam e\mlam \kcapp\app f\app (\kcpair\app M_1\app (\kcpair\app M_2\app e))))\app 
              K\app M' \limply$\\
\>\>$(\sfall f,e,x_1. \kcgen\app M_2$\\
\>\>\>\>$(x_2\mlam \kclet\app (\kalloc\app (\ks\app (\ks\app \kz)))$\\
\>\>\>\>$\quad(p_1\mlam\kclet\app (\kmove\app p_1\app \kz\app x_2)$\\
\>\>\>\>$\quad(v_1\mlam\kclet\app (\kmove\app p_1\app (\ks\app \kz)\app e)$\\
\>\>\>\>$\quad(v_2\mlam\kclet\app (\kalloc\app (\ks\app (\ks\app \kz)))$\\
\>\>\>\>$\quad(p_2\mlam\kclet\app (\kmove\app p_2\app \kz\app x_1)$\\
\>\>\>\>$\quad(v_3\mlam\kclet\app (\kmove\app p_2\app (\ks\app \kz)\app p_1)$\\
\>\>\>\>$\quad(v_4\mlam\kclet\app (\kcapp\app f\app p_2)\app (v\mlam K\app v))))))))$\\
\>\>\>$(M_3\app f\app e\app x_1)) \scomma$\\
\>\>$\kcgen\app M_1\app (x_1\mlam \kclet\app (\kload\app x_1\app \kz)\app
                        (f\mlam \kclet\app (\kload\app x_1\app (\ks\app \kz))\app
                        (e\mlam M_3\app f\app e\app x_1)))$.
\end{tabbing}
In the clause encoding $\cgenpair$, we use the terms of \kalloc,
\kmove embedded in let expressions to represent the sequence of
instructions for allocating memory cells for a pair and initializing
the cells with elements of the pair. Similarly in the clause
encoding $\cgenfst$ and $\cgensnd$, we use terms of \kload to
represent the instructions for reading values from the cells for
the pairs. The clauses encoding $\cgenclos$ and $\cgenopen$ are direct
translations from the corresponding rules. As usual, the freshness
side conditions of these rules are captured via universal goals.

We identify the following predicate constants to represent the
transformation relations $\cgenf \rho F {F'}$ and $\cgenp \rho M {P}$:
\begin{tabbing}
\qquad\=\kill
\>$\kecgen : \klist\app \kctm \to \klist\app \kctm \to \omic.$\\
\>$\kbcgen : \kctm \to \kctm \to \omic.$\\
\>$\khcgen : \kctm \to \kctm \to \omic.$
\end{tabbing}
Their definitions are given as follows:
\begin{tabbing}
\qquad\=\quad\=\quad\=\kill
\>$\kecgen\app \knil\app \knil.$\\
\>$\kecgen\app 
     ((\kcabs\app l\mlam \kcabs\app x\mlam F\app l\app x) \cons \FE)\app
     ((\kcabs\app l\mlam \kcabs\app x\mlam F'\app l\app x) \cons \FE') \limply$\\
\>\>$(\sfall l,x. 
        (\forallx k {\kcgen\app l\app k\app (k\app l)}) \simply 
        (\forallx k {\kcgen\app x\app k\app (k\app x)}) \simply$\\
\>\>\>$\kcgen\app (F\app l\app x)\app (x\mlam x)\app (F'\app l\app x)) \scomma
          \kecgen\app \FE\app \FE'.$
\\[1ex]
\>$\kbcgen\app (\khbase\app M)\app (\khbase\app M') \limply \kcgen\app M\app (x\mlam x)\app M'.$\\
\>$\kbcgen\app (\khabs\app R)\app (\khabs\app R') \limply
    \forallx {x} {(\forallx k {\kcgen\app x\app k\app (k\app x)})} \simply 
        \kbcgen\app (R\app x)\app (R'\app x).$
\\[1ex]
\>$\khcgen\app (\khtm\app \FE\app M)\app (\khtm\app \FE'\app M') \limply
     \kecgen\app \FE\app \FE' \scomma \kbcgen\app M\app M'.$
\end{tabbing}
As usual, we use universal and hypothetical goals to preform recursion
over binding operators and to introduce the rules for transforming
variables.
By definition $\khcgen\app M\app P$ holds if $M$ encodes code that has
the form of outputs of code hoisting and $P$ encodes the result of apply code generation
to $M$. The constant \kecgen represents the transformation of the
top-level functions and \kbcgen represents the transformation of the
body of $M$. Note that in the base case of \kbcgen we use $\kcgen$
with the initial context $(x \mlam x)$ to generate the target code
from the body of the program. Similarly, we use \kcgen in the definition of
\kecgen to generate code from the bodies of functions.

\subsection{Informal verification of the transformation}
\label{subsec:informal_verify_ch}

Because code generation closely corresponds to the CPS transformation,
its verification is carried out in a way similar to the CPS
transformation. The only interesting part of the verification is to
identify an operational semantics for the target language of code
generation. In the rest of this section, we give the rule-based
description of this semantics and its encoding and briefly talk about
the verification of code generation using this semantics.

We first describe the representation of and the operations on
heaps. We use $H$ to denote heaps. A heap with no allocated cells is
written as $\emptyset$. A heap with cells indexed from $0$ to $n$
allocated such that $V_i$ is the content of the $i$-th cell is written
as $\heap{0 \to V_0, 1 \to V_1, \ldots, n \to V_n}$. We define the
following operations for allocation, modification and query of the
contents of a heap:
\begin{itemize}
\item
  The function $allocate$ for allocating memory such that $H' =
  \allocate{H}{i}$ if $n$ is the smallest index of the unallocated
  cells in $H$ and $H'$ is $H$ extended with the mappings $\heap{n \to
    \unit, \ldots, n+i-1 \to \unit}$. That is, $\allocate{H}{i}$
  allocates $i$ fresh memory cells in $H$ and set their default
  values to $\unit$.
\item
  The function $update$ for updating the content of a memory cell such
  that $H' = \update{H}{i}{V}$ if $H$ contains a mapping for $i$ and
  $H'$ is obtained from $H$ by replacing this mapping with $(i \to
  V)$.
\item
  The function $lookup$ for reading the content of a memory cell such
  that $V = \lookup{H}{i}$ if $H$ contains the mapping $(i \to V)$.
\end{itemize}

\pagelabel{txt:cg_eval}
We identify the following classes of terms for describing evaluation
where $V$ denotes values, $B$ denotes values or variables and $D$ and
$U$ respectively denote the intermediate forms of expressions and
statements in the evaluation process.
\begin{tabbing}
  \qquad\qquad\qquad\qquad\=$T$ \quad\=$::=$ \quad\=\kill
  \>$V$ \>$::=$ \>$n \sep \unit \sep \loc n \sep \abs x U$\\
  \>$B$ \>$::=$ \>$V \sep x$\\
  \>$D$ \>$::=$ \>$B \sep \pred B \sep B_1 + B_2 \sep
  \ifz {B} {U_1} {U_2} \sep$ \\
  \>\>\>$(B_1\app B_2) \sep \alloc n \sep \move {B_1} n {B_2} \sep \load {B} n$
  \\[1ex]
  \>$U$ \>$::=$ \>$D \sep \letexp x U U$
\end{tabbing}
We now describe an operational semantics based on a left-to-right,
call-by-value evaluation strategy and in a small-step form. One step
of evaluation takes a statement $U$ and a heap $H$ as input, evaluates
the first instruction in $U$ based on the content of $H$ and outputs
the modified statement and heap. The changes to the heap reflects the side
effects incurred by the evaluation.
We write $H, M \step{1} H', M'$ where $M$ is either an intermediate
expression or statement to denote that $M$ evaluates to $M'$ in one
step and the evaluation changes the heap $H$ to $H'$. We write $H, M
\step{n} H', M'$ where $n > 0$ to denote the $M$ evaluates to $M'$ in
$n$ steps and $H$ is changed to $H'$ by the evaluation and $P \eval H,
V$ to denote that a program $P$ evaluates to the value $V$ starting
with an empty heap and the final memory state is represented by $H$.
Figure\cspc\ref{fig:cg_eval_rules} depicts the rules defining these
relations.

\begin{figure}[!ht]
  \begin{gather*}
    \infer[]{
      H, \pred n \step{1} H, n'
    }{
      n' \text{ is the predecessor of } n
    }
    \qquad
    \infer[]{
      H, n_1 + n_2 \step{1} H, n_3
    }{
      n_3 \text{ is the sum of } n_1 \text{ and } n_2
    }
    \\
    \infer[]{
      H, \ifz 0 {M_1} {M_2} \step{1} H, M_1
    }{}
    \;\;
    \infer[]{
      H, \ifz n {M_1} {M_2} \step{1} H, M_2
    }{n > 0}
    \\
    \infer[]{
      H, \alloc n, H \step{1} H', \loc i
    }{
      i = \mbox{the number of allocated cells in $H$}
      &
      H' = \allocate H n
    }
    \\
    \infer[]{
      H, \move {(\loc n)} i V \step{1}  H', \unit
    }{
      H' = \update H {(n+i)} V
    }
    \qquad
    \infer[]{
      H, \load {(\loc n)} i \step{1} H, V
    }{
      V = \lookup H {(n+i)}
    }
    \\
    \infer[]{
      H, \letexp x {M_1} {M_2} \step{1} H', \letexp x {M_1'} {M_2}
    }{
      M_1, H \step{1} M_1', H'
    }
    \\
    \infer[]{
      H, \letexp x {V} {M} \step{1} H, M[V/x]
    }{}
    \qquad
    \infer[]{
      H, {(\abs x U)}\app V \step{1} H, U[V/x]
    }{}
    \\[2ex]
    \infer{
      H, M \step{0} H, M
    }{}
    \qquad
    \infer{
      H, M \step{n+1} H'', M''
    }{
      H, M \step{1} H', M'
      &
      H', M' \step{n} H'', M''
    }
    \\
    \infer{
      \letfun {\vec{f} = \vec{F}} {S} \eval H, V
    }{
      \emptyset, S[\vec{F}/\vec{f}] \step{n} H, V
    }
  \end{gather*}
  \caption{Evaluation Rules for the Target Language of Code Generation}
  \label{fig:cg_eval_rules}
\end{figure}

\pagelabel{txt:cg_sem_pres}
Given the evaluation semantics, we can prove that code generation
preserves semantics by following a development similar to that for the
CPS transformation. In the end, we can prove the following theorem
that code generation on statements preserves semantics:
\begin{mythm}\label{thm:cgen_sem_pres_atom}
  If $\stseq M : \tnat$, $\cgen \emptyset M K {M'}$ and $M \eval V$,
  then $\emptyset, M' \step{n} H, (\capp K V)$ for some $n$ and $H$.
\end{mythm}
\noindent From this theorem, it is easy to derive the following theorem which
states code generation on programs preserves semantics:
\begin{mythm}\label{thm:cgenp_sem_pres_atom}
  If $\stseq P : \tnat$, $\cgenp \emptyset P {P'}$ and $P \eval V$,
  then $P' \eval H, V$ for some $H$.
\end{mythm}

\subsection{Verifying the \LProlog implementation in \Abella}
\label{subsec:formal_verify_ch}

We first formalize the representation of and operations on memories in
\LProlog. We represent a heap as a list of the type $(\klist\app
(\kmap\app \knat\app \kctm))$. It contains the mappings from indexes
to the values for the allocated cells. We use the type \kstate to
represent the memory state. The sole constructor of \kstate is $\kst :
\knat \to (\klist\app (\kmap\app \knat\app \kctm)) \to \kstate$ which
takes a heap $H$ and the smallest index of the allocated cells in $H$
to form a memory state. We then identify the following predicate
constants to represent the allocation, update and look-up operations
on heaps:
\begin{tabbing}
\qquad\=\quad\=\kill
\>$\kallocate : \knat \to \knat \to (\klist\app (\kmap\app \knat\app \kctm)) \to 
                (\klist\app (\kmap\app \knat\app \kctm)) \to \omic.$\\
\>$\kupdateheap : (\klist\app (\kmap\app \knat\app \kctm)) \to \knat \to 
                   \kctm \to (\klist\app (\kmap\app \knat\app \kctm)) \to \omic.$\\
\>$\klookupheap :  (\klist\app (\kmap\app \knat\app \kctm))\app \to \knat \to \kctm \to \omic.$
\end{tabbing}
They are defined as follows:
\begin{tabbing}
\qquad\=\quad\=\kill
\>$\kallocate\app N\app \kz\app H\app H.$\\
\>$\kallocate\app N\app (\ks\app S)\app H\app H' \limply
     \kallocate\app (\ks\app N)\app S\app (\kmap\app N\app \kunit \cons H)\app H'.$\\
\>$\kupdateheap\app (\kmap\app L\app V \cons H)\app L\app V'\app (\kmap\app L\app V' \cons H).$\\
\>$\kupdateheap\app (M \cons H)\app L\app V'\app (M \cons H') \limply 
     \kupdateheap\app H\app L\app V'\app H'.$\\
\>$\klookupheap\app (\kmap\app L\app V \cons H)\app L\app V.$\\
\>$\klookupheap\app (M \cons H)\app L\app V \limply \klookupheap\app H\app L\app V.$
\end{tabbing}

\pagelabel{ecd:cg_eval}
We then formalize the evaluation of expressions and
statements. We identify the following constants to represent the
evaluation relations:
\begin{tabbing}
\qquad\=\quad\=\kill
\>$\kccstep : \kstate \to \kctm \to \kstate \to \kctm \to \omic$ \\
\>$\kccnstep : \knat \to \kstate \to \kctm \to \kstate \to \kctm \to \omic.$\\
\>$\kccceval : \ktm \to \kstate \to \kctm \to \omic.$
\end{tabbing}
The constant \kccstep is used to represent the one-step evaluation
relation such that $\kccstep\app \St\app M\app \St'\app M'$ if and
only if the term $M$ evaluates to $M'$ in the state $\St$ and $\St$ is
changed to $\St'$ by the evaluation, \kccnstep is used to represent
the $n$-step evaluation relation, and \kccceval is used to represent
the evaluation of programs such that $\kccceval\app P\app H\app V$
holds if the program $P$ in an empty heap evaluates to $V$ and the
final heap is $H$. We summarize the clauses defining these constants which
are transparently translated from the rules in
Figure\cspc\ref{fig:cg_eval_rules} follows:
\begin{tabbing}
\qquad\=\quad\=\kill
\>$\kccstep\app \St\app (\kcpred\app (\kcnat\app N)) \St\app (\kcnat\app N') 
     \limply \knpred\app N\app N'.$\\
\>$\kccstep\app \St\app (\kcplus\app (\kcnat\app N_1)\app 
                (\kcnat\app N_2))\app \St\app (\kcnat\app N) \limply \kadd\app N_1\app N_2\app N.$\\
\>$\kccstep\app \St\app (\kcifz\app (\kcnat\app \kz)\app M_1\app M_2)\app \St\app M_1.$\\
\>$\kccstep\app \St\app (\kcifz\app (\kcnat\app (\ks\app N))\app M_1\app M_2) \St\app M_2.$\\
\>$\kccstep\app (\kst\app N\app H)\app (\kalloc\app S)\app (\kst\app N'\app H')\app (\kloc\app N) \limply$\\
\>\>$\kadd\app N\app S\app N' \scomma \kallocate\app N\app S\app H\app H'.$\\
\>$\kccstep\app (\kst\app N\app H)\app (\kmove\app (\kloc\app L)\app S\app V)\app 
     (\kst\app N\app H')\app \kunit \limply$\\
\>\>$\kadd\app L\app S\app L' \scomma \kupdateheap\app H\app L'\app V\app H'.$\\
\>$\kccstep\app (\kst\app N\app H)\app (\kload\app (\kloc\app L)\app S)\app (\kst\app N\app H)\app V \limply$\\
\>\>$\kadd\app L\app S\app L' \scomma \klookupheap\app H\app L'\app V.$\\
\>$\kccstep\app \St\app (\kclet\app M\app R)\app \St\app' 
                (\kclet\app M'\app R) \limply \kccstep\app \St\app M \St\app' M'.$\\
\>$\kccstep\app \St\app (\kclet\app V\app R)\app \St\app (R\app V) \limply \kccval\app V.$\\
\>$\kccstep\app \St\app (\kcapp\app (\kcabs\app R)\app V)\app \St\app (R\app V) \limply \kccval\app V.$
\\[1ex]
\>$\kccnstep\app \kz\app \St\app M\app \St\app M.$\\
\>$\kccnstep\app (\ks\app N)\app \St\app M\app \St''\app M'' \limply$\\
\>\>$\kccstep \St\app M\app \St'\app M' \scomma \kccnstep\app N\app \St\app' M'\app \St\app'' M''.$
\\[1ex]
\>$\kccceval\app (\khtm\app \knil\app (\khbase\app M))\app \St\app V 
     \limply \kccnstep\app N\app (\kst\app \kz\app \knil)\app M\app \St\app V \scomma \kccval\app V.$\\
\>$\kccceval\app (\khtm\app (F \cons \FE)\app (\khabs\app R))\app \St\app V 
     \limply \kccceval\app (\khtm\app \FE\app (R\app F))\app \St\app V.$
\end{tabbing}
Here the predicate constant $\kccval : \kctm \to \omic$ is used to
identify values in the target language.

\pagelabel{ecd:cg_sem_pres}
We formalize the informal development of semantics preservation proofs
in \Abella by following a way similar to that in which we develop the formal proof of the CPS
transformation. In the end, Theorem\cspc\ref{thm:cgen_sem_pres_atom}
is proved as the following theorem in \Abella:
\begin{tabbing}
\qquad\=\quad\=\kill
\>$\rfall M,K,M',V. \oseq{\kcof\app M\app \ktnat} \rimp \oseq{\kcgen\app M\app K\app M'} \rimp
   \oseq{\kceval\app M\app V} \rimp$\\
\>\>$\rexistsx {N,\St} {\oseq{\kccstep\app N\app (\kst\app \kz\app \knil)\app M'\app \St\app (K\app V)}}.$
\end{tabbing}
From this theorem it is easy to prove the following formalized version
of Theorem\cspc\ref{thm:cgenp_sem_pres_atom}:
\begin{tabbing}
\qquad\=\quad\=\kill
\>$\rfall P,P',V. \oseq{\kccof\app P\app \ktnat} \rimp \oseq{\khcgen\app P\app P'} \rimp
   \oseq{\kcceval\app P\app V} \rimp$\\
\>\>$\rexistsx {\St} {\oseq{\kccceval\app P'\app \St\app V}}.$
\end{tabbing}

\section{The Correctness Proof for the Full Compiler}
\label{sec:full_correctness}

We designate the constant $\kcompile : \ktm \to \kctm \to \omic$ to
represent the full compilation process. It is defined by the following
program clause in \LProlog:
\begin{tabbing}
\qquad\=\quad\=\kill
\>$\kcompile\app M\app M' \limply$\\
\>\>$\kcps\app M\app (x\mlam x)\app M_1 \scomma
     \kcc\app \knil\app \knil\app M_1\app M_2 \scomma$\\
\>\>$\kch\app M_2\app (\khtm\app \FE\app M_3) \scomma
     \khcgen\app (\khtm\app \FE\app M_3)\app M'.$
\end{tabbing}

To prove the correctness of the full compiler, we need the type and
semantics preservation theorems of its individual transformations. We
summarize the type preservation theorems corresponding to
Corollaries\cspc\ref{coro:cps_typ_pres_closed_id},
\ref{coro:cc_typ_pres_closed} and \ref{coro:ch_typ_pres_closed} as
follows:
\begin{tabbing}
\qquad\=\quad\=\kill
\>$\rfall M,M'.
     \oseq{\kof\app M\app \ktnat} \rimp
     \oseq{\kcps\app M\app (x\mlam x)\app M'} \rimp
     \oseq{\kof\app M'\app \ktnat}$.\\
\>$\rfall M, M', T.
      \oseq{\kof\app M\app T} \rimp \oseq{\kcc\app \knil\app \knil\app M\app M'} 
      \rimp \oseq{\kcof\app M'\app T}.$\\
\>$\rforallx {M,M',T} {
   \oseq{\kcof\app M\app T} \rimp 
   \oseq{\kch\app M\app M'} \rimp
   \oseq{\kccof\app M'\app T}}.$
\end{tabbing}
Note that the code generation transformation does not preserve types
because its target language is typeless. We summarize the semantics
preservation theorems corresponding to
Corollaries\cspc\ref{coro:cps_sem_pres_atom_id},
\ref{coro:cc_sem_pres_atom}, \ref{coro:ch_sem_pres_atom} and
\ref{thm:cgenp_sem_pres_atom} as follows:
\begin{tabbing}
\qquad\=\quad\=\kill
\>$\rfall M, M', V.
     \oseq{\kof\app M\app \ktnat} \rimp \oseq{\kcps\app M\app (x\mlam x)\app M'} \rimp$\\
\>\>$\oseq{\keval\app M\app V} \rimp \oseq{\keval\app M'\app V}.$\\
\>$\rfall K, T, M, M'. \oseq{\kof\app M\app \ktnat} \rimp 
    \oseq{\kcc\app \knil\app \knil\app M\app M'} \rimp$\\
\>\>$\oseq{\keval\app M\app (\knat\app N)} \rimp 
     \oseq{\kceval\app M'\app (\kcnat\app N)}$\\
\>$\rfall M, K, M', V.
     \oseq{\kcof\app M\app \ktnat} \rimp \oseq{\kch\app M\app (\khtm\app \FE\app M')} \rimp$\\
\>\>$\oseq{\kceval\app M\app (\kcnat\app N)} \rimp 
     \oseq{\kcceval\app (\khtm\app \FE\app M')\app (\kcnat\app N)}$\\
\>$\rfall P,P',V. \oseq{\kccof\app P\app \ktnat} \rimp \oseq{\khcgen\app P\app P'} \rimp
   \oseq{\kcceval\app P\app V} \rimp$\\
\>\>$\rexistsx {\St} {\oseq{\kccceval\app P'\app \St\app V}}.$
\end{tabbing}

The correctness theorem of the full compiler is stated as follows:
\begin{tabbing}
\qquad\=\quad\=\kill
\>$\rfall M,M',N. 
              \oseq{\kof\app M\app \ktnat} \rimp
              \oseq{\kcompile\app M\app M'} \rimp
              \oseq{\keval\app M\app (\knat\app N)} \rimp$\\
\>\>$\rexistsx {\St} {\oseq{\kccceval\app M'\app \St\app (\kcnat\app N)}}$
\end{tabbing}
It is proved by analyzing $\oseq{\kcompile\app M\app M'}$ and
interleaving the applications of the above type and semantics
preservation theorems in an obvious way.

%%%%%%%%%%%%%%%%%%%%%%%%%%%%%%%%%%%%%%%%%%%%%%%%%%%%%%%%%%%%%%%%%%%%%%%%%%%%%%%%

% Related work
%%%%%%%%%%%%%%%%%%%%%%%%%%%%%%%%%%%%%%%%%%%%%%%%%%%%%%%%%%%%%%%%%%%%%%%%%%%%%%%
% related.tex: Related work
%%%%%%%%%%%%%%%%%%%%%%%%%%%%%%%%%%%%%%%%%%%%%%%%%%%%%%%%%%%%%%%%%%%%%%%%%%%%%%%%
\chapter{Related Work}
\label{ch:related}

Compiler verification is an old topic, interest in which can be traced back to
1960s~\cite{mccarthy67}. The past decade has witnessed impressive
developments on mechanizing compiler verification, due partly to the
maturation of formal verification tools. Many of these developments
have focused on implementing and verifying compilers for imperative
programming languages such as C, C++
and Java; see \cite{dave03sigsoft} for a catalog of these
developments. Among these
efforts, the most notable and influential has been that of the
CompCert project that has developed a verified multi-pass compiler for a
subset of C using the Coq theorem prover~\cite{leroy09cacm}. The
correctness of
the CompCert compiler has been proved by establishing the permutability of
evaluation and the compilation process as we have described in
Section~\ref{sec:sem_pres_nuances}. As such, it only makes sense for
compilation of full programs. Recently, researchers have also begun
investigating the separate compilation of program modules and a
composible way to verify such compilation~\cite{stewart15popl}.
The CompCert project has shown that verification of non-trivial compilers
for realistic programming languages is feasible with the state-of-art
verification tools. This project has also provided the impetus for
other efforts
such as the Verified Software Toolchain project~\cite{appel14book}
related to overall program verification.

Our focus in this thesis has been on the implementation and formal
verification of compilers for functional programming languages.
In contrast to the compilation of imperative languages,
%% This compared to the
%% verified compilation of imperative languages is much more complicated
%% because of the necessity to transform highly abstracted functional
%% programs into low-level executable code and the requirement for
%% representing, manipulating, analyzing and reasoning about binding
%% structure in compiler transformations. Despite of these difficulties,
this task requires the transformation of more abstract
programs into low-level executable code and brings with it the need to
represent, manipulate, analyze and reason about binding
structure in compiler transformations. While there are new
difficulties in this area, there have also been
efforts targeted at overcoming them. These efforts have ranged from
verifying individual compiler transformations to ones with a much more
ambitious scope such as the verification of complete compilers for
realistic functional languages. We discuss these efforts in the rest of
this chapter, paying particular attention to how they deal with binding
structure and how this impacts on the development of proofs.
Because there are significant differences in scope and focus
and in the theories and tools used in implementation
and verification, a direct comparison of our work with many of the
projects described below is neither feasible nor sensible.
%
%% However, we do try
%% to compare one project closely related to our work from some sensible
%% perspectives.
%
However, we do try to make comparative assessments where this seems
possible and sensible.

A large part of the existing work on verified compilation of
functional languages makes use of general theorem provers such as
Coq~\cite{bertot04book}, Isabelle~\cite{nipkow02book} and
HOL~\cite{gordon91tphol}.
In~\cite{dargaye09phd} Dargaye describes a verified compiler for a
subset of ML, which extends the simply typed $\lambda$-calculus with
$n$-ary functions, recursive functions, data types and pattern
matching. The compilation process for this language is similar to
ours: it transforms source programs through multiple passes, including
the CPS transformation, closure conversion and code generation, into the
Cminor intermediate language of CompCert. The compiler transformations
are verified by following the approach adopted by CompCert,
\ie, by showing the permutability between evaluation and
transformation. The verification of the full compiler is obtained by
utilizing the result previously established within CompCert of the
correctness of the translation from Cminor to low-level code.
In~\cite{chlipala07pldi}, Chlipala develops a verified multi-pass
compiler for the simply typed $\lambda$-calculus in Coq by using
logical relations as the notion of semantics preservation. In a manner
similar to what we did in Section~\ref{sec:full_correctness}, Chlipala
composed the correctness proofs of individual transformations to
prove that closed programs at atomic types have the same behavior
before and after compilation.
In~\cite{benton09icfp}, Benton and Hur have implemented a verified
single-pass compiler that takes programs in an extension of
PCF as input and outputs virtual code executable an SECD
machine~\cite{landin64}. Their notion of semantics preservation is based on a
step-indexing logical relation. Because logical relations satisfy the
modularity property described in Section~\ref{sec:sem_pres_nuances},
the semantics preservation property of this compiler makes sense for
program modules with external references. Moreover, Benton and Hur
have not only proved semantics preservation, they have also proved
that the compiler is \emph{fully abstract}, namely two pieces of
source programs are contextual equivalent if and only if their
compiled versions are. As a result, properties proved about program
equivalence at the source level are still valid after the compilation.

Because general theorem provers do not have built-in support for
bindings, users have to implement it explicitly or use a library based
on some first-order or higher-order approach to treating bindings. All
of the above work uses standard de Bruijn indexes or its variant for
representing bindings, bearing the burden of explicitly representing,
manipulating and reasoning about binding related notions such as
substitution and renaming. These difficulties are best illustrated by
considering the treatment of the CPS transformation in the Danvy and
Filinski style. Chlipala has used standard de Bruijn indexes to represent both
the administrative and dynamic abstractions in the transformation and
explicitly implemented all the binding related operations and proved
their properties~\cite{chlipala07pldi}. To alleviate the
effort, Dargaye has used two kinds of de Bruijn indexes, one for
representing administrative abstractions, another for dynamic
abstractions~\cite{dargaye09phd,dargaye07lpar}. Still she needs to
explicitly implement and reason about such binding representations. In
\cite{minamide03merlin}, Minamide and Okuma have also formally
verified a Danvy Filinski style CPS transformation using
Isabelle/HOL. They argue that de Bruijn indexes are not effective for
formalizing the CPS transformation because they are sensitive to the
changes to the context. For example, the reduction of an
administrative $\beta$-redex will cause the shifting of de Bruijn
indexes. To avoid this problem they use named variables to represent
abstractions and realize $\alpha$-equivalence via explicit renaming.
In contrast, by adopting an \HOAS approach we are able to use meta-level
abstractions to represent administrative abstractions, which
eliminates the necessity to explicitly encode renaming and
substitution and to reason about such encoding. This is also observed
in \cite{tian06cats} by Tian who has used \Twelf to encode a CPS
transformation and has provided a simple correctness proof for the
transformation via the \HOAS approach (We shall discuss compiler
verification work done using \Twelf more thoroughly later in the
chapter.)

The most comprehensive work on verified compilers for functional
languages we know of is the CakeML compiler which supports a
substantial subset of Standard ML~\cite{kumar14popl}. CakeML is
designed to serve as a platform for running verified software with a
reduced trusted computing base. The compiler is implemented as an
interactive read-eval-print loop that compiles Standard ML programs
into x86-64 machine code through a sequence of transformations. Its
verification is done by using HOL4 and essentially based on showing
permutability of evaluation and compilation. A distinguish
characteristic of CakeML is that it is bootstrapped: the compiler is
fed as a source program into itself to generate an x86-64
implementation of it. Moreover, it is proved that the correctness of
the source compiler is preserved by the compilation. Therefore we
get a verified x86-64 implementation of CakeML. CakeML represents
bindings through named variables in its source language and through de
Bruijn indexes in its intermediate language. As a result, all the
properties about bindings must be proved explicitly based on those
representations. For instance, the closure conversion transformation
in CakeML represents an environment for a function as a list of de
Bruijn indexes pointing to the abstractions binding the free variables
in the function body. The encoding of closure conversion replaces the
de Bruijn indexes of the free variables in the function body with
pointers to the elements in the environment. Such manipulation of de
Bruijn indexes must be reasoned about explicitly in the correctness
proof of closure conversion. All this takes significant effort that is
obviously orthogonal to the actual task that is of interest.

In \cite{chlipala08icfp} Chlipala has tried to alleviate the
difficulties in dealing with bindings in general theorem provers by
introducing the \emph{Parametric Higher Order Abstract Syntax} or
\emph{PHOAS} and applied PHOAS to re-implement the verified compiler
for the \STLC introduced in \cite{chlipala07pldi}. PHOAS is a
further development of an approach for treating bindings known as the
\emph{weak higher-order abstract syntax} or \emph{weak
  HOAS}\cite{despeyroux95tlca,honsell01icalp}. In weak \HOAS, we
designate a type $\kvar$ for representing variables which is separate
from the type $\ktm$ for representing terms and we use abstractions
over $\kvar$ in the meta-language to encode object-level abstractions
over variables. For instance, to encode the \STLC we can introduce a
constructor $\kvar : \kvar \to \ktm $ for encoding variables, $\kapp :
\ktm \to \ktm \to \ktm$ for encoding applications and $\kabs : (\kvar
\to \ktm) \to \ktm$ for encoding object-level abstractions. By
embedding those constructors in an inductive definition for the type
$\ktm$, we partially derive the benefits of the \HOAS approach, such
as that renaming is modeled by $\alpha$-conversion. However, because
variables are separated from ordinary terms, substitutions in weak
\HOAS cannot be represented elegantly through $\beta$-conversion like
in \HOAS; more specifically, the term $(M\app N)$ where $M$ has the
type $(\kvar \to
\ktm)$ and $N$ has the type $\ktm$ is not well-typed. Instead, we must
explicitly define substitutions as a relation or function that
traverses a term and replaces variables with terms. Moreover,
properties about substitutions must be proved explicitly using their
representation. PHOAS inherits those characteristics of weak \HOAS and
further \emph{parameterizes} the type $\ktm$ and its constructors with
the type of variables instead of using the fixed type $\kvar$. As a
result, it is possible to choose different representations for
variables to suit the need of users. Chlipala demonstrates the
usefulness of this parameterization in implementing and verifying a
compiler for the \STLC in \cite{chlipala07pldi}. For verifying 
compiler transformations that need to check identity of variables 
such as closure conversion, he instantiates the variable type with
the type of natural numbers for facilitating the identity check. 
For this, he needs to define the well-formedness of the 
terms of the instantiated types and to prove that the every 
parametric term is well-formed under this instantiation. 
This incurs an extra level of complexity 
in reasoning about terms with bindings.

Some comparison of our compiler is possible with the above compiler
for the \STLC by Chlipala.  Chlipala's implementation of closure
conversion comprises about 400 lines of Coq code, in contrast to about
70 lines of \LProlog code that are needed in our implementation.
Chlipala's proof of correctness comprises about 270 lines but it
benefits significantly from the automation framework that was the
focus of his work; that framework is built on top of the already
existing Coq libraries and consists of about 1900 lines of code. The
\Abella proof script runs about 1600 lines.
We note that \Abella has virtually no automation and no library. We
also note that,
in contrast to Chlipala's work, our development treats a version of
the \STLC that includes recursion. This necessitates the use of a
step-indexed logical relation which makes the overall proof more
complex.

There has also been some work on implementing and verifying compiler
transformations using the functional higher-order approach; it is to
be noted that the  ``verification'' part of such work has typically
been quite weak or even non-existent.
%GN Do you really know that this is the reason? Don't throw in things
%that are not substantiated or difficult to authenticate.
%% because of the difficulties in analysis of binding structure using
%% this approach.
Guillemette has encoded a CPS transformation in Haskell
using this style~\cite{guillemette07plpv}. He argues that the type checking in
Haskell ensures typing is preserved by the transformation. Because of
the incapability of analyzing variables in Haskell, closure conversion
cannot be encoded by using the functional higher-order
approach. Instead, he falls back to de Bruijn indexes for encoding
closure conversion and proves that the transformation is type
preserving~\cite{guillemette07haskell}.  Because of the inability to
analyze higher-order objects in the functional higher-order
approach, none of the work above has proved semantics preservation of
the encoded transformations. Hickey and Nogin have proposed an
implementation of closure conversion in the MetaPRL logical framework
that is based on the functional higher-order
approach~\cite{hickey06hosc}. Rather than analyzing the body of a
function to compute what variables appear free in it and thereby to
form a suitable environment, they take the simplistic approach of
constructing the environment based on all the abstractions the function
term appears under. They also do not prove the correctness of their
implementation because the MetaPRL
framework offers no capabilities for verification.
%

%We now examine the existing work on verified compilation of functional
%languages using the \HOAS approach.
The earliest work related to using the \HOAS approach in compiler
verification seems to be that of
Hannan and Pfenning in which they use \Twelf (called Elf at that time)
as the specification and reasoning vehicle~\cite{hannan92lics}. In
that paper they have implemented
and verified a compiler from the \STLC to a variant of the Categorical
Abstract Machine~\cite{cousineau87}. The compiler consists of some
very simple transformations such as conversion of $\lambda$-terms into
their de Bruijn forms. Since this work, there have been investigations of more
complicated compiler transformations for functional languages using
\Twelf. In \cite{murphy08modal} Murphy has used \Twelf to
verify the type preservation of the CPS transformation and closure conversion 
for a programming language for distributed computing based on a modal logic.
In \cite{tian06cats} Tian implemented and
verified a Danvy-Filinski style CPS transformation for a slight
extension of the \STLC. All of the described efforts exploit the support for the
\HOAS approach that \Twelf provides to simplify the implementation and
verification of binding related operations in the
transformations. 

The studies of compiler transformations on functional languages using
\Twelf have all been based on the permutability of evaluation and
transformation. As we have discussed in
Section\cspc\ref{sec:sem_pres_nuances}, permutability is not very
flexible as a notion of semantics preservation. Recently, researchers
have proposed many different notions of semantics preservation that
possess most or all of the desired properties such as modularity,
flexibility and transitivity as described in
Section\cspc\ref{sec:sem_pres_nuances} (\eg, see
\cite{neis15icfp,benton09icfp,perconti14esop}). These notions are
either more complex forms of logical relations or have similar
characteristics as logical relations. This suggests that the ability
to effectively treat logical relations in a verification framework
that supports HOAS may greatly simplify the verification of compilers
for functional languages based on those more powerful notions of
semantic preservation.
However, this idea has not been explored using \Twelf, perhaps because
%the way in which logical properties can be expressed in Twelf
%is restricted; in particular, it is not easy to encode a logical
it is not easy to encode a logical relation-style definition within it
(see \cite{schurmann08lics} for discussions about a way to achieve this.)

The \Beluga system~\cite{pientka10ijcar}, which implements a
functional programming language based on contextual modal type theory
and also supports \HOAS~\cite{nanevski08tocl}, overcomes some of the
shortcomings of \Twelf. Reasoning in \Beluga is based on the idea that
by type checking programs and by ensuring that they satisfy certain
coverage and termination conditions, we can obtain a guarantee of their
correctness with respect to the properties expressed by the types.
Coupling this approach with an expressive type system that is able
to express rich properties of programs results in a framework that has
the potential for proving deep properties, such as semantics
preservation, for the programs we write. 
%% idea: rich properties of programs can be embedded in types in \Beluga;
%% by constructing a program of a particular type and checking that it
%% satisfies certain coverage and termination conditions, we obtain a
%% program that satisfies the property encoded by that type.
Belanger
\etal\ have shown how this idea can be used to ensure
type preservation for implementations of CPS transformation, closure
conversion and code hoisting in Beluga~\cite{belanger13cpp}.
%GN These statements sound a little silly: what can you compare
%directly with your work then? More to the point, there are
%interesting common points to the two approaches and these sentences
%obscures that issue.
%% A direct comparison between this
%% work and our work is not very fruitful because, first, it does not
%% deal with semantics preservation which is the central part of our verification work
%% and, second, there are fundamental
%% differences between the approaches to implementation and reasoning in
%% Beluga and our framework: For implementation, \Beluga is based on
%% functional programming and \LProlog is based on logic programming; For
%% reasoning, the program is correct (type preserving in the case of \cite{belanger13cpp}) by
%% construction in \Beluga and verification is carried out automatically
%% while the proofs in \Abella must be explicitly developed in an
%% interactive style.
There are many similarities between the use made of \HOAS in the cited
work and that described in this thesis, although the property
considered there is weaker than the semantics preservation that we
have been concerned with here.  
Recently, Cave and Pientka have shown that stronger properties, such
as those based on logical relations, can be reflected into the types
in Beluga and shown to hold of programs through the methods it
provides\cspc\cite{cave15lfmtp}. However, these ideas have not yet
been applied to developing semantics preservation proofs
of the kind discussed in this thesis in \Beluga.

There has been recent work towards developing approaches to verified
compilation that satisfy modularity, flexibility and transitivity at
the same time. One such approach was proposed by Perconti and Ahmed in
\cite{perconti14esop} based on what is called \emph{multi-language
  semantics}. In particular, a ``big-tent'' language that encompasses
all the source, target and intermediate languages of the compiler is
defined and the program equivalence between different languages is
defined as contextual equivalence of terms in the big-tent
language. They then use logical relations to prove that the source
program module which may not be closed (\ie, it is a module containing
external references) is contextually equivalent to its compiled code
when embedded into the big-tent language. The correctness argument
using this approach currently only exists on paper. Another promising
approach, which we have already discussed in
Section~\ref{sec:sem_pres_nuances}, is that based on a notion of
semantics preservation called \emph{Parametric Inter-Languages
  Simulation} or \emph{PILS}\cspc\cite{neis15icfp}. Using PILS, Neis
\etal have developed a verified multi-pass compiler called Pilsner
that satisfies all the desired properties. Pilsner and its correctness
proof are formalized in Coq where explicit names are used to represent
bindings. At the beginning of the compilation, $\alpha$-renaming is
applied to establish the invariant that every variable is bound at
most once. This invariant is carefully maintain by the rest of the
transformations, thereby avoiding the problem of variable
capturing. We conjecture that using a \HOAS approach may enable a more
flexible approach to compilation and may also simplify the
verification effort even with a formulation of semantic correctness in
the style of PILS.

Formalizing binding related notions is not unique to verified
compilation of functional programs. It has been long recognized that
it is an important part of mechanizing the meta-theories of formal
systems involving types, programs, formulas and proofs. The
% GN Don't use words like ``famous'' to qualify projects, especially
% not POPLmark.
POPLMark challenge~\cite{aydemir05tphols} proposed a set of problems
for measuring the strength of formalization systems in dealing with
bindings. The posted solutions to those problems~\cite{poplmark15url}
make use of various kinds of approaches to representing binding
structure. We have discussed some of these approaches in
Section~\ref{sec:intro_binding} in the introductory chapter.

%Conclusion
%%%%%%%%%%%%%%%%%%%%%%%%%%%%%%%%%%%%%%%%%%%%%%%%%%%%%%%%%%%%%%%%%%%%%%%%%%%%%%%%
% conclusion.tex:
%%%%%%%%%%%%%%%%%%%%%%%%%%%%%%%%%%%%%%%%%%%%%%%%%%%%%%%%%%%%%%%%%%%%%%%%%%%%%%%%
\chapter{Conclusion and Future Work}
\label{ch:conclusion}

%% The section titled ``Verified Compiler Optimizations'' should be
%% given the title ``Verified Implementation of Realistic Compilers''
%% and moved to the second place in this chapter. It should point out
%% that 

%% The last section should focus on what we have learned from our
%% formalization work and what Abella still lacks to effectively
%% carried out such work. We then discuss how to address these
%% shortcomings in the future.

This thesis has demonstrated the effectiveness of the
Higher-Order Abstract Syntax approach in implementing and verifying
compilers for functional programming languages. In particular, it has
shown how this approach simplifies the treatment of binding related
notions which is an essential part of the task. In the demonstration,
we have used a
framework consisting of \LProlog, a language suitable for specifying
formal systems in a rule-based and relational style such that the
specifications are also executable programs, and \Abella, a theorem
proving system that provides rich mechanisms for reasoning about
\LProlog programs. Both \LProlog and \Abella support a realization of
the HOAS approach known as the $\lambda$-tree syntax approach. We have
proposed the following methodology for implementing and verifying
compilers for functional programming languages: we implement the
compiler transformations as \LProlog programs and we formulate and
prove the correctness of these implementations using
\Abella. Using this methodology we have developed a verified compiler
for a representative functional programming language. In doing so we
have demonstrated that the \HOAS approach provided by \LProlog and
\Abella significantly simplifies the treatment of bindings in both the
implementation and verification of the compilation of functional
programs, leading to formal correctness proofs that closely follow the
informal ones.

In the course of the above work, we have also addressed some
shortcomings of the \Abella system. First, we have extended \Abella so
that it can be used to reason about the full class of specifications
in \LProlog. Second, we have built support for a schematic form of
polymorphism into \Abella, thereby allowing us to use more modular
\LProlog implementations and also to modularize proofs of
correctness.

The work in this thesis leaves several interesting questions
unanswered that could provide the topics for further investigations.
We discuss some of these in the sections below.

\section{Verified Compilers for Realistic Functional Languages}
\label{sec:verify_real_fun_lang}

One natural continuation of our work is to apply our methodology to
verify compiler transformations for richer functional programming
languages. Such transformations often involve more complicated
manipulation of bindings. For example, in \cite{minamide96popl},
Minamide \etal have presented a description of the closure conversion
transformation for a polymorphic functional language and provided a
pencil-and-paper correctness proof based on logical relations. To
separate the code of a polymorphic function from its context, they
abstract the function not only over an environment binding the free
term variables of the function but also an environment binding the
free type variables. Moreover, to make the code shareable across
different sites where the function is called, they require the type
environment to have \emph{translucent types}~\cite{harper94popl}. It
is interesting to investigate if \HOAS can be used to simplify the
implementation and formal verification of such a transformation.

A more ambitious project would be to construct a verified compiler for
a realistic functional programming language, such as a substantial
subset of Standard ML or OCaml that supports features like data
structures, polymorphism, exceptions and modules. A possible and
interesting experiment would be to formalize the development of the
TIL compiler in Morrisett's Ph.D.  thesis~\cite{morrisett95phd} using
our framework. The TIL compiler transforms a subset of Standard ML to
Alpha assembly code through a sequence of transformations.  Its key
property is that it uses typed intermediate languages at all but the
lowest levels of compilation. TIL is able to generate efficient code
by exploiting the type information which is difficult to do if untyped
languages were used instead. For example, a common technique used by
compilers with untyped intermediate language for representing data
structures of types instantiated from polymorphic types is to tag the
data objects together with the instantiating types. To use such tagged
values, they must be unpacked first. The tagging and unpacking
operations are quite expensive and have significant impact on the
performance of the generated code. By using a technique called
\emph{dynamic type dispatch} that exploits the type information of
intermediate languages, TIL avoids tagging and unpacking data
structures. It would be
interesting to see how our framework can be applied to formally
develop the implementation and verification of a compiler like TIL and
if the \HOAS approach can benefit this formal development.

\section{Verified Implementation of Realistic Compilers}
\label{sec:verified_real_compiler}

The work of this thesis is just a starting point towards showing the
effectiveness of HOAS in verified implementation of realistic
compilers. We have implemented and verified the core transformations
for compiling functional languages to bring out the effectiveness
argument.
However, our compiler does not perform any kind of
optimization. Consequently, the code generated by it is not
very efficient. A realistic compiler should contain optimization
phases besides the core transformations for generating efficient code.
In the past programming language researchers have
developed various optimization transformations for functional
programs. Such transformations often involve complicated analysis of
the structure of functional programs, especially their binding
structure. An example would be \emph{lambda shrinking} proposed by
Appel and Jim~\cite{appel97jfp}. Lambda shrinking inlines functions or
values bound by let expressions when this operation does not increase
the code size, \ie, when the variable binding the value or function
occurs at most once in the program (excluding the binding site) and 
the arguments the function applies to are variables. Such
an inlining operation may expose further opportunities for inlining
because it may remove occurrences of variables bound at other
places. Lambda shrinking works roughly as follows: For every binding
variable it computes the number of times the variable occurs in the
program; It maintain and update this information as the inlining
operations go until no more inlining is possible. The analysis and
update carried out in lambda shrinking is quite complicated when a
first-order representation of bindings is used, as indicated in
\cite{appel97jfp}.
As we have shown in the thesis, our framework is able to carry out
non-trivial analysis on binding structure. An interesting research
topic would be to investigate how our methodology could help simplify
the mechanization of the optimization transformations for compiling
functional languages like lambda shrinking.

\section{Stronger Notions of Semantics Preservation}
\label{sec:deep_correctness}

In Section~\ref{sec:sem_pres_nuances} we have discussed three
criteria proposed by Neis \etal in \cite{neis15icfp} for measuring the
strength of notions of semantics preservation: modularity, flexibility
and transitivity. We have also discussed why logical relations do not
satisfy all three properties in that section. We would like to base
our compiler verification on richer notions of semantics preservation
that satisfy all these properties so that we get a stronger safety
guarantee.
One such notion is the Parametric Inter-Languages Simulation (PILS)
proposed in \cite{neis15icfp}.
%We have already exposed PILS in
%Section~\ref{sec:sem_pres_nuances} and Chapter~\ref{ch:related}.
The formulation of PILS in \cite{neis15icfp} has a
lot of similarities to logical relations: PILS is initially defined on
closed programs using their operational semantics and then extended to
open terms such that they are related by PILS if and only if the
results of applying any related and closed substitutions to them are
related. Neis \etal have developed two compilers, Pilsner and Zwickel,
to demonstrate the effectiveness of PILS, which according to their
paper ``has been a significant undertaking, involving several
person-years of work and over 55,000 lines of Coq''.  We suspect that
part of the difficulties come from how the binding related notions
are handled in their formal development. For example, they maintain an
invariant that every variable is bound at most once throughout the
compilation and they have used operational semantics which maintain
explicit environments containing bindings for free variables in
evaluation instead of the more natural operational semantics that base
evaluation on substitutions. It would be interesting to see how our
methodology can be used to alleviate their effort.
Eventually, we would like to leverage our understanding of this kind
of rich notions of semantics preservation to benefit the formal
verification of compilers for functional languages in realistic
settings.

\section{Further Extensions to the Framework}
\label{sec:further_exts}

While the framework comprising \LProlog and \Abella has significant
benefits in the verified implementation of compiler transformations
for functional languages, its current realization has some limitations
that make it insufficient for compiler verification in the more
realistic settings. By removing those limitations we can get a more
flexible system suitable for carrying out the tasks we
described in the previous sections.

One limitation of \Abella is that it lacks a module system for
managing proofs in large scales. This causes extra effort in managing
proofs, and more importantly, results in proofs that are not modular or
portable. We can already see this problem manifested in the
verification work in this thesis. First, we have to carefully assign
distinct names to definitions and theorems for individual compiler
transformations to avoid clashing of names. Second, even though some
of the definitions and theorems are only used locally for the type or
semantics preservation proofs for a particular transformation, they
are exposed to proofs for other transformations because there is no mechanism
for hiding such definitions and theorems; this leads to pollution of
name space and less abstracted proofs. 
We are interested in solving those problems by borrowing ideas from
other established interactive theorem provers such as Isabelle and Coq
to develop a module system for \Abella.

\Abella also has some practical limitations that lead to a larger proof
development effort than seems necessary.
One such limitation is the need to make explicit every step in the
process of constructing interactive proofs.
The effect of this requirement is especially
felt with respect to lemmas about contexts that arise routinely in the
$\lambda$-tree syntax approach: such lemmas have fairly obvious proofs
but, currently, the user must provide them to complete the overall
verification task. In the Twelf and Beluga systems, such lemmas are
obviated by absorbing them into the meta-theoretic framework. There
are reasons related to the validation of verification that lead us to
prefer explicit proofs (\eg, it is easier to generate proof 
certificates from explicit proofs). However, as shown in
\cite{savary-belanger14lfmtp}, it is often possible to generate these
proofs automatically, thereby allowing the user to focus on the less
obvious aspects. In ongoing work, we are exploring the impact of using
such ideas on reducing the overall proof effort.

%%%%%%%%%%%%%%%%%%%%%%%%%%%%%%%%%%%%%%%%%%%%%%%%%%%%%%%%%%%%%%%%%%%%%%%%%%%%%%%%

%%%%%%%%%%%%%%%%%%%%%%%%%%%%%%%%%%%%%%%%%%%%%%%%%%%%%%%%%%%%%%%%%%%%%%%%%%%%%%%%

%%%%%%%%%%%%%%%%%%%%%%%%%%%%%%%%%%%%%%%%%%%%%%%%%%%%%%%%%%%%%%%%%%%%%%%%%%%%%%%%

%%%%%%%%%%%%%%%%%%%%%%%%%%%%%%%%%%%%%%%%%%%%%%%%%%%%%%%%%%%%%%%%%%%%%%%%%%%%%%%%
% Bibliography, uncomment for final version
%%%%%%%%%%%%%%%%%%%%%%%%%%%%%%%%%%%%%%%%%%%%%%%%%%%%%%%%%%%%%%%%%%%%%%%%%%%%%%%%
%\bibliography{references/master.bib,local.bib}
\bibliography{root}
%%%%%%%%%%%%%%%%%%%%%%%%%%%%%%%%%%%%%%%%%%%%%%%%%%%%%%%%%%%%%%%%%%%%%%%%%%%%%%%%

%%%%%%%%%%%%%%%%%%%%%%%%%%%%%%%%%%%%%%%%%%%%%%%%%%%%%%%%%%%%%%%%%%%%%%%%%%%%%%%%
% Appendices
%%%%%%%%%%%%%%%%%%%%%%%%%%%%%%%%%%%%%%%%%%%%%%%%%%%%%%%%%%%%%%%%%%%%%%%%%%%%%%%%
\appendix
\chapter{Tables}
\label{ch:tables}

The two tables below contain references to the notions related to the
source, intermediate and target languages, the transformations in our
compiler and the type and semantics preservation proofs for the
transformations. In these tables, ``CPS'' stands for ``the CPS
transformation'', ``CC'' stands for ``closure conversion'', ``CH''
stands for ``code hoisting'' and ``CG'' stands for ``code
generation''.

\begin{table}[!ht]
  \center
  \scriptsize
  \begin{tabular}{p{1.9cm} ||p{1.3cm} |p{1.3cm} |p{1.5cm} |p{1.3cm} |p{1.5cm} |p{1.3cm}}
    \hline
    \multirow{2}{*}{\emph{Language}}
              & \multicolumn{2}{|c}{Syntax} & \multicolumn{2}{|c}{Typing} & \multicolumn{2}{|c}{Evaluation}
    \\\cline{2-7}
           & On Paper & \LProlog   & Rule-Based & \LProlog &  Rule-Based & \LProlog    
    \\\hline\hline
    Source
           & Figure~\ref{fig:cps_src_lang} & Page~\pageref{ecd:cps_src_lang}  
           & Figure~\ref{fig:cps_src_typing} & Page~\pageref{ecd:cps_src_typing}
           & Page~\pageref{txt:cps_eval} & Page~\pageref{ecd:cps_eval}
    \\\hline
      Target of CPS
           & \multicolumn{6}{c}{(Same as Source)}
    \\\hline
      Target of CC
           & Figure~\ref{fig:cc_targ_lang} & Page~\pageref{ecd:cc_targ_lang}
           & Page~\pageref{txt:cc_targ_typing} & Page~\pageref{ecd:cc_targ_typing}
           & Page~\pageref{txt:cc_eval} & Page~\pageref{ecd:cc_eval}
    \\\hline
      Target of CH
           & Page~\pageref{txt:ch_targ_lang}
           & Page~\pageref{ecd:ch_targ_lang}
           & Page~\pageref{txt:cc_targ_typing}
           & Page~\pageref{ecd:ch_targ_typing}
           & Page~\pageref{txt:cc_eval}
           & Page~\pageref{ecd:cc_eval}
    \\\hline
      Target of CG
           & Figure~\ref{fig:cg_targ_lang}
           & Pages~\pageref{ecd:cg_targ_lang}
           & \multicolumn{2}{c|}{(untyped)}
           & Page~\pageref{txt:cg_eval}
           & Page~\pageref{ecd:cg_eval}
     \\\hline
  \end{tabular}

  %% \begin{quote}
  %% {\small * Note that the typing contexts are implicit in the encoding (see the discussion in Section~\ref{subsec:ecd_bindings})}
  %% \end{quote}
  \caption{Summary of the Source, Intermediate and Target Languages}
\end{table}

\begin{table}[!ht]
  \center
  \scriptsize

\begin{tabular}{p{0.5cm} ||p{1.5cm} |p{1.3cm} |p{1.3cm} |p{1cm} |p{1.3cm} |p{1cm} |p{1.3cm} |p{1cm}}
    \hline
    \multirow{2}{*}{}
              & \multicolumn{2}{|c}{Transformation} 
              & \multicolumn{2}{|c}{Type Preservation}
              & \multicolumn{2}{|c}{Simulation} 
              & \multicolumn{2}{|c}{Semantics Preservation}
    \\\cline{2-9}
           & Rule-Based & \LProlog  & On Paper & \Abella   & On Paper & \Abella &  On Paper & \Abella
    \\\hline\hline
      CPS
           & Figure~\ref{fig:cps_rules} & Page~\pageref{ecd:cps_rules}  
           & Theorem~\ref{thm:cps_typ_pres_open} & Page~\pageref{ecd:cps_typ_pres_open}
           & Figure~\ref{fig:cps_logical_relations} & Page~\pageref{ecd:cps_logical_relations}
           & Theorem~\ref{thm:cps_sem_pres_open} & Page~\pageref{ecd:cps_sem_pres_open}
    \\\hline
      CC
           & Figure~\ref{fig:cc_rules} & Page~\pageref{ecd:cc_rules}
           & Theorem~\ref{thm:cc_typ_pres_open} & Page~\pageref{ecd:cc_typ_pres_open}
           & Figure~\ref{fig:cc_logical_relations} & Page~\pageref{ecd:cc_logical_relations}
           & Theorem~\ref{thm:cc_sem_pres_open} & Page~\pageref{ecd:cc_sem_pres_open}
    \\\hline
      CH
           & Figure~\ref{fig:ch_rules} & Page~\pageref{ecd:ch_rules}
           & Theorem~\ref{thm:ch_typ_pres_open} & Page~\pageref{ecd:ch_typ_pres_open}
           & Figure~\ref{fig:ch_logical_relations} & Page~\pageref{ecd:ch_logical_relations}
           & Theorem~\ref{thm:ch_sem_pres_open} & Page~\pageref{ecd:cc_sem_pres_open}
    \\\hline
      CG
           & Figure~\ref{fig:cg_rules} & Page~\pageref{ecd:cg_rules}
           & --- & ---
           & --- & ---
           & Theorem~\ref{thm:cgenp_sem_pres_atom} & Page~\pageref{ecd:cg_sem_pres}
     \\\hline
  \end{tabular}

  \caption{The Transformations and Their Type and Semantics Preservation Proofs}
\end{table}

\end{document}
%%%%%%%%%%%%%%%%%%%%%%%%%%%%%%%%%%%%%%%%%%%%%%%%%%%%%%%%%%%%%%%%%%%%%%%%%%%%%%%%